\documentclass[11pt]{article}
\pdfoutput=1
\usepackage{jcapmod}
\usepackage{shorthand}
\usepackage[scaled]{helvet}

\usepackage[toc,page]{appendix}
\usepackage{mathtools}
\usepackage{booktabs}
\usepackage[english]{babel}
\usepackage{amsmath,amssymb,amsbsy,amstext, amsthm, simplewick, amsfonts}
\usepackage{hyperref}
\usepackage{graphicx}
\usepackage[small]{caption}
\usepackage{siunitx}
\usepackage{upgreek}
\usepackage{framed}
\usepackage{wrapfig}
\usepackage{multirow}
\usepackage{bbm}
\usepackage[svgnames,dvipsnames,x11names]{xcolor}
\usepackage[utf8x]{inputenc}
\usepackage{selinput}
\usepackage{bm}
\usepackage{float}
\usepackage{geometry}
\usepackage{yfonts}
\usepackage{caption}
\usepackage{subcaption}
\usepackage{sidecap}
\usepackage{longtable}
\usepackage{anyfontsize}
\usepackage{dsfont}
\usepackage{tikz}
\usepackage{relsize}
\usepackage{tcolorbox}
\usepackage{shuffle}
\usetikzlibrary{matrix}
\usepackage{bigstrut}
\usepackage{fnpct}
\usetikzlibrary{arrows.meta}
\usetikzlibrary{decorations.pathreplacing,calligraphy}

\usetikzlibrary{decorations.markings, decorations.pathmorphing, decorations.pathreplacing, decorations.shapes}

\usepackage{xcolor}
\usepackage{xparse}

\usepackage{fontawesome5}

\makeatletter
\newcommand{\github}[1]{%
   \href{#1}{\faGithub}%
}
\makeatother

\NewDocumentCommand{\colornucleus}{omme{_^}}{%
  \begingroup\colorlet{currcolor}{.}%
  \IfValueTF{#1}
   {\textcolor[#1]{#2}}
   {\textcolor{#2}}
    {%
     #3
     \IfValueT{#4}{_{\textcolor{currcolor}{#4}}}
     \IfValueT{#5}{^{\textcolor{currcolor}{#5}}}
    }%
  \endgroup
}

\usepackage{array}
\newcolumntype{L}[1]{>{\raggedright\let\newline\\\arraybackslash\hspace{0pt}}m{#1}}
\newcolumntype{C}[1]{>{\centering\let\newline\\\arraybackslash\hspace{0pt}}m{#1}}
\newcolumntype{R}[1]{>{\raggedleft\let\newline\\\arraybackslash\hspace{0pt}}m{#1}}

\usepackage[framemethod=default]{mdframed}
\newmdenv[skipabove=7pt,
skipbelow=7pt,
rightline=true,
leftline=true,
topline=true,
bottomline=true,
backgroundcolor=gray!10,
linecolor=black,
innerleftmargin=5pt,
innerrightmargin=5pt,
innertopmargin=5pt,
innerbottommargin=5pt,
leftmargin=0cm,
rightmargin=0cm,
linewidth=1pt]{eBox}

\definecolor{Red}{RGB}{214, 39, 40}
\definecolor{Blue}{RGB} {31, 119, 180}
\definecolor{Orange}{RGB}{255, 153, 51}
\definecolor{Purple}{RGB}{178, 102, 255}
\definecolor{Green}{RGB}{44, 160, 44}
\definecolor{regal}{RGB}{90,0,120}
\definecolor{darkblue}{rgb}{0.15,0.35,0.55}
\definecolor{reddish}{rgb}{0.65, 0.2, 0.2}
\definecolor{darkgreen}{RGB}{50,150,0}
\definecolor{greyish}{rgb}{.90,.90,.90}
\definecolor{greyish2}{rgb}{.96,.96,.96}
\definecolor{greyish3}{rgb}{.37,.37,.37}
\definecolor{darkblue2}{rgb}{0.3,0.4,0.9}
\definecolor{Blue3}{RGB}{31, 119, 180}

\definecolor{blue3}{RGB}{31, 119, 180}
\definecolor{red3}{RGB}{	214, 39, 40}
\definecolor{orange3}{RGB}{255, 127, 14}
\definecolor{green3}{RGB}{44, 160, 44}
\definecolor{repBlue}{RGB}{31, 119, 180}
\definecolor{repRed}{RGB}{	214, 39, 40}
\definecolor{repGreen}{RGB}{44, 160, 44}

\newcommand{\blue}[1]{\textcolor{blue3}{#1}}
\newcommand{\green}[1]{\textcolor{green3}{#1}}
\newcommand{\red}[1]{\textcolor{red3}{#1}}

\newcommand{\orange}[1]{\textcolor{orange3}{#1}}

\definecolor{vio}{RGB}{19, 130, 164}
\definecolor{vioo}{RGB}{89, 2, 155}
\newcommand{\Comment}[1]{{}}
\usepackage[linktocpage=true]{hyperref}
\hypersetup{
colorlinks=true,
citecolor=darkblue,
linkcolor=reddish,
urlcolor=darkblue,
pdfauthor={},
pdftitle={},
pdfsubject={}
}

\def\Lap{\raisebox{-3pt}{\includegraphics{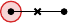}}}
\def\Lam{\raisebox{-3pt}{\includegraphics{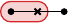}}}
\def\Lbp{\raisebox{-3pt}{\includegraphics{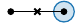}}}
\def\Lbm{\raisebox{-3pt}{\includegraphics{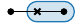}}}

\def\Labg{\raisebox{-3pt}{\includegraphics{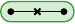}}}
\def\Labr{\raisebox{-3pt}{\includegraphics{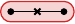}}}
\def\Labb{\raisebox{-3pt}{\includegraphics{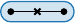}}}

\def\gpsi{\raisebox{-3pt}{\includegraphics{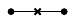}}}
\def\gF{\raisebox{-3pt}{\includegraphics{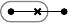}}}
\def\gFt{\raisebox{-3pt}{\includegraphics{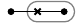}}}
\def\gZ{\raisebox{-3pt}{\includegraphics{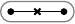}}}

\def\LLap{\raisebox{-3pt}{\includegraphics{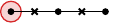}}}
\def\LLam{\raisebox{-3pt}{\includegraphics{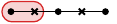}}}
\def\LLcp{\raisebox{-3pt}{\includegraphics{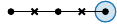}}}
\def\LLcm{\raisebox{-3pt}{\includegraphics{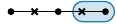}}}

\def\LLbpp{\raisebox{-3pt}{\includegraphics{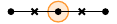}}}
\def\LLbmp{\raisebox{-3pt}{\includegraphics{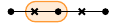}}}
\def\LLbpm{\raisebox{-3pt}{\includegraphics{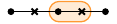}}}
\def\LLbmm{\raisebox{-3pt}{\includegraphics{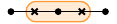}}}

\def\LLabp{\raisebox{-3pt}{\includegraphics{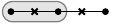}}}
\def\LLabm{\raisebox{-3pt}{\includegraphics{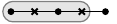}}}
\def\LLbcp{\raisebox{-3pt}{\includegraphics{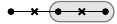}}}
\def\LLbcm{\raisebox{-3pt}{\includegraphics{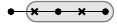}}}
\def\LLabc{\raisebox{-3pt}{\includegraphics{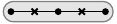}}}

\def\LLabpo{\raisebox{-3pt}{\includegraphics{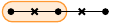}}}
\def\LLabpr{\raisebox{-3pt}{\includegraphics{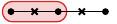}}}
\def\LLabmo{\raisebox{-3pt}{\includegraphics{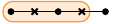}}}
\def\LLabmr{\raisebox{-3pt}{\includegraphics{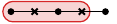}}}

\def\LLbcpo{\raisebox{-3pt}{\includegraphics{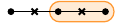}}}
\def\LLbcpb{\raisebox{-3pt}{\includegraphics{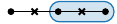}}}

\def\LLbcmb{\raisebox{-3pt}{\includegraphics{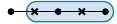}}}

\def\LLabco{\raisebox{-3pt}{\includegraphics{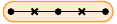}}}
\def\LLabcb{\raisebox{-3pt}{\includegraphics{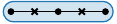}}}

\def\ggF{\raisebox{-3pt}{\includegraphics{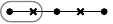}}}
\def\ggFt{\raisebox{-3pt}{\includegraphics{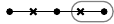}}}
\def\ggQa{\raisebox{-3pt}{\includegraphics{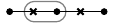}}}
\def\ggQb{\raisebox{-3pt}{\includegraphics{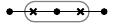}}}
\def\ggQc{\raisebox{-3pt}{\includegraphics{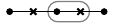}}}

\def\ggf{\raisebox{-3pt}{\includegraphics{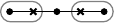}}}
\def\ggqa{\raisebox{-3pt}{\includegraphics{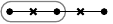}}}
\def\ggqb{\raisebox{-3pt}{\includegraphics{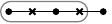}}}
\def\ggqc{\raisebox{-3pt}{\includegraphics{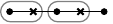}}}
\def\ggqta{\raisebox{-3pt}{\includegraphics{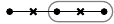}}}
\def\ggqtb{\raisebox{-3pt}{\includegraphics{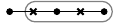}}}
\def\ggqtc{\raisebox{-3pt}{\includegraphics{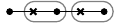}}}

\def\ggg{\ \raisebox{-3pt}{\includegraphics{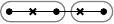}}}
\def\gggt{\ \raisebox{-3pt}{\includegraphics{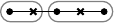}}}
\def\ggZ{\ \raisebox{-3pt}{\includegraphics{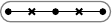}}}

\def\gstarQ{\raisebox{-15pt}{\includegraphics{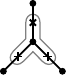}}}
\def\gstarQa{\raisebox{-15pt}{\includegraphics{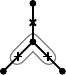}}}
\def\gstarQb{\raisebox{-15pt}{\includegraphics{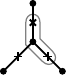}}}
\def\gstarQc{\raisebox{-15pt}{\includegraphics{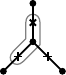}}}

\usepackage{colortbl}
\definecolor{lightgreen}{cmyk}{0.2, 0, 0.2, 0.2}
\definecolor{lightgray2}{cmyk}{0.1,0.1,0,0.1}
\definecolor{Red2}{RGB}{214, 39, 40}
\definecolor{Blue2}{RGB} {31, 119, 180}
\definecolor{Orange2}{RGB}{255, 127, 14}
\definecolor{Green2}{RGB}{44, 160, 44}

\makeatletter
\newlength{\apb@width}
\newcommand{\autoparbox}[2][c]{\settowidth{\apb@width}{#2}\parbox[#1]{\apb@width}{#2}}

\makeatother


\def\hs{\hskip 1pt}

\def\beq{\begin{equation}}
\def\eeq{\end{equation}}
\def\be{\begin{equation}}
\def\ee{\end{equation}}

\def\k{\vec k}

\newcommand{\ud}{{\rm d}}

\def\e{\varepsilon}

\allowdisplaybreaks[1]
\newcommand\BC{\cellcolor{Blue!20}}
\newcommand\RC{\cellcolor{Red!20}}
\newcommand\GC{\cellcolor{Green!20}}
\newcommand\OC{\cellcolor{Orange!20}}

\newcommand\sqmatrix[2][c]{%
  \fixTABwidth{T}%
  \setbox0=\hbox{$\tabbedCenterstack{#2}$}%
  \setstackgap{L}{\dimexpr\maxTAB@width+\tabbed@gap}%
  \tabbedCenterstack[#1]{#2}%
}

\usetikzlibrary{shapes.misc}

\tikzset{cross/.style={cross out, draw=black, minimum size=2*(#1-\pgflinewidth), inner sep=0pt, outer sep=0pt},
cross/.default={1pt}}

\newcommand{\Cross}{$\mathbin{\tikz [x=1.4ex,y=1.4ex,line width=.2ex, black] \draw (0,0) -- (0.45,0.45) (0,0.45) -- (0.45,0);}$}%
\newcommand{\CrossR}{$\mathbin{\tikz [x=1.4ex,y=1.4ex,line width=.2ex, black] \draw (0,-0.318) -- (0,0.318) (-0.318,0) -- (0.318,0);}$}%

\usetikzlibrary{calc}

\begin{document}


\newgeometry{top=2cm, bottom=2cm, left=2cm, right=2cm}

\begin{titlepage}
\setcounter{page}{1} \baselineskip=15.5pt 
\thispagestyle{empty}

\begin{center}
{\fontsize{18}{18} \bf Differential Equations for Cosmological Correlators}
\end{center}

\vskip 10pt
\begin{center}
\noindent
{\fontsize{12}{18}\selectfont 
Nima Arkani-Hamed,$^{1}$
Daniel Baumann,$^{2,3,4}$ Aaron Hillman,$^{5,6}$ \\[4pt]   
Austin Joyce,$^{7,8}$  Hayden Lee,$^{8}$ and Guilherme L.~Pimentel\hskip 1pt$^{9}$}
\end{center}

\vskip16pt
\begin{center}
\textit{$^1$ Institute for Advanced Study, Princeton, NJ 08540, USA}

  \vskip8pt
\textit{$^2$ 
Leung Center for Cosmology and Particle Astrophysics, Taipei 10617, Taiwan}

  \vskip8pt
\textit{$^3$ Center for Theoretical Physics, National Taiwan University, Taipei 10617, Taiwan}

  \vskip8pt
\textit{$^4$ Institute of Physics, University of Amsterdam, Amsterdam, 1098 XH, The Netherlands}

\vskip 8pt
\textit{$^5$ Walter Burke Institute for Theoretical Physics, Caltech, Pasadena, CA 91125, USA}

\vskip 8pt
\textit{$^6$ Department of Physics, Jadwin Hall, Princeton University, NJ 08540, USA}

\vskip 8pt
\textit{$^7$ 
Department of Astronomy and Astrophysics,
University of Chicago, Chicago, IL 60637, USA}

\vskip 8pt
\textit{$^8$ Kavli Institute for Cosmological Physics, 
University of Chicago, Chicago, IL 60637, USA}

\vskip 8pt
\textit{$^9$ Scuola Normale Superiore and INFN, Piazza dei Cavalieri 7, 56126, Pisa, Italy}
\end{center}

\vspace{0.4cm}
\begin{center}{\bf Abstract}
\end{center}
\noindent
Cosmological fluctuations retain a memory of the physics that generated them in their spatial correlations. The strength of correlations varies smoothly as a function of external kinematics, which is encoded in differential equations satisfied by cosmological correlation functions. In this work, we provide a broader perspective on the origin and structure of these differential equations. As a concrete example, we study conformally coupled scalar fields in a power-law cosmology. The wavefunction coefficients in this model have integral representations, with the integrands being the product of the corresponding flat-space results and ``twist factors" that depend on the cosmological evolution. Similar twisted integrals arise for loop amplitudes in dimensional regularization, and their recent study has led to the discovery of rich mathematical structures and powerful new tools for computing multi-loop Feynman integrals in quantum field theory. The integrals of interest in cosmology are also part of a finite-dimensional basis of master integrals, which satisfy a system of first-order differential equations. We develop a formalism to derive these differential equations for arbitrary tree graphs. The results can be represented in graphical form by associating the singularities of the differential equations with a set of graph tubings. Upon differentiation, these tubings grow in a local and predictive fashion. In fact, a few remarkably simple rules allow us to predict---by hand---the equations for all tree graphs. While the rules of this ``kinematic flow" are defined purely in terms of data on the boundary of the spacetime, they reflect the physics of bulk time evolution. We also study the analogous structures in ${\rm tr}\,\phi^3$ theory, and see some glimpses of hidden structure in the sum over planar graphs. This suggests that there is an autonomous combinatorial or geometric construction from which cosmological correlations, and the associated spacetime, emerge.

\end{titlepage}
\restoregeometry

\newpage
\setcounter{tocdepth}{2}
\setcounter{page}{2}

\linespread{0.75}
\tableofcontents
\linespread{1.}

\newpage
\section{Introduction}

There are many indications that the spacetime concept must be replaced by deeper principles as part of the next upheavals in fundamental physics. Nowhere is the need for an understanding of emergent space-{\it time} more pressing than in cosmology, where the birth of spacetime and the Universe itself are intimately connected at the Big Bang singularity.
Moreover, actual observations of the Universe are static, correlating structures at different spatial locations, but fixed moments in time.  We invoke a cosmological history to explain these spatial patterns, but ultimately “cosmological time” is an auxiliary concept, not present in the observables themselves. This calls for a new ``timeless" understanding of cosmology, reproducing the approximate notion of time evolution where appropriate, but deforming away from it when the need becomes exigent.

\vskip 4pt
Descending from these perhaps dangerously lofty conceptual heights, there is also a more practical reason to seek such a timeless description of cosmology. The conventional textbook methods for computing cosmological correlators\footnote{In recent years, there has been remarkable progress in understanding features of the quantum field theory wavefunctionals of various theories, both in flat space and in cosmology~\cite{Maldacena:2002vr,Anninos:2014lwa,Ghosh:2014kba,Arkani-Hamed:2017fdk,Albayrak:2018tam,Albayrak:2019yve,Hillman:2019wgh,Baumann:2020dch,Goodhew:2020hob,Cespedes:2020xqq,Baumann:2021fxj,Goodhew:2021oqg,Jazayeri:2021fvk,Hillman:2021bnk,Meltzer:2021zin,Bittermann:2022nfh,Bonifacio:2022vwa,Salcedo:2022aal,Lee:2022fgr,Stefanyszyn:2023qov,Albayrak:2023hie,Cespedes:2023aal}, and of correlation functions directly~\cite{Chen:2009zp,Maldacena:2011nz,Mata:2012bx,Creminelli:2012ed,Assassi:2012zq,Hinterbichler:2013dpa,Kundu:2014gxa,Arkani-Hamed:2015bza,Lee:2016vti,Arkani-Hamed:2018kmz,Baumann:2019oyu,Sleight:2019mgd,Sleight:2019hfp,Pajer:2020wxk,DiPietro:2021sjt,Hogervorst:2021uvp,McFadden:2009fg,Bzowski:2012ih,Sleight:2021plv,Bzowski:2022rlz,Baumgart:2019clc,Gorbenko:2019rza,Cohen:2020php,Cohen:2021fzf,Cohen:2021jbo,Pimentel:2022fsc,Qin:2022fbv,Wang:2022eop,Penedones:2023uqc,Loparco:2023akg,Loparco:2023rug,Qin:2023nhv,DuasoPueyo:2023viy,Xianyu:2023ytd}. In many cases, these correlations are most naturally studied in momentum space, and there has been commensurate progress in the study of momentum-space conformal field theory~\cite{Bzowski:2013sza,Bzowski:2015pba,Bzowski:2017poo,Bzowski:2019kwd,Bzowski:2020kfw,Dymarsky:2014zja,Gillioz:2018mto,Gillioz:2019lgs,Gillioz:2020mdd}.}  (and also the underlying cosmological wavefunction) are very complicated.  One of the main reasons  is that all Feynman diagram calculations involve time integrals from past infinity to the present for all interaction vertices. This leads to tremendous complexity in intermediate stages of  the computations, mirroring a similar explosion of complexity familiar in the study of scattering amplitudes in asymptotically flat space. And yet, as in the case of scattering amplitudes, the final expressions are vastly simpler. This provides a much more down-to-earth motivation for finding a new understanding of time evolution, one purely formulated in terms of spatial variables defining the kinematic dependence of the cosmological correlators, without a trace of integration over time coordinates. Echoing developments in scattering amplitudes over the past decade, we can hope to find entirely new sorts of mathematical questions in kinematic space, to which cosmological correlators are the answers, with the interpretation in terms of time evolution arising only as a derived concept. 

\vskip 4pt
In this paper, we will give the first complete example of such a description of the cosmological wavefunction for a simple class of toy models, working at tree level. We will consider the wavefunction  for conformally coupled scalars with general polynomial interactions, evolving in an FRW cosmology with scale factor $a(\eta) = (\eta/\eta_0)^{-(1 + \e)}$. Equivalently, after a conformal rescaling, we have a theory of a massless scalar field in flat space, with time-dependent interactions. 

\vskip 4pt
As a first step, we will consider the contributions to the wavefunction associated with individual Feynman diagrams. Representing the time-dependent couplings in the frequency domain allows all time integrations to be performed, determining the cosmological wavefunction in terms of the flat-space wavefunction as 
\begin{equation}
\Psi_{{\rm FRW}}(E_v,E_I) = \int_0^\infty \prod_v \ud \omega_v \left(\prod_v \omega_v\right)^\e \Psi_{{\rm flat}}(E_v + \omega_v, E_I)\,,
\label{equ:PsiCosmo}
\end{equation}
where $E_v $ and $E_I $ are the ``energies" associated with the vertices and the internal edges of the graph, respectively.  The flat-space wavefunction is a rational function of the energies $E_v$ and $E_I$, with simple poles when the sum of the energies entering any subgraph vanishes. The cosmological wavefunction (\ref{equ:PsiCosmo}) is then an integral over the deformed flat-space wavefunction, weighted by a ``twist factor" $\left(\prod_v \omega_v\right)^\e$.  Already this integral gives a formula for the wavefunction with no explicit reference to integrations over time, but this is clearly only a cosmetic difference---we have simply gone to Fourier space, with the integrals over vertex energies being conjugate to those over the vertex times. 

\vskip 4pt
However, there is a much more interesting way in which this integral representation opens the door to a ``timeless" description of cosmological correlators. Our cosmological integrals are  special cases of a wide class of integrals of the form 
\begin{equation}
I(C,D;n;\e) = \int_0^\infty \ud x_1 \cdots \ud x_m \,P(x) \prod_I (C_{Ij} x_j + D_I)^{-n_I + \e_I}\,,
\end{equation}
where $P(x)$ is some polynomial in the $x$ variables, and the singularities of the integrand are powers of linear factors, raised to integer powers $n_I$ possibly ``twisted" by fractional parameters~$\e_I$. There are naively an infinite number of these integrals parameterized by the general integers $n_I$. However, there are also an infinite number of linear relations between these integrals generated by integration-by-parts identities. A complete understanding of the vector space of all independent integrals of this type is offered by the study of the so-called ``twisted cohomology'' of the ``hyperplane arrangement'' attached to the linear half-spaces $(C \cdot x + D) > 0$. Quite beautifully, there is a {\it finite-dimensional} space of linearly independent integrals, whose dimension is given by the number of bounded regions carved out by the hyperplanes. We can therefore choose a basis $I_a$ of these integrals, and every other integral with arbitrary powers $n_I$ can be written as a linear combination of these basis integrals, with coefficients that are rational functions of the parameters $(C,D,n,\e)$. This in turn implies that the total differential of any of the basis functions $I_a$ with respect to the parameters $(C,D)$ must satisfy a differential equation. The reason is simply that the total differential of any integral in this class must be a linear combination of basis integrals, so that
\begin{equation}
\ud I_a = A_{ab} I_b \,, 
\label{equ:DEQ}
\end{equation}
where $A_{ab}$ is a matrix of one-forms depending on the data $(C,D)$. 

\vskip 4pt
When applied to the cosmological wavefunction (\ref{equ:PsiCosmo}), we learn a first important fact: 
\begin{center}
{\it The cosmological wavefunction satisfies a differential equation, \\
which governs how it changes as the external kinematics are varied.} 
\end{center}
We will see that the particular solution of this linear system of differential equations is completely fixed by a natural set of boundary conditions, enforcing  the absence of unphysical ``folded singularities" in the wavefunction, as dictated by the choice of the adiabatic vacuum in the far past, together with a single ``factorization" condition near certain singularities in energy space.  The factorization limit distinguishes the wavefunction from the actual correlation function (which satisfies the same differential equation).

\vskip 4pt
\newpage
We stress that this differential equation is  formulated purely in the space of kinematic variables, but encodes all the dynamical content of cosmological time evolution. This is an extension of something we are familiar with in inflationary cosmology, namely that the overall scale of momenta gives a coarse measure for inflationary time, with longer wavelength perturbations generated earlier in the inflationary epoch. Similar differential equations also play a starring role in the ``cosmological bootstrap"~\cite{Baumann:2022jpr}, where they arise as the conformal Ward identities obeyed by  the boundary correlators in an approximate de Sitter spacetime~\cite{Arkani-Hamed:2015bza, Arkani-Hamed:2018kmz}.  In this paper, we will provide a broader perspective on the origin and mathematical properties of these differential equations. In particular, we will show that a similar set of differential equations also arises in general FRW spacetimes, without any conformal symmetry. 

\vskip 4pt
The fact that the cosmological wavefunction satisfies a differential equation in kinematic space is a tantalizing clue for finding an autonomous, ``timeless" theory of cosmological observables, but does not by itself provide us with such a theory. At the most practical level, there is no especially canonical choice of a basis for the twisted cohomology, and hence no preferred form of the connection one-form $A$ in (\ref{equ:DEQ}). This means that the  differential equations have to be studied on a case-by-case basis, which is problematic, since, for example for $\phi^3$ theory  the size of the connection matrices grows exponentially with the number of external legs for the wavefunction. For instance, at five points, the naive size of the basis given by a standard treatment of the cohomology is 25-dimensional, and even the more correct understanding tailored to the cosmological wavefunction is 16-dimensional.  For the general $n$-point problem, the size of the basis scales as $4^{n-3}$. Thus, while the analysis of the differential equation can be carried out manually for small $n$, it quickly becomes entirely impractical for larger $n$. 

\vskip 4pt
Clearly, some new organizing principle is needed to be able to tame this exponential jungle of complexity, but the general properties of twisted integrals are not enough to provide this guidance. Any hope for understanding the general structure of these differential equations, say even just for arbitrary tree graphs, must rely on some special property of the twisted integrals specifically associated with our cosmological problem, rather than properties of completely generic integrals, which are all we have appealed to so far. 

\vskip 4pt
This brings us to the second, and much more non-trivial and remarkable, observation of our paper: 
\begin{center}
{\it There is an ab-initio, autonomous description of the differential equations satisfied by \\
 the cosmological wavefunction, vividly captured by a ``flow'' in kinematic space}. 
\end{center}
This kinematic flow is revealed when we represent the singularities of the differential equations~(\ref{equ:DEQ}) as graph tubings. We find that these tubings grow in a local and predictive fashion. In fact, a few remarkably simple rules govern the equations for all tree graphs.  Although the complexity of these equations increases dramatically for larger graphs the underlying rules remain amazingly simple.

\vskip 4pt
\newpage
This picture extends from a single graph to the sum over all graphs, for the special case of the cosmological wavefunction of ${\rm tr}\,\phi^3$ theory, much studied in the amplitudes literature. The final important result of our paper is:
\begin{center}
{\it The cosmological wavefunction of ${\rm tr}\,\phi^3$ theory satisfies a differential equation\\ determined by ``kinematic flow" defined directly on the polygons of external momenta}. 
\end{center}
The singularities of the differential equations are now represented by shaded subpolygons, and the kinematic flow describes how these shadings  ``evolve" as we take derivatives. We find that the
differential equations connect basis functions for different graphs (channels) suggesting the existence of a unifying geometric object capturing the sum over graphs. 

\vskip 4pt
The existence of this ``kinematic flow" description of the cosmological differential equations was by no means obvious a priori. Indeed, our own investigations of this subject were spurred at first simply by the existence of the differential equations guaranteed on the general grounds described above, which led us to study simple graphs with $n=4,5$ external lines in some detail. In doing so, we also encountered the familiar frustration of ``no canonical choice of basis.'' It was only after switching tactics from somewhat arbitrary organizing principles involving notions like the  ``sparseness" of the connection matrix $A$, to drawing suggestive pictures for the differential equations, that the kinematic flow structure finally revealed itself. As we hope the reader will see, kinematic flow is an example of a sort of ``magic": hidden structure in the cosmological wavefunction not guaranteed a priori, but capturing bulk time evolution in an entirely new way, in purely boundary terms. 

\vskip 4pt
Our goal in this paper is to explain the general existence of the differential equations for cosmological correlators, as well the surprising emergence of the kinematic flow picture determining all these differential equations at tree level. Our exposition will be entirely self-contained, with no previous knowledge of the relevant mathematical background being assumed or needed. We will proceed systematically by first deriving the differential equations, and then seeing how a natural organization of basis integrals, together with a pictorial representation of the differential equations brings the kinematic flow to life. For this reason, our exposition will be as long as it is systematic and, hopefully, pedagogical in character. The culmination of this presentation will be a systematic set of rules for kinematic flow, defined either on a given graph 
or on the full polygon of momentum kinematics. 

\vskip 4pt
In keeping with our ultimate goal of formulating a timeless theory of cosmological correlators, we can just as well tell the story backwards, beginning  by simply stating the rule of the kinematic flow, using them to predict the equation $\ud I_a = A_{ab} I_b$ and seeing how the crucial consistency condition $\ud^2 = 0$ is implied.  The result is a first step to back-constructing the energy/time-integral expression for the wavefunction from a purely boundary perspective. We will take up this more adventurous presentation of our results in a companion Letter~\cite{Arkani-Hamed:2023bsv}.

\newpage
\subsection{Road Map and Summary}

It has not escaped our notice that this paper is rather long. This section therefore serves as a high-level overview of the salient points of the analysis, and as a roadmap for the developments in the rest of the text. We will sketch the logic that leads from the time/energy integrals arising in bulk perturbation theory, to the differential equations that they satisfy, and finally to an alternative viewpoint where these differential equations originate from combinatorial ideas applied to graphs and kinematic polygons associated to the cosmological wavefunction. The story will be told through the lens of the simplest nontrivial example---the cosmological four-point function---with pointers to where details and generalizations can be found in the text.

\paragraph{Toy model} Throughout, we will focus on the particular model of a conformally coupled scalar field (with polynomial self-interactions), described by the action 
\beq
S = \int \ud^4 x \sqrt{-g} \left[-\frac{1}{2}( \partial\phi)^2-\frac{1}{12} R \phi^2 - \sum_{p>2} \frac{ \lambda_p}{p!} \phi^p \right]  .
\label{eq:ccscalartoy}
\eeq
We will study correlation functions in an FRW cosmology, with a power-law scale factor $a(\eta) = (\eta/\eta_0)^{-(1+\e)}$. The utility of this toy model is that it can be simply related to a flat-space quantum field theory with particular time-dependent couplings, allowing us to parameterize the features of various cosmologies in a uniform way.
As a practical matter, we will study correlations in this model by computing the vacuum wavefunctional.
We briefly review the basic features of the wavefunction and its connection to correlation functions in flat space in~Section~\ref{ssec:FlatSpace}.

\paragraph{Twisted integrals} The fact that the action~\eqref{eq:ccscalartoy} can be cast as a flat-space field theory with time-dependent couplings allows us to relate the  elementary building blocks of the wavefunction (``wavefunction coefficients")  to their flat-space counterparts integrated over energies, as in~\eqref{equ:PsiCosmo}. In Section~\ref{ssec:FRW}, we explain this correspondence in detail, while in Section~\ref{sec:integralselection} we provide many explicit examples.

\vskip 4pt
An illustrative example is the single-exchange process at tree level:  
\vspace{.1cm}
\beq
 \raisebox{-33pt}{
\begin{tikzpicture}[line width=1. pt, scale=2]
\draw[fill=black] (0,0) -- (1,0);
\draw[lightgray, line width=1.pt] (0,0) -- (-0.25,0.55);
\draw[lightgray, line width=1.pt] (0,0) -- (0.25,0.55);
\draw[lightgray, line width=1.pt] (1,0) -- (0.75,0.55);
\draw[lightgray, line width=1.pt] (1,0) -- (1.25,0.55);
\draw[lightgray, line width=2.pt] (-0.45,0.55) -- (1.45,0.55);
\draw[fill=black] (0,0) circle (.03cm);
\draw[fill=black] (1,0) circle (.03cm);
\node[scale=1] at (0,-.15) {$X_1$};
\node[scale=1] at (1,-.15) {$X_2$};
\node[scale=1] at (0.5,-.12) {$Y$};
\end{tikzpicture}
} 
\nonumber
\eeq
In flat space, the number of external lines of this ``two-site chain" is irrelevant (though we have drawn it for $p=3$), and the result depends only on  the sum of the external energies flowing into the two vertices---which we denote by 
$X_1$ and $X_2$---as well as the internal energy $Y$ flowing through the diagram.
The wavefunction coefficient associated to this diagram is 
\beq
\psi_{\rm flat} =  \frac{1}{(X_1+X_2)(X_1+Y)(X_2+Y)}\,.
\label{equ:flat-space}
\eeq
Notice that the result is a rational function with three poles: the ``total-energy singularity" at $X_1+X_2=0$ and two ``partial-energy singularities" at $X_1+Y = 0$ and $X_2+Y = 0$. The locations of these singularities encode aspects of the spacetime evolution that generated these correlations, but interestingly an autonomous, spacetime-independent definition of this wavefunction exists in terms of the volume of a ``cosmological polytope"~\cite{Arkani-Hamed:2017fdk} (see Section~\ref{ssec:CosmoPolytope}).

\vskip 4pt
We obtain the wavefunction in the power-law {\it FRW background} by shifting the external energies as $X_a \mapsto X_a + x_a$ and integrating against the twist factor $(x_1x_2)^\e$:
\beq
\psi_{\rm FRW} \propto \int_0^\infty \ud x_1 \int_0^\infty \ud x_2\,  \frac{(x_1x_2)^\e}{(X_1+X_2 + x_1 + x_2)(X_1 + x_1+Y)(X_2 + x_2+Y)}\,.
\label{equ:Integral}
\eeq
For $\e=0$---i.e.~without the twist factor---this corresponds to the wavefunction of conformally coupled scalars in {\it de Sitter space}. In that case, performing the integral explicitly leads to a solution in terms of logs and dilogs~\cite{Arkani-Hamed:2015bza}.  In Section~\ref{sec:frwintegral}, we study the integral for general finite~$\e$.\footnote{We also consider the de Sitter limit of the system explicitly in Section~\ref{ssec:dS}.}

\paragraph{Differential equations}  If one is interested in obtaining an analytic understanding of integrals of the form~\eqref{equ:Integral}---including their analytic continuation, singularities, etc.---the only practical way to proceed is to derive the differential equations that these integrals satisfy.
The integral~\eqref{equ:Integral} is similar to integrals that arise in the computation of one-loop scattering amplitudes in dimensional regularization (in that context, $2\e = 4-D$ is the deviation from the canonical spacetime dimension).  
The basic reason that the tree-level wavefunction is similar to loop amplitudes is that energy is not conserved in cosmology, and therefore must be integrated over.
 This connection allows us to import tools from the treatment of loop amplitudes to the study of tree-level correlators.
 In particular, an important insight is that loop amplitudes can be written in terms of a set of master integrals which span a finite-dimensional vector space. The finiteness of this vector space guarantees that the master integrals themselves satisfy interesting differential equations with respect to the kinematic data, which then provides a powerful way to compute the original Feynman integrals~\cite{Bern:1993kr,Kotikov:1990kg,Remiddi:1997ny,Gehrmann:1999as,Henn:2013pwa} (see~\cite{Henn:2014qga,Abreu:2022mfk} for nice reviews). 
In this paper, we will see how the  differential equation method  
can be translated to the cosmological context.\footnote{Recently, this approach has also been applied to multi-loop Feynman integrals of relevance to the black hole binary inspiral problem~\cite{Parra-Martinez:2020dzs,Kalin:2020fhe}.} 

 \begin{figure}[t!]
\centering
\includegraphics[scale=0.8]{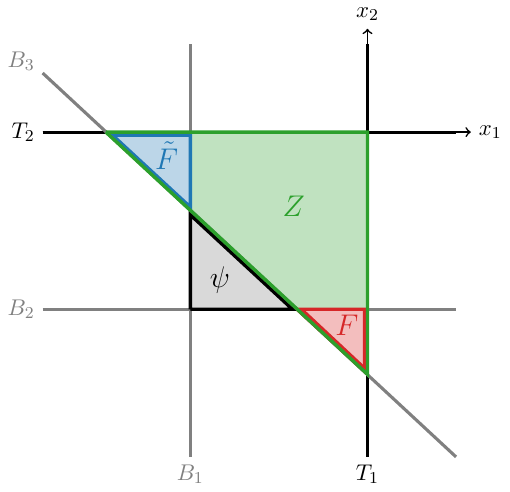}
\caption{Illustration of the singular lines in the integrand of (\ref{equ:Integral}). We see that there are four bounded regions which implies a four-dimensional vector space of basis integrals. The shaded regions correspond to a specific choice of basis.}
\label{fig:regionsX}
\end{figure}

\vskip4pt
The integrals of interest are naturally related to hyperplane arrangements (see Appendix~\ref{app:hyperplanes}).
For example, the integrand of~\eqref{equ:Integral} contains 5 singular lines, 
\beq
\begin{aligned}
T_1 &\equiv x_1\,,\qquad\quad B_1 \equiv  X_1 + x_1 +  Y\,, \\
T_2 &\equiv x_2\,, \qquad\quad B_2 \equiv  X_2 + x_2+    Y\,, \qquad\quad B_3 \equiv  X_1+X_2 + x_1 +x_2\, .
\end{aligned}
\eeq
As we see in Figure~\ref{fig:regionsX}, these lines define four bounded region in the $x_1$--$x_2$ plane. 
It is useful to think of~\eqref{equ:Integral} as one member of a family of integrals, where each 
singular line is raised to an arbitrary integer power: 
\beq
I_{N} = \int (x_1 x_2)^\e\, \Omega_N\,, \qquad\quad \Omega_N \equiv \frac{\ud x_1\wedge \ud  x_2}{T_{1}^{m_1} T_{2}^{m_2} B_1^{n_1} B_2^{n_2} B_3^{n_3}} \,,
\label{equ:family0}
\eeq
where $N\equiv (m_1,m_2, n_1,n_2,n_3)$ is a multi-index.
 Despite appearances, not all these integrals are independent because many are related through integrations by parts (integrands that differ by a total derivative lead to the same integrated function).  
A beautiful result from ``twisted cohomology" (see Appendix~\ref{app:twistedcohomology}) states that the number of independent basis integrals is equal to the number of bounded regions defined by the singular lines of the integrand. In the case of interest, the basis is therefore four-dimensional. The fact that there is a finite basis immediately implies that the basis integrals will satisfy first-order differential equations: taking derivatives of integrals of the form~\eqref{equ:family0} just shifts the integers $m_a, n_a$, but keeps the integral in the same family. It must then be possible to re-write the result back in terms of the basis integrals, leading to a matrix differential equation.

\vskip4pt
A preferred integral basis can be defined using the ``canonical forms" of each bounded region.\footnote{Canonical forms are the unique forms with logarithmic singularities
on the boundaries of the polytopes (and nowhere else), and whose residues on these facets are canonical forms of lower-dimensional regions; see Appendix~\ref{app:canforms} for a pedagogical introduction.} 
Denoting the four-dimensional vector of basis integrals by $\vec{I} \equiv (I_1,I_2,I_3,I_4)^T$ and defining the total differential $\ud \equiv \sum_I \partial_{Z_I} \ud Z_I $ (where $Z_I$ includes both the external energies $X_{1,2}$ and the exchange energy $Y$), the differential equation of interest can be written as
\be
\ud\, \vec{I} =  \e A\, \vec{I}\, ,
\label{equ:DE}
\eeq
where $A$ is a matrix-valued one-form, which defines an Abelian flat connection over the space of integrated functions because it
satisfies 
$\ud A =0$ and $A \wedge A=0$. 
The matrix $A$ can then be written as a sum of dlog forms 
\beq
A = \sum_i \alpha_i\, \ud \log \Phi_i(Z)\,,
\eeq 
where $\alpha_i$ are constant matrices and the functions $\Phi_i(Z)$ are called ``letters.”
The set of all letters is the ``alphabet" and it determines the possible singularities of functions in the family~\eqref{equ:family0}, and plays an important role in describing the nature of the differential equation. We derive the system of differential equations satisfied by the two-site graph in detail in Section~\ref{sec:frwintegral} (see also~\cite{De:2023xue}).

\paragraph{Graphical representation}
The explicit form of the connection $A$ of course depends on the choice of basis integrals.
Choosing the basis associated to the bounded regions depicted in Figure~\ref{fig:regions}, we can write~\eqref{equ:DE} in the form  
\begin{align}
\ud \psi
&\ =  \  \e\, \Big[(\psi-F)
\  \Lap
 \ + \   F
\  \Lam
 \ + \   (\psi- \tilde F)
\  \Lbp
\ \ + \ \  \tilde F
\   \Lbm \Big] \label{equ:TwoSite-dPsi-X} \\[-4pt]
\cline{1-2}
\ud F 
&\ =  \   \e\, \Big[ F 
\  \Lam 
 \ +  \  (F-Z)\ \Lbp 
\ + \ 
Z\  \Labb \Big] \label{equ:TwoSite-dF-X}
  \\[5pt]
\ud \tilde F 
&\ =\  \e\, \Big[ \tilde F 
\ \Lbm 
 \ +  \  (\tilde F-Z) \  
\Lap  
 \ + \ 
Z\  \Labr  \Big]
  \\[-4pt]
  \cline{1-2}
\ud Z 
&\ =  \  2\e\,Z\  \Labg   
\label{equ:TwoSite-dZ-X}
\end{align}
where we have introduced a graphical representation of the dlog forms: 
\beq
\begin{aligned}
	&\Lap \ \equiv\ \ud\log(X_1+Y)\,,  &&\Lam\ \equiv\ \ud\log(X_1-Y)\,,\\
	&\Lbp \ \equiv\ \ud\log(X_2+Y)\,, &&\Lbm\ \equiv\ \ud\log(X_2-Y)\,,\\
	&\Labg \ \equiv\ \ud\log(X_1+X_2)\,,
\end{aligned}
\label{eq:summaryletters}
\eeq
and we have given the various basis integrals $I_a$ their own names, $\psi, F,\tilde F, Z$. Once we have this system of differential equations, we can effectively evaluate the original integrals of interest by solving them. For the two-site chain, the explicit solution for~\eqref{equ:Integral} can be written in terms of power laws like $(X_n+Y)^\e$ and the Gauss hypergeometric function~${}_2 F_1$.

\vskip4pt
The representation~\eqref{equ:TwoSite-dPsi-X}--\eqref{equ:TwoSite-dZ-X} contains the same information as the matrix equation~\eqref{equ:DE}, but the graphical form of the equation will turn out to be very powerful. 
We see that the letters of the differential equations can be represented by connected ``tubings" of ``marked graphs", with the markings denoting sign flips of the internal energies.
What is less apparent in this single example is that the structure of differentials themselves can also be deduced in terms of some dynamics of these tubes. 
Importantly, in this and other examples, the relevant graph of interest is very closely related to the original Feynman diagram that we would use to compute the cosmological observable directly.

\paragraph{More complex examples} For the two-site chain, the master integrals (\ref{equ:family0}) are two-dimensional and their singularities can easily be visualized as lines in the plane. However, we will also be interested in more complicated graphs with additional exchanges, like
\beq
 \raisebox{-8pt}{
 \begin{tikzpicture}[baseline=(current  bounding  box.center)]
 \coordinate (1) at (-1.4,0);
 \coordinate (2) at (0,0);
 \coordinate (3) at (1.4,0);
  \coordinate (1p) at (-1.4,-0.75);
 \coordinate (2p) at (0,-0.75);
 \coordinate (3p) at (1.4,-0.75);
  \coordinate (4p) at (2.8,-0.75);
  \draw[fill] (3) circle (.5mm);
   \draw[fill] (2) circle (.5mm);
  \draw[fill] (1) circle (.5mm);
      \draw[fill] (4p) circle (.5mm);
    \draw[fill] (3p) circle (.5mm);
   \draw[fill] (2p) circle (.5mm);
  \draw[fill] (1p) circle (.5mm);
  \draw[thick] (1) -- (2) -- (3);
    \draw[thick] (1p) -- (2p) -- (3p) -- (4p);
\end{tikzpicture}}
\hspace{1.5cm}
\begin{tikzpicture}[baseline=(current  bounding  box.center)]
 \coordinate (4) at (0,0);
 \coordinate (3) at (0,1.4);
  \coordinate (1) at (-1.1,-1.1);
 \coordinate (2) at (1.1,-1.1);
 \draw[fill] (4) circle (.5mm);
  \draw[fill] (3) circle (.5mm);
   \draw[fill] (2) circle (.5mm);
  \draw[fill] (1) circle (.5mm);
  \draw[thick] (4) -- (3);
    \draw[thick] (4) -- (2);
   \draw[thick] (4) -- (1); 
\end{tikzpicture}
\hspace{1.5cm}
\begin{tikzpicture}[baseline=(current  bounding  box.center)]
  \coordinate (1p) at (-1.4,-0.75);
 \coordinate (2p) at (0,-0.75);
 \coordinate (3p) at (1.4,-0.75);
  \coordinate (4p) at (2.8,-0.75);
   \coordinate (5) at (0,0.65);
 \coordinate (6) at (1.4,0.65);
       \draw[fill] (5) circle (.5mm);
    \draw[fill] (6) circle (.5mm);
      \draw[fill] (4p) circle (.5mm);
    \draw[fill] (3p) circle (.5mm);
   \draw[fill] (2p) circle (.5mm);
  \draw[fill] (1p) circle (.5mm);
    \draw[thick] (1p) -- (2p) -- (3p) -- (4p);
      \draw[thick] (2p) -- (5);
       \draw[thick] (3p) -- (6);
\end{tikzpicture}
\nonumber
\eeq
In those cases, the integrals are higher-dimensional---a graph with $n-1$ exchanges can be written as an $n$-dimensional integral. The singularities of the integrands lie along codimension-1 hyperplanes in this space. The higher-dimensional setting makes it difficult to visualize the geometry of intersections of these planes and the bounded regions that they define, and so a more algebraic approach is required (see Section~\ref{sec:MoreComplex}). 

\vskip4pt
The strategy to analyze these more complicated examples is to enumerate all possible $n$-tuples of planes, which along with the ``plane at infinity" define simplices in projective $n$-space. In Section~\ref{ssec:simplex}, we show that by defining a suitable boundary operation, we can then find the linear combinations of these simplices in ${\mathbb P}^n$ that are also simplices in ${\mathbb R}^n$ (those which are closed). The number of such linear combinations is the same as the number of bounded regions of the hyperplane arrangement, and therefore they define a basis of possible integrands with these singularities. It is then a mechanical (if tedious) task to differentiate these basis functions and re-write them back in terms of the original basis, mirroring the approach taken for the two-site chain. The systematics of taking differentials is explained in Section~\ref{ssec:diffeqssec4}, and many examples of the procedure are illustrated in Section~\ref{ssec:sec4-examples}.

\vskip4pt
Remarkably, what we find by carrying out this procedure is that the resulting differential equations can always be given a graphical representation, quite similar to that of the two-site example~\eqref{equ:TwoSite-dPsi-X}--\eqref{equ:TwoSite-dZ-X}. Given the original Feynman graph that generates the energy integral of interest, both the functions that appear in the differential system and the letters of the differential equation can be associated to certain tubings of a marked graph.
More strikingly, the structure of the differentials continues to be interpretable in terms of simple dynamical rules that these tubings obey. This suggests that it should be possible to understand the differential equations in terms of some sort of autonomous  ``flow" directly in the space of kinematics. Indeed, one of the main results of this paper is an efficient graphical method to derive the differential equations for arbitrary tree graphs (see Section~\ref{sec:GraphicalRules}).

\paragraph{Kinematic flow} The analysis of concrete examples in an algorithmic fashion suggests a shift of perspective that reveals a hidden structure underlying the differential equations satisfied by the cosmological wavefunction. 

\vskip4pt
The fundamental starting point is a tree graph with marked internal edges.
We begin by enumerating the basis functions that will appear in the differential system; each
is associated to a (possibly disconnected) complete tubing of the graph (without any nested tubes).  
For the two-site chain, these tubings are
 \beq
 \begin{tikzpicture}[baseline=(current  bounding  box.center)]
  \node at (-0.85,0)  {$\psi$};
\node[inner sep=0pt] at (0,0)
   {\includegraphics[scale=1]{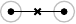}}; 
 \node at (1.3,0)  {$F$};
 \node[inner sep=0pt] at (2.2,0)
   {\includegraphics[scale=1]{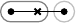}}; 
    \node at (1.3,-0.45)  {$\tilde F$};
    \node[inner sep=0pt] at (2.2,-0.5)
   {\includegraphics[scale=1]{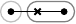}}; 
    \node at (3.4,0)  {$Z$};
 \node[inner sep=0pt] at (4.3,0)
   {\includegraphics[scale=1]{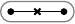}}; 
\end{tikzpicture}
\nonumber
\eeq
In this case, there are only four possible tubings, but for more complicated graphs the number of distinct complete tubings---and hence the size of the basis---grows. We can similarly associate the letters of the system to the connected tubings of the same graph, as in~\eqref{eq:summaryletters}. Note that the translation between tubings and letters relies on the association with the original Feynman graph, which specifies how each vertex and edge depends on the energies involved in the problem. We describe the general features of this graphical representation in Sections~\ref{ssec:graphical} and~\ref{ssec:rules}.

\vskip 4pt
Given this graphical representation of letters and basis functions, there is then a completely systematic algorithm to write down the associated differential equations. 
We start with the graph tubing associated to a ``parent function" of interest and generate a ``family tree" of its ``descendants" according to a set of simple and universal rules. For the case of the two-site chain, the relevant trees are 
 \begin{align}
 & \raisebox{10pt}{\begin{tikzpicture}[baseline=(current  bounding  box.center)]
  \node at (-1.2,0)  {$\psi$\,:};
\node[inner sep=0pt] at (0,0)
   {\includegraphics[scale=1]{Figures/Tubings/two/psi/twopsi}}; 
\draw [color=gray,thick,-stealth] (0.8,0.15) -- (1.8,0.6);
\draw [color=gray,thick,-stealth] (0.8,-0.15) -- (1.8,-0.6);
\node[inner sep=0pt] at (2.65,0.7)
    {\includegraphics[scale=1]{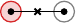}};
\node[inner sep=0pt] at (2.65,-0.7)
    {\includegraphics[scale=1]{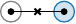}};   
    \draw [color=gray,thick,-stealth] (3.6,0.7) -- (4.6,0.7);
    \draw [color=gray,thick,-stealth] (3.6,-0.7) -- (4.6,-0.7);
    \node[inner sep=0pt] at (5.5,0.7)
    {\includegraphics[scale=1]{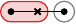}};
    \node at (6.4,0.7)  {$F$};
       \node[inner sep=0pt] at (5.5,-0.7)
    {\includegraphics[scale=1]{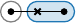}};
    \node at (6.4,-0.7)  {$\tilde F$};
 \draw [color=gray,thick,line width=0.5pt, dashed] (1.95,-0.95) -- (1.95,0.95) --  (3.35,0.95) -- (3.35,-0.95) -- (1.95,-0.95); 
  \node[above] at (2.65,.95)  {$\psi$};
\end{tikzpicture}} \label{equ:psi-Tree}\\[16pt]
&\raisebox{-6pt}{\begin{tikzpicture}[baseline=(current  bounding  box.center)]
  \node at (-1.2,0)  {$F$\,:};
\node[inner sep=0pt] at (0,0)
   {\includegraphics[scale=1]{Figures/Tubings/two/F/twoF}}; 
\draw [color=gray,thick,-stealth] (0.8,0.15) -- (1.8,0.6);
\draw [color=gray,thick,-stealth] (0.8,-0.15) -- (1.8,-0.6);
\node[inner sep=0pt] at (2.65,0.7)
    {\includegraphics[scale=1]{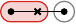}};
\node[inner sep=0pt] at (2.65,-0.7)
    {\includegraphics[scale=1]{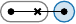}};   
    \draw [color=gray,thick,-stealth] (3.6,-0.7) -- (4.6,-0.7);
       \node[inner sep=0pt] at (5.5,-0.7)
    {\includegraphics[scale=1]{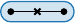}};
    \node at (6.4,-0.7)  {$Z$};
 \draw [color=gray,thick,line width=0.5pt, dashed] (1.95,-0.95) -- (1.95,0.95) --  (3.35,0.95) -- (3.35,-0.95) -- (1.95,-0.95); 
  \node[below] at (2.65,-.95)  {$F$};
\end{tikzpicture}} \\
&\begin{tikzpicture}[baseline=(current  bounding  box.center)]
  \node at (-1.2,0)  {$Z$\,:};
\node[inner sep=0pt] at (0,0)
   {\includegraphics[scale=1]{Figures/Tubings/two/Z/twoZ}}; 
\draw [color=gray,thick,-stealth] (0.8,0) -- (1.8,0);
\node[inner sep=0pt] at (2.65,0.0)
    {\includegraphics[scale=1]{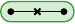}};
    \node at (3.7,0)  {$2Z$};
\end{tikzpicture}
\label{equ:Z-Tree}
\end{align}
The precise procedure to generate these trees is explained in Section~\ref{ssec:rules}, but it should be apparent that it is related to a set of rules for how tubes can grow and combine with each other. These flowcharts encode the differential of the function represented by the original graph tubing. There is then a procedure for reading off the differential. Schematically, each shaded tubing corresponds to a letter in the differential equation, and the
coefficient of each letter is given by the function associated to the relevant complete tubing minus the functions associated to its immediate descendants. For the two-site chain, applying this to (\ref{equ:psi-Tree})--(\ref{equ:Z-Tree}) leads to the equations~\eqref{equ:TwoSite-dPsi-X}--\eqref{equ:TwoSite-dZ-X}. In this case, the equations are relatively simple, and the power of the graphical method may not be fully apparent. For more complicated graphs, however, the differential equations quickly become much more complex. Remarkably, the rules for predicting the structure of the equations stay simple and uniform. 

\vskip4pt
Perhaps the most intriguing feature of this perspective is that it is possible to state this procedure in a way that is completely independent from the original perturbation theory algorithm that produced the integrals we wanted to compute. Instead, we could imagine starting with a graph (not necessarily thinking of it as a Feynman diagram), associating functions and letters to its tubings, and defining a differential that is closed ($\ud^2 = 0$).\footnote{See Appendix~\ref{app:locality}, for more details about the relation between the graphical rules and the nilpotency of the exterior derivative.} It would then be natural to study the properties of the functions that appear in this system. 
The fact that these objects have an interpretation as pieces of the wavefunction in cosmological spacetimes would come as quite a surprise from this point of view. The idea that we can trade the usual procedure of tracking time evolution in cosmological perturbation theory for a sort of ``kinematic flow" directly in the configuration space of cosmological correlators is rather tantalizing. We have been led to this perspective systematically by transforming the bulk calculation into the space of boundary kinematics, but these features suggest that there is also an alternative starting point where this combinatorial question is the natural one to ask, while bulk time evolution would be a fully derived concept.

\paragraph{Beyond single graphs} For the most part, we study the differential equations for individual Feynman graphs, with a fixed ordering of the external legs. Similarly, most work in the cosmological bootstrap is still organized graph by graph~\cite{Baumann:2022jpr}. In contrast, for scattering amplitudes the most stunning simplifications are seen when one sums over all graphs contributing to a process. This is especially true for gravity and gauge theories, where the individual graphs aren't even uniquely defined. It is therefore an important challenge to find some irreducible structures inside the cosmological wavefunction that emerge when summing over graphs.

\vskip 4pt
A minimal situation where the sum over graphs is expected to be interesting is
${\rm tr}\,\phi^3$ theory, and we study its cosmological wavefunction in Section~\ref{sec:Channels}.  
In this theory, the external fields carry flavor charges, so that permutations of external particles are inequivalent.
It is convenient to describe the kinematics of an $n$-point function in terms of
the $n$-gon associated to the $n$ (spatial) momentum vectors, which sum to zero: $\sum_a\vec{k}_a =0$. 
The different ``channels" contributing to a given flavor-ordered wavefunction then correspond to the different triangulations of the $n$-gon. 
For example, the four-point function is described by a closed quadrilateral, which has two different triangulations corresponding to $s$ and $t$-channel exchange:
\begin{equation*}
 \raisebox{5pt}{
 \begin{tikzpicture}[baseline=(current  bounding  box.center)]
 \draw[lightgray, line width=1.pt] (0,0) -- (-0.25,0.75);
\draw[lightgray, line width=1.pt] (0,0) -- (0.25,0.75);
\draw[lightgray, line width=1.pt] (1.4,0) -- (1.15,0.75);
\draw[lightgray, line width=1.pt] (1.4,0) -- (1.65,0.75);
\draw[lightgray, line width=2.pt] (-0.5,0.75) -- (1.9,0.75);
\node[above] at (-0.25,0.75) {\scriptsize $1$};
\node[above] at (0.25,0.75) {\scriptsize $2$};
\node[above] at (1.15,0.75) {\scriptsize $3$};
\node[above] at (1.65,0.75) {\scriptsize $4$};
\node[below] at (.7,-.15) {\small $s$-channel};
\draw[thick] (0, 0) -- (1.4, 0);
\draw[fill,color=Red] (0, 0) circle (.5mm);
\draw[fill,color=Blue] (1.4, 0) circle (.5mm);
   \draw[black, line width=0.5pt] (3,-0.5) -- (4.25,0.75);
 \draw[Red, line width=1.pt] (3,-0.5) -- (3,0.75) -- (4.25,0.75);
  \draw[Blue, line width=1.pt] (3,-0.5) -- (4.25,-0.5) -- (4.25,0.75);
  \node[left,color=Red] at (3,0.125) {\scriptsize $k_1$};
  \node[above,color=Red] at (3.625,0.75) {\scriptsize $k_2$};
    \node[right,color=Blue] at (4.25,0.125) {\scriptsize $k_3$};
  \node[below,color=Blue] at (3.625,-0.5) {\scriptsize $k_4$};
\end{tikzpicture}} 
\hspace{1.6cm}
 \raisebox{5pt}{
 \begin{tikzpicture}[baseline=(current  bounding  box.center)]
 \draw[lightgray, line width=1.pt] (0,0) -- (-0.25,0.75);
\draw[lightgray, line width=1.pt] (0,0) -- (0.25,0.75);
\draw[lightgray, line width=1.pt] (1.4,0) -- (1.15,0.75);
\draw[lightgray, line width=1.pt] (1.4,0) -- (1.65,0.75);
\draw[lightgray, line width=2.pt] (-0.5,0.75) -- (1.9,0.75);
\node[above] at (-0.25,0.75) {\scriptsize $1$};
\node[above] at (0.25,0.75) {\scriptsize $4$};
\node[above] at (1.15,0.75) {\scriptsize $2$};
\node[above] at (1.65,0.75) {\scriptsize $3$};
\node[below] at (.7,-.15) {\small $t$-channel};
\draw[thick] (0, 0) -- (1.4, 0);
\draw[fill,color=Red] (0, 0) circle (.5mm);
\draw[fill,color=Blue] (1.4, 0) circle (.5mm);
   \draw[black, line width=0.5pt] (3,0.75) -- (4.25,-0.5);
 \draw[Red, line width=1.pt] (3,0.75)  -- (3,-0.5) -- (4.25,-0.5);
  \draw[Blue, line width=1.pt] (3,0.75) -- (4.25,0.75) -- (4.25,-0.5);
  \node[left,color=Red] at (3,0.125) {\scriptsize $k_1$};
  \node[above,color=Blue] at (3.625,0.75) {\scriptsize $k_2$};
    \node[right,color=Blue] at (4.25,0.125) {\scriptsize $k_3$};
  \node[below,color=Red] at (3.625,-0.5) {\scriptsize $k_4$};
\end{tikzpicture}} 
\end{equation*}
We can therefore think of the graphs studied earlier as corresponding to these kinematic polygons plus a choice of triangulation.
(In general, the number of triangulations of an $n$-gon is given by the Catalan number $C_{n-2}$, see Section~\ref{sec:triangulations} for details.) 
It is natural to suspect that we will find additional structures in the larger kinematic space that combines together different compatible exchange channels.

\vskip 4pt
We are interested in the differential equations satisfied by these $n$-point functions when taking derivatives with respect to the external side lengths $k_a$ (without specifying a bulk process/triangulation). It is again enlightening to represent the result graphically.
Similar to before, we can associate letters and functions to geometric objects related to the original polygon, which is explained in Section~\ref{ssec:letterssec5}.
Each letter is now a connected ``shading" of a sub-polygon, which represent the different kinematic limits that can be taken.
To account for sign flips in the internal energies, we allow the internal lines to be solid or dashed, with the dashed lines representing internal energies with flipped signs. (This is the analogue of tubes enclosing markings on internal lines in the graph case.) 
The letters are then given by the perimeters of the shaded sub-polygon, with sign flips for any dashed lines.
For example, three letters (dlog forms) appearing in the differential equation for the four-point function are
\beq
\raisebox{-8pt}{\includegraphics[scale=0.8]{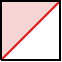}}
=\ud\log(k_{12}+s)\,,
\hspace{.5cm}
\raisebox{-8pt}{\includegraphics[scale=0.8]{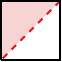}}=\ud\log(k_{12}-s)\,,
\hspace{.5cm}
\raisebox{-8pt}{\includegraphics[scale=0.8]{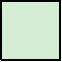}}=\ud\log(k_{1234})\,,
\eeq
where  $k_{n_1\cdots n_j} \equiv |\vec{k}_{n_1}| +\cdots+ |\vec{k}_{n_j}|$ and $s\equiv |\vec{k}_1+\vec{k}_2|$. There are also letters associated to the other triangle in this triangulation, as well as the triangles involved in the other possible triangulation of the square.

\vskip4pt
The functions that appear when we take derivatives are associated to a second type of shading of the triangulated polygon (see Section~\ref{ssec:polygondifferential}). Much like in the single-graph case, the shadings associated to the basis functions can in general be disconnected. 
In the four-point function example, there are $7$ functions, corresponding to the following shadings
\begin{equation*}
 \begin{tikzpicture}[baseline=(current  bounding  box.center),scale=0.85]
          \draw[gray, line width=1.pt] (-2.75,3.5) -- (-1.75,4.5); 
 \draw[black, line width=1pt] (-2.75,3.5) -- (-2.75,4.5) -- (-1.75,4.5) -- (-1.75,3.5) -- (-2.75,3.5); 
\node[below] at (-2.25,3.5) {$\psi^{(s)}$};
     \draw[gray, line width=1.pt] (2.75,3.5) -- (1.75,4.5); 
 \draw[black, line width=1pt] (2.75,3.5) -- (2.75,4.5) -- (1.75,4.5) -- (1.75,3.5) -- (2.75,3.5); 
\node[below] at (2.25,3.5) {$\psi^{(t)}$};
    \fill[gray!20] (-3.5,1.5) -- (-3.5,2.5) -- (-2.5,2.5) -- (-3.5,1.5);
    \draw[gray, line width=1.pt, dashed] (-2.5,2.5) -- (-3.5,1.5); 
  \draw[black, line width=1pt] (-3.5,1.5) -- (-3.5,2.5) -- (-2.5,2.5) -- (-2.5,1.5) -- (-3.5,1.5); 
  \node[below] at (-3,1.5) {$F^{(s)}$};
    \fill[gray!20] (3.5,1.5) -- (3.5,2.5) -- (2.5,2.5) -- (3.5,1.5);
    \draw[gray, line width=1.pt, dashed] (2.5,2.5) -- (3.5,1.5); 
  \draw[black, line width=1pt] (3.5,1.5) -- (3.5,2.5) -- (2.5,2.5) -- (2.5,1.5) -- (3.5,1.5); 
  \node[below] at (3,1.5) {$\tilde F^{(t)}$};
    \fill[gray!20] (-2,1.5) -- (-1,1.5) -- (-1,2.5) -- (-2,1.5);
    \draw[gray, line width=1.pt, dashed] (-1,2.5) -- (-2,1.5); 
  \draw[black, line width=1pt] (-2,1.5) -- (-2,2.5) -- (-1,2.5) -- (-1,1.5) -- (-2,1.5); 
  \node[below] at (-1.5,1.5) {$\tilde F^{(s)}$};
    \fill[gray!20] (2,1.5) -- (1,1.5) -- (1,2.5) -- (2,1.5);
    \draw[gray, line width=1.pt, dashed] (1,2.5) -- (2,1.5); 
  \draw[black, line width=1pt] (2,1.5) -- (2,2.5) -- (1,2.5) -- (1,1.5) -- (2,1.5); 
  \node[below] at (1.5,1.5) {$F^{(t)}$};
\draw[black, fill=gray!20, line width=1pt] (-0.5,-0.5) -- (-0.5,0.5) -- (0.5,0.5) -- (0.5,-0.5) -- (-0.5,-0.5); 
\node[below] at (0,-0.5) {$Z$};
\end{tikzpicture}
\end{equation*}
The un-shaded triangulations are each associated to a distinct source function, which is the wavefunction coefficient in the corresponding channel.

\vskip4pt
The kinematic flow rules then determine how the shadings of a function ``evolve" when we take derivatives; see Section~\ref{ssec:polygondifferential}.
The novelty of this situation is that there are typically several paths through the space of shadings that end up at the same configuration. Consistency demands that these different paths lead to the same functions, and this requires that certain functions are ``shared" between different channels. As a simple example of this phenomenon, we can consider the differential of the function $F^{(s)}$---which is the function $F$ appearing in~\eqref{equ:TwoSite-dF-X}---and that of its $t$-channel analogue $F^{(t)}$. These two differentials are determined by the following tree of shadings
 \beq
 \begin{tikzpicture}[baseline=(current  bounding  box.center)]
\node[inner sep=0pt] at (0.1,1.4)
    {\includegraphics[scale=0.8]{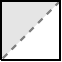}}; 
    \node[inner sep=0pt] at (0.1,-1.4)
    {\includegraphics[scale=0.8]{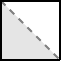}}; 
    \draw [color=gray,thick,-stealth] (0.6,1.5) -- (1.45,1.9);
     \draw [color=gray,thick,-stealth] (0.6,1.3) -- (1.45,0.9);
     \draw [color=gray,thick,-stealth] (0.6,-1.5) -- (1.45,-1.9);
     \draw [color=gray,thick,-stealth] (0.6,-1.3) -- (1.45,-0.9);
 \draw [color=gray,thick,line width=0.5pt, dashed] (1.6,0.3) -- (1.6,2.5) --  (2.6,2.5) -- (2.6,0.3) -- (1.6,0.3); 
   \node[above] at (2.1,2.55) {$F^{(s)}$};
    \draw [color=gray,thick,line width=0.5pt, dashed] (1.6,-0.3) -- (1.6,-2.5) --  (2.6,-2.5) -- (2.6,-0.3) -- (1.6,-0.3); 
   \node[below] at (2.1,-2.55) {$F^{(t)}$};
\node[inner sep=0pt] at (2.1,2)
    {\includegraphics[scale=0.8]{Figures/Shadings/Square/sqFsr}};
\node[inner sep=0pt] at (2.1,0.8)
    {\includegraphics[scale=0.8]{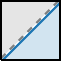}};
\node[inner sep=0pt] at (2.1,-2)
    {\includegraphics[scale=0.8]{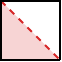}};
    \node[inner sep=0pt] at (2.1,-0.8)
    {\includegraphics[scale=0.8,angle=90]{Figures/Shadings/Square/sqFsZ}};
   \draw [color=gray,thick,-stealth] (2.8,0.7) -- (3.7,0.2); 
   \draw [color=gray,thick,-stealth] (2.8,-0.7) -- (3.7,-0.2);
\node[inner sep=0pt] at (4.2,0)
    {\includegraphics[scale=0.8]{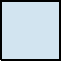}}; 
   \node[right] at (4.65,0) {$Z$};    
\end{tikzpicture}
\nonumber
\eeq
We see that the {\it same} source function $Z$ (associated to the fully shaded kinematic square) appears in both differentials,  which requires a certain compatibility between the different exchange channels. It is of course possible to split apart the function $Z$ and associate half of it to each exchange channel, but this is in a sense artificial.

\vskip4pt
The sharing of functions between different channels becomes much more frequent and nontrivial at higher points. For example, we discuss situations at five and six points in Section~\ref{ssec:polygondifferential}, and describe how aspects of the shared letters and functions can be understood using  the associahedron in Section~\ref{sec:associahedron}. This is a first hint of some simplicity obtained by asking the right question of the sum over individual graphs, but it is clear that there are still deeper structures to uncover. For scattering amplitudes in ${\rm tr}\,\phi^3$ theory, the compatibility between different channels is related to the kinematic associahedron~\cite{Arkani-Hamed:2017mur}, which serves as a positive geometry that encodes the amplitude geometrically. In the cosmological context, there are now tantalizing glimpses of similar structures. A positive geometry that computes the full sum over graphs remains elusive, but there are interesting features in the compatibility between different channels that hint at something similar waiting to be found.

\subsection*{Outline} 

The outline of the paper is as follows: In~Section~\ref{sec:WF}, we review the wavefunction approach to computing cosmological correlators. We derive explicit results for wavefunction coefficients in flat space and then present a simple model of a conformally coupled scalar in an FRW background. We show that the wavefunction coefficients in this model can be written as twisted integrals of the corresponding flat-space results. In Section~\ref{sec:TwoSite}, we study in detail the integral (\ref{equ:Integral}) for the two-site chain (see also~\cite{De:2023xue}). We define the integral basis and derive the corresponding differential equation. Imposing boundary conditions at the singularities of the equation, we present an explicit solution to the problem. In Section~\ref{sec:MoreComplex}, we extend the treatment to more complex tree graphs.  We derive the differential equations for all $n$-site chains up to $n=6$ and for the $4$-site star topology.  In Section~\ref{sec:GraphicalRules}, we identify simple graphical rules to predict the structure of the equations for arbitrary tree graphs. 
 In Section~\ref{sec:Channels}, we move beyond single graphs and study the differential equations satisfied by sums of Feynman graphs. 
 Finally, in Section~\ref{sec:Outlook}, we present our conclusions and give an outlook on future directions.

\vskip 4pt
Five appendices contain additional background material and technical details: In Appendix~\ref{app:Maths}, we give a pedagogical overview of some of the mathematics that is relevant to this work. In Appendix~\ref{app:flow-examples}, we present further examples for the graphical representation of the differential equations and explain how these arise from the rules for the kinematic flow.  In Appendix~\ref{app:locality}, we describe the connection between integrability of the differential equations and the physics of locality.  In Appendix~\ref{app:functions}, we discuss the space of functions arising as solutions to the differential equations. Finally, in Appendix~\ref{app:symbology}, we present the relation between this work and the symbol approach to the differential equations of conformal scalars in de Sitter space~\cite{Hillman:2019wgh}.

\vskip 4pt
The details of the computations can be found in supplemental {\sc Mathematica} notebooks on a github repository for this paper. We will provide links to the relevant notebooks using the github icon \href{https://github.com/haydenhylee/kinematic-flow}{\faGithub}.

 \newpage
\section{Correlators as Twisted Integrals}
\label{sec:WF}

We begin by introducing the physical object of interest: the wavefunction for cosmological fluctuations~\cite{LectureNotes, Benincasa:2022gtd}.  It is a somewhat more primitive quantity than actual correlation functions, and as such it is simpler to compute and interpret. Already in the time-independent setting of flat-space field theory  the wavefunction possesses a rich and interesting structure~\cite{Arkani-Hamed:2017fdk}. 
More remarkably for our purposes, the flat-space wavefunction can serve as a seed from which the wavefunction in cosmological spacetimes will germinate. In particular, the wavefunction of conformally coupled scalars in a power-law cosmology can be written as a twisted integral over the corresponding flat-space wavefunction.

\subsection{Preliminary Definitions}
\label{ssec:Definitions}

The wavefunction encodes the quantum-mechanical state of a system on a given time slice. In terms of the fundamental fields of the theory, $\phi$, the wavefunctional is the overlap between a given state $\rvert \uppsi\rangle$ and the basis of field eigenstates $\rvert \varphi(\vec x) \rangle $:\footnote{Here, the field eigenstates satisfy $\hat\phi(t,\vec x)\lvert \varphi(\vec x)\rangle = \varphi(\vec x)\lvert \varphi(\vec x)\rangle$, where $\hat\phi$ is a field operator (in the Heisenberg picture) and $\varphi(\vec x)$ is a spatial field profile.}
\beq
\Psi[ \varphi(\vec x)] \equiv \langle \varphi(\vec x) \rvert \uppsi\rangle\,.
\label{eq:WFdef}
\eeq
Any state of a system can be expressed in this way, and the wavefunction~\eqref{eq:WFdef} contains all  information about correlations in this state. For example, equal-time correlation functions can be extracted by squaring the wavefunction and integrating against some operator insertions:
\beq
\langle\varphi(\vec{x}_1)\cdots \varphi(\vec{x}_n)\rangle =  \frac{\displaystyle\int{\cal D} \varphi\,\varphi(\vec{x}_1)\cdots \varphi(\vec{x}_n) \left\lvert\Psi[\varphi]\right\rvert^2}{\displaystyle\int{\cal D} \varphi\hskip 1pt \left\lvert\Psi[\varphi]\right\rvert^2}\, .
\eeq
In the following, we will be interested in the structure of the wavefunction of the vacuum, $ \rvert \uppsi\rangle \to\, \rvert 0 \rangle$, which is particularly simple.

\vskip 4pt
In perturbation theory, it is convenient to expand the wavefunction in powers of the field fluctuations:
\beq
\begin{aligned}
\Psi[\varphi] &\,=\, \exp\bigg(\!-\frac{1}{2}\int{\rm d}^3x_1{\rm d}^3x_2\, \Psi_2(\vec x_1,\vec x_2)\,\varphi(\vec x_1)\varphi({\vec x_2}) \\
&\hspace{1.6cm} +\sum_{n=3}^\infty \int{\rm d}^3x_1\cdots {\rm d}^3x_n\,\Psi_n(\underline{\vec x}) \, \varphi(\vec x_1)\cdots\varphi({\vec x_n})
\bigg)\,,
\end{aligned}
\label{eq:WFexp}
\eeq
where $\underline{\vec x} \equiv \{\vec x_1, \cdots, \vec x_n \}$ is a set of spatial coordinates, and the kernels $ \Psi_n(\underline{\vec x})$ are the so-called {\it wavefunction coefficients}. The functions $\Psi_n$ characterize the statistics of the field fluctuations.
The systems that we will consider are spatially translation invariant, so it is convenient to work in Fourier space so that
\beq
\int{\rm d}^3x_1\cdots {\rm d}^3x_n\, \Psi_n(\underline{\vec x}) \, \varphi(\vec x_1)\cdots\varphi({\vec x_n}) = \int\frac{{\rm d}^3k_1\cdots {\rm d}^3k_n}{(2\pi)^{3n}}\, \Psi_n(\underline{\vec k}) \, \varphi_{\vec k_1}\cdots\varphi_{\vec k_n}\, ,
\eeq
where the momentum-space wavefunction coefficients are proportional to momentum-conserving delta functions
\beq
\Psi_n(\underline{\vec k}) = (2\pi)^3  \delta(\vec k_1+\cdots+\vec k_n)\, \psi_n(\underline{\vec k}) \, .
\eeq
We will mostly focus directly on the functions $\psi_n(\underline{\vec k})$, which we will also refer to as wavefunction coefficients.

\vskip 4pt
To compute the wavefunction, we start from its 
path-integral representation 
\beq
\Psi[\varphi] \ =  \hspace{-0.2cm} \int\limits_{\substack{\phi(0) \,=\,\varphi\\ \hspace{-0.45cm}\phi(-\infty)\,=\,0}} 
\hspace{-0.45cm} \raisebox{-.05cm}{ ${\cal D} \phi\, e^{iS[\phi]}\,,$ }
\label{eq:wfPI}
\eeq
where the integration is over all field configurations that interpolate between the free vacuum in the far past, $t=- \infty(1-i\epsilon)$, and some field configuration $\varphi(\vec x)$ at a later time $t \equiv 0$. 
 The tree-level approximation of this path integral is given by its saddle point, $\Psi[ \varphi] \approx \exp\left(iS[ \phi_{\rm cl}]\right)$, where $\phi_{\rm cl}$ is the classical solution satisfying the same boundary conditions as the path integral~\eqref{eq:wfPI}. Fluctuations about this saddle capture quantum corrections to the wavefunction coefficients.
 
\subsection{Flat-Space Wavefunction}
\label{ssec:FlatSpace}

We will illustrate the wavefunction approach with the simple example of a massless scalar field with polynomial self-interactions in flat space~\cite{Arkani-Hamed:2017fdk, Benincasa:2022gtd}:
\beq
 S = \int \ud^4 x  \left[-\frac{1}{2}( \partial\phi)^2- \sum_{p>2} \frac{\lambda_p}{p!} \phi^p \right] . 
 \label{eq:flatphinaction}
\eeq
The correlators of this theory will also provide the seeds from which the correlators of conformally coupled scalars in FRW spacetime can be generated; see Section~\ref{ssec:FRW}.

\paragraph{Feynman rules:} Perturbatively, the evaluation of the path integral~\eqref{eq:wfPI} can be given a simple diagrammatic interpretation, paralleling the Feynman rules for the computation of scattering amplitudes in flat space. The rules for computing the $n$-point coefficient $\psi_n$ are:\\[4pt]
$\bullet$ Draw all diagrams with $n$ lines ending on the late-time surface at $t=0$. \\[2pt]
$\bullet$ Assign a vertex factor $iV$ to each bulk interaction, derived as usual from the action~\eqref{eq:flatphinaction}.\\[2pt]
$\bullet$ Assign a bulk-to-boundary propagator to each external line:
\beq
K(k,t) = e^{ikt}\,,
\label{eq:bbdyprop}
\eeq
where $k \equiv |\k|$.
This is a solution to the wave equation, $(\partial_t^2 +k^2)\hskip 1pt  K(k,t) = 0$, with $K(k,t=0) = 1$ and $K(k,t=-\infty(1-i\epsilon))=0$.\\[2pt]
$\bullet$ Assign a bulk-to-bulk propagator to each internal line:
\beq
G(k;t,t') = \frac{1}{2k} \left[e^{-ik(t-t')} \theta(t-t') + e^{+ik(t-t')} \theta(t'-t) - e^{ik(t+t')}\right] , \label{eq:Gbb}
\eeq
which is a Green's function for the wave operator in Fourier space, $(\partial_t^2+k^2)\hskip 1pt G(k;t,t') =-i\delta(t-t')$. The non-time-ordered piece in (\ref{eq:Gbb}) is required by the boundary condition $G(k;t,0) = 0$.\\[2pt]
$\bullet$ Finally, integrate over the time insertions of all bulk vertices and over any undetermined loop momenta.

\paragraph{Wavefunction coefficients:} The simplest example is the wavefunction coefficient associated to a contact interaction.
We call this a ``one-site graph" and denote the corresponding wavefunction coefficient by $\psi^{(1)}_{\rm flat}$.
Using the above Feynman rules, we get
\beq
\psi^{(1)}_{\rm flat} = \raisebox{-29pt}{
\begin{tikzpicture}[line width=1. pt, scale=2]
\draw[lightgray, line width=1.pt] (0,0) -- (-0.25,0.55);
\draw[lightgray, line width=1.pt] (0,0) -- (0.0,0.55);
\draw[lightgray, line width=1.pt] (0,0) -- (0.25,0.55);
\draw[lightgray, line width=2.pt] (-0.5,0.55) -- (0.5,0.55);
\draw[fill=black] (0,0) circle (.03cm);
\node[scale=1] at (0,-.15) {$X$};
\end{tikzpicture}
} 
= i \int_{-\infty}^0 \ud t\, e^{i X t} = \frac{1}{X}\, ,
\label{equ:contact}
\eeq
where we have set the coupling constant $\lambda$ to unity.
The number of external lines is arbitrary, as only the total energy $X \equiv \sum_a |\k_a|$ enters in the computation. 

\vskip 4pt
A more nontrivial example is the single exchange of a particle at tree level. 
The wavefunction coefficient associated to this ``two-site graph" is  
 \begin{equation}
\psi^{(2)}_{\rm flat}=
\scalebox{1.0}{
 \raisebox{-26pt}{
\begin{tikzpicture}[line width=1. pt, scale=2]
\draw[fill=black] (0,0) -- (1,0);
\draw[fill=black] (0,0) -- (1,0);
\draw[line width=1.pt,lightgray] (0,0) -- (-0.25,0.55);
\draw[line width=1.pt,lightgray] (0,0) -- (0,0.55);
\draw[line width=1.pt,lightgray] (0,0) -- (0.25,0.55);
\draw[line width=1.pt,lightgray] (1,0) -- (0.75,0.55);
\draw[line width=1.pt,lightgray] (1,0) -- (1,0.55);
\draw[line width=1.pt,lightgray] (1,0) -- (1.25,0.55);
\draw[lightgray, line width=2.pt] (-0.5,0.55) -- (1.5,0.55);
\draw[fill=Red,Red] (0,0) circle (.03cm);
\draw[fill=Blue,Blue] (1,0) circle (.03cm);
\node[scale=1] at (0,-.15) {$X_1$};
\node[scale=1] at (1,-.15) {$X_2$};
\node[scale=1] at (0.5,-.12) {$Y$};
\end{tikzpicture}
} 
}
= - \int_{-\infty}^0 \ud t  \int_{-\infty}^0 \ud t'\, e^{i X_1t}\, \,G(Y; t,t')\, \, e^{i X_2t'} \,,
\end{equation}
where $Y$ is the energy of the internal particle. 
Note that the result depends only to the total energies flowing into each vertex, which we have denoted by $X_1$ and $X_2$. Performing the time integrals, we find
\beq
\psi^{(2)}_{\rm flat}(X_1,X_2,Y) = \frac{1}{(X_1+X_2){\color{Red}(X_1 +Y)}{\color{Blue} (X_2+Y)}} \equiv \frac{1}{X {\color{Red}X_L} {\color{Blue} X_R}}\, .
\label{eq:flatWF}
\eeq
The result has simple poles at vanishing ``total energy", $X \equiv X_1+X_2$, and the two ``partial energies" $X_L \equiv X_1 + Y$ and $X_R \equiv X_2+Y$.

\paragraph{Combinatorial definition:} Wavefunction coefficients corresponding to more complicated diagrams can be computed in the same way, but the degree of complexity of the time integrals involved increases quickly with the number of internal lines. To simplify the calculation of higher-order wavefunction coefficients, we can take advantage of some of the hidden structure underlying the flat-space wavefunction. In particular, there is a simple combinatorial prescription for generating the wavefunction coefficient associated to a graph. 
To explain this, we must first introduce the concept of {\it graph tubings}, which will play a critical role in the rest of this paper.

\vskip 4pt
Given a graph, we define a {\it tube} to be a subset of vertices of the graph which form a connected subgraph of the original graph. It is convenient to denote the tube by circling the relevant subgraph. For example, the following is a tube of the three-site graph: 
\be
\begin{tikzpicture}[baseline=(current  bounding  box.center)]
\draw[fill=black] (-0.75,0) -- (0.75,0);
\draw[fill=black] (0,0) circle (.5mm);
\draw[fill=black] (-0.75,0) circle (.5mm);
\draw[fill=black] (0.75,0) circle (.5mm);
\node [
        draw, color=black, line width=0.8pt,
        rounded rectangle,
        minimum height = 1.1em,
        minimum width = 4.1em,
        rounded rectangle arc length = 180,
    ] at (0.38,0)
    {};
\end{tikzpicture}
\ee
For our purposes, the only compatibility condition needed for tubes is that they do {\it not} intersect. (Two tubes intersect if their intersection is 
non-empty and neither is a subset of the other.) Finally, a {\it complete tubing} is a maximal set of compatible (non-overlapping) tubes of a graph.\footnote{We should warn the reader that a different notion of compatibility is used in the mathematical literature~\cite{CARR20062155}. There, tubes are considered to be compatible only if they neither intersect nor are adjacent. For us, the restriction that tubes not be adjacent would be rather artificial. An interesting avenue for investigation is to explore the mathematical properties of the compatibility conditions that are natural from the cosmological perspective.}

\vskip 4pt
To compute the flat-space wavefunction coefficients, we first generate the set of compatible complete tubings of a given graph.  For each tube, we sum up all the energies of vertices enclosed by the tube, along with the energy of any lines that intersect the tube. 
(This is the energy associated to the subgraph enclosed by the tube.) 
We then assign a rational function to the complete tubing which is the inverse of the product of all of the energy sums associated to the constituent tubings, and sum over all possible complete tubings of the graph. The resulting rational function is precisely the wavefunction coefficient associated to the graph~\cite{Arkani-Hamed:2017fdk}.

\vskip4pt
It is simplest to understand this algorithm through examples. Above we derived the wavefunction of the two-site graph, which has a single complete tubing:
\be\label{equ:2pt-tubing}
\psi^{(2)}_{\rm flat} \ = \  \raisebox{2pt}{
 \begin{tikzpicture}[baseline=(current  bounding  box.center)]
\draw[fill] (4, 0) circle (.5mm);
\draw[fill] (5, 0) circle (.5mm);
\draw[thick] (4, 0) -- (5, 0);
\node [
        draw, color=black, line width=0.6pt,
        rounded rectangle,
        minimum height = 1.3em,
        minimum width = 4.9em,
        rounded rectangle arc length = 180,
    ] at (4.5,0)
    {};
\draw[color=Red, line width=0.6pt] (4, 0) circle (2mm);
\draw[color=Blue, line width=0.6pt] (5, 0) circle (2mm);
\end{tikzpicture}}
\,\,=\,\,  \frac{1}{{\color{black}(X_1+X_2)}{\color{Red}(X_1+Y)}{\color{Blue}(X_2+Y)}} \,,
\ee
where on the right-hand side we have written the product of the sums of energies associated to each tube of the same color. We see that this reproduces exactly the wavefunction coefficient~\eqref{eq:flatWF}. A more nontrivial example is provided by the three-site graph. In this case, there are two inequivalent complete tubings. Summing them, we obtain the wavefunction coefficient
\begin{align}
\label{equ:3pt-tubing}
\psi^{(3)}_{\rm flat} &= \
 \raisebox{2pt}{
 \begin{tikzpicture}[baseline=(current  bounding  box.center)]
\draw[fill=black] (-0.75,0) -- (0.75,0);
\draw[fill=black] (0,0) circle (.5mm);
\draw[fill=black] (-0.75,0) circle (.5mm);
\draw[fill=black] (0.75,0) circle (.5mm);
\draw[color=Red, line width=0.6pt] (-0.75,0) ellipse (.15cm and .15cm);
\draw[color=Blue, line width=0.6pt] (0.75,0) ellipse (.15cm and .15cm);
\draw[color=Orange, line width=0.6pt] (0,0) ellipse (.15cm and .15cm);
\node [
        draw, color=Purple, line width=0.8pt,
        rounded rectangle,
        minimum height = 1.1em,
        minimum width = 4.0em,
        rounded rectangle arc length = 180,
    ] at (-0.385,0)
    {};
    \node [
        draw, color=black, line width=0.6pt,
        rounded rectangle,
        minimum height = 1.4em,
        minimum width = 6.2em,
        rounded rectangle arc length = 180,
    ] at (-0.1,0)
    {};
\end{tikzpicture}}
\,\,
+
 \raisebox{2pt}{
 \begin{tikzpicture}[baseline=(current  bounding  box.center)]
\draw[fill=black] (-0.75,0) -- (0.75,0);
\draw[fill=black] (0,0) circle (.5mm);
\draw[fill=black] (-0.75,0) circle (.5mm);
\draw[fill=black] (0.75,0) circle (.5mm);
\draw[color=Red, line width=0.6pt] (-0.75,0) ellipse (.15cm and .15cm);
\draw[color=Blue, line width=0.6pt] (0.75,0) ellipse (.15cm and .15cm);
\draw[color=Orange, line width=0.6pt] (0,0) ellipse (.15cm and .15cm);
\node [
        draw, color=brown, line width=0.8pt,
        rounded rectangle,
        minimum height = 1.1em,
        minimum width = 4.1em,
        rounded rectangle arc length = 180,
    ] at (0.38,0)
    {};
        \node [
        draw, color=black, line width=0.6pt,
        rounded rectangle,
        minimum height = 1.4em,
        minimum width = 6.2em,
        rounded rectangle arc length = 180,
    ] at (0.1,0)
    {};
\end{tikzpicture}}\\\nonumber
&=\ \frac{1}{(X_1+X_2+X_3){\color{Red}(X_1+Y)}{\color{Orange}(X_2+Y+Y^\prime)}{\color{Blue}(X_3+Y^\prime)} }\bigg(\frac{1}{\color{Purple}X_1+X_2+Y^\prime}+\frac{1}{\color{brown}X_2+X_3+Y} \bigg)\, ,
\end{align}
which matches the result of a bulk computation. It is straightforward to consider more complicated graphs and derive their wavefunction coefficients in the same way.

 \subsection{Wavefunction from Polytopes}
 \label{ssec:CosmoPolytope}
 
The flat-space wavefunction also has an interesting geometric origin~\cite{Arkani-Hamed:2017fdk} which we will sketch in the following. Our description will be purely qualitative.  
 
\vskip 4pt
To each graph ${\cal G}$ we assign a polytope ${\cal P}_{\cal G}$, whose facets are defined by the energy singularities of ${\cal G}$. For example, the polytope associated to the two-site graph is simply a triangle
 \begin{align}
 \raisebox{2pt}{\begin{tikzpicture}[baseline=(current  bounding  box.center)]
\draw[line width=1pt] (2, 0) -- (0, 0);
\draw[fill,color=Red] (0, 0) circle (0.75mm);
\draw[fill,color=Blue] (2, 0) circle (0.75mm);
\end{tikzpicture}}
&\qquad \ \ \Longleftrightarrow \qquad
 \raisebox{-4pt}{
 \begin{tikzpicture}[baseline=(current  bounding  box.center)]
 \draw[fill = lightgray!50,opacity=0.7] (4., 0) -- (5,1.5) -- (6., 0) -- (4.,0);
\draw[line width=1.25pt] (4., 0) -- (6., 0);
\draw[line width=1.25pt,color=Red] (4., 0) -- (5,1.5);
\draw[line width=1.25pt,color=Blue] (6., 0) -- (5,1.5);
\node at (5, -0.3) {\small $X=0$};
\node at (3.7, 0.8) {\small \color{Red}$X_L=0$};
\node at (6.3, 0.8) {\small \color{Blue}$X_R=0$};
\end{tikzpicture}}
\nonumber
\end{align}
 The sides of the triangle are given by the three energy singularities of the exchange diagram; cf.~\eqref{eq:flatWF}.
More complicated graphs correspond to intersections of triangles $\triangle_i$ subject to certain rules (intersections occur at the midpoints of the triangles' edges and only on edges corresponding to vertices of subgraphs; and we require the collection of intersecting triangles to be connected).
 For example, the following intersection of two triangles corresponds to the three-site graph:
  \begin{align}
 \hspace{-1.5cm} 
 \raisebox{2pt}{\begin{tikzpicture}[baseline=(current  bounding  box.center)]
\coordinate (A) at (-1.25, 0);
\coordinate (B) at (1.25, 0);
\coordinate (C) at (0,1.8);
\coordinate (D) at (-0.75,-0.5);
\coordinate (E) at (0.75,0.5);
\coordinate (F) at (0.4,-1.65);
\draw[line width=1pt] (-1.5, 0) -- (1.5, 0);
\draw[fill,color=Red] (-1.5, 0) circle (0.75mm);
\draw[fill,color=Orange] (0, 0) circle (0.75mm);
\draw[fill,color=Blue] (1.5, 0) circle (0.75mm);
\end{tikzpicture}}
&\qquad \ \ \Longleftrightarrow \qquad \ \
 \scalebox{.905}{
 \raisebox{4pt}{
 \begin{tikzpicture}[baseline=(current  bounding  box.center)]
  \draw[line width = 0.5pt] (A) -- (E)  -- (B);
    \draw[line width = 0.5pt] (E)  -- (C);
 \draw[fill = lightgray!50,opacity=0.7] (D) -- (E) -- (F) -- (D);
 \draw[line width=1.25pt,color=Orange] (D) -- (E);
\draw[line width=1.25pt,color=black] (E) -- (F);
\draw[line width=1.25pt,color=Blue] (D) -- (F);
 \draw[fill = lightgray!50,opacity=0.7] (A) -- (B) -- (C) -- (A);
\draw[line width=1.25pt,color=Orange] (A) -- (B);
\draw[line width=1.25pt,color=Red] (A) -- (C);
\draw[line width=1.25pt,color=black] (B) -- (C);
\draw[fill,color=Orange] (0, 0) circle (0.75mm);
  \draw[line width = 0.5pt] (A) -- (D)  -- (B);
    \draw[line width = 0.5pt] (D)  -- (C);
      \draw[line width = 0.5pt, dashed] (B)  -- (F);
\end{tikzpicture}} 
}
\nonumber
\end{align}
 The polytope ${\cal P}_{\cal G}$ is then the {\it convex hull} of the vertices of all the intersecting triangles $\triangle_i$. For the two-site graph, this is simply the original triangle, while, for the three-site graph, it is a ``double square pyramid" living in the four-dimensional projective space $\mathbb{P}^4$.

\vskip 4pt
The polytopes ${\cal P}_{\cal G}$ are autonomous geometrical objects, which provide an ab initio definition of the wavefunction without invoking any spacetime structure. 
In particular, the {\it canonical form} defined on each polytope, $\Omega({\cal P}_{\cal G})$, gives the  wavefunction coefficient $\psi_{\cal G}$ for the associated graph.
These canonical forms are the unique forms with logarithmic singularities
on the boundaries of the polytopes (and nowhere else). 
They have a fascinating recursive structure: evaluating the canonical form on the facets of the polytopes gives the canonical form of the facet.  This recursive facet structure reflects the factorization properties of the wavefunction on its energy singularities.
Canonical forms will play  a central role in our work and we describe their properties in more detail in Appendix~\ref{app:Maths}, where we also show how to compute canonical forms for various examples.

\subsection{Cosmological Wavefunction}
\label{ssec:FRW}

The polytopes introduced in the previous section provide a purely combinatorial description of the flat-space wavefunction, without any reference to time evolution in the bulk spacetime.
In this paper, we will study a natural way of integrating the flat-space wavefunction (or, equivalently, the canonical form of the polytope) to obtain the wavefunction in a cosmological spacetime. These ``twisted integrals" provide a boundary-centric definition of the cosmological wavefunction and will give a new mathematical perspective on the question of time in cosmology. In this section, we will explain from a bulk perspective how these integrals arise. 

\vskip4pt
To begin, we consider a {\it conformally coupled scalar} with (non-conformal) polynomial self-interactions 
\beq
S = \int \ud^4 x \sqrt{-g} \left[-\frac{1}{2}( \partial\phi)^2-\frac{1}{12} R \phi^2 - \sum_{p>2} \frac{ \lambda_p}{p!} \phi^p \right] , \label{equ:S}
\eeq
where the metric is ${\rm d} s^2 = a^2(\eta) \left(- {\rm d}\eta^2+{\rm d}\vec x^{\hskip 1pt 2}\right)$.
We will assume that the scale factor takes the form of a power law
 \beq
 a(\eta) = \left(\frac{\eta}{\eta_0}\right)^{-(1+\e)}\,,
 \label{equ:power}
 \eeq
which captures many cases of interest, such as $\e = 0$ (de Sitter), $\e \approx 0$ (inflation), $\e =-1$ (Minkowski), $\e=-2$ (radiation) and $\e=-3$ (matter). 
 Correlators will be evaluated at a fixed time $\eta =0$, which in accelerating cosmologies is  the future boundary of the spacetime. We will be interested in correlators where scalars are evolved from $-\infty$ to 0 in $\eta$, with Bunch--Davies initial conditions imposed in the far past.\footnote{Later, we will observe that the differential equations governing the correlators are unaffected by this assumption, and solving for correlators that arise from generic initial conditions amounts to imposing different boundary conditions.} For non-accelerating cosmologies with $\e<-1$, this will be equivalent to considering a contracting phase of the universe~\cite{Chen:2011zf, Chen:2014cwa}.

\vskip 4pt
The model (\ref{equ:S}) is particularly simple because the quadratic action is invariant under the Weyl transformation $g_{\mu\nu}\mapsto a^2 g_{\mu\nu}$, $\phi\mapsto a^{-1}\phi$, which allows us to write the action as that of a {\it massless scalar in flat space}, but now with time-dependent couplings:
 \beq
 S = \int \ud^4 x \left[-\frac{1}{2}( \partial\phi)^2 - \sum_{p>2} \frac{\lambda_p(\eta)}{p!} \phi^p \right] .
 \label{equ:S2}
 \eeq
The time dependence of the couplings is determined by the scale factor as
 \beq
 \lambda_p(\eta) \equiv  \lambda_p\, a(\eta)^{4-p} \propto \eta^{(p-4)(1+\e)}\,,
 \label{equ:lambda}
 \eeq
 where in the last equality we have specialized to the power-law cosmologies defined in~\eqref{equ:power}.

\vskip 4pt
Given the close relationship between~\eqref{equ:S2} and~\eqref{eq:flatphinaction}, it is not surprising that the perturbative construction of the wavefunction for these power-law cosmologies closely parallels the flat-space case. In particular, since the quadratic action is the same, the bulk-to-boundary and bulk-to-bulk propagators are still given by~\eqref{eq:bbdyprop} and~\eqref{eq:Gbb}, respectively (only written in terms of conformal time~$\eta$ rather than coordinate time $t$). The Feynman rules are also the same, the only difference being that now all vertex factors involve time-dependent couplings. This complicates the evaluation of the time integrals involved in the calculation of the wavefunction, and is the primary technical challenge involved in computing the cosmological wavefunction, as compared to flat space. However, an elegant way to deal with these time-dependent couplings is to consider their frequency representations \cite{Arkani-Hamed:2017fdk}: 
\beq
\lambda_p(\eta) = \int_0^\infty \ud \omega\, \tl\lambda_p(\omega)\,e^{i\omega \eta} \, .
\label{equ:lambda2}
\eeq
For the power-law scaling in \eqref{equ:lambda}, we get
\beq
\tl\lambda_p(\omega) \propto \omega^{\alpha}\,, \quad {\rm with} \quad \alpha \equiv (4-p)(1+\e) - 1\, .
\eeq
Working in frequency space, the time integrals become identical to their flat-space counterparts, where the energies involved are shifted by the frequencies associated to the couplings. To obtain the full cosmological wavefunction coefficients, we must integrate these shifted flat-space wavefunctions over {\it energies}. Strictly speaking, the frequency integral \eqref{equ:lambda2} only converges for $\alpha>-1$, which 
excludes some physical cosmologies of interest (e.g.~convergence requires $\e>-1$ when $p=3$). 
 However, we will see that the differential equations satisfied by the energy integrals, and their solutions, can be analytically continued in $\e$, and that we can extract the physical wavefunctions by taking suitable limits. This is a conceptual advantage of focusing on the differential equations satisfied by the wavefunction coefficients rather than the integrals per se.

\vskip 4pt
This is most simply illustrated with examples. 
First, consider a three-point contact interaction. The associated wavefunction coefficient is
 \begin{equation}
 \psi_{(1)} =
\scalebox{1.0}{
 \raisebox{-26pt}{
\begin{tikzpicture}[line width=1. pt, scale=2]
\draw[line width=1.pt] (0,0) -- (-0.3,0.55);
\draw[line width=1.pt] (0,0) -- (0,0.55);
\draw[line width=1.pt] (0,0) -- (0.3,0.55);
\node[scale=1,above] at (-0.3,0.55) {$k_1$};
\node[scale=1,above] at (0.,0.55) {$k_2$};
\node[scale=1,above] at (0.3,0.55) {$k_3$};
\draw[lightgray, line width=2.pt] (-0.6,0.55) -- (0.6,0.55);
\draw[fill=black] (0,0) circle (.03cm);
\node[scale=1] at (0,-.15) {$X$};
\end{tikzpicture} } }
= i \int_{-\infty}^0 \ud \eta \, e^{i X\eta}\, \lambda_3(\eta) \,,
\label{equ:BulkIntegral0}
\end{equation}
where $X \equiv k_1+k_2+k_3$. Substituting the frequency representation of the couplings \eqref{equ:lambda}, where  $\tl\lambda_3(\omega) \propto \omega^\e$, we find
\beq
\psi_{(1)} \propto \int_0^\infty \ud \omega\, \omega^\e \ \underbrace{ i \int \ud \eta\, e^{i (X+\omega)\eta}}_{\displaystyle \psi^{(1)}_{\rm flat}(X+\omega)} =  \int_0^\infty \ud \omega\, \frac{\omega^\e}{X+\omega}  \, ,
\eeq
where we have identified the integrand with the shifted flat-space wavefunction, $\psi^{(1)}_{\rm flat}(X+\omega)$, times a twist factor, $\omega^\e$.
Performing the integral, we get
\be
\psi_{(1)} = -\pi \csc(\pi\e) \,X^\e\, . \label{equ:One}
\ee
We therefore see that the solution of the one-site graph in a general FRW spacetime is a simple power law.
In de Sitter space, with $\e =0$, we instead have $\psi_{(1)} \propto \log(iX)$.

\vskip4pt
Next, consider the four-point wavefunction coefficient arising from a cubic interaction:
 \begin{equation}
 \psi_{(2)} =
\scalebox{1.0}{
 \raisebox{-26pt}{
\begin{tikzpicture}[line width=1. pt, scale=2]
\draw[fill=black] (0,0) -- (1,0);
\draw[fill=black] (0,0) -- (1,0);
\draw[line width=1.pt] (0,0) -- (-0.25,0.55);
\draw[line width=1.pt] (0,0) -- (0.25,0.55);
\draw[line width=1.pt] (1,0) -- (0.75,0.55);
\draw[line width=1.pt] (1,0) -- (1.25,0.55);
\node[scale=1,above] at (-0.25,0.55) {$k_1$};
\node[scale=1,above] at (0.25,0.55) {$k_2$};
\node[scale=1,above] at (0.75,0.55) {$k_3$};
\node[scale=1,above] at (1.25,0.55) {$k_4$};
\draw[lightgray, line width=2.pt] (-0.6,0.55) -- (1.6,0.55);
\draw[fill=black] (0,0) circle (.03cm);
\draw[fill=black] (1,0) circle (.03cm);
\node[scale=1] at (0,-.15) {$X_1$};
\node[scale=1] at (1,-.15) {$X_2$};
\node[scale=1] at (0.5,-.12) {$Y$};
\end{tikzpicture}
} 
}
= -\int_{-\infty}^0 \ud \eta \hskip 1pt \ud \eta'\, e^{i X_1\eta}\, \lambda_3(\eta) \,G(Y; \eta,\eta')\, \lambda_3(\eta')\, e^{i X_2\eta'} \,,
\label{equ:BulkIntegral}
\end{equation}
where we have defined $X_1 \equiv k_1+k_2$, $X_2 \equiv k_3+k_4$ and $Y \equiv |\k_1 +\k_2|$.\footnote{Due to the Weyl rescaling $\phi\mapsto a^{-1}\phi$, the original FRW wavefunction coefficient evaluated at $\eta=\eta_*$ has an overall factor of $a^4(\eta_*)$ relative to \eqref{equ:BulkIntegral}. We will ignore this constant rescaling.} Substituting the frequency representation of the couplings \eqref{equ:lambda}, we get
\beq
\psi_{(2)} \propto \int_0^\infty \ud \omega_1 \ud \omega_2\, \omega_1^\e \omega_2^\e  \underbrace{ \int \ud \eta \hskip 1pt \ud \eta'\, e^{i (X_1+\omega_1)\eta} \,G(Y; \eta,\eta')\,  e^{i (X_2+\omega_2)\eta'}}_{\displaystyle \psi^{(2)}_{\rm flat}(X_1+\omega_1,X_2+\omega_2,Y)}\, .
\eeq
We see that the wavefunction coefficient can be expressed as a two-dimensional frequency integral, where the integral kernel is precisely the flat-space wavefunction coefficient $\psi^{(2)}_{\rm flat}$ in~\eqref{eq:flatWF}, with external energies shifted by the frequencies $\omega_a$. We then have
\begin{eBox}
\vspace{-0.3cm}
\beq
	\psi_{(2)} = - \int_0^\infty \ud x_{1} \ud x_{2}\,(x_{1}x_{2})^{\e}\frac{2Y}{(X_{1}+X_{2}+x_{1}+x_{2}) (X_{1} +x_{1}+Y)(X_{2}+x_{2}+Y)}\, , \label{equ:Master0}
\eeq
\end{eBox}
where we have relabelled the integration variables $x_a\equiv \omega_a$ and set the overall normalization factor to $-2Y$ for later convenience.
This reduces the challenge of computing the FRW wavefunction coefficient associated to the exchange~\eqref{equ:BulkIntegral} to the task of evaluating this two-dimensional integral involving a rational integrand ``twisted" by powers of the integration variables. Despite appearances, this integral is actually quite simple and rich in structure. Before developing the formalism to evaluate integrals of this type, we enumerate some other explicit examples.

\subsection{Selected Integrals}
\label{sec:integralselection}
By following the procedure outlined in the previous section, it is possible to cast all wavefunction coefficients of a conformally coupled scalar in a power-law FRW background as Euler-type twisted integrals of rational functions. In the following, we give some more complicated examples of such integrals, which will serve as useful test cases for the formalism developed in the rest of the paper. These integrals can be obtained systematically from the time-integral representation, but the integrand associated to a given graph is always the flat-space wavefunction coefficient, with the external energies shifted by integration variables, times a twist factor.  Consequently, one can easily generate many examples by employing the combinatorial construction of the flat-space wavefunction in terms of graph tubings and then shifting the energies and integrating.

\paragraph{Three-site chain:}  The next-simplest example is the wavefunction coefficient involving two exchanges. 
The exponential form of the  bulk-to-boundary propagator for the conformally coupled scalar implies that the answer only depends on the total energy flowing out of a given vertex to the boundary, and so we suppress the external lines that connect to the boundary surface. The relevant graph is therefore the {\it three-site chain}
\begin{equation*}
\begin{tikzpicture}[line width=1. pt, scale=2]
\draw[fill=black] (0,0) -- (2,0);
\draw[fill=black] (0,0) circle (.03cm);
\draw[fill=black] (1,0) circle (.03cm);
\draw[fill=black] (2,0) circle (.03cm);
\node[scale=1] at (0,-.15) {$X_1$};
\node[scale=1] at (1,-.15) {$X_2$};
\node[scale=1] at (2,-.15) {$X_3$};
\node[scale=1] at (0.5,.15) {$Y$};
\node[scale=1] at (1.5,.15) {$Y^{\prime}$};
\end{tikzpicture}
\end{equation*}
where we have labeled the external energies by $X_a$ and the internal energies as $Y,Y'$. The wavefunction coefficient associated to this graph is
\beq
\psi_{(3)} =\int_0^\infty (x_1x_2 x_3)^{\e} \, \Omega_\psi\,, \quad\text{where} \quad \Omega_\psi \equiv \frac{4 YY'}{B_1 B_2 B_3 B_4} \left(\frac{1}{B_5}+\frac{1}{B_6}\right) \ud x_1 \wedge \ud x_2 \wedge \ud x_3\,,\label{equ:ThreeSite}
\eeq
with the linear factors defined by
\beq
\begin{aligned}
B_1 &\equiv X_1+x_1 + Y\,, & \quad B_4 &\equiv  X_1+X_2  + X_3+ x_1+x_2+x_3 \,,\\
B_2 &\equiv X_2+ x_2 + Y+Y'\,,&B_5&\equiv  X_1+X_2 + x_1+x_2 +Y'\, ,
 \\
B_3&\equiv X_3 + x_3  +Y'\, , & B_6&\equiv  X_2+X_3+ x_2+x_3 +Y\, .
\end{aligned}
\label{equ:Singularities2}
\eeq
Comparing this integral to~\eqref{equ:3pt-tubing}, we see that the integrand is exactly the flat-space wavefunction associated to this graph, written in dimensionless form and shifted by the integration variables.

\paragraph{Four-site chain:} We will also consider the {\it four-site chain}
\begin{equation*}
\begin{tikzpicture}[line width=1. pt, scale=2]
\draw[fill=black] (0,0) -- (3,0);
\draw[fill=black] (0,0) circle (.03cm);
\draw[fill=black] (1,0) circle (.03cm);
\draw[fill=black] (2,0) circle (.03cm);
\draw[fill=black] (3,0) circle (.03cm);
\node[scale=1] at (0,-.15) {$X_1$};
\node[scale=1] at (1,-.15) {$X_2$};
\node[scale=1] at (2,-.15) {$X_3$};
\node[scale=1] at (3,-.15) {$X_4$};
\node[scale=1] at (0.5,.15) {$Y$};
\node[scale=1] at (1.5,.15) {$Y^{\prime}$};
\node[scale=1] at (2.5,.15) {$Y^{\prime\prime}$};
\end{tikzpicture}
\end{equation*}
The associated wavefunction coefficient can be written as
\beq
\label{equ:FourSiteChain}
\psi_{(4)}^{(c)} =\int_0^\infty   (x_1x_2 x_3x_4)^{\e} \, \Omega_\psi^{(c)}\,, \quad {\rm where} \quad
 \Omega_\psi^{(c)} \equiv \hat \Omega_\psi^{(c)}\, \ud x_1 \wedge \ud x_2 \wedge \ud x_3\wedge \ud x_4\,\, .
\eeq
The function in the integrand is the flat-space wavefunction for the four-site chain (with shifted arguments)
\beq
\hat \Omega_\psi^{(c)} \equiv \psi_{\rm flat}^{(4,c)}(X_a+x_a) = \frac{8 YY'Y''}{B_1 B_2 B_3 B_4B_5} \left[\frac{1}{B_6 B_8}+\frac{1}{B_9}\left(\frac{1}{B_6}+\frac{1}{B_7}\right)+\frac{1}{B_{10}}\left(\frac{1}{B_7}+\frac{1}{B_8}\right)\right] ,
\label{equ:psi4c}
\eeq
where the linear factors are
\beq
\begin{aligned}
B_1 &\equiv X_1+x_1 + Y\,, & \quad  B_5 &\equiv X_1+X_2  + X_3+X_4+ x_1+x_2+x_3+x_4\,, \\ 
B_2 &\equiv X_2+ x_2 + Y+Y'\,,& B_6 &\equiv X_1+X_2+x_1 +x_2+ Y' \,, \\
B_3&\equiv X_3 + x_3  +Y'+Y''\, , & B_7&\equiv  X_2+X_3+x_2 +x_3+Y+ Y'' \, , \\ 
B_4 &\equiv X_4+ x_4 + Y''\,,& B_8&\equiv  X_3+X_4+x_3 +x_4+Y' \,, \\ 
&& B_9&\equiv  X_1+X_2  + X_3+ x_1+x_2+x_3+Y''\, , \\
& & B_{10}&\equiv  X_2+X_3  + X_4+ x_2+x_3+x_4+Y\, .
\end{aligned}
\label{equ:Singularities4}
\eeq
The form of the function (\ref{equ:psi4c}) is easily determined from the graph tubings described in Section~\ref{ssec:FlatSpace}.

\paragraph{Four-site star:} An interesting new phenomenon happens for graphs involving four sites (or three exchanges). 
In addition to the chain topology, we can also arrange the graph into a star by connecting three of the vertices to a single central vertex with valence three:
\begin{equation*}
\begin{tikzpicture}[baseline=(current  bounding  box.center)]
 \coordinate (4) at (0,0);
 \coordinate (3) at (0,1.8);
  \coordinate (1) at (-1.4,-1.4);
 \coordinate (2) at (1.4,-1.4);
 \draw[fill] (4) circle (.5mm);
  \draw[fill] (3) circle (.5mm);
   \draw[fill] (2) circle (.5mm);
  \draw[fill] (1) circle (.5mm);
  \draw[thick] (4) -- (3);
    \draw[thick] (4) -- (2);
   \draw[thick] (4) -- (1); 
   \node[right] at (0,0.1) {$X_4$};
      \node[right] at (0,1.8) {$X_3$};
   \node[right] at (1.4,-1.4) {$X_2$};
 \node[left] at (-1.4,-1.4) {$X_1$};
  \node[left] at (0,0.9) {$Y''$};
    \node[left] at (-0.7,-0.7) {$Y$};
        \node[right] at (0.8,-0.7) {$Y'$};
\end{tikzpicture}
\end{equation*}

\vskip 4pt
\noindent
The associated wavefunction coefficient takes the same form as~\eqref{equ:FourSiteChain}, but the integrand is now
\beq
 \hat\Omega_\psi^{(s)} \equiv \  \frac{8 YY' Y''}{B_1 B_2 B_3 B_4 B_5} \bigg[\frac{1}{B_9}\left(\frac{1}{B_6}+\frac{1}{B_7}\right) \ + \frac{1}{B_{10}}\left(\frac{1}{B_6}+\frac{1}{B_8}\right) + \frac{1}{B_{11}}\left(\frac{1}{B_7}+\frac{1}{B_8}\right)\bigg] \,, 
 \label{equ:4Site-Star}
\eeq
with 
\beq
\begin{aligned}
B_1 &\equiv X_1+  x_1 +  Y\,, & \quad B_5 &\equiv X_1+X_2  + X_3+X_4 + x_1+x_2+x_3 +x_4 \,, \\ 
B_2 &\equiv  X_2+x_2 +Y'\,,&  B_6&\equiv X_1+X_4 + x_1+x_4 + Y'+Y''\,, 
 \\
B_3&\equiv  X_3 +x_3 +Y''\, , & B_7&\equiv X_2+X_4 + x_2+x_4 +  Y+Y''\, , \\
B_4&\equiv X_4 +  x_4 +  Y+Y'+Y''\,,  & B_8&\equiv  X_3+X_4 + x_3+x_4 + Y+Y' \, ,   \\
& & B_9&\equiv  X_1+X_2+X_4 + x_1+x_2+x_4 + Y'' \,, \\
    & & B_{10}&\equiv  X_1+X_3+X_4 +x_1+x_3+x_4 + Y'\,, \\
    & & B_{11}&\equiv  X_2+X_3+X_4 +  x_2+x_3+x_4 + Y\,,
\end{aligned}\label{equ:4Site-StarB}
\eeq
The function (\ref{equ:4Site-Star})---the flat-space wavefunction coefficient for the four-site star $ \hat\Omega_\psi^{(s)}  \equiv \psi_{\rm flat}^{(4,s)}(X_n+x_n)$---is again easily derived from the graph tubings described in Section~\ref{ssec:FlatSpace}.

\vskip4pt
An important commonality of all the above integrals is that they are integrals of rational functions, whose singularities are associated to linear factors. The singularities of these integrals therefore lie along hyperplanes in the integration space; i.e.~each integral is associated to a {\it hyperplane arrangement}. There is a rich mathematical theory of hyperplane arrangements (see Appendix~\ref{app:Maths}), which will aid us in developing the formalism to evaluate integrals of this type.

\section{Case Study: Two-Site Chain}
\label{sec:TwoSite}

In the previous section, we showed that the wavefunction coefficients of a conformally coupled scalar in an FRW spacetime can be written as twisted integrals of the corresponding flat-space results. Specifically, for the two-site chain, we found 
\beq
	\psi = -\int_0^\infty \ud x_{1} \ud x_{2}\,(x_{1}x_{2})^{\e}\frac{2Y}{(X_{1}+X_{2}+x_{1}+x_{2}) (X_{1} +x_{1}+Y)(X_{2}+x_{2}+Y)}\, .\label{equ:FRW-2site}
\eeq
Similar integrals arise for loop amplitudes in dimensional regularization. In that case, powerful mathematical tools have been used to evaluate these integrals, such as differential equations~\cite{Henn:2014qga} and twisted cohomology~\cite{Mizera:2019ose}. We will now show that a similar approach is very effective for evaluating FRW correlators. In this section, we will present the analysis for the two-site chain as a case study,\footnote{While this work was being written up, related analyses of the two-site chain have appeared in~\cite{De:2023xue, Brunello}.} while, in the next section, we extend this to arbitrary tree graphs, bypassing many of the  steps outlined here.

\subsection{Differential Equation}

\begin{figure}[tb]
\centering
\includegraphics[scale=0.875]{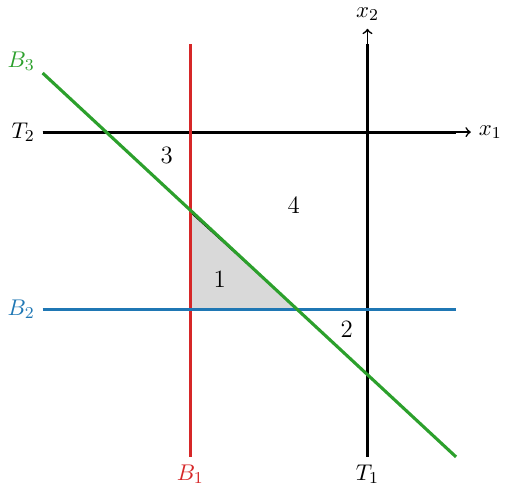}
\caption{Locations of the singular lines of the integrand in (\ref{equ:FRW-2site}). We see that there are four bounded regions which implies a four-dimensional vector space of integrals in (\ref{equ:family}). The gray shaded region is associated to the original integral~\eqref{equ:FRW-2site}. }
\label{fig:regions}
\end{figure}

The key idea is to enlarge the problem and consider a {\it family of integrals} with the same singularities as the original integral:
\beq
I_{N} = \int (x_1 x_2)^\e\, \Omega_N\,, \quad \Omega_N \equiv \frac{ \ud x_1\wedge \ud  x_2}{T_{1}^{m_1} T_{2}^{m_2} B_1^{n_1} B_2^{n_2} B_3^{n_3}}\,,
\label{equ:family}
\eeq
where $N \equiv (m_1,m_2, n_1,n_2,n_3)$ is a set of integers and we have defined
\beq
\begin{aligned}
T_1 &\equiv x_1\,,\qquad\quad B_1 \equiv  X_1 + x_1 + Y \,, \\
T_2 &\equiv x_2\,, \qquad\quad B_2 \equiv  X_2 + x_2+ Y \,, \qquad\quad B_3 \equiv  X_1+X_2 + x_1 +x_2\, .
\end{aligned}
\label{eq:2siteboundaries}
\eeq
The original integral~\eqref{equ:FRW-2site} is one member of this family, namely $\psi =  -2Y \times I_{00111}$. Remarkably, the twisted integrals in (\ref{equ:family}) form a finite-dimensional vector space. This is not obvious from their definition, since the exponents $(m_i, n_i)$ have an infinite range. A finite basis of {\it master integrals} exists because all other integrals can be transformed into linear combinations of basis elements by a combination of partial fraction identities and integrations by parts, and so the integrals belong to equivalence classes. From a more abstract perspective, these equivalence classes are 
determined by the (twisted) cohomology induced by the exterior derivative (see Appendix~\ref{app:Maths}). Twisted cohomology also gives a simple way to predict the size of the vector space~\cite{aomoto1975vanishing, mastrolia2019feynman}:
\begin{center}
{\it The number of independent master integrals is equal to the number\\
 of bounded regions defined by the divisors of the integrand.}
\end{center}
Figure~\ref{fig:regions} shows the arrangement of the singular lines in the $(x_1,x_2)$-plane. There are four bounded regions which implies that the vector space of integrals in (\ref{equ:family}) is four dimensional.

\vskip 4pt
It is convenient to collect the four independent master integrals into a four-component vector
\beq
\vec{I}\equiv  \left[ \begin{array}{c} I_1 \\ I_2 \\ I_3 \\ I_4 \end{array} \right]  = \int (x_1 x_2)^\varepsilon  \left[ \begin{array}{c} \Omega_1 \\ \Omega_2 \\ \Omega_3 \\ \Omega_4 \end{array} \right]  ,
\eeq
where $\Omega_{a}$ are four independent two-forms. The existence of a finite-dimensional vector space has important consequences. In particular, it means that the master integrals must satisfy a set of coupled differential equations.

\vskip 4pt
Consider taking derivatives with respect to the kinematic variables
$Z_I \equiv\{X_1,X_2,Y\}$. Whatever the result, it must be writable as a linear combination of the basis integrals.
This leads to differential equations of the form
\beq
\frac{\partial }{\partial Z_I} I_a = \sum_{b=1}^4 A_{I,ab} I_b\,, \label{equ:Eqn0}
\eeq
where the coefficients $A_{I,ab}$ are rational functions of the variables $Z_I$.
For each $I=1,2,3$, we have a $4\times 4$ matrix $A_I$ with entries $A_{I,ab}$.
Defining the total differential $\ud I_a \equiv \sum_I  (\partial_{Z_I} I_a)\hs {\rm d} Z_I$ and the matrix-valued one-form $A \equiv  \sum_I A_I \hskip 1pt{\rm d} Z_I$, equation (\ref{equ:Eqn0}) can be written compactly as
\beq
\ud \vec{I} = A \, \vec{I}\,.
\label{equ:Eqn1} 
\eeq
The form of $A$ will depend on our particular choice of basis, which we will discuss momentarily. Acting with $\ud$ on (\ref{equ:Eqn1}) gives the constraint $\ud A + A \wedge A = 0$. 

\vskip 4pt
A convenient choice of basis integrals is defined in terms of the {\it canonical forms} of the bounded regions (see box below).   This choice is still not unique, since we have some freedom in choosing different bounded regions to tessellate the region in Figure~\ref{fig:regions}. For all such choices, however, the differential equation~(\ref{equ:Eqn1}) takes the so-called ``$\e$-form"~\cite{Henn:2013pwa}, where the right-hand side becomes proportional to the twist parameter:
 \beq\label{eq:diffeqcanonical}
\ud \vec{I} = \e \tilde A \,\vec{I}\,.
\eeq
The fact that the source of this differential equation is linear in $\e$ has important consequences.
First, a truncated Dyson series (or Chen iterated integral) gives a reliable approximation to the solution at small $\e$ (which is the case of interest for inflationary correlators). Moreover, the condition 
$\ud A + A \wedge A = 0$
becomes stronger for $\tilde A$: it must separately satisfy
$\ud  \tilde A=0$ and $ \tilde A \wedge  \tilde A=0$, i.e.~$\tilde A$ is an {\it Abelian flat connection}. The matrix $\tilde A$ can then be written as a sum of dlog forms 
\beq\label{equ:tildeAPhi}
\tilde A = \sum_i \alpha_i\, \ud \log \Phi_i(Z)\,,
\eeq 
where the functions $\Phi_i(Z)$ are the ``letters" of the differential equation.  

 \vspace{0.25cm}
\begin{eBox}
\noindent
{\small
{\bf Canonical forms:} \ The canonical form of a polytope is the unique differential form with unit logarithmic singularities on all its boundaries, and nowhere else, and which reduces to the canonical form of its boundaries when taking residues.  
In this section, we only need the canonical forms in a few special cases, but more background is given in Appendix~\ref{app:Maths}.

\vskip 4pt
Consider a one-dimensional interval between $x=a$ and $x=b$:
\begin{center}
\begin{tikzpicture}
\draw[black,->,line width=0.75pt] (-4.5,0) -- (1.75,0);
\draw[black] (1.75,0) node[right] {$x$};

\draw[Blue!40,line width=3pt] (-2.5,0) -- (0,0);

\filldraw[Red] (0,0) circle (3pt);
\node[Red,below] at (0,-0.05) {$b$};

\filldraw[Blue] (-2.5,0) circle (3pt);
\node[Blue,below] at (-2.5,-0.07) {$a$};
\end{tikzpicture}
\end{center} 
The canonical form of this interval is
\beq
\Omega = \frac{\ud x}{x-a} - \frac{\ud x}{x-b} = \frac{(a-b)\, \ud x}{(x-a)(x-b)}= \ud \log\left(\frac{x-a}{x-b}\right) ,
\label{equ:line}
\eeq
where the relative minus sign in the first expression ensures the absence of a singularity at infinity.

\vskip 4pt
In two dimensions, we can have a triangle defined by the lines $\{L_i, L_j, L_k\} =0$: 
\begin{center}
\begin{tikzpicture}

\draw[black,->,line width=0.75pt] (-3,1) -- (2,1);
\draw[black] (2,1) node[right] {$x_1$};

\draw[black,->,line width=0.75pt] (-2.5,0.5) -- (-2.5,3);
\draw[black] (-2.5,3) node[above] {$x_2$};

\filldraw[Blue!20] (-2.5,1) -- (0,2.5) -- (1,1) -- (-2.5, 1);

\draw[Blue!50,line width=1.5pt] (-3,0.7)  -- (0.5,2.8);
\draw[Red!50,line width=1.5pt]  (-0.3,3) -- (1.3,0.5);
\draw[Green!50,line width=1.5pt] (-3, 1) -- (1.5,1);

\draw[Blue,line width=2.5pt]  (-2.5,1)  -- (0,2.5);
\draw[Red,line width=2.5pt] (1,1) -- (0,2.5);
\draw[Green,line width=2.5pt] (-2.5,1) -- (1,1);

\draw[Red] (0.6,1.8) node[right] {$L_j$};
\draw[Blue] (-1,2.1) node[left] {$L_i$};
\draw[Green] (-0.75,0.9) node[below] {$L_k$};

\filldraw (-2.5,1) circle (3pt);
\filldraw (0,2.5) circle (3pt);
\filldraw (1,1) circle (3pt);

\end{tikzpicture}
\end{center}
In Appendix~\ref{app:Maths}, we show that the canonical form 
of a general triangle is 
\beq
\label{equ:triangle}
\Omega = \ud\log\left(\frac{L_i}{L_k}\right)\wedge \ud\log\left(\frac{L_j}{L_k}\right) .
\eeq
The expressions (\ref{equ:line}) and (\ref{equ:triangle}) are all that is needed in this section.}
\end{eBox}
 
 \vspace{0.25cm}
\subsection{De Sitter Limit}
\label{ssec:dS}

 \begin{figure}[tb]
\centering
\includegraphics[scale=0.875]{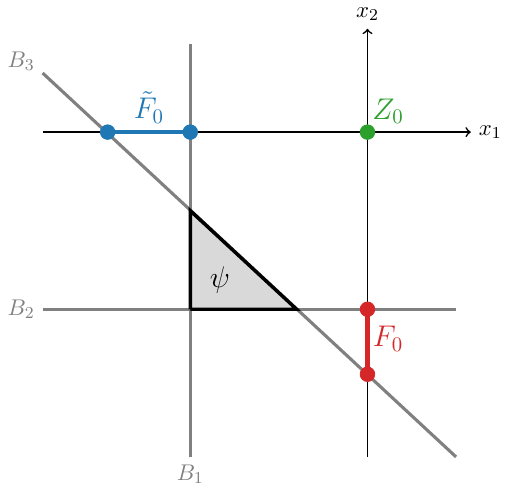}
\caption{Illustration of the geometric meaning of the source functions arising in the differential equations for the de Sitter wavefunction.}
\label{fig:dS-regions}
\end{figure}

Before studying the most general form of the twisted integral~\eqref{equ:FRW-2site}, it is instructive to first consider the special case $\e=0$. The corresponding untwisted integral gives the wavefunction in de Sitter space: 
\be
	\psi_{0}=\int\,\frac{-2Y}{B_1 B_2 B_3}\, \ud x_1 \wedge \ud x_2 \equiv  \int \Omega_\psi \, ,\label{equ:deSitter-2site}
\ee
where $\Omega_\psi \equiv \hat \Omega_\psi \, \ud x_1 \wedge \ud x_2 $ is the canonical form of the triangle bounded by $\{B_1,B_2,B_3\} = 0$.

\vskip 4pt
Importantly, the integrand in (\ref{equ:deSitter-2site}) depends only on the combinations $X_1+x_1$ and $X_2+x_2$. 
 When we take a derivative with respect to the external energy $X_{1}$, we therefore have  $\partial_{X_1}  \hat \Omega_\psi  
= \partial_{x_1} \hat \Omega_\psi$, so that we get a total derivative in the integration variable $x_1$. Evaluating this integral on its boundary, we find 
\beq
\partial_{X_1} \psi_0 = \int  \ud x_2 \,(- \hat \Omega_\psi)\Big|_{x_1=0} = f(X_1) \int \ud \log \left(\frac{B_2}{B_3}\right) \bigg|_{x_1=0} \equiv f(X_1) F_0\,,
\label{equ:dPsi0}
\eeq
where we have written the result in terms of the unique one-form on the line $x_1=0$ that shares a nontrivial residue with $(-\hat \Omega_\psi)\big|_{x_1=0}$, i.e.~the form associated to the red interval in Figure~\ref{fig:dS-regions}. 
Matching residues at $(x_1=0,B_2=0)$ on both sides of (\ref{equ:dPsi0}) gives $ f(X_1) =2Y/(X_1^2-Y^2)$, so that
\beq
\partial_{X_1} \psi_0 = \left( \frac{1}{X_1-Y} - \frac{1}{X_1 +Y}\right) F_0\, .
\eeq
Similarly, taking derivatives with respect to $X_2$ and $Y$ gives
\beq
\begin{aligned}
\partial_{X_2} \psi_0 &= \left( \frac{1}{X_2-Y} - \frac{1}{X_2 +Y}\right) \tilde F_0\,,\\
\partial_{Y} \psi_0 &= -\left( \frac{1}{X_1-Y} + \frac{1}{X_1 +Y}\right) F_0-\left(\frac{1}{X_2-Y} + \frac{1}{X_2 +Y}\right) \tilde F_0\,,
\end{aligned}
\eeq
where $\tilde F_0$ is associated to the canonical form of the blue interval on $x_2=0$ in Figure~\ref{fig:dS-regions}.

\vskip 4pt
Next, we take derivatives of the source functions $F_0$ and $\tilde F_0$.  Starting from the definition of $F_0$ in (\ref{equ:dPsi0}), its $X_1$-derivative is
\beq
\begin{aligned}
\partial_{X_1} F_0  = \int_0^\infty \ud x_2\, \partial_{x_{2}}\left(- \frac{1}{B_3}\right)\bigg|_{x_1=0} = \frac{1}{B_3}\bigg|_{x_1=x_2=0} &= \frac{1}{X_1+X_2} \\
&\equiv \frac{1}{X_1+X_2}\, Z_0\,,
\end{aligned}
\eeq
where we have defined the ``function" $Z_0 \equiv 1$ for later convenience.  Similarly, for the $X_2$-derivative, we get
\beq
\begin{aligned}
\hspace{1cm} \partial_{X_2} F_0  = \int_0^\infty \ud x_2\, \partial_{x_{2}}\left(\frac{1}{B_2}- \frac{1}{B_3}\right)\bigg|_{x_1=0} &= -\left(\frac{1}{B_2} - \frac{1}{B_3}\right)\bigg|_{x_1=x_2=0} \\[4pt]
&\equiv \left(- \frac{1}{X_2+Y} + \frac{1}{X_1+X_2}\right)\, Z_0\,.
\end{aligned}
\eeq
Finally, the $Y$-derivative gives
\beq
\partial_Y F_0 =\int_0^\infty \ud x_2\, \partial_{x_{2}}\frac{1}{B_2}= - \frac{1}{X_2+Y} Z_0\, .
\eeq
By permutation symmetry, $\partial_{X_1} \tilde F_0$ and $\partial_{X_2} \tilde F_0$ satisfy the same equations with $X_1$ and $X_2$ interchanged, while $\partial_{Y} \tilde F_0$  is equal to $\partial_{Y} F_0$ with $X_1$ and $X_2$ interchanged. 

\vskip 4pt
Defining the total differential, $\ud \equiv \sum_I \partial_{Z_I} \ud Z_I $ (where $Z_I$ includes both the external energies $X_{1,2}$ and the exchange energy $Y$), we can combine these equations as
\begin{eBox}
\vspace{-0.3cm}
\beq
\begin{aligned}
\ud \psi_0 &= \big[ \ud \log(X_1-Y) - \ud \log(X_1+Y)  \big] F_0 +  \big[ \ud \log(X_2-Y) - \ud \log(X_2+Y) \big] \tilde F_0  \,,\\[4pt]
\ud F_0 &= \big[ \ud \log(X_1+X_2) -  \ud \log(X_2+Y) \big] Z_0 \,,\\[4pt]
\ud \tilde F_0 &=  \big[ \ud \log(X_1+X_2) -  \ud \log(X_1+Y) \big] Z_0\,,\\[4pt]
\ud Z_0 &= 0\,.
\end{aligned}
\label{equ:dS-Eqns}
\eeq
\end{eBox}
\vspace{0.25cm}
\noindent
In this system of equations, taking a differential produces ``simpler" functions, which is a characteristic of polylogarithmic functions. 
The information encoded in (\ref{equ:dS-Eqns}) can also be repackaged using the {\it symbol} of a transcendental function (for details, see Appendix~\ref{app:symbology}). The symbol $\cal S$ of a function $F^{(n)}$ of transcendental weight $n$ is defined recursively by the relation  
\beq
\ud F^{(n)}= \sum_i F_i^{(n-1)}\, \ud \log R_i \quad \Rightarrow \quad {\cal S}(F^{(n)}) \equiv \sum_i {\cal S}(F_i^{(n-1)}) \otimes R_i\,,
\eeq
where $R_i$ are rational functions of the variables associated to the differential. This is consistent with the structure of the differential equations \eqref{equ:dS-Eqns}, with $R_i$ involving ratios of the letters $\Phi_j/\Phi_l$.
From this definition, we see that the symbol of the de Sitter wavefunction is
\beq
\label{eq:twositesymbol}
{\cal S}(\psi_0) = \frac{X_1+X_2}{X_2+Y} \otimes \frac{X_1-Y}{X_1+Y} + \frac{X_1+X_2}{X_1+Y} \otimes \frac{X_2-Y}{X_2+Y}\,.
\eeq
The symbol can also be determined by other recursive methods, using either bulk perturbation theory~\cite{Hillman:2019wgh} or cosmological polytopes \cite{Arkani-Hamed:2017fdk}. It is interesting that the same building blocks that appear in the symbol here arise in the differential equations satisfied by  the master integrals. This is highly nontrivial, and a more detailed discussion can be found in Appendix~\ref{app:symbology}. 

\newpage
\subsection{FRW Integral}
\label{sec:frwintegral}

 \begin{figure}[t!]
\centering
\includegraphics[scale=0.865]{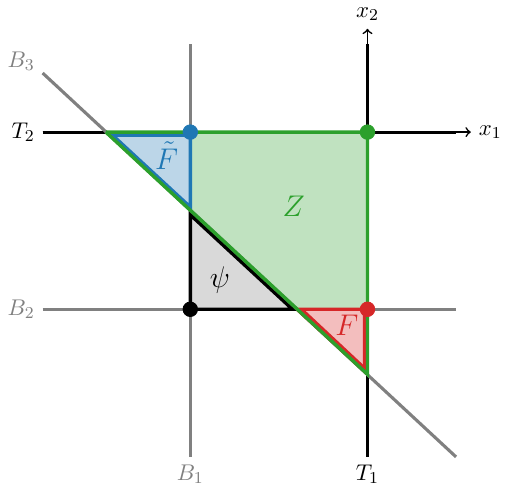}
\caption{Illustration of the geometry behind the functions defined in this section. Associated to each bounded region is a two-form. Each triangle contains one vertex that isn't shared with any other triangle, so any two-form can be decomposed into this basis by matching residues at these special points. }
\label{fig:regions2}
\end{figure}

After this warmup in de Sitter space, we now consider the twisted case:
\be
\psi=\int\, (x_1 x_2)^\e\, \Omega_\psi\, .\label{equ:FRW-2siteX}
\ee
We define a specific integral basis in terms of the canonical forms associated to the four shaded triangles in Figure~\ref{fig:regions2}.\footnote{The selection of these boundary regions is of course not unique, and in principle any four linearly independent combinations of the canonical forms in \eqref{equ:CanonicalForms} can function as a basis. Alternative choices of the basis yield different matrix forms, with varied coefficients and levels of sparsity, etc. Nevertheless, as we will elucidate in later sections, this specific integral basis possesses a canonical character, enabling the reinterpretation of the corresponding differential equations in combinatorial terms.}  Using (\ref{equ:triangle}), we have 
\beq
\begin{aligned}
\Omega_\psi &=  \ud \log\left(\frac{B_1}{B_3} \right) \wedge \ud \log\left(\frac{B_2}{B_3} \right) =  \frac{-2Y}{B_1 B_2 B_3} \, \ud x_1 \wedge \ud x_2 \,,\\[4pt]
\Omega_F &= \ud \log\left(\frac{T_1}{B_3} \right) \wedge \ud \log\left(\frac{B_2}{B_3} \right) =  \frac{X_1-Y}{T_1 B_2 B_3} \, \ud x_1 \wedge \ud x_2 \,,\\[4pt]
\Omega_{\tilde F} &= \ud \log\left(\frac{B_1}{B_3} \right) \wedge \ud \log\left(\frac{T_2}{B_3} \right) =  \frac{X_2-Y}{B_1 T_2 B_3} \, \ud x_1 \wedge \ud x_2\,, \\[4pt]
\Omega_Z &= \ud \log\left(\frac{T_1}{B_3} \right) \wedge \ud \log\left(\frac{T_2}{B_3} \right)=  \frac{X_1+X_2}{T_1 T_2 B_3}\,\ud x_1 \wedge \ud x_2\,.
\end{aligned}
\label{equ:CanonicalForms}
\eeq
Given this choice of bounded regions and associated canonical forms,
 the basis of integrals is 
\be
\vec{I}\equiv  \left[ \begin{array}{c} \psi \\ F \\ \tilde F \\ Z \end{array} \right]  = \int (x_1 x_2)^\e  \left[ \begin{array}{c} \Omega_\psi \\ \Omega_F \\ \Omega_{\tilde F} \\ \Omega_Z \end{array} \right]  .
\label{equ:Psi-Basis}
\ee
To derive the differential equation, we take derivatives of these basis integrals with respect to the external energies. As before, we have $\partial_{X_1} \hat \Omega_\psi = \partial_{x_1} \hat \Omega_\psi$. This time, however, integration by parts will pick up a contribution from the twist factor:
\begin{align}
\partial_{X_1} \psi =\int (x_1 x_2)^\varepsilon  \,\partial_{x_1}\hat \Omega_\psi \, \ud x_1 \wedge \ud x_2 
&= \varepsilon\, \int (x_1 x_2)^\varepsilon  \, \left(-\frac{1}{ T_1}\,\hat \Omega_\psi \right) \ud x_1 \wedge \ud x_2\nonumber\\
&= \varepsilon\, \int (x_1 x_2)^\varepsilon \,  \left(g(X_1)\, \hat \Omega_\psi + f(X_1) \,\hat \Omega_F \right)\ud x_1 \wedge \ud x_2 \, ,
\end{align}
where, in the second line, we have expanded the result into the basis of two-forms.
The only two-forms that share nontrivial residues with the integrand $(-\hat \Omega_\psi/T_1)$, are those associated to the gray and red triangles in Figure~\ref{fig:regions2}. Matching residues at $(B_1=0, B_2=0)$ and $(T_1=0, B_2=0)$, respectively, we find
$g(X_1) =  1/(X_1+Y)$ and $f(X_1) =  2Y/(X_1^2 - Y^2)$, so that
\beq\partial_{X_1} \psi = \frac{\e}{X_1+Y} (\psi- F) + \frac{\e}{X_1-Y} F\, .
\eeq
By symmetry, $\partial_{X_2} \psi$ gives the same equation with $X_1 \to X_2$ and $F \to \tilde F$. Finally, since $Y$ appears only as $X_{1,2}\pm Y$, 
the $\partial_Y$ equations don't produce any new functions, like in the untwisted case.

\vskip 4pt
The rest of the calculation then proceeds in the same way. We take derivatives of the source functions and expand the result in terms of the basis functions, where the coefficient functions are determined by matching residues. This leads to the following total differentials
\begin{eBox}
\vspace{-0.5cm}
\begin{align}
&\ud \psi = \e \, \big[(\psi - F)\,\ud \log(X_1+Y) + F\, \ud \log(X_1-Y) +  (\psi - \tilde F)\, \ud \log(X_2+Y) +\tilde F\,\ud \log(X_2-Y)  \big]\,,\nonumber \\[4pt]
\begin{split}
\ud F &= \e\, \big[ F\,\ud \log(X_1-Y)  + (F-  Z)\, \ud \log(X_2+Y) + Z\, \ud \log(X_1+X_2) \big] \,,  \\[4pt]
\ud \tilde F &=  \e\, \big[ \tilde F\,\ud \log(X_2-Y)  + (\tilde F-  Z)\, \ud \log(X_1+Y) + Z\, \ud \log(X_1+X_2) \big]\,,  \\[4pt]
\end{split} \label{equ:2SiteEqns}\\
&\ud Z = 2 \e\, Z\, \ud \log(X_1+X_2)\,.\nonumber
\end{align}
\vspace{-0.75cm}
\end{eBox}
All integrals have weight $2\e$ under rescalings of the energies~$Z_I \equiv\{X_1,X_2,Y\}$.
In other words, the dilation operator acts as 
\beq
\sum_{I}Z_I \frac{\partial}{\partial Z_I} \ \vec{I} = 2 \e \, \vec{I}\,.
\eeq
An important consequence of this constraint is that 
the sum of all coefficients in each differential equals $2\e$ times the parent function: all source functions cancel in pairs. For example, for the wavefunction $\psi$, we have
\be
\sum_{I}Z_I \frac{\partial}{\partial Z_I}  \psi = \e \, \big[(\psi - F) + F +  (\psi - \tilde F) +\tilde F  \big]=2\e\, \psi\, ,
\ee
where the sum indeed involves all coefficients of the dlog forms appearing in $\ud \psi$. 

\vskip 4pt
Before considering the solution of (\ref{equ:2SiteEqns}), let us show that these equations become the correct de Sitter equations (\ref{equ:dS-Eqns}) in the limit $\e \to 0$. Before taking this limit, we define $Z^\prime \equiv \e^{2} Z$, $F^\prime \equiv \e F$, $\tilde F^\prime \equiv \e \tilde F$, so that
\beq
\begin{aligned}
\ud \psi &= \e \,  \big[ \ud \log(X_1+Y)  + \ud \log(X_2+Y)  \big]\,\psi \\
&\quad +  \big[ \ud \log(X_1-Y) - \ud \log(X_1+Y)  \big]  \hs F^\prime + \big[ \ud \log(X_2-Y) - \ud \log(X_2+Y) \big]\hs \tilde F^\prime  \,,
  \\[8pt]
\ud F^\prime &= \e\, \big[ \ud \log(X_1-Y)  + \ud \log(X_2+Y)\big]\hs F^\prime + \big[ \ud \log(X_1+X_2) -  \ud \log(X_2+Y) \big] \hs Z^\prime \,,\\[4pt]
\ud \tilde F^\prime &=  \e\, \big[ \ud \log(X_2-Y)  +  \ud \log(X_1+Y)  \big]\hs \tilde F^\prime+ \big[ \ud \log(X_1+X_2) -  \ud \log(X_2+Y) \big] \hs Z^\prime \,,
  \\[4pt]
\ud Z^\prime &= 2 \e\, \ud \log(X_1+X_2)\hs Z^\prime \,.
\end{aligned}
\eeq
Sending $\e \to 0$ then amounts to deleting the diagonal entries of the matrix, reproducing~\eqref{equ:dS-Eqns}.

\vskip 4pt
The above system of differential equations can be recast as a matrix differential equation, $\ud \vec{I} = \e \tilde A\, \vec{I}$, with $\tilde A$ taking the form
\beq
\tilde A = \left[\begin{array}{cccc}
\phantom{\hskip 3pt} \ell_1+ \ell_2\phantom{\hskip 3pt} & \phantom{\hskip 3pt} \ell_3- \ell_1 \phantom{\hskip 3pt}&  \phantom{\hskip 3pt}\ell_4- \ell_2\phantom{\hskip 3pt} & 0 \\[5pt]
 0 &  \ell_2+ \ell_3 & 0 &  \ell_5- \ell_2 \\[5pt]
 0 & 0 &  \ell_1+ \ell_4 & \phantom{\hskip 3pt} \ell_5- \ell_1 \phantom{\hskip 3pt}\\[5pt]
 0 & 0 & 0 & 2  \ell_5
\end{array}\right],
\eeq
where we have defined
\beq
\begin{aligned}\label{eq:dlogell}
	\ell_1 &\equiv \ud\log(X_1+Y)\,,\qquad \ell_3 \equiv \ud\log(X_1-Y)\,,\\
	\ell_2 &\equiv \ud\log(X_2+Y)\,,\qquad\ell_4 \equiv \ud\log(X_2-Y)\,, \qquad \ell_5 \equiv \ud\log(X_1+X_2)\,.
\end{aligned}
\eeq
It can be readily checked that the integrability condition is satisfied, $\tilde A\wedge\tilde A=0$,\footnote{In order to check this equation, we use $\ell_{1,2}\wedge \ell_{4,3}+\ell_{4,3}\wedge \ell_5+\ell_5\wedge \ell_{1,2}=0$, which follows from $(X_1\pm Y) + (X_2 \mp Y) = X_1+X_2$. This will be important later to show consistency of $\tilde A$ for more complicated diagrams.} while $\ud\tilde A=0$ is trivially satisfied because the matrix is given by a sum of exact forms with constant coefficients.
We also see that the matrix is upper triangular. The matrix differential equation is therefore amenable to an iterative solution, ascending through rows, each time solving a first-order equation, and then using the solution as a source in the subsequent row's equation. 
Interestingly, $\tilde A$ has ``folded singularities" as $X_{1,2} \to Y$, which are singularities that arise when two (or more) momenta become collinear. 
These singularities shouldn't exist in physical correlation functions (for the usual adiabatic vacuum initial conditions), and enforcing the absence of these singularities in the solutions is an important boundary condition (see below).

\paragraph{Explicit solution:} With the matrix $\tilde A$ in hand, let us now solve the system of differential equations. The easiest entries are $Z$ and the linear combination $\psi_D\equiv \psi-F-\tilde F+Z$:\footnote{This combination doesn't have a singularity in the total energy, hence the subscript $D$ for ``disconnected." Explicitly, the integral is
\beq
\psi_D=\int \ud x_1 \ud x_2\, (x_1 x_2)^\e \frac{(X_1+Y)(X_2+Y)}{x_1 x_2 (x_1+X_1+Y)(x_2+X_2+Y)} \, ,
\eeq
where the integrand is the canonical form associated to the square bounded by $\{T_1,T_2,B_1,B_2\} =0$.}
\begin{align}
\ud  Z &= 2 \e \,\ud \log(X_1+X_2) \, Z\,, \label{eq:Zeq}\\
\ud  \psi_D &= \e\, \Big[\ud \log(X_1+Y) + \ud \log(X_2+Y)\Big]\, \psi_D\,.
\label{eq:Deq}
\end{align}
It is easy show that the solutions to these equations are
\begin{align}
Z &= c_Z\, (X_1+X_2)^{2\e}\, ,
\label{equ:Z} \\
\psi_D &= c_D\, (X_1+Y)^\e (X_2+Y)^{\e}\, .
\end{align}
The normalizations of these solutions can be fixed by performing the integrals explicitly for specific kinematics (or by imposing appropriate boundary conditions on the full solution; see Appendix~\ref{app:functions}).
Evaluating $Z$ for $X_1=X_2=1/2$ and $\psi_D$ for $X_1=X_2= 0$, $Y=1$ gives 
\beq
\begin{aligned}
c_Z &=  4^{-\e} \sqrt{\pi}\csc(\pi \e)\, \Gamma(\e) \Gamma(\tfrac{1}{2}-\e)\, , \\
c_D &=  \pi^2 \csc^2(\pi \e)\,.
\end{aligned}
\label{equ:cZcD}
\eeq
Next, we consider the equation for $F$, which is not fully decoupled, but depends on our previous result for $Z$ as a source. Explicitly, we have
\beq
\begin{aligned}
\ud F = &\ \e\, \Big[\ud \log (X_1-Y) + \ud \log(X_2+Y) \Big] \,F +\e\, \Big[\ud \log (X_1+X_2) - \ud \log(X_2+Y) \Big]  \, Z\,.
\end{aligned}
\eeq
Substituting the solution~\eqref{equ:Z} for $F$, we find 
\beq
\begin{aligned}
\frac{\partial F}{\partial X_1} &=  \frac{\e}{X_1-Y} \, F + \e\, c_Z (X_1+X_2)^{2\e -1}\,, \\
\frac{\partial F}{\partial X_2} &= \frac{\e}{X_2+Y} \,  F - \e\, c_Z (X_1+X_2)^{2\e}\left(\frac{1}{X_2+Y} - \frac{1}{X_1+X_2}\right) .
\end{aligned}
\label{equ:XXX}
\eeq
These equations are solved by
\beq
F= c_F\,(X_1-Y)^\e (X_2+Y)^\e +\frac{c_Z}{2}\, \left(\frac{X_1-Y}{X_2+Y}\right)^\e(X_1+X_2)^{2\e} 
\,{}_2F_1\left[\begin{array}{c}
\e\,,\,2\e\\[-2pt]
1+2\e
\end{array}\Big\rvert \,\frac{X_1+X_2}{Y+X_2}\,\right] ,
\eeq
where ${}_2F_1$ is the Gauss hypergeometric function.
The normalization of the solution can again be fixed by performing the integral  for specific kinematics.
Evaluating $F$ for $X_1=X_2= 0$  yields
\beq
c_F = -\pi^2\csc(\pi\e)\csc(2\pi\e) \, .
\eeq
A similar analysis for $\tilde F$ leads to the same result with $X_1$ and $X_2$ interchanged. 
By shifting the indices of the hypergeometric function, we can absorb the terms proportional to $c_F$. 

\vskip4pt
Putting everything together, we finally obtain
\begin{eBox}
\vskip -10pt
\beq
\begin{aligned}
\psi = &\ c_D(\e)\, (X_1+Y)^\e (X_2+Y)^\e \\
&+ c_Z(\e)\, (X_1+X_2)^{2\e} \left(1 - \,{}_2F_1\left[\begin{array}{c}
1\,,\,\e\\[-2pt]
1-\e
\end{array}\Big\rvert \,\frac{Y-X_2}{Y+X_1}\,\right]
 -  \,{}_2F_1\left[\begin{array}{c}
1\,,\,\e\\[-2pt]
1-\e
\end{array}\Big\rvert \,\frac{Y-X_1}{Y+X_2}\,\right]  \right) ,
\end{aligned}
\label{equ:Two}
\eeq
\end{eBox}
where $c_Z$ and $c_D$ are given in (\ref{equ:cZcD}). Even though the original energy integral in \eqref{equ:FRW-2site} only converged for $\e>-1$, this solution to the differential equations contains the physical wavefunctions for any $\e$. The result~\eqref{equ:Two} is a meromorphic function of $\e$ with poles at integer and positive half-integer values of $\e$, as can be seen from the normalization factors in \eqref{equ:cZcD}. The physical wavefunctions for these (half-)integer values can then be extracted by taking the coefficient of the leading pole in the singular limits.

\newpage
\subsection{Cosmology-Shift Relations}
\label{sec:cosmoshift}

For the two-site process, we have the explicit solution (\ref{equ:Two}) for any value of $\e$, but we will not have this luxury for more complicated correlation functions. It turns out that knowing the wavefunction at a particular value of $\e$ gives us a way to obtain it for other cosmologies that differ in $\e$ by integer units. This is due to {\it shift relations} between the integrals. Similar shift relations are standard in scattering amplitudes, where a shift in $\e$ relates the amplitudes computed in different spacetime dimensions.

\vskip 4pt
The proof of the existence of shift relations is simple: a shift in $\e$ of all master integrals is just a change of basis in the same vector space. Hence, they must be related by a rotation matrix:
\beq
\vec{I}(\e-1)=M(Z,\e)\,\vec{I}(\e)\,,
\eeq
where $M$ satisfies $\ud M+\tilde A M=\e[\tilde A,M]$ and is invertible for $\e\ne \tfrac{1}{2},1$. 
If one insists on relating directly the wavefunctions, instead of the whole basis, that is also possible. We simply use the flat connection $A$ to write the other basis members in terms of derivatives of the wavefunction. In that case, the cosmology-shift relation is performed by a differential operator.\footnote{In fact, the origin of such cosmology-shifting differential operators is easy to see from the time integral representation. For cubic interactions, lowering $\e$ by one unit amounts to introducing an extra factor of $\eta_i$ for each vertex $i$. This is easily achieved with the differential operator $D_- = \prod_{i}\partial_{X_i}$, which acts on the product of the bulk-to-boundary propagators $\prod_{i} e^{iX_i\eta_i}$ and effectively shifts $\e$. Similar differential operators that shift the masses of scalar fields were studied in flat space~\cite{Benincasa:2019vqr} and de Sitter space~\cite{Arkani-Hamed:2018kmz, Baumann:2019oyu}.}

\vskip 4pt
The matrix $M$ can be found using partial fractions and integrations by part. We obtain
\begin{align}
&\psi_{\e-1}=\frac{1}{(X_1+Y)(X_2+Y)}\left(\psi_{\e}+\frac{2Y}{X_1-1}F_{\e}+\frac{2Y}{X_2-1}\tilde F_{\e}-\frac{2Y}{X_1+X_2}Z_{\e}\right) ,\nonumber \\
\begin{split}
F_{\e-1}&=\frac{1}{(X_1-Y)(X_2+Y)}\,F_{\e}+\left(\frac{1-2\e}{(1-\e)(X_1+X_2)^2}-\frac{1}{(X_2+Y)(X_1+X_2)}\right)Z_{\e}\, ,\\
\tilde F_{\e-1}&=\frac{1}{(X_1+Y)(X_2-Y)}\,\tilde F_{\e}+\left(\frac{1-2\e}{(1-\e)(X_1+X_2)^2}-\frac{1}{(X_1+Y)(X_1+X_2)}\right)Z_{\e}\, ,\\
\end{split}\\
&Z_{\e-1}=\frac{2(1-2\e)}{(1-\e)(X_1+X_2)^2}\,Z_{\e}\, . \nonumber
\end{align}
We see that the shift relation is algebraic, and hence care must be taken when using it at certain (half-)integer values of $\e$.
For example, in the de Sitter limit $\e\to 0$, only $\lim_{\e\to 0} Z_{\e} = 1/\e^2$ survives on the right-hand side. 
Keeping only $Z_\e$, we see that the shift relation gives the correct expression for the flat-space wavefunction $\psi_{-1}$ in terms of the basis functions for $\e=0$, despite the fact that the de Sitter wavefunction is a transcendental function. 
On the other hand, the relation is invalid for $\e=\frac{1}{2},1$ because the basis functions on the right-hand side are rational functions, while the wavefunctions for $\e=-\frac{1}{2},0$ are transcendental functions.

\subsection{Locality and Time Evolution}

The wavefunction $\psi$, and more generally the family of twisted integrals we have encountered, is rather special from a mathematical perspective. This is maybe not too surprising, because we expect that objects with a physical interpretation have a somewhat distinguished status in the space of all possible mathematical objects with similar properties. One obvious special feature of the integral family~(\ref{equ:family})  is that only two out of the five singular lines are twisted. This seemingly innocuous fact has a dramatic consequence: the physical wavefunction of interest satisfies a {\it local} differential equation. As we will see, this is a non-generic situation, and more generic twists do not lead to local differential equations. 

\paragraph{Local evolution:}
Since the integral basis is four dimensional, each member of the basis will satisfy at most a fourth-order ordinary homogeneous differential equation in each of $X_1$ and~$X_2$.  
A small surprise is that we instead get a third-order equation. 
Explicitly, the wavefunction satisfies 
\begin{align}
\Big[(X_1+X_2)(X_1^2-Y^2)\,\partial^3_{X_1} &- \big[(1-2\e)Y^2 -2 X_1 X_2 (2-\e)-X_1^2(5-4\e)\big]\,\partial^2_{X_1}\label{eq:odese} \\&+(1-\e)\big(X_1(4-5\e)+X_2(2-\e)\big)\,\partial_{X_1}-\e(1-\e)(1-2\e)\Big]\, \psi =0\,,\nonumber
\end{align}
and, by symmetry, an equivalent equation holds in $X_2$. 

\vskip 4pt
Note that the coefficients in~\eqref{eq:odese} depend on both $X_1$ and $X_2$, so that the equation looks ``non-local," in the sense that it seems to depend on the energies of both interaction vertices from the bulk perspective.  However, the equation can also be written as
\beq
\partial_{X_1}\Big[(X_1+X_2)^{1-2\e}\big((X_1^2-Y^2)\partial_{X_1}^2+2(1-\e)X_1\partial_{X_1}-\e(1-\e)\big)\Big]\,\psi=0 \,,
\label{eq:odefactor}
\eeq
which can then be integrated to give a {\it local}, inhomogeneous second-order equation\footnote{In the mathematical literature, equations like (\ref{eq:odefactor}) are also referred to as {\it Picard--Fuchs equations}; moreover, the differential operator annihilating the wavefunction function in (\ref{eq:odese}) is called the  {\it Picard--Fuchs operator}. There are interesting algorithms to derive the Picard--Fuchs equations directly without going through the procedure described in this paper~\cite{Muller-Stach:2012tgj,Agostini:2022cgv,Lairez:2022zkj}.} 
\begin{eBox}
\be
\Big[(X_1^2-Y^2)\partial^2_{X_1}+2(1-\e) X_1\partial_{X_1}-\e(1-\e)\Big] \,
\psi= g \left(\frac{1}{X_1+X_2}\right)^{1-2\e}\, ,
\label{equ:2ndOrder}
\ee
\end{eBox}
where $g$ is a constant of integration. 
A simple way to arrive at the same result is by solving (\ref{equ:2SiteEqns}) iteratively.
Integrating the equation for $Z$ is trivial and leads to $Z = c_Z (X_1+X_2)^{2\e}$. The remaining equations can then be written as
\begin{align}
	\label{equ:2siteX1s}
	\frac{X_1^2-Y^2}{2}\left(\partial_{X_1}-\frac{\e}{X_1+Y} \right)\psi &=\e  F\,,\\
		\left(\partial_{X_1}-\frac{\e}{X_1-Y} \right)F &= \e\, c_Z\,\left(\frac{1}{X_1+X_2}\right)^{1-2\e}\,,
\end{align}
which can easily be combined into the form (\ref{equ:2ndOrder}).

\vskip 4pt
Importantly, the differential operator on the left-hand side of  (\ref{equ:2ndOrder}) depends only on $X_1$ (and not on $X_2$).
This reflects the fact that the wavefunction comes from local evolution in the bulk spacetime. To see this, we start from the bulk time integral~(\ref{equ:BulkIntegral}):
 \begin{equation}
 \psi \propto \int \ud \eta \hskip 1pt \ud \eta'\, (\eta \eta')^{-(1+\e)}\, e^{i X_1\eta}  \,G(Y; \eta,\eta')\, e^{i X_2\eta'} \,,
\label{equ:BulkIntegral2}
\end{equation}
where the bulk-to-bulk propagator satisfies 
\be
\big(\partial_\eta^2 + Y^2\big) G(Y; \eta,\eta')  = -i \delta(\eta-\eta')\, .
\label{eq:greensfeq}
\ee
The fact that $G$ obeys this differential equation can be used to derive a differential equation for the function~$\psi$. We simply consider the auxiliary object 
\beq
F\equiv  \int \ud \eta \hskip 1pt \ud \eta'\, (\eta \eta')^{-(1+\e)}\, e^{iX_{1}\eta} e^{iX_2 \eta'} \,\eta^2 \big(\partial_\eta^2 + Y^2\big) G(Y;\eta,\eta')\,,
\label{eq:intermediatedsdiff}
\eeq
and evaluate it in two different ways. First, we use~\eqref{eq:greensfeq} to compute the integral directly. The delta function $\delta(\eta-\eta')$ reduces this to a one-dimensional time integral leading to the contact term on the right-hand side of \eqref{equ:2ndOrder}.
Second, we integrate the time derivatives in~\eqref{eq:intermediatedsdiff} by parts and trade factors of $\eta$ for $-i\partial_{X_{1}}$. Pulling the energy derivatives outside of the time integrals produces the left-hand side of \eqref{equ:2ndOrder}. This illustrates how the inhomogeneous second-order equation~\eqref{equ:2ndOrder} reflects local evolution in the bulk spacetime. In some sense, we can think of the existence of a local differential operator that simplifies the singularity structure of $\psi$ as a reflection of its origin from local spacetime physics.

\paragraph{When is it local?}
We found that the integral $\psi$ obeys two separate second-order inhomogeneous differential equations in each variable $X_1$ and $X_2$.  This is a simpler state of affairs than we could have hoped for, since a generic arrangement of five hyperplanes and twists yields a fourth-order homogeneous differential equation with coefficients depending on both $X_1$ and $X_2$.  Of course, the reason for this simplification is 
that our integral comes from a local spacetime process mediated by the exchange of a particle with a two-derivative equation of motion.  However, more abstractly, we can ask what mathematical property of the arrangement of hyperplanes and twists gives rise to a pair of local differential equations. In other words, if we were given a random set of exponents and hyperplanes, how could we determine whether the resulting integral describes local evolution in spacetime? This is a deep question, for which we do not have a complete answer, but in the following we explore it in the context of our specific problem.  

\vskip 4pt
Let us consider the most general twisting of the hyperplane arrangement depicted in Figure~\ref{fig:regions}, with five twist parameters $\e_i$. The relevant integral family then is
\be\label{eq:Igentwist}
I_N \equiv \int \ud x_1 \ud x_2\, \frac{1}{T_1^{m_1-\varepsilon_1} T_2^{m_2-\varepsilon_2} \colornucleus{Red}{B_1}^{n_1+\varepsilon_3} \colornucleus{Blue}{B_2}^{n_2+\varepsilon_4} \colornucleus{Green}{B_3}^{n_3+\varepsilon_5}}\,,
\ee
which again only has four basis elements. We take the basis to be defined by the same canonical forms as before.
The first-order differential equation obeyed by this basis then again takes the form $\ud \vec{I} = A\,\vec{I}$, with 
\beq
\hskip -7pt A = \left[\begin{array}{cccc}
 \e_1 \ell_1+ \e_2 \ell_2-\e_{345}^+\ell_6 &  \e_1(\ell_3- \ell_1) &  \e_2(\ell_4- \ell_2) & 0 \\[5pt]
 \e_3 (\ell_1-\ell_6) &  \e_2 \ell_2-\e_3\ell_1-\e_{451}^-\ell_3 & 0 &  \e_2(\ell_5- \ell_2) \\[5pt]
 \e_4(\ell_2-\ell_6) & 0 &  \e_1\ell_1 - \e_4\ell_2-\e_{352}^- \ell_4 &  \e_1(\ell_5- \ell_1) \\[5pt]
 0 & \e_4(\ell_2-\ell_3) & \e_3(\ell_1-\ell_4) & \e_{125}^- \ell_5-\e_3\ell_1-\e_4\ell_2
\end{array}\right]\!,
\eeq
where $\e_{ijk}^\pm \equiv  \e_i+\e_j\pm \e_k$ and $\ell_i$ denote the forms defined in \eqref{eq:dlogell}, with the new letter $\ell_6\equiv \ud \log Y$. 
Notice that it reduces to $\e \tilde A$ for $\e_1=\e_2=\e$ and $\e_3=\e_4=\e_5=0$. As expected, the matrix is linear in the twists, and the coefficients of each twist parameter are $\pm 1$.

\vskip4pt
First, we discuss the case with distinct twists $\e_1\ne \e_2$ on $x_1$ and $x_2$, and no twists on the other lines,
$\e_3=\e_4=\e_5=0$.
The most notable  features of the special case $\e_1 = \e_2 \equiv \e$ are preserved: the matrix $A$ is still triangular and leads to a third-order homogeneous differential equation (one order lower than expected in the generic case).  Furthermore, this homogeneous differential equation still implies a local second-order equation, which now takes the form
\beq
\Delta_{X_1}^{(\e_1)}\psi= g^2 \left(\frac{1}{X_1+X_2}\right)^{1-(\e_1+\e_2)} \,,
\eeq
where $\Delta_{X_1}^{(\e_1)} \equiv (X_1^2-Y^2)\partial_{X_1}^2+2(1-\e_1)X_1\partial_{X_1}-\e_1(1-\e_1)$ is the same operator as in~\eqref{equ:2ndOrder}. 
It is not surprising that $\e_1\ne \e_2$ still leads to a local equation, since we could implement this form of twisting by having two different cubic couplings with some relative time dependence in the bulk spacetime. Hence, there is still a natural bulk spacetime interpretation of these objects. 

\vskip 4pt 
Next, we explore under which conditions twisting of the lines $\{B_1,B_2,B_3\}$ leads to local equations. We first impose that each basis integral satisfies
a third-order homogeneous differential equation (and not a fourth-order equation).  We find that this requires $\e_3 = \e_4 = 0$; i.e.~any twisting of the lines associated to the partial energy singularities would lead to non-locality.
On the other hand, a twisting of the total energy line, $\e_5\ne 0$, is allowed and leads to 
\beq\label{eq:DeltaXe5}
\bigg(\Delta_{X_1}^{(\e_1)} + \e_5 \Big[ (Y+X_1) \partial_{X_1}  - \e_1 \Big]\bigg)\,\psi \ =\ g^2 \left(\frac{1}{X_1+X_2}\right)^{1-(\e_1+\e_2-\e_5)} \,.
\eeq
In this case, there is indeed a two-derivative operator collapsing the exchange diagram to a contact solution.  However, this operator is qualitatively different from the previous case. The single-derivative term, proportional to $\e_5$, has a different weight in $X_1$ and does not amount to a mere redefinition of the scalar coefficients appearing in the second-order equation~\eqref{equ:2ndOrder}.  
The spacetime interpretation is a term linear in $\eta$ in the equation of motion for the bulk-to-bulk propagator: 
\be
\left(\partial_\eta^2 +\frac{\e_5}{\eta}(iY - \partial_\eta) + Y^2 \right) G(Y; \eta,\eta')  = -i \delta(\eta-\eta')\, .\label{equ:prope5}
\ee
In flat space, this modified propagator gives the wavefunction
\begin{align}
\psi_{\rm flat} = \frac{1}{(X_1+Y)(X_2+Y)(X_1+X_2)^{1+\e_5}}\,,
\end{align}
consistent with \eqref{eq:Igentwist}. It can readily be checked that this satisfies \eqref{eq:DeltaXe5} for $\e_1=\e_2=-1$. However, it is evident that the term  linear in $Y$  in \eqref{equ:prope5} is spatially non-local (recall that $Y\equiv|\k_1+\k_2|$, so this is non-analytic in $\vec k_{1,2}$). Thus, only the twisting of the coordinate planes corresponds to the physical wavefunction arising from a local process in the bulk spacetime.

\newpage
\section{An Algorithm for All Trees}
\label{sec:MoreComplex}

In the previous section, we studied in detail the differential equations for the two-site chain, where the integrals were two-dimensional and their singularities could easily be visualized as lines in the plane.
We will also be interested, however, in more complicated graphs like  
\beq
 \raisebox{-8pt}{
 \begin{tikzpicture}[baseline=(current  bounding  box.center)]
 \coordinate (1) at (-1.4,0);
 \coordinate (2) at (0,0);
 \coordinate (3) at (1.4,0);
  \coordinate (1p) at (-1.4,-0.75);
 \coordinate (2p) at (0,-0.75);
 \coordinate (3p) at (1.4,-0.75);
  \coordinate (4p) at (2.8,-0.75);
  \draw[fill] (3) circle (.5mm);
   \draw[fill] (2) circle (.5mm);
  \draw[fill] (1) circle (.5mm);
      \draw[fill] (4p) circle (.5mm);
    \draw[fill] (3p) circle (.5mm);
   \draw[fill] (2p) circle (.5mm);
  \draw[fill] (1p) circle (.5mm);
  \draw[thick] (1) -- (2) -- (3);
    \draw[thick] (1p) -- (2p) -- (3p) -- (4p);
\end{tikzpicture}}
\hspace{1.5cm}
\begin{tikzpicture}[baseline=(current  bounding  box.center)]
 \coordinate (4) at (0,0);
 \coordinate (3) at (0,1.4);
  \coordinate (1) at (-1.1,-1.1);
 \coordinate (2) at (1.1,-1.1);
 \draw[fill] (4) circle (.5mm);
  \draw[fill] (3) circle (.5mm);
   \draw[fill] (2) circle (.5mm);
  \draw[fill] (1) circle (.5mm);
  \draw[thick] (4) -- (3);
    \draw[thick] (4) -- (2);
   \draw[thick] (4) -- (1); 
\end{tikzpicture}
\hspace{1.5cm}
\begin{tikzpicture}[baseline=(current  bounding  box.center)]
  \coordinate (1p) at (-1.4,-0.75);
 \coordinate (2p) at (0,-0.75);
 \coordinate (3p) at (1.4,-0.75);
  \coordinate (4p) at (2.8,-0.75);
   \coordinate (5) at (0,0.65);
 \coordinate (6) at (1.4,0.65);
       \draw[fill] (5) circle (.5mm);
    \draw[fill] (6) circle (.5mm);
      \draw[fill] (4p) circle (.5mm);
    \draw[fill] (3p) circle (.5mm);
   \draw[fill] (2p) circle (.5mm);
  \draw[fill] (1p) circle (.5mm);
    \draw[thick] (1p) -- (2p) -- (3p) -- (4p);
      \draw[thick] (2p) -- (5);
       \draw[thick] (3p) -- (6);
\end{tikzpicture}
\nonumber
\eeq
In those cases, the integrals are higher-dimensional and therefore much harder to visualize. 
To analyze such examples, we need a more algebraic approach.

\vskip 4pt
In this section, we will introduce a general algorithm to derive the differential equations for arbitrary tree graphs.  Although this approach will be completely systematic, 
for sufficiently complicated graphs it can become rather tedious and the results often aren't very illuminating. However, a closer inspection of the differential equations reveals something remarkable. As we will show in Section~\ref{sec:GraphicalRules}, the structure of the differential equations obeys very simple graphical rules, which will allow us to understand the differential equations for all tree graphs.  Readers who don't want to be
distracted by technical aspects are therefore invited to skip this section and look directly at the efficient graphical representation of the results in Section~\ref{sec:GraphicalRules}.

\vskip 4pt
The algorithm has two essential features. For an $n$-dimensional integral, we first enumerate a basis of integrals by considering all possible projective simplices involving $n$ planes, along with the plane at infinity. We then construct linear combinations 
 that are boundary-less (in an appropriately defined sense), so that they correspond to bounded regions. The canonical forms of these regions then provide a specific basis of master integrals. The derivatives of the canonical forms associated to the simplices, with respect to external data, can be written as a linear combination of the basis of simplices---cf.~\eqref{equ:Formula}.
  It is then a matter of algebra to take derivatives of a given integral and decompose the result into the chosen basis. In this section, we will describe the detailed implementation of this strategy for the two- and three-site graphs, but the approach is sufficiently algorithmic that it can easily be implemented for arbitrary graphs, and we discuss some features of these more complicated situations. Further examples are presented 
  in a {\sc Mathematica} notebook, which is available at a github repository for this paper~\href{https://github.com/haydenhylee/kinematic-flow}{\faGithub}. 

\subsection{A Simplex Basis}
\label{ssec:simplex}
We have seen that the FRW wavefunction associated to a general graph can be written as a twisted integral of a rational function. At tree level, the wavefunction arising from a graph with $n-1$ exchanges can be written as an $n$-dimensional integral, and the singularities of the integrand define a hyperplane arrangement in ${\mathbb R}^n$. Denoting these hyperplanes involved by $L_I \equiv \{T_a, B_a\}$, we can construct a convenient basis of integrals by considering the projective simplices involving all possible $n$-tuples of planes:
\beq
[L_1\cdots L_n] \equiv \ud \log L_1 \wedge \cdots \wedge \ud \log L_n\,.
\label{eq:simplexbracket}
\eeq
Note that this is an $n$-form in the space of integration variables, $x_a$, and so the exterior derivative in this equation is with respect to these coordinates.
This dlog form can be thought of as the canonical form of the (projective) simplex formed from the intersection of $n$ hyperplanes with the plane at infinity $L_\infty$. (For a brief review of the salient features of projective geometry, see Appendix~\ref{app:Maths}.) In general, there are far more possible $n$-tuples of intersecting planes than the dimension of the master integral basis. This is because a typical choice of planes will not bound a finite region. Instead, we must take linear combinations of these projective simplices to obtain bounded regions.\footnote{It may also happen that some of the planes in~\eqref{eq:simplexbracket} are parallel, in which case the bracket will simply vanish.}

\vskip4pt
To determine the combinations of brackets~\eqref{eq:simplexbracket} of interest, we introduce the {\it boundary operator} 
\beq
\partial[L_1 \cdots L_{n}] \equiv \sum_{J=1}^{n} (-1)^{J+1} [L_1 \cdots \widehat L_{J} \cdots L_{n}]\,,
\label{eq:simplexbdyoperator}
\eeq
where $~\,\widehat{}\,~$ indicates the omission of that hyperplane. We then
 search for linear combinations that have vanishing boundary. The vector space of cycles with respect to this operation is the space of integrands defining our master integrals.\footnote{In fact, given a hyperplane arrangement, this seems to be one of the more efficient ways to count the number of bounded regions.}

\paragraph{Two-site chain:} Let us illustrate this procedure for the case of the two-site chain. The relevant hyperplane arrangement was given in Figure~\ref{fig:regions} and consists of the five lines~\eqref{eq:2siteboundaries}: 
 \beq
\begin{aligned}
T_1 &\equiv x_1\,,\qquad B_1 \equiv  X_1 + x_1 + Y\,, \\
T_2 &\equiv x_2\,, \qquad B_2 \equiv  X_2 + x_2+ Y\,, \qquad B_3 \equiv  X_1+X_2 + x_1 +x_2\, .
\end{aligned}
\eeq
Since the arrangement is not generic, these lines intersect in only 8 places (as opposed to 10 for a generic configuration). We can write a general linear combination as
\be
\Omega = c_1 [B_1B_2]+ c_2 [B_1B_3]+ c_3 [B_1T_2]+ c_4 [B_2B_3]+ c_5 [B_2T_1]+ c_6 [B_3T_1]+ c_7 [B_3T_2]+ c_8 [T_1T_2]\,.
\ee
There is a four-dimensional space of combinations that are annihilated by $\partial$, and a convenient choice of basis is
\beq
\begin{aligned}
\Omega_\psi  &=  [B_1 B_2] + [B_2 B_3] + [B_3 B_1]\,,\\
\Omega_F &=  [T_1 B_2] + [B_2 B_3] + [B_3 T_1] = \Omega_\psi(B_1 \to T_1) \,,\\
\Omega_{\tilde F} &=  [B_1 T_2] + [T_2 B_3] + [B_3 B_1] = \Omega_\psi(B_2 \to T_2) \,,\\
\Omega_Z &= \,[T_1 T_2] + [T_2 B_3] \,+ [B_3 T_1] = \Omega_\psi(B_1 \to T_1, B_2 \to T_2) \,.
\end{aligned}
\label{eq:2sitebasissec4}
\eeq
It is straightforward to check that these forms are boundary-less. The simplest way is to note that the flat-space wavefunction can be written as the boundary of a higher-dimensional object
\be
\Omega_\psi =  \partial[B_1 B_2 B_3]\,,
\ee
with similar expressions for the other functions (and recall that $\partial^2 = 0$). Notice also that the wavefunction is the unique form that is both annihilated by $\partial$ and which has vanishing residues on $T_1=0$ and $T_2=0$.
This basis is particularly convenient because $\Omega_\psi$ is the form that appears in the integrand of the two-site wavefunction itself, and the other forms are related to $\Omega_\psi$ by simple replacements of the boundary lines $B_1,B_2$ with the twisted lines $T_1,T_2$.

\paragraph{Three-site chain:} As a more nontrivial example, we consider the integral associated to the three-site chain~\eqref{equ:ThreeSite}:
\beq
\psi_{(3)} =\int_0^\infty \ud x_1\, \ud x_2 \, \ud x_3\, (x_1x_2 x_3)^{\e} \frac{4 YY'}{B_1 B_2 B_3 B_4} \left(\frac{1}{B_5}+\frac{1}{B_6}\right) .
\label{eq:sec43site}
\eeq
Recall that this integral is singular on the following six (untwisted) hyperplanes
\beq
\begin{aligned}
B_1 &= X_1+x_1 + Y\,, & \qquad B_4 &=  X_1+X_2  + X_3+ x_1+x_2+x_3 \,,\\
B_2 &= X_2+ x_2 + Y+Y'\,,&B_5&=  X_1+X_2 + x_1+x_2 +Y'\, ,
 \\
B_3&= X_3 + x_3  +Y'\, , & B_6&=  X_2+X_3+ x_2+x_3 +Y\, ,
\end{aligned}
\eeq
along with the three twisted planes $T_{1,2,3} = x_{1,2,3}$.
The arrangement defined by these 9 planes has 25 bounded regions, but already in this example directly visualizing the arrangement to count is challenging.\footnote{Aside from the approach taken in this section, another way to determine the number of bounded regions is to consider these hyperplanes not in $\mathbb{R}^d$, but in some finite field $\mathbb{F}_p^d$, with $p$ being a prime number.  This is a discrete lattice of $p^d$ points.  We then count the number of points not on the hyperplanes.  This count $c(p)$ is a polynomial in $p$ of degree $d$ and $|c(1)|$ is the number of bounded regions.  To determine $c(p)$, one counts the number of points which do not vanish mod $p$ for $d+1$ different primes $p$ (for example, by evaluating the product of the $L_I$ on all $p^d$ points in $\mathbb{F}_p^d$).  This fixes the coefficients and we can then evaluate $c(1)=25$.  The intersecting surfaces therefore bound 25 regions. See~Appendix~\ref{app:Maths} for more details.} 

\vskip4pt
There are naively $9$ choose $3$ $= 84$ intersections of planes that can be used to form the brackets~\eqref{eq:simplexbracket}. However, the hyperplane arrangement in this example is highly non-generic, because there are linear relations between the functions $B_a$. This means that certain planes don't intersect.
(For example, the linear relation $B_1 + B_6 = B_4 + 2Y$ implies that $[B_1B_4B_6] = 0$.  Similar relations hold for the set of planes $B_3,B_4,B_5$, along with $B_1,B_2,B_5$ and $B_2,B_3,B_6$.) Additionally, there is an interesting four-term relation $B_2 + B_4 = B_5 + B_6$. Taking all of these relations into account, there are $50$ nonzero brackets and a 
 general linear combination of them is
\beq
\Omega = \sum c_{IJK} [IJK]\,,
\label{equ:Omega-Ansatz}
\eeq
where $I,J,K$ run over all the planes.
We find a $25$-dimensional space of forms annihilated by the boundary operator~\eqref{eq:simplexbdyoperator}. The integrand of the wavefunction appearing in~\eqref{eq:sec43site} is the specific linear combination
\beq
\begin{aligned}
\Omega_\psi = \ & [B_1 B_2 B_3] - [B_1 B_2 B_6] - [B_1 B_3 B_4] + [B_1 B_3 B_5] + [B_1 B_3 B_6]  \\
& - [B_1 B_4 B_5] - [B_2 B_3 B_5] +[B_2 B_4 B_5] - [B_2 B_4 B_6] + [B_3 B_4 B_6] \, .
\end{aligned}
\label{eq:3siteWFsimplex}
\eeq
This form is the unique boundary-less object that does not involve any of the twisted planes.
We will not choose a specific basis for the other 24 integrals at this stage, since we will shortly be working with a smaller basis tailored to our problem. 
An interesting feature of the form~\eqref{eq:3siteWFsimplex} is that it is completely determined by its pattern of vanishing residues, much like the wavefunction integrand for the two-site graph.\footnote{This pattern appears to end here---the four-site chain wavefunction is not uniquely specified by demanding a certain set of zeroes.}

\vskip4pt
It is straightforward to consider more complicated examples following the same procedure. Given the possible singularities of the integrand of interest, we can consider the arrangement cut out by these hyperplanes and search for combinations of intersections of the planes that are annihilated by~\eqref{eq:simplexbdyoperator}. This then serves as a basis for all the possible integrands that can appear when we differentiate the wavefunction. We now consider this second step.

\subsection{Differential Equations}
\label{ssec:diffeqssec4}
When we differentiate the wavefunction, we necessarily generate new functions whose integrands involve the twisted planes. These functions can always be decomposed into the basis obtained in the previous step, so the challenge is essentially to figure out what linear combinations of functions appear. This is made quite simple by considering the differential of the individual terms in~\eqref{eq:simplexbracket}. As proven in the box below, the following identity holds inside of a twisted integral 
\begin{eBox}
\beq
\ud [L_1 \cdots L_n] \,= \,\e\, \sum_a \ud \log \langle \hat T_a\hat L_1 \cdots \hat L_n  \rangle\times \partial[T_aL_1 \cdots L_n ]\, ,
\label{equ:Formula}
\eeq
\end{eBox}
where the total differential $\ud$ is with respect to the kinematic variables $Z_I \equiv\{X_a,Y_{ab}\}$, and we have defined $L_a \equiv \hat L_a \cdot P$, where $P^I \equiv (1,x_1,\cdots, x_n)$.
Note that $\hat L_{a,I}$ 
is the expression for a given hyperplane in projective coordinates. The boundary operation is defined as in~\eqref{eq:simplexbdyoperator}, while the angle bracket is the determinant
\be
\langle \hat L_1 \cdots \hat L_{n+1} \rangle \equiv \det(\hat L_1, \cdots, \hat L_{n+1})\,.
\ee
With this formula, we can easily convert derivatives with respect to external kinematics into linear combinations of simplex forms, but now involving the twisted planes. 

\vskip 4pt
We emphasize that the simple formula \eqref{equ:Formula} is based on the physical properties of the wavefunction and is therefore not applicable to generic simplices.
Recall that the hyperplane arrangements associated with the FRW wavefunction are highly non-generic due to the linear relations they satisfy. 
Notably, these hyperplanes have the property that each coordinate $x_a$ always comes in pairs with an external variable $X_a$, allowing us to interchange~$\partial_{X_a}$ with~$\partial_{x_a}$. 
It is this special characteristic that enables us to easily differentiate projective simplices in the integrands with respect to the external kinematics, and subsequently elevate these to the differentials of the {\it integrated} functions.

\vskip2pt
\begin{eBox}
\noindent
{\small {\bf Proof:} The following is a proof of~\eqref{equ:Formula}. \\[4pt]
\noindent
Recall that $[L_1\cdots L_n]$ is a projective simplex form involving 
the plane at infinity $L_\infty= \hat L_\infty\cdot P$, with $\hat L_{\infty,I}=(1,0,\cdots,0)$. 
It will be useful restore the $L_\infty$ dependence and express \eqref{eq:simplexbracket} in a gauge-invariant way as \cite{Arkani-Hamed:2017tmz}
\beq
\begin{aligned}
	[L_1\cdots L_n] 
	&= \partial[L_\infty L_1 \cdots L_n]= \frac{ \langle \hat L_\infty\hat L_1\cdots \hat L_n \rangle \langle P\, \ud^n P\rangle}{(\hat L_\infty\cdot P)(\hat L_1\cdot P)\cdots (\hat L_n\cdot P)}\,,
\end{aligned}
\eeq
where $\langle P\, \ud^n P\rangle$ is the standard measure on projective space
\beq
	\langle P\, \ud^n P\rangle 
	= \sum_{I=0}^n\, (-1)^I P^{I}\, \ud P^{0} \wedge\cdots \wedge \widehat{\ud P^I} \wedge\cdots \wedge \ud P^{n}\, .
\eeq
For our boundary planes, we have that $\langle \hat L_\infty\hat L_1\cdots \hat L_n \rangle \equiv \eta $ is an integer. Similarly, the boundary of the simplex form involving the twisted plane $T_a=\hat T_a\cdot P =x_a$ in \eqref{equ:Formula} can be expressed as
\begin{align}
	\partial[T_a L_1 \cdots L_n ] 
	&=\frac{\langle \hat T_a \hat L_1\cdots \hat L_n  \rangle \langle P \, \ud^n P\rangle}{(\hat T_a\cdot P)(\hat L_1\cdot P)\cdots (\hat L_n\cdot P)}\,.
\end{align}
Now, consider the twisted integral of the simplex form
\begin{align}
	I = \int (x_1\cdots x_n)^\e\, [L_1\cdots L_n] = \int (x_1\cdots x_n)^\e\, \frac{\eta \langle P\, \ud^n P\rangle}{(\hat L_1\cdot P)\cdots (\hat L_n\cdot P)}\,.
\end{align}
To facilitate integration by parts on the right-hand side of~\eqref{equ:Formula}, we would like to express kinematic derivatives in terms of derivatives of the integration variables. The relation $\partial_{X_a}[L_1\cdots L_n] = \partial_{x_a}[L_1\cdots L_n]$ only holds when $T_a$ is not a member of $\{L_i\}$; i.e.~when the twisted line $T_a$ is not a line in the simplex form. 
The trick is to express all kinematic derivatives $\partial_{Z_I}$ on the simplex form in terms of only $\partial_{X_a}$, for which $\partial_{X_a} = \partial_{x_a}$ holds, i.e.~those $a$ for which $T_a \notin \{ L_i\}$.  

\vskip 3pt
It is a straightforward linear algebra exercise to deduce the following determinant expression relating derivatives acting on $[L_1\cdots L_n]$:
\begin{align}
	\partial_{Z_I}[L_1\cdots L_n] = \sum\limits_a \frac{\partial_{Z_I}\langle\hat T_a \hat L_1\cdots\hat L_n\rangle}{\partial_{X_a}\langle\hat T_a \hat L_1\cdots\hat L_n\rangle}\partial_{X_a}[L_1\cdots L_n]\,,
\end{align}
where it is understood that for $a$ such that $T_a$ is already a member of the $\{L_i\}$, the determinants vanish and the term doesn't contribute.  
This guarantees that any kinematic derivative is expressed in terms of only derivatives associated to $X_a$ without any twisted lines in the simplex form.  
Using  that $\partial_{X_a}\langle\hat T_a \hat L_1\cdots\hat L_n\rangle = -\eta$,
for $T_a \notin \{ L_i\}$, we find the following expression for the total differential:
\begin{align}
	\ud [L_1\cdots L_n] = -\frac{1}{\eta} \sum_a \ud \langle\hat T_a \hat L_1\cdots\hat L_n\rangle \partial_{X_a} [L_1\cdots L_n]\,.
\end{align}
We are now free to use the fact that $\partial_{X_a}=\partial_{x_a}$ when acting on the untwisted simplex form, and integration by parts then gives 
\beq
\begin{aligned}
	\ud I &= - \int (x_1\cdots x_n)^\e \sum_a \ud \langle\hat T_a \hat L_1\cdots\hat L_n\rangle \partial_{x_a} \frac{\langle P\, \ud^n P\rangle}{(\hat L_1\cdot P)\cdots (\hat L_n\cdot P)}\\
	&= \e   \int (x_1\cdots x_n)^\e \sum_a \frac{\ud \langle\hat T_a \hat L_1\cdots\hat L_n\rangle}{\langle \hat T_a \hat L_1\cdots \hat L_n  \rangle} \frac{\langle \hat T_a \hat L_1\cdots \hat L_n  \rangle \langle P\, \ud^n P\rangle}{(\hat T_a\cdot P)(\hat L_1\cdot P)\cdots (\hat L_n\cdot P)}\\
	&= \e \int (x_1\cdots x_n)^\e \sum_a \ud \log \langle \hat T_a\hat L_1 \cdots \hat L_n  \rangle\times\partial[T_a L_1 \cdots L_n ]\,,
\end{aligned}
\eeq
where we have multiplied and divided by $\langle \hat T_a \hat L_1\cdots \hat L_n  \rangle$ in the second line. This proves the desired formula \eqref{equ:Formula}.}
\end{eBox}

\vskip 6pt
We can use~\eqref{equ:Formula} to differentiate each of the terms appearing in the wavefunction. After that, we just have to decompose the result back into whatever basis we choose for the integral family. This linear algebra problem can be solved in many ways, but it is often most convenient to choose a basis where as few elements as possible share singularities and then match residues on these singularities. We now show how this works in some cases.

\subsection{Selected Examples}
\label{ssec:sec4-examples}
Deriving the differential equation for the wavefunction $\psi$ is now completely algorithmic. We first enumerate a basis of canonical forms written in terms of simplices, and then use~\eqref{equ:Formula} to differentiate each dlog form that appears in the wavefunction. We then introduce new functions as necessary to write the result back in terms of the original wavefunction, plus new pieces. We then differentiate these new pieces and keep going until the system closes. Remarkably, proceeding in this way typically requires fewer functions than occur for totally generic twists.

\vskip 6pt
\paragraph{Two-site chain:} We first revisit the example of the two-site chain.
In the simplex basis, the canonical form associated to the wavefunction was given in \eqref{eq:2sitebasissec4}:
 \beq
 \Omega_\psi=  [B_1 B_2] + [B_2 B_3] + [B_3 B_1]\, .
 \eeq
Using~\eqref{equ:Formula}, we have 
\beq
\begin{aligned}
\ud [B_1 B_2] &= \e\Big(\ud \log(X_1+Y)\, \partial [T_1 B_1 B_2 ]  + \ud \log(X_2+Y)\, \partial [T_2 B_1 B_2]\Big) \,, \\
\ud [B_2 B_3] &= \e\Big(\ud \log(X_1-Y) \, \partial [T_1 B_2 B_3] + \ud \log (X_2+Y) \, \partial [T_2 B_2 B_3] \Big)\,,\\
 \ud [B_3 B_1] &= \e\Big(\ud \log(X_1+Y) \, \partial [T_1 B_3 B_1] + \ud \log (X_2 -Y) \, \partial [T_2 B_3 B_1]\Big) \,.
 \end{aligned}
 \eeq
This means that we can write the total differential of the wavefunction integrand as
\beq
\begin{aligned}
\ud \Omega_\psi = \e\, \Big( &\, \ud \log(X_1+Y) \,\partial \big([T_1 B_1 B_2] +  [T_1 B_3 B_1]\big) + \ud \log(X_1-Y) \,\partial [T_1 B_2 B_3]  \\
+  &\ \ud \log(X_2+Y) \,\partial \big([T_2 B_1 B_2] +  [T_2 B_2 B_3]\big) + \ud \log(X_2-Y) \,\partial [T_2 B_3 B_1] \Big)\, . 
\end{aligned}
\label{equ:dXOmega}
\eeq
It is convenient to identify the coefficients of the dlog forms involving the letters $X_1-Y$ and $X_2-Y$ as the integrands of new functions, which we call $\Omega_F$ and $\Omega_{\tilde F}$, respectively.
Using $\partial [a b c] = [ab] + [bc] + [ca]$, we find 
\beq
\begin{aligned}
\Omega_F &=  [T_1 B_2] + [B_2 B_3] + [B_3 T_1] = \Omega_\psi(B_1 \to T_1) \,,\\
\Omega_{\tilde F} &=  [B_1 T_2] + [T_2 B_3] + [B_3 B_1] = \Omega_\psi(B_2 \to T_2) \,.
\end{aligned}
\eeq
We see that the new forms $\Omega_F$ and $\Omega_{\tilde F}$ are the ones that we chose as part of our basis in~\eqref{eq:2sitebasissec4} with some foresight. They can be obtained from the original form $\Omega_\psi$ by replacing boundary lines $B_a$ with twisted lines $T_a$.  
With these definitions, the result in~\eqref{equ:dXOmega} can then be written as 
\beq
\begin{aligned}
\ud \Omega_\psi = \e\,\Big( &\, (\Omega_\psi- \Omega_F) \,\ud \log(X_1+Y) + \Omega_F \,\ud \log(X_1-Y)  \\
+ &\ (\Omega_\psi- \Omega_{\tilde F}) \,\ud \log(X_2+Y) + \Omega_{\tilde F} \,\ud \log(X_2-Y) \Big)\,,
\end{aligned}
\label{equ:OmegaPsi}
\eeq
where  $\Omega_F$ and $\Omega_{\tilde F}$ appear as sources. 

\vskip4pt
We now repeat the above procedure to obtain the derivatives of these functions.
To evaluate $\ud \Omega_F$, we again use~\eqref{equ:Formula} and write the total differentials of the simplex basis forms as
\beq
\begin{aligned}
\ud [T_1 B_2] &=  \e\, \ud \log(X_2+Y)\, \partial [T_2 T_1 B_2] \,, \\
\ud [B_2 B_3] &= \e \Big(\ud \log(X_1-Y) \, \partial [T_1 B_2 B_3] + \ud \log (X_2+Y) \, \partial [T_2 B_2 B_3] \Big)\,,\\
 \ud [B_3 T_1] &= \e\,  \ud \log (X_1+X_2) \, \partial [T_2 B_3 T_1] \,,
 \end{aligned}
 \eeq
so that the differential of $\Omega_F$ is
\beq
\begin{aligned}
\ud \Omega_F =  \e\, \Big(&\, \ud \log(X_1-Y) \, \partial [T_1 B_2 B_3]  + \ud \log(X_2+Y) \,\partial \big([T_2 T_1 B_2] +  [T_2 B_2 B_3]\big)  \\
+ &\ \ud \log (X_1+X_2) \, \partial [T_2 B_3 T_1] \Big)\,.
\end{aligned}
\label{equ:dXOmegaF}
\eeq
As before, it makes sense to call the coefficient of $ \ud \log (X_1+X_2)$ a new function, as this is the first appearance of this letter. We call this form $\Omega_Z$, and it is easy to see that it is related to $\Omega_F$ by the replacement $B_2 \to T_2$.
With this definition, the result~\eqref{equ:dXOmegaF} becomes
\beq
\ud \Omega_F = \e\, \Big(\, \Omega_F\, \ud \log(X_1-Y) + (\Omega_F  - \Omega_Z)\, \ud \log(X_2+Y) + \Omega_Z\, \ud \log(X_1 +X_2)\Big)\,.
\label{equ:dXOmegaF2}
\eeq
By symmetry, a similar result holds for $\tilde F$, with $X_1 \leftrightarrow X_2$ and the same function $\Omega_Z$ (which can also be obtained from $\Omega_{\tilde F}$ by replacing $B_1\to T_1$).

\vskip 4pt
Lastly, we evaluate $\ud \Omega_Z$.  Using~\eqref{equ:Formula} again, we have the differentials
\beq
\begin{aligned}
\ud [T_1 T_2] &= 0 \,, \\
\ud [T_2 B_3] &= \e \, \ud \log(X_1+X_2) \, \partial [T_1 T_2 B_3]  \,,\\
 \ud [B_3 T_1] &= \e \, \ud \log (X_1+X_2) \, \partial [T_2 B_3 T_1] \,.
 \end{aligned}
 \eeq
In this case, no new functions appear. Instead, we just get back $\Omega_Z$ and the system closes:
\beq
\ud \Omega_Z = 2\e\, \Omega_{Z}\, \ud \log(X_1+X_2) \, .
\label{equ:OmegaZ}
\eeq
The equations for the actual functions (as opposed to the integrands) are obtained by simply integrating~\eqref{equ:OmegaPsi},~\eqref{equ:dXOmegaF2}, and~\eqref{equ:OmegaZ} against the twist factor. This gives
\beq
\begin{aligned}\label{eq:twositefull}
\ud \psi &= \e \, \big[(\psi - F)\, \ud \log(X_1+Y) + F\, \ud \log(X_1-Y) \\
&\quad \ +  (\psi - \tilde F)\, \ud \log(X_2+Y) + \tilde F\,\ud \log(X_2-Y)  \big] \,,
  \\[8pt]
\ud F &= \e\, \big[ F\,\ud \log(X_1-Y)  + (F-  Z)\, \ud \log(X_2+Y) + Z\, \ud \log(X_1+X_2) \big] \,,\\[4pt]
\ud \tilde F &=  \e\, \big[ \tilde F\,\ud \log(X_2-Y)  + (\tilde F-  Z)\, \ud \log(X_1+Y) + Z\, \ud \log(X_1+X_2) \big]\,,
  \\[8pt]
\ud Z &= 2 \e\, Z\, \ud \log(X_1+X_2)\,,
\end{aligned}
\eeq
which is precisely the same as~\eqref{equ:2SiteEqns}. 

\vskip 4pt
The advantage of this approach is that it is completely algorithmic, and does not require us to understand the geometric details of the hyperplane arrangement. Moreover, we are led to a natural basis of functions: as each new letter appears when we take derivatives, we call its coefficient a new function and continue on. This systematic procedure readily generalizes to more complicated examples, as we will now see.

\vskip 6pt
\paragraph{Three-site chain:} 
Above we found that the form associated to the three-site chain is
\beq
\begin{aligned}
\Omega_\psi = \ & [B_1 B_2 B_3] - [B_1 B_2 B_6] - [B_1 B_3 B_4] + [B_1 B_3 B_5] + [B_1 B_3 B_6]  \\
& - [B_1 B_4 B_5] - [B_2 B_3 B_5] +[B_2 B_4 B_5] - [B_2 B_4 B_6] + [B_3 B_4 B_6] \, .
\end{aligned}
\label{equ:WF-simplex}
\eeq
The procedure is now the same as it was for the two-site chain.
To evaluate $\ud \Omega_\psi$, 
we apply (\ref{equ:Formula}) to each term $[IJK]$ on the right-hand side of (\ref{equ:WF-simplex}) and identify the coefficients multiplying the resulting ``letters" as new functions.
We then act with $\ud$ on each of these functions and repeat the procedure until the system closes.  The method is completely systematic and is implemented in a {\sc Mathematica} notebook available at \href{https://github.com/haydenhylee/kinematic-flow}{\faGithub}. 

\vskip 4pt
To illustrate this, let us first introduce the shorthand notations $[I\cdots \tilde J\cdots]\equiv [B_I\cdots T_J\cdots]$ with $\ \tilde{}\ $ denoting the twisted planes, and
\beq
\begin{aligned}\label{threesiteX}
	X_1^\pm &\equiv X_1 \pm Y\,,	&&\quad \ X_{12}^\pm \equiv X_1+X_2\pm Y'\,,\\
	X_2^{\pm \pm} &\equiv X_2 \pm Y \pm Y'\,,	&&\quad \ X_{23}^\pm \equiv X_2+X_3\pm Y\,, \\
	X_3^\pm &\equiv X_3 \pm Y'\,,&&\quad X_{123} \equiv X_1+X_2+X_3\,,
\end{aligned}
\eeq
for the letters. 
Taking the total differential of $\Omega_\psi$ using \eqref{equ:Formula}, we get 
\beq
\begin{aligned}
	\ud \Omega_\psi =  \e\, \Big(&a_1^+\, \ud\log X_1^+ + a_1^-\,\ud\log X_1^- +a_3^+\, \ud\log X_3^+ + a_3^{-}\, \ud\log X_3^-\\
	&\quad +a_2^{-+}\, \ud\log X_2^{-+}+ a_2^{--}\, \ud\log X_2^{--}+ a_2^{+-} \,\ud\log X_2^{+-}+ a_2^{++}\, \ud\log X_2^{++}\Big)\,,
\end{aligned}
\label{equ:dPsi3}
\eeq
where the coefficients of the dlog forms are given by the following simplex forms
\beq
\begin{aligned}
	a_1^- &\equiv \partial(-[ \tilde{1}235] +[ \tilde{1}245] -[ \tilde{1}246] +[ \tilde{1}346]) \,,\\
	a_1^+ &\equiv \partial([ \tilde{1}123] -[ \tilde{1}126] -[ \tilde{1}134] +[ \tilde{1}135]
   +[ \tilde{1}136] -[ \tilde{1}145])\,,\\
	a_2^{-+} &\equiv \partial([ \tilde{2}135] -[ \tilde{2}145])\,,\\
	a_2^{--} &\equiv -\partial[ \tilde{2}134]\,,\\
	a_2^{+-} &\equiv \partial([ \tilde{2}136] +[ \tilde{2}346])\,,\\
	a_2^{++} &\equiv \partial([ \tilde{2}123] -[ \tilde{2}126] -[ \tilde{2}235] +[ \tilde{2}245]
   -[ \tilde{2}246])\,,\\	
	a_3^-&\equiv \partial(-[ \tilde{3}126] -[ \tilde{3}145] +[ \tilde{3}245] -[ \tilde{3}246])\,,\\
	a_3^+&\equiv \partial([ \tilde{3}123] -[ \tilde{3}134] +[ \tilde{3}135] +[ \tilde{3}136]
   -[ \tilde{3}235] +[ \tilde{3}346])\,.
\end{aligned}
\label{eq:diffpsi}
\eeq
We now want to split these up in a natural way. First notice that some new letters (which have relative minus signs between energies) appear in~\eqref{eq:diffpsi} which are not associated to lines appearing in $\Omega_\psi$. These letters are $X_1^-$, $X_2^{-+}$, $X_2^{--}$, $X_2^{+-}$, and $X_3^+$.
 As in the two-site case, it is convenient to identify the coefficients of these letters as new sources: 
\beq
\begin{aligned}
a_1^- &\equiv \Omega_F\,,\\
a_2^{-+} &\equiv \Omega_{Q_1}\,,\\
a_2^{--} &\equiv \Omega_{Q_2}\,,\\
a_2^{+-} &\equiv \Omega_{Q_3}\,,\\
a_3^- &\equiv \Omega_{\tilde F}\,.
\end{aligned}
\eeq
Explicit formulas for these forms can be obtained by taking the boundary of the simplex forms appearing in~\eqref{eq:diffpsi}
\beq
\begin{aligned}
	\Omega_F & = -[235]+[23\tilde 1]+[245]-[246]-[26\tilde 1]+[346]-[34\tilde 1]+[35\tilde 1]+[36\tilde 1]-[45\tilde 1]\,,\\
	\Omega_{Q_1} & = [135]-[13\tilde 2]-[145]+[14\tilde 2]-[35\tilde 2]+[45\tilde 2]\,,\\
	\Omega_{Q_2} & = -[134]+[13\tilde 2]-[14\tilde 2]+[34\tilde 2]\,,\\
	\Omega_{Q_3} & = [136]-[13\tilde 2]+[16\tilde 2]+[346]-[34\tilde 2]-[46\tilde 2]\,,\\
	\Omega_{\tilde F} & = -[126]+[12\tilde 3]-[145]+[14\tilde 3]-[15\tilde 3]-[16\tilde 3]+[245]-[246]+[25\tilde 3]+[46\tilde 3]\,,
\end{aligned}
\label{eq:sourcedefs}
\eeq
which allows us to write the remaining coefficient functions in terms of these functions and $\Omega_\psi$:
\beq
\begin{aligned}
a_1^+ &\equiv \Omega_\psi - \Omega_F\,,\\
a_2^{++} &\equiv \Omega_\psi - (\Omega_{Q_1}+ \Omega_{Q_2}+\Omega_{Q_3})\,,\\
a_3^+ &\equiv \Omega_\psi - \Omega_{\tilde F}\,.
\end{aligned}
\eeq
Interestingly, the sources~\eqref{eq:sourcedefs} can be obtained directly from $\Omega_\psi$ through various replacements of the boundary planes with the twisted ones as
\beq
\begin{aligned}\label{equ:replace1}
	\Omega_F & =\Omega_\psi (1\to \tilde 1)\,,\\
	\Omega_{Q_1} & =\Omega_\psi (2,6\to \tilde 2)-\Omega_\psi (2,5,6\to \tilde 2)\,,\\
	\Omega_{Q_2} & = \Omega_\psi (2,5,6\to \tilde 2)\,,\\
	\Omega_{Q_3} & = \Omega_\psi (2,5\to \tilde 2)-\Omega_\psi (2,5,6\to \tilde 2)\,,\\
	\Omega_{\tilde F} & = \Omega_\psi (3\to \tilde 3)\,.
\end{aligned}
\eeq
In the two-site example, we saw the same relation between $\Omega_\psi$ and the sources $\Omega_F$ and $\Omega_{\tilde F}$ associated to the end vertices. 
A new feature in the three-site case is that the letters associated to the middle vertex $X_2$, connected by two edges, can have multiple sources depending on the relative signs of the internal energies $Y$ and $Y'$.

\vskip 4pt
We then proceed in the same way and take the differential of the source functions. First, consider $\Omega_F$ and $\Omega_{\tilde F}$. Taking their differentials gives
\beq
\begin{aligned}\label{equ:dXF3site}
	\ud \Omega_F = \e\,\Big(& \Omega_F \, \ud\log X_1^-  + \Omega_f\,\ud\log X_3^- +(\Omega_F-\Omega_f) \,\ud \log X_3^+ +\Omega_{q_1}\,\ud\log X_{12}^{+} \\
	&\quad + \Omega_{q_2}\,\ud\log X_{12}^- +\Omega_{q_3}\,\ud\log X_{2}^{+-} + (\Omega_F-\Omega_{q_1}-\Omega_{q_2}-\Omega_{q_3})\,\ud\log X_2^{++} \Big)\,,\\[4pt]
	\ud \Omega_{\tilde F} = \e\,\Big( &\Omega_{\tilde F} \, \ud\log X_3^-  + \Omega_f\,\ud\log X_1^- +(\Omega_{\tilde F}-\Omega_f)\, \ud \log X_1^+ + \Omega_{\tilde q_1}\,\ud\log X_{23}^{+}\\
	&\quad  + \Omega_{\tilde q_2}\,\ud\log X_{12}^- +\Omega_{\tilde q_3}\,\ud\log X_2^{-+} + (\Omega_{\tilde F}-\Omega_{\tilde q_1}-\Omega_{\tilde q_2}-\Omega_{\tilde q_3})\,\ud\log X_2^{++} \Big)\,, 
\end{aligned}
\eeq
where we have expressed the coefficients of the dlog forms in terms of the newly defined sources
\beq
\begin{aligned}
	\Omega_{f} &\equiv [245]-[246]+[25\tilde 3]-[26\tilde 1]-[2\tilde 1\tilde 3]-[45\tilde 1]+[46\tilde 3]-[4\tilde 1\tilde 3]+[5\tilde 1\tilde 3]+[6\tilde 1\tilde 3] \,,\\
	\Omega_{q_1} &\equiv [35\tilde 1]-[35\tilde 2]+[3\tilde 1\tilde 2]-[45\tilde 1]+[45\tilde 2]-[4\tilde 1\tilde 2] \,,\\
	\Omega_{q_2} &\equiv  -[34\tilde 1]+[34\tilde 2]-[3\tilde 1\tilde 2]+[4\tilde 1\tilde 2] \,,\\
	\Omega_{q_3} &\equiv  [346]-[34\tilde 2]+[36\tilde 1]+[3\tilde 1\tilde 2]-[46\tilde 2]-[6\tilde 1\tilde 2]\,,\\
	\end{aligned}
\eeq	
along with
\beq
\begin{aligned}
	\Omega_{\tilde q_1} &\equiv  -[145]+[14\tilde 2]-[15\tilde 3]+[1\tilde 2\tilde 3]+[45\tilde 2]-[5\tilde 2\tilde 3]\,,\\
	\Omega_{\tilde q_2}&\equiv  -[14\tilde 2]+[14\tilde 3]-[1\tilde 2\tilde 3]+[4\tilde 2\tilde 3]\,,\\
	\Omega_{\tilde q_3}&\equiv  [16\tilde 2]-[16\tilde 3]+[1\tilde 2\tilde 3]-[46\tilde 2]+[46\tilde 3]-[4\tilde 2\tilde 3]\,.
\end{aligned}
\eeq	
The equations \eqref{equ:dXF3site} involve four new letters, $X_{12}^+$, $X_{12}^-$, $X_{23}^+$, and $X_{23}^-$, and we have identified their coefficients as new sources, which we call $\Omega_{q_1}$, $\Omega_{q_2}$, $\Omega_{\tilde q_1}$, and $\Omega_{\tilde q_2}$, respectively. 
While the letters $X_1^-$, $X_3^-$, $X_2^{-+}$, $X_2^{+-}$ are not new, some of their dlog form coefficients involve two twisted planes and therefore cannot be represented in terms of the previous sources. Instead, we have assigned new sources $\Omega_{q_3}$, $\Omega_{\tilde q_3}$, and $f$ to these coefficients. 
These new sources can also be obtained from the previous ones through replacements of the boundary planes as
\beq
\begin{aligned}
	\Omega_{f} &= \Omega_F(3\to\tilde 3)=\Omega_{\tilde F}(1\to\tilde 1)\,,\\
	\Omega_{q_1} &= \Omega_{Q_1}(1\to\tilde 1) =  \Omega_F(2,6\to\tilde 2)-\Omega_F(2,5,6\to\tilde 2) \,,\quad	 \\
	\Omega_{q_2} &=\Omega_{Q_2}(1\to\tilde 1) = \Omega_F(2,5,6\to\tilde 2)\,,\\
	\Omega_{q_3} &= \Omega_{Q_3}(1\to\tilde 1) = \Omega_F(2,5\to\tilde 2)-\Omega_F(2,5,6\to\tilde 2)\,,\\
	\Omega_{\tilde q_1} &= \Omega_{Q_3}(3\to\tilde 3)= \Omega_{\tilde F}(2,5\to\tilde 2)-\Omega_{\tilde F}(2,5,6\to\tilde 2)\,,\\
	\Omega_{\tilde q_2} &= \Omega_{Q_2}(3\to\tilde 3)=\Omega_{\tilde F}(2,5,6\to\tilde 2)\,,\\
	\Omega_{\tilde q_3}&=\Omega_{Q_1}(3\to\tilde 3)=\Omega_{\tilde F}(2,6\to\tilde 2)-\Omega_{\tilde F}(2,5,6\to\tilde 2)\,.
\end{aligned}
\eeq	
Notice that these are exactly the same replacements that were used to obtain the previous sources from $\Omega_\psi$ in \eqref{equ:replace1}. Due to the commutative nature of the replacement operations, these sources with two twisted planes can be obtained via two different paths. 

\vskip 4pt
Taking the differential of $\Omega_{Q_{1,2,3}}$, we find
\beq
\begin{aligned}\label{equ:dXQ3site}
	\ud\Omega_{Q_1} = \e\,\Big(& \Omega_{Q_1} \ud\log X_2^{-+} +\Omega_{q_1} \ud\log X_{12}^+ + (\Omega_{Q_1}-\Omega_{q_1}) \,\ud\log X_1^+ \\
	&\quad  -\Omega_{\tilde q_2}\, \ud\log X_{23}^- + (\Omega_{\tilde q_1}+\Omega_{\tilde q_2})\,\ud\log X_1^+ + (\Omega_{Q_1}-\Omega_{\tilde q_1}) \,\ud\log X_3^+\Big)\,,\\[4pt]
	\ud\Omega_{Q_2} = \e\,\Big( &\Omega_{Q_2}\, \ud\log X_2^{--} +\Omega_{q_2}\, \ud\log X_{12}^- +(\Omega_{Q_2}-\Omega_{q_2})\,\ud\log X_1^+   \\
	&\quad +\Omega_{\tilde q_2} \,\ud\log X_{23}^- + (\Omega_{Q_2}-\Omega_{\tilde q_2})\,\ud\log X_3^+ \Big)\,,\\[4pt]
	\ud\Omega_{Q_3} = \e\,\Big( &\Omega_{Q_3}\, \ud\log X_2^{+-} +\Omega_{\tilde q_3}\, \ud\log X_{23}^+ + (\Omega_{Q_3}-\Omega_{\tilde q_3})\, \ud\log X_3^+ \\
	&\quad  -\Omega_{ q_2}\, \ud\log X_{12}^- + (\Omega_{q_2}+\Omega_{q_3})\,\ud\log X_1^- + (\Omega_{Q_3}-\Omega_{q_3}) \,\ud\log X_1^+\Big)\,.
\end{aligned}
\eeq
No new sources are necessary this time, since the source assignments of these equations are completely fixed in terms of the sources already defined.

\vskip 4pt
We now move on to the differential equations for the sources with two twisted planes. First, the differential of $\Omega_f$ can be expressed as
\beq
\begin{aligned}
	\ud \Omega_f = \e\, \Big( &\Omega_f\, \ud\log X_1^- +\Omega_f\, \ud\log X_3^- +\Omega_g\,\ud\log X_{12}^+ + \Omega_{\tilde g}\,\ud\log X_{23}^+\\
	&\quad\ +\Omega_Z\, \ud\log X_{123} + (\Omega_f-\Omega_g-\Omega_{\tilde g}-\Omega_Z) \,\ud\log X_2^{++}\Big)\,,
\end{aligned}
\eeq
in terms of three new sources
\beq
\begin{aligned}
	\Omega_g &\equiv  -[45\tilde 1]+[45\tilde 2]-[4\tilde 1\tilde 2]+[5\tilde 1\tilde 3]-[5\tilde 2\tilde 3]+[\tilde 1\tilde 2\tilde 3] \,,\\
	\Omega_{\tilde g} &\equiv  -[46\tilde 2]+[46\tilde 3]+[4\tilde 2\tilde 3]-[6\tilde 1\tilde 2]+[6\tilde 1\tilde 3]+[\tilde 1\tilde 2\tilde 3]\,,\\
	\Omega_Z &\equiv 	[4\tilde 1\tilde 2]-[4\tilde 1\tilde 3]+[4\tilde 2\tilde 3]-[\tilde 1\tilde 2\tilde 3]\,.
\end{aligned}
\eeq
Perhaps unsurprisingly, these can be obtained using the same replacement rules used earlier as
\beq
\begin{aligned}
	\Omega_g & = \Omega_{q_1}(3\to\tilde 3)=\Omega_{\tilde q_3}(1\to\tilde 1) = \Omega_f(2,6\to \tilde 2)-\Omega_f(2,5,6\to \tilde 2)\,,\\
	\Omega_{\tilde g} & = \Omega_{\tilde q_1}(3\to\tilde 3)=\Omega_{q_3}(1\to\tilde 1) = \Omega_f(2,6\to \tilde 2)-\Omega_f(2,5,6\to \tilde 2)\,,\\
	\Omega_Z & = \Omega_{q_2}(3\to\tilde 3)=\Omega_{\tilde q_2}(1\to\tilde 1) = \Omega_f(2,5,6\to \tilde 2)\,,
\end{aligned}
\eeq
via three different commuting paths. These sources then fully determine the coefficients appearing in the differentials of $\Omega_{ q_{1,2,3}}$ as
\beq
\begin{aligned}\label{equ:dXq3site}
	\ud \Omega_{q_1} &= \e\,\Big( 2\Omega_{q_1}\, \ud\log X_{12}^+ + (\Omega_{q_1}-\Omega_g)\,\ud\log X_3^+ + (\Omega_g+\Omega_Z)\,\ud\log X_3^- - \Omega_Z\, \ud\log X_{123}\Big)\,,\\[4pt]
	\ud \Omega_{q_2} &= \e\,\Big(2\Omega_{q_2}\,\ud\log X_{12}^- + \Omega_Z\,\ud\log X_{123} + (\Omega_{q_2}-\Omega_Z)\,\ud\log X_3^+\Big)\,,\\[4pt]
	\ud \Omega_{q_3} &= \e\,\Big( \Omega_{q_3}\,\ud\log X_2^{+-} - \Omega_{q_2}\,\ud\log X_{12}^- +(\Omega_{q_3}+\Omega_{q_2})\,\ud\log X_1^- \\
	&\quad\qquad\ + \Omega_{\tilde g}\,\ud\log X_{23}^+ + (\Omega_{q_3}-\Omega_{\tilde g})\,\ud\log X_3^+\Big)\,.
\end{aligned}
\eeq
The differentials of $\Omega_{\tilde q_{1,2,3}}$ can be obtained from these by permutation symmetry.

\vskip 4pt
Finally, taking the differential of the sources with three twisted planes gives
\beq
\begin{aligned}\label{equ:dXg3site}
	\ud\Omega_g &= \e\,\Big( 2\Omega_g\,\ud\log X_{12}^+ -\Omega_Z\,\ud\log X_{123} + (\Omega_g+\Omega_Z)\,\ud\log X_3^-\Big)\,,\\
	\ud\Omega_{\tilde g} &= \e\,\Big( 2\Omega_{\tilde g}\,\ud\log X_{23}^+ -\Omega_Z\,\ud\log X_{123} + (\Omega_{\tilde g}+\Omega_Z)\,\ud\log X_1^-\Big)\,,\\
	\ud\Omega_Z &=  3\e\,\Omega_Z\,\ud\log X_{123}\,.
\end{aligned}
\eeq
We see that no additional sources are required this time, and the system therefore closes.

\vskip 4pt
Integrating the basis three-forms against the twist factor yields the corresponding equations for the actual functions, which maintain the same form. We will therefore not reiterate the equations, but instead simply illustrate below the nonzero elements of the matrix $\tilde A$ in the $\ud {\cal I} = \e \tilde A \,{\cal I}$ equation:
 \be
\tilde A =  \left[\renewcommand{\arraystretch}{0.7} {\footnotesize{\begin{array}{cccccccccccccccc}
 \RC & \OC & \OC & \OC & \OC & \OC & \phantom{0} & \phantom{0} & \phantom{0} & \phantom{0} & \phantom{0} & \phantom{0} & \phantom{0} & \phantom{0} & \phantom{0} & \phantom{0}\\
 \phantom{0} & \OC & \phantom{0} & \phantom{0} & \phantom{0} & \phantom{0} & \GC & \GC & \GC & \GC & \phantom{0} & \phantom{0} & \phantom{0} & \phantom{0} & \phantom{0} & \phantom{0} \\
 \phantom{0} & \phantom{0} & \OC & \phantom{0} & \phantom{0} & \phantom{0} & \GC & \phantom{0} & \phantom{0} & \phantom{0} & \GC & \GC & \GC & \phantom{0} & \phantom{0} & \phantom{0} \\
 \phantom{0} & \phantom{0} & \phantom{0} & \OC & \phantom{0} & \phantom{0} & \phantom{0} & \GC & \GC & \phantom{0} & \phantom{0} & \phantom{0} & \GC & \phantom{0} & \phantom{0} & \phantom{0} \\
 \phantom{0} & \phantom{0} & \phantom{0} & \phantom{0} & \OC & \phantom{0} & \phantom{0} & \phantom{0} & \GC & \phantom{0} & \phantom{0} & \GC & \phantom{0} & \phantom{0} & \phantom{0} & \phantom{0} \\
 \phantom{0} & \phantom{0} & \phantom{0} & \phantom{0} & \phantom{0} & \OC & \phantom{0} & \phantom{0} & \phantom{0} & \GC & \GC & \GC & \phantom{0} & \phantom{0} & \phantom{0} & \phantom{0} \\
 \phantom{0} & \phantom{0} & \phantom{0} & \phantom{0} & \phantom{0} & \phantom{0} & \GC & \phantom{0} & \phantom{0} & \phantom{0} & \phantom{0} & \phantom{0} & \phantom{0} & \BC & \BC & \BC \\
 \phantom{0} & \phantom{0} & \phantom{0} & \phantom{0} & \phantom{0} & \phantom{0} & \phantom{0} & \GC & \GC & \phantom{0} & \phantom{0} & \phantom{0} & \phantom{0} & \phantom{0} & \BC & \phantom{0} \\
 \phantom{0} & \phantom{0} & \phantom{0} & \phantom{0} & \phantom{0} & \phantom{0} & \phantom{0} & \phantom{0} & \GC & \phantom{0} & \phantom{0} & \phantom{0} & \phantom{0} & \phantom{0} & \phantom{0} & \BC \\
 \phantom{0} & \phantom{0} & \phantom{0} & \phantom{0} & \phantom{0} & \phantom{0} & \phantom{0} & \phantom{0} & \phantom{0} & \GC & \phantom{0} & \phantom{0} & \phantom{0} & \BC & \phantom{0} & \BC \\
 \phantom{0} & \phantom{0} & \phantom{0} & \phantom{0} & \phantom{0} & \phantom{0} & \phantom{0} & \phantom{0} & \phantom{0} & \phantom{0} & \GC & \GC & \phantom{0} & \BC & \phantom{0} & \phantom{0} \\
 \phantom{0} & \phantom{0} & \phantom{0} & \phantom{0} & \phantom{0} & \phantom{0} & \phantom{0} & \phantom{0} & \phantom{0} & \phantom{0} & \phantom{0} & \GC & \phantom{0} & \phantom{0} & \phantom{0} & \BC \\
 \phantom{0} & \phantom{0} & \phantom{0} & \phantom{0} & \phantom{0} & \phantom{0} & \phantom{0} & \phantom{0} & \phantom{0} & \phantom{0} & \phantom{0} & \phantom{0} & \GC & \phantom{0} & \BC & \BC \\
 \phantom{0} & \phantom{0} & \phantom{0} & \phantom{0} & \phantom{0} & \phantom{0} & \phantom{0} & \phantom{0} & \phantom{0} & \phantom{0} & \phantom{0} & \phantom{0} & \phantom{0} & \BC & \phantom{0} & \BC \\
 \phantom{0} & \phantom{0} & \phantom{0} & \phantom{0} & \phantom{0} & \phantom{0} & \phantom{0} & \phantom{0} & \phantom{0} & \phantom{0} & \phantom{0} & \phantom{0} & \phantom{0} & \phantom{0} & \BC & \BC \\
 \phantom{0} & \phantom{0} & \phantom{0} & \phantom{0} & \phantom{0} & \phantom{0} & \phantom{0} & \phantom{0} & \phantom{0} & \phantom{0} & \phantom{0} & \phantom{0} & \phantom{0} & \phantom{0} & \phantom{0} & \BC \\
\end{array}}}  \begin{matrix} \phantom{\hskip 0pt} \\[110pt] \phantom{\hskip 0pt}\end{matrix} \right] ,
\ee
with the basis vector defined as
\begin{align}
	\vec I = ( \red{\psi},\,\orange{F},\,\orange{\tilde F},\,\orange{Q_3},\,\orange{Q_2},\,\orange{Q_1},\,\green{f},\,\green{q_3},\,\green{q_2},\,\green{q_1},\,\green{\tilde q_3},\,\green{\tilde q_2},\,\green{\tilde q_1},\,\blue{g},\,\blue{\tilde g},\,\blue{Z})\,.
\end{align}
The different colors indicate the four layers of the differential equations or equivalently the number of twisted planes that define the functions. The components of the matrix can be read off from the equations \eqref{equ:dPsi3}, \eqref{equ:dXF3site}, \eqref{equ:dXQ3site}, \eqref{equ:dXq3site}, and \eqref{equ:dXg3site}, or the {\sc Mathematica} notebook available at~\href{https://github.com/haydenhylee/kinematic-flow}{\faGithub}. 
It can be checked that this matrix indeed satisfies the integrability condition $\tilde A\wedge \tilde A=0$. 

\vskip 4pt
Some reordering of the functions is needed to put the matrix in a manifestly triangular form, 
due to the mixing between functions within the same layer. 
For instance, $q_2$ (or $\tilde q_2$) appears as a source in the equation for $q_3$ (or $\tilde q_3$), while $Z$ appears as a source in the equations for both $g$ and $\tilde g$. 
Relatedly, taking the de Sitter limit $\e\to 0$ no longer amounts to simply deleting the diagonal entries as in the two-site case, since the rescaling in $\e$ depends on the number of twisted planes. 

\vskip 4pt
As in the two-site case, we can solve the differential equations line-by-line and obtain the physical solution for $\psi$ after imposing the correct boundary conditions. We provide the details of this procedure in Appendix~\ref{app:functions}.
For $\e=0$, the solution is given in terms of classical polylogarithms up to transcendental weight three~\cite{Hillman:2019wgh}. Moreover, it can be checked that the above matrix in the $\e\to 0$ limit correctly reproduces the symbol of the de Sitter wavefunction (see Appendix~\ref{app:symbology} for details).  
For negative integer $\e$, which includes the flat-space case $\e=-1$, the solution is given by a rational function.
For generic $\e$, the solution is given by an integral over a ${}_2F_1$, which gives a two-variable generalization of the hypergeometric series.

\subsection{Source Functions}\label{sec:source}

The previous examples suggest that all sources in the differential equations may be obtained through suitable ``replacement rules" starting from the wavefunction integrand and replacing untwisted planes $B_a$ by twisted ones $T_a$.
It turns out that this is indeed the case, and this gives a systematic way to generate all source functions for any tree graph.

\vskip 4pt
For the two-site chain, there were two replacement rules, one for each vertex: 
\beq
\begin{aligned}
1&: \quad F = \psi(B_1\to T_1)\,, \quad Z=\tilde F(B_1\to T_1)\,,\\
2&: \quad \tilde F = \psi(B_2 \to T_2)\,, \quad Z=F(B_2\to T_2)\,.
\label{equ:replacement}
\end{aligned}
\eeq
Note that the functions $F$ and $\tilde F$ in \eqref{eq:twositefull} were associated to the letters $X_1-Y$ and $X_2-Y$ with flipped signs of the internal energies. These correspond to folded singularities that are not present in the Bunch--Davies wavefunction, whose physical energy singularities were represented by various tubings of a graph, cf.~\eqref{equ:2pt-tubing} and \eqref{equ:3pt-tubing}.
It turns out that there is a simple graphical way to capture the dependence on folded singularities with a slight modification of the previous tubing picture, in terms of a ``marked graph''. 
In this dressed version of a graph, we add a cross for every edge of a graph as an extra label for internal energies, and the nodes indicate external energies as usual. 
Folded singularities are then indicated by tubings that enclose these crosses.
This modified tubing picture is necessary to capture all singularities and basis functions that are present in the differential equations before imposing boundary conditions.
The relation between the parent function $\psi$ and its source functions can then be graphically depicted as
\beq
 \begin{tikzpicture}[baseline=(current  bounding  box.center)]
\node at (-1.3,0) {$\psi$};
\node at (-0.5,0) {\gpsi};
\node[right] at (2.1,1) {$F$};
\node at (3.7-.4,1)
    {\gF}; 
\node[right, color=Red] at (1,0.8) {\small $1$};
\node[right, color=Blue] at (1,-0.8) {\small $2$};
\node[right, color=Blue] at (1+3.7,0.8) {\small $2$};
\node[right, color=Red] at (1+3.7,-0.8) {\small $1$};
\draw [color=Red,thick,-stealth] (0.3,0.1) -- (2.1,0.9);
\draw [color=Blue,thick,-stealth] (0.3,-0.1) -- (2.1,-0.9);
\node[right] at (2.1,-1) {$\tilde F$};
\node at (3.7-.4,-1)
    {\gFt}; 
\draw [color=Red,thick,-stealth] (0.3+4.0-.2,0.1-1) -- (2.1+4.0-.2,0.9-1);
\draw [color=Blue,thick,-stealth] (0.3+4.0-.2,.9) -- (2.1+4.0-.2,0.1);
\node[right] at (2.1+4.0-.2,0) {$Z$};
\node at (3.7+4.0-.6,0)
    {\gZ}; 
\end{tikzpicture}
\label{equ:twositesplit}
\eeq
Here, each source function is associated to a tubing of a marked graph, containing at least one cross (except the wavefunction, which corresponds to a graph without any tubings).
Each substitution of a twisted plane $T_i$ is represented by a tube enclosing the vertex $i$. 
When the two tubings corresponding to $F$ and $\tilde F$ overlap, i.e.~when they share a cross, their union is represented by a merged tubing. This new tubing is associated to the function $Z$.

\vskip 4pt
For the three-site chain, the source functions were obtained from $\psi$ using five different replacement rules: one for each of the end vertices $X_1$ and $X_3$, and three for the middle vertex $X_2$. 
In the equation for the wavefunction, these were
\begin{align}
\begin{split}
1&: \quad F = \psi(B_1\to T_1)\,,\\
2_1&: \ \ \,  Q_1  = \psi(B_{2,6} \to T_2)- \psi(B_{2,5,6} \to T_2)\,,\\
2_2&:\ \ \,  Q_2 = \psi(B_{2,5,6} \to T_2)\,,\\
2_3&: \ \ \, Q_3  = \psi(B_{2,5} \to T_2)-\psi(B_{2,5,6} \to T_2)\,,\\
3&: \quad \tilde F = \psi(B_3 \to T_3)\,,
\label{equ:replacement}
\end{split}
\end{align}
where $i_r$ refers to the $r$-th replacement operation for vertex $i$ (so that e.g.~$2_2$ is the second replacement rule for vertex $2$).
In the differential equation, these source functions were associated with new letters involving at least one sign flip of the internal energies for lines a vertex is connected to.
The number of replacement rules for each vertex is therefore given by $2^{e-1}$ for a vertex connected to $e$ edges.  
As in the two-site case, the relations between $\psi$ and its source functions can be graphically represented as
\beq
 \begin{tikzpicture}[baseline=(current  bounding  box.center),scale=0.98]
\node at (0,0) {$\psi$};
\node[right] at (2.1,2.9) {$F
$};
\node at (3.7,2.9)
    {\ggF};  
\node[right,color=Red] at (1,1.9) {\small $1$};
\node[right] at (2.1,1) {$Q_1$};
\node at (3.7,1)
    {\ggQa}; 
\node[right] at (2.1,0) {$Q_2$};
\node at (3.7,0)
    {\ggQb}; 
\node[right, color=Orange] at (1,0.8) {\small $2_1$};
\node[right, color=Orange] at (1,0.2) {\small $2_2$};
\node[right, color=Orange] at (1,-0.3) {\small $2_3$};
\draw [color=Red,thick,-stealth] (0.2,0.2) -- (2.1,2.7);
\draw [color=Orange,thick,-stealth] (0.3,0.1) -- (2.1,0.9);
\draw [color=Orange,thick,-stealth] (0.3,0) -- (2.1,0);
\draw [color=Orange,thick,-stealth] (0.3,-0.1) -- (2.1,-0.9);
\draw [color=Blue,thick,-stealth] (0.2,-0.2) -- (2.1,-2.7);
\node[right] at (2.1,-1) {$Q_3$};
\node at (3.7,-1)
    {\ggQc}; 
\node[right] at (2.1,-2.9) {$\tilde F$};
\node at (3.7,-2.9)
    {\ggFt}; 
\node[right,color=Blue] at (1,-1.9)  {\small $3$};
\draw [color=Orange,thick,-stealth] (0.3+4.5,0.1+2.9) -- (2.1+4.5,0.9+2.9);
\draw [color=Orange,thick,-stealth] (0.3+4.5,0+2.9) -- (2.1+4.5,0+2.9);
\draw [color=Orange,thick,-stealth] (0.3+4.5,-0.1+2.9) -- (2.1+4.5,-0.9+2.9);
\draw [color=Red,thick,-stealth] (0.2+4.5,0.2+1) -- (2.1+4.5,2.7+1);
\draw [color=Red,thick,-stealth] (0.2+4.5,0.2) -- (2.1+4.5,2.7);
\draw [color=Red,thick,-stealth] (0.2+4.5,0.2-1) -- (2.1+4.5,2.7-1);
\node[right] at (2.1+4.5,1+2.9) {$q_1$};
\node at (3.7+4.5,1+2.9)
    {\ggqa}; 
\node[right] at (2.1+4.5,0+2.9) {$q_2$};
\node at (3.7+4.5,0+2.9)
    {\ggqb}; 
\node[right] at (2.1+4.5,-1+2.9) {$q_3$};
\node at (3.7+4.5,-1+2.9)
    {\ggqc}; 

\draw [color=Orange,thick,-stealth] (0.3+4.5,0.1-2.9) -- (2.1+4.5,0.9-2.9);
\draw [color=Orange,thick,-stealth] (0.3+4.5,0-2.9) -- (2.1+4.5,0-2.9);
\draw [color=Orange,thick,-stealth] (0.3+4.5,-0.1-2.9) -- (2.1+4.5,-0.9-2.9);
\draw [color=Blue,thick,-stealth] (0.2+4.5,-0.2+1) -- (2.1+4.5,-2.7+1);
\draw [color=Blue,thick,-stealth] (0.2+4.5,-0.2) -- (2.1+4.5,-2.7);
\draw [color=Blue,thick,-stealth] (0.2+4.5,-0.2-1) -- (2.1+4.5,-2.7-1);
\draw [color=Blue,thick,-stealth] (0.2+4.5,-0.2+2.9) -- (2.1+4.5,-2.7+2.9);
\draw [color=Red,thick,-stealth] (0.2+4.5,0.2-2.9) -- (2.1+4.5,+2.7-2.9);
\node[right] at (2.1+4.5,1-2.9) {$\tilde q_3$};
\node at (3.7+4.5,1-2.9)
    {\ggqtc}; 
\node[right] at (2.1+4.5,0-2.9) {$\tilde q_2$};
\node at (3.7+4.5,0-2.9)
    {\ggqtb}; 
\node[right] at (2.1+4.5,-1-2.9) {$\tilde q_1$};
\node at (3.7+4.5,-1-2.9)
    {\ggqta}; 
\node[right] at (2.1+4.5,0) {$f$};
\node at (3.7+4.5,0)
    {\ggf}; 
\node[right] at (2.1+4.5+4.5,0) {$Z$};
\node at (3.7+4.5+4.5,0)
    {\ggZ}; 
 \node[right] at (2.1+4.5+4.5,0-2.9) {$\tilde g$};
\node at (3.7+4.5+4.5,0-2.9)
    {\gggt}; 
    \node[right] at (2.1+4.5+4.5,0+2.9) {$g$};
\node at (3.7+4.5+4.5,0+2.9)
    {\ggg}; 
\draw [color=Orange,thick,-stealth] (0.3+9,0) -- (2.1+9,0);
\draw [color=Blue,thick,-stealth] (0.2+9,-0.2+2.9) -- (2.1+9,-2.7+2.9);
\draw [color=Orange,thick,-stealth] (0.2+9,-0.2) -- (2.1+9,-2.7-.1);
\draw [color=Orange,thick,-stealth] (0.2+9,0.2) -- (2.1+9,2.7+.1);
\draw [color=Blue,thick,-stealth] (0.3+9,-0.1+2.9+1) -- (2.1+9,-0.9+2.9+1);
\draw [color=Red,thick,-stealth] (0.3+9,0.1-2.9-1) -- (2.1+9,0.9-2.9-1);
\draw [color=Red,thick,-stealth] (0.2+9,0.2-2.9) -- (2.1+9,+2.7-2.9);
\draw [color=Red,thick,-stealth] (0.2+9,0.2-2.9+1) -- (2.1+9,+2.7-.2);
\draw [color=Blue,thick,-stealth] (0.2+9,-0.2+2.9-1) -- (2.1+9,-2.7+.2);
\end{tikzpicture}
\label{equ:splitting}
\eeq
We see that there are a total of 16 basis functions in the differential system including $\psi$ itself.

\vskip 4pt
This graphical representation of basis functions can be straightforwardly generalized to arbitrary tree graphs.
Since the substituted twisted planes do not need to be adjacent to each other, they can be depicted as a union of disconnected tubings of a marked graph. 
There are four independent tubings for a graph with one edge (i.e.~the two-site chain), and a simple inductive argument implies that the total number of independent tubings for a graph with $e$ edges is $4^e$, irrespective of its geometry.\footnote{This counting is also given by the normalized volume of cosmological polytopes~\cite{Kuhne:2022wze, Juhnke-Kubitzke:2023nrj}.}
This establishes a one-to-one correspondence between the basis functions in the differential equations and the disconnected tubings of a marked graph. 
The basis functions are therefore elegantly characterized by the combinatorics of a marked graph, a topic that will be explored in greater detail in the next section.

\paragraph{Function count:}
We can also easily count the number of basis functions in different {\it layers}, where a layer refers to the number of vertices enclosed in the tubings (or, equivalently, the number of substituted twisted planes).
For a graph $G$, the number of disconnected tubings $t_\ell$ at layer $\ell$ is 
\beq
	t_\ell(G) = \sum_{x_1+\cdots+x_{r_{\rm max}}=\ell}\ \prod_{r=1}^{r_{\rm max}} \binom{a_r}{x_r}(2^r-1)^{x_r}\, , \label{eq:tl}
\eeq
where $r=1,2,\cdots,r_{\rm max}$ is the number of neighboring edges for a vertex, $a_r$ is the number of vertices of $G$ having exactly $r$ edges, 
and $x_r$ is the number of vertices enclosed in tubings that have exactly $r$ edges for a given layer.  Interestingly, the alternating sum of the number of functions in each layer  always vanishes, regardless of its geometry:
\begin{align}
	\sum_{\ell} (-1)^\ell\, t_\ell = 0\,.
\end{align}
This feature holds in general and suggests the existence of some underlying geometric object. 

\vskip 4pt
Let us give a few examples. For the $n$-site chain, we have $r_{\rm max}=2$, with $a_1=2$ and $a_2=n-2$. 
The number of functions in the $\ell$-th layer of the differential equation then is
\begin{align}
	t_\ell(\text{$n$-site chain}) = \binom{n-2}{\ell}\,3^\ell+2\binom{n-2}{\ell-1}\,3^{\ell-1}+\binom{n-2}{\ell-2}\,3^{\ell-2}\,.
\end{align}
Explicitly, this gives $\{t_0,t_1,t_2\}=\{1,2,1\}$ for the two-site chain, $\{t_0,t_1,t_2,t_3\}=\{1,5,7,3\}$ for the three-site chain as we have found above, and $\{t_0,t_1,t_2,t_3,t_4\}=\{1,8,22,24,9\}$ for the four-site chain. For a pure star geometry, we have $a_1=n-1$, $a_{n-1}=1$ and $a_i=0$, for $i\notin \{1,n-1\}$. This gives
\begin{align}
	t_\ell(\text{$n$-site star})  = \binom{n-1}{\ell}+\binom{n-1}{\ell-1}(2^{n-1}-1)\,.
\end{align}
The simplest graph with a star geometry is the four-site star, which has $\{t_0,t_1,t_2,t_3,t_4\}=\{1,10,24,22,7\}$.

\paragraph{Replacement rules:}
Let us now describe a procedure to derive the replacement rules. 
We saw above that the replacement rule involving the middle vertex of the three-site chain was slightly nontrivial. To gain more insight, it is best to demonstrate the procedure for more nontrivial examples. 
Consider the wavefunction for the four-site chain, whose wavefunction integrand and boundary planes were given in \eqref{equ:psi4c} and \eqref{equ:Singularities4}.  
The end vertices, 1 and 4, are each connected to a single edge, so obtaining the source functions associated to these vertices amounts to the simple replacements $B_1\to T_1$ or $B_4\to T_4$. 
Vertex 2, on the other hand, is connected to two edges, and the three source functions that appear in the differential of $\psi$ can be obtained as
\beq
\begin{aligned}
2_2:\quad	\raisebox{2pt}{
 \begin{tikzpicture}[baseline=(current  bounding  box.center)]
\draw[fill=black] (-0.75,0) -- (1.5,0);
\draw[fill=black] (0,0) circle (.5mm);
\draw[fill=black] (-0.75,0) circle (.5mm);
\draw[fill=black] (0.75,0) circle (.5mm);
\draw[fill=black] (1.5,0) circle (.5mm);
\node at (0.375,0)  {\Cross};
\node at (0.375+.75,0)  {\Cross};
\node at (0.375-.75,0)  {\Cross};
\node [
        draw, color=gray, line width=0.6pt,
        rounded rectangle,
        minimum height = 0.9em,
        minimum width = 3.5em,
        rounded rectangle arc length = 180,
    ] at (0,0)
    {};
\end{tikzpicture}} \ &=\ \psi(B_{2,6,7,10}\to T_2)\,,\\
2_1:\quad \raisebox{2pt}{
 \begin{tikzpicture}[baseline=(current  bounding  box.center)]
\draw[fill=black] (-0.75,0) -- (1.5,0);
\draw[fill=black] (0,0) circle (.5mm);
\draw[fill=black] (-0.75,0) circle (.5mm);
\draw[fill=black] (0.75,0) circle (.5mm);
\draw[fill=black] (1.5,0) circle (.5mm);
\node at (0.375,0)  {\Cross};
\node at (0.375+.75,0)  {\Cross};
\node at (0.375-.75,0)  {\Cross};
\node [
        draw, color=gray, line width=0.6pt,
        rounded rectangle,
        minimum height = 0.9em,
        minimum width = 2.5em,
        rounded rectangle arc length = 180,
    ] at (-.2,0)
    {};
\end{tikzpicture}}\ &= \ \psi(B_{2,7,10}\to T_2) \ -\ 	\raisebox{2pt}{
 \begin{tikzpicture}[baseline=(current  bounding  box.center)]
\draw[fill=black] (-0.75,0) -- (1.5,0);
\draw[fill=black] (0,0) circle (.5mm);
\draw[fill=black] (-0.75,0) circle (.5mm);
\draw[fill=black] (0.75,0) circle (.5mm);
\draw[fill=black] (1.5,0) circle (.5mm);
\node at (0.375,0)  {\Cross};
\node at (0.375+.75,0)  {\Cross};
\node at (0.375-.75,0)  {\Cross};
\node [
        draw, color=gray, line width=0.6pt,
        rounded rectangle,
        minimum height = 0.9em,
        minimum width = 3.5em,
        rounded rectangle arc length = 180,
    ] at (0,0)
    {};
\end{tikzpicture}}\,\ ,\\
2_3:\quad\raisebox{2pt}{
 \begin{tikzpicture}[baseline=(current  bounding  box.center)]
\draw[fill=black] (-0.75,0) -- (1.5,0);
\draw[fill=black] (0,0) circle (.5mm);
\draw[fill=black] (-0.75,0) circle (.5mm);
\draw[fill=black] (0.75,0) circle (.5mm);
\draw[fill=black] (1.5,0) circle (.5mm);
\node at (0.375,0)  {\Cross};
\node at (0.375+.75,0)  {\Cross};
\node at (0.375-.75,0)  {\Cross};
\node [
        draw, color=gray, line width=0.6pt,
        rounded rectangle,
        minimum height = 0.9em,
        minimum width = 2.5em,
        rounded rectangle arc length = 180,
    ] at (.2,0)
    {};
\end{tikzpicture}} \ & = \ \psi(B_{2,6}\to T_2)\hskip 10.5pt\  -\  	\raisebox{2pt}{
 \begin{tikzpicture}[baseline=(current  bounding  box.center)]
\draw[fill=black] (-0.75,0) -- (1.5,0);
\draw[fill=black] (0,0) circle (.5mm);
\draw[fill=black] (-0.75,0) circle (.5mm);
\draw[fill=black] (0.75,0) circle (.5mm);
\draw[fill=black] (1.5,0) circle (.5mm);
\node at (0.375,0)  {\Cross};
\node at (0.375+.75,0)  {\Cross};
\node at (0.375-.75,0)  {\Cross};
\node [
        draw, color=gray, line width=0.6pt,
        rounded rectangle,
        minimum height = 0.9em,
        minimum width = 3.5em,
        rounded rectangle arc length = 180,
    ] at (0,0)
    {};
\end{tikzpicture}}\,\ .
\end{aligned}
\eeq
Note that this is similar to how the functions $Q_{1,2,3}$ were obtained in the three-site case, involving a subtraction of the function involving both crosses enclosed in the tubing. A new feature is that the replacement $2_2$ involves two planes $B_7$ and $B_{10}$ that contain both $X_2$ and $X_3$, with $B_{10}$ also containing $X_4$.
The source functions for vertex 3 can be obtained in a similar way.

\vskip 4pt
A slightly more nontrivial example is the four-site star, whose wavefunction integrand and boundary planes were given in \eqref{equ:4Site-Star} and \eqref{equ:4Site-StarB}. The source functions associated to the three outer vertices are  again generated through trivial replacements $B_i\to T_i$, for $i=1,2,3$. 
For the central vertex, there are $7$ source functions which appear in the differential of the wavefunction:
\beq
\begin{aligned}\label{starreplacemets}
4_7:\quad\	\gstarQ \ &  =\ \psi(B_{4,6,7,8,9,10,11}\to T_4)\, ,\hskip -15pt\\
4_6:\quad\ \gstarQc \ & = \ \psi(B_{4,6,7,8,9,11}\to T_4)   &&-\ \gstarQ\,\ ,\\
4_5:\quad\	\gstarQb \ & = \ \psi(B_{4,6,7,8,9,10}\to T_4)  &&-\ \gstarQ\,\ ,\\
4_4:\quad\	\gstarQa \ & = \  \psi(B_{4,6,7,8,10,11}\to T_4)  &&-\ \gstarQ\,\ ,\\
4_3:\quad	\raisebox{3pt}{
 \begin{tikzpicture}[baseline=(current  bounding  box.center)]
 \coordinate (4) at (0,0);
 \coordinate (3) at (0,0.65);
  \coordinate (1) at (-0.5,-0.5);
 \coordinate (2) at (0.5,-0.5);
   \coordinate (y1) at (-0.25,-0.25);
  \coordinate (y2) at (0.25,-0.25);
   \coordinate (y3) at (0,0.325);
 \draw[gray,line width=0.6pt]  ($(4) + (-0.12,0)$)  to[out=-90,in=-90,looseness=2]  ($(4) + (0.12,0)$) -- ($(y3) + (0.12,0)$)  to[out=90,in=90,looseness=2]  ($(y3) + (-0.12,0)$)   --  ($(4) + (-0.12,0)$)  ; 
 \draw[fill] (4) circle (.5mm);
  \draw[fill] (3) circle (.5mm);
   \draw[fill] (2) circle (.5mm);
  \draw[fill] (1) circle (.5mm);
  \draw[thick] (4) -- (3);
    \draw[thick] (4) -- (2);
   \draw[thick] (4) -- (1); 
\node at (0,0.3)  {\Cross};
\node at (-0.25,-0.25) {\CrossR};
\node at (0.25,-0.25) {\CrossR};
\end{tikzpicture}} \ & = \ \psi(B_{4, 6, 7, 9}\to T_4)  &&- \ \gstarQb\ -\  \gstarQc\   +\ \gstarQ\,\ ,\\
4_2:\quad	\raisebox{3pt}{
 \begin{tikzpicture}[baseline=(current  bounding  box.center)]
 \coordinate (4) at (0,0);
 \coordinate (3) at (0,0.65);
  \coordinate (1) at (-0.5,-0.5);
 \coordinate (2) at (0.5,-0.5);
   \coordinate (y1) at (-0.25,-0.25);
  \coordinate (y2) at (0.25,-0.25);
   \coordinate (y3) at (0,0.325);
 \draw[gray,line width=0.6pt]  ($(y2) + (-0.085,-0.085)$)  to[out=-45,in=-45,looseness=2]  ($(y2) + (0.085,0.085)$) -- ($(4) + (0.085,0.085)$)  to[out=135,in=135,looseness=2]  ($(4) + (-0.085,-0.085)$) -- ($(y2) + (-0.085,-0.085)$); 
 \draw[fill] (4) circle (.5mm);
  \draw[fill] (3) circle (.5mm);
   \draw[fill] (2) circle (.5mm);
  \draw[fill] (1) circle (.5mm);
  \draw[thick] (4) -- (3);
    \draw[thick] (4) -- (2);
   \draw[thick] (4) -- (1); 
\node at (0,0.3)  {\Cross};
\node at (-0.25,-0.25) {\CrossR};
\node at (0.25,-0.25) {\CrossR};
\end{tikzpicture}} \ & = \ \psi(B_{4, 6, 8, 10}\to T_4)  &&-\ \gstarQa\ -\  \gstarQb\  +\ \gstarQ\,\ ,\\
4_1:\quad	\raisebox{3pt}{
 \begin{tikzpicture}[baseline=(current  bounding  box.center)]
 \coordinate (4) at (0,0);
 \coordinate (3) at (0,0.65);
  \coordinate (1) at (-0.5,-0.5);
 \coordinate (2) at (0.5,-0.5);
   \coordinate (y1) at (-0.25,-0.25);
  \coordinate (y2) at (0.25,-0.25);
   \coordinate (y3) at (0,0.325);
 \draw[gray,line width=0.6pt]  ($(y1) + (0.085,-0.085)$)  to[out=-135,in=-135,looseness=2]  ($(y1) + (-0.085,0.085)$) -- ($(4) + (-0.085,0.085)$)  to[out=45,in=45,looseness=2]  ($(4) + (0.085,-0.085)$) -- ($(y1) + (0.085,-0.085)$); 
 \draw[fill] (4) circle (.5mm);
  \draw[fill] (3) circle (.5mm);
   \draw[fill] (2) circle (.5mm);
  \draw[fill] (1) circle (.5mm);
  \draw[thick] (4) -- (3);
    \draw[thick] (4) -- (2);
   \draw[thick] (4) -- (1); 
\node at (0,0.3)  {\Cross};
\node at (-0.25,-0.25) {\CrossR};
\node at (0.25,-0.25) {\CrossR};
\end{tikzpicture}} \ & = \ \psi(B_{4,7,8,11}\to T_4)  &&- \ \gstarQa\  -\  \gstarQc\  +\ \gstarQ\,\ .
\end{aligned}
\eeq
First, note that the seven boundary planes replaced in the rule $4_7$ are those that involve $X_4$ and at least one internal energy. 
For the rule $4_6$, there is a single boundary plane $B_{10}$ that contains $X_4$ but no $Y,Y''$. To generate this function, we thus do not replace $B_{10}$, and then subtract off the function obtained through $4_7$.
The functions $4_5$ and $4_4$ can be obtained in a similar way. 
For~$4_3$, whose tubing encloses a single cross, we do not replace the planes that contain $X_4$ but not $Y'$. 
We then subtract off all functions that contain $X_4$ and $Y'$. 
The remaining functions can be obtained following the same procedure.

\vskip 4pt
This procedure can be made systematic as follows. Let $y_{i_r}$ denote a non-empty subset of the set of internal energies associated to vertex $i$. We have $r=1,\cdots,2^{e_i}-1$, where $e_i$ is the number of edges connected to vertex $i$. For example, for vertex 4 of the four-site star, $y_{4_r}$ would be an element of the set $\{ \{Y\},\{Y'\},\{Y''\}, \{Y,Y'\},\{Y,Y''\},\{Y',Y''\},\{Y,Y',Y''\} \}$.
The basis functions associated with each $y_{i_r}$ in the $\ud\psi$ equation can then be obtained recursively as
\begin{align}
	F_{i_r} = \psi({\cal B}_{i_r}\to T_i) - \sum_{y_{i,u}\,\supset\, y_{i,r}} (-1)^{|y_{i,u}|-|y_{i,r}|} F_{i_{u}}\,,\label{eq:replace}
\end{align}
where ${\cal B}_{i_r}$ is the set of boundary planes that contain $X_i$ and at least one element of $y_{i_r}$, which get replaced with $T_i$. The last term subtracts (with alternating signs) the functions that are associated to bigger subsets of internal energies. All other functions can be obtained by successive iterations of these replacement operations.

\paragraph{Algorithm:}
Given that we have a procedure to generate the various source functions, we are now ready to state a complete algorithm for deriving the differential equations of the wavefunction for a given graph:
\begin{itemize}
	\item For a given graph $G$, write down all independent replacement rules for each vertex using~\eqref{eq:replace}. The total number of replacement rules is given by $t_1(G)$ in \eqref{eq:tl}. 
	\item From $\psi$ defined in terms of simplex forms \eqref{eq:simplexbracket}, use the replacement rules to generate all source functions. The total number of basis functions for an $n$-site graph is $4^{n-1}$.
	\item Take the differential of the basis functions using \eqref{equ:Formula}. Express the coefficient of each letter as a linear combination of the basis functions.
\end{itemize}
The last step is a linear algebra problem, which we implemented using computer algebra~\href{https://github.com/haydenhylee/kinematic-flow}{\faGithub}.

\newpage
\section{Time Evolution as Kinematic Flow}
\label{sec:GraphicalRules}

In the previous section, we presented a systematic algorithm to derive the differential equations satisfied by the wavefunction for arbitrary tree graphs. 
The output of this analysis, however, is typically not very 
 illuminating. Fortunately, a closer inspection of the results reveals a striking hidden simplicity in the complex equations. 
 This hidden structure becomes manifest when we represent the singularities of the differential equations graphically as tubings of marked graphs.
 Upon differentiation, these tubings grow in a local and predictive fashion. In fact, a few remarkably simple graphical rules will allow us to predict the equations for all tree graphs.\footnote{At this point, the reader may (correctly) object that the Feynman rules presented in Section~\ref{ssec:FlatSpace} are also very simple rules underlying all tree-level wavefunctions. However, it is important to emphasize that---unlike the Feynman rules---the rules presented in this section apply to the final wavefunction {\it after} performing the time (or energy) integrals.} 
The complexity of the differential equations is then reduced to a few universal rules.  While these rules are defined purely in terms of the kinematic data on the boundary of the spacetime, they reflect the physics of bulk time evolution.

\subsection{Graphical Representation}
\label{ssec:graphical}

In Sections~\ref{sec:TwoSite} and \ref{sec:MoreComplex}, we derived the differential equations for the two and three-site chains. We begin by translating these equations into graphical form.

\subsubsection{Two-Site Chain}
\label{ssec:2site-graphical}

We start by introducing a graphical representation for the letters of the differential equations.
To capture all letters, we add a marking (cross) to each internal line.
The  letters are then represented by ``connected tubes" (circlings of the vertices and crosses) of these marked graphs. The tubes must contain at least one vertex. To each tubing, we then assign a letter given by the sum of vertex energies enclosed by the tube and the energies of the internal lines piercing the tube. For tubes that intersect an internal line and enclose the corresponding cross, we flip the sign of the internal energy.

\vskip 4pt
The letters\footnote{Strictly speaking, the tubings in (\ref{equ:Letters-2Site})
 represent dlog forms with the letters appearing as their arguments. For simplicity, we will continue to use this somewhat imprecise language.} of the two-site chain then are
\beq
\begin{aligned}
	&\Lap \ \equiv\ \ud\log(X_1+Y)\,,  &&\Lam\ \equiv\ \ud\log(X_1-Y)\,,\\
	&\Lbp \ \equiv\ \ud\log(X_2+Y)\,, &&\Lbm\ \equiv\ \ud\log(X_2-Y)\,,\\
	&\Labg \ \equiv\ \ud\log(X_1+X_2)\,.
\end{aligned}\label{equ:Letters-2Site}
\eeq
The letters in the left column correspond to the known energy singularities of the graph, while those on the right come from a sign flip of the internal energy. 
Using this graphical representation, the differential equations for the two-site chain, eq.~\eqref{equ:2SiteEqns}, can be written as
\begin{align}
\ud \psi
&\ =  \  \e\, \Big[(\psi-F)
\  \Lap
 \ + \   F
\  \Lam
 \ + \   (\psi- \tilde F)
\  \Lbp
\ \ + \ \  \tilde F
\   \Lbm \Big] \label{equ:TwoSite-dPsi} \\[-4pt]
\cline{1-2}
\ud F 
&\ =  \   \e\, \Big[ F 
\  \Lam 
 \ +  \  (F-Z) \ \Lbp 
\ + \ 
Z \ \Labb \Big] \label{equ:TwoSite-dF}
  \\[5pt]
\ud \tilde F 
&\ =\  \e\, \Big[ \tilde F 
\ \Lbm 
 \ +  \  (\tilde F-Z) \  
\Lap  
 \ + \ 
Z \ \Labr  \Big]
\label{equ:TwoSite-dtF}
  \\[-4pt]
  \cline{1-2}
\ud Z 
&\ =  \  2\e\,Z\ \Labg   
\label{equ:TwoSite-dZ}
\end{align}
 We see that each letter is represented by a graph tubing and that these tubings ``grow" as we take derivatives. The system closes when then entire graph is encircled by a tube.
 This growth is a boundary manifestation of bulk time evolution. 
 In the following, we will see that this pattern generalizes to arbitrary tree graphs.

\subsubsection{Three-Site Chain}
 
 The two-site chain is too small to capture all features of a generic tree graph. A second illustrative example is the case of the three-site chain, whose differential equations were derived in  Section~\ref{ssec:sec4-examples}.
 In the following, we present these equations in graphical form. 
 
\vskip 4pt
 The letters for the three-site chain are given by the following tubings of the marked graph:
\beq
\begin{aligned}
	&\LLap \ \equiv\ \ud\log(X_1+Y) \,, &&\LLam\ \equiv\ \ud\log(X_1-Y)\,,\\
	&\LLcp \ \equiv\ \ud\log(X_3+Y')\,, &&\LLcm\ \equiv\ \ud\log(X_3-Y')\,,\\
	&\LLbpp \ \equiv\ \ud\log(X_2+Y+Y') \,,&&\LLbmp\ \equiv\ \ud\log(X_2-Y+Y')\,,\\
	& &&\LLbmm\ \equiv\ \ud\log(X_2-Y-Y')\,,\\
	& &&\LLbpm\ \equiv\ \ud\log(X_2+Y-Y')\,,\\
	&\LLabp\ \equiv\ \ud\log(X_1+X_2+Y')\,, &&\LLabm\ \equiv\ \ud\log(X_1+X_2-Y')\,,\\
	&\LLbcp\ \equiv\ \ud\log(X_2+X_3+Y)\,, \quad &&\LLbcm\ \equiv\ \ud\log(X_2+X_3-Y')\,,\\
	&\LLabc\ \equiv\ \ud\log(X_1+X_2+X_3)\,. 
\end{aligned}\label{equ:3Site-Letters}
\eeq
Again, the letters in the left column are the known energy singularities, while those on the right come from all possible ways of flipping the signs of the internal energies.
This predicts the 13 letters appearing in the differential equations for the three-site chain.

\vskip 4pt
Using \eqref{equ:3Site-Letters}, the differentials of all basis functions can be written as
\begin{align}
\begin{array}{rrrr}
  \ud \psi \,=\, \e\, \Big[    & (\psi- F)\ \LLap  &\hskip 8pt  +\ (\psi-\tilde F) \LLcp\phantom{\Big]} & +\ (\psi-{\textstyle \sum}Q_i) \LLbpp\phantom{\Big]}\\
  &	+\ F\ \LLam &   +\ \tilde F\ \LLcm\phantom{\Big]} & +\ Q_1\ \LLbmp\phantom{\Big]}\\
  &	&  & +\ Q_2\ \LLbmm\phantom{\Big]}\\
  &	&  & +\ Q_3\ \LLbpm \Big]
  \end{array}\label{equ:dPsi-3S}\\
  \cline{1-2}
\begin{array}{rrrr}
  \ud F \,=\, \e\, \Big[    & \hskip 8pt F\ \LLam  &\hskip 12pt +\ (F-f) \LLcp\phantom{\Big]} & +\ (F-{\textstyle \sum}q_i) \LLbpp\phantom{\Big]}\\
  &	&   +\ f\ \LLcm\phantom{\Big]} & +\ q_1\ \LLabpo\phantom{\Big]}\\
  &	&  & +\ q_2\ \LLabmo\phantom{\Big]}\\
  &	&  & +\ q_3\ \LLbpm \Big]
  \end{array}\label{equ:FR-1}\\
\begin{array}{rrrr}
  \ud Q_1 \,=\, \e\, \Big[ &\hskip 3pt Q_1\ \LLbmp & +\ (Q_1-q_1)\ \LLap\phantom{\Big]} & +\ (Q_1-\tilde q_3) \LLcp\phantom{\Big]}\\
  &	&  +\ q_1\ \LLabpr\phantom{\Big]} & \hskip 9.5pt +\ (\tilde q_3+\tilde q_2) \LLcm\phantom{\Big]}\\
  &	& & -\ \tilde q_2\ \LLbcmb\Big]
  \end{array}\label{equ:dQ2}\\
\begin{array}{rrrr}
  \ud Q_2 \,=\, \e\, \Big[ &\hskip 3pt Q_2\ \LLbmm & +\ (Q_2-q_2)\ \LLap\phantom{\Big]} &\hskip 6pt +\ (Q_2-\tilde q_2) \LLcp\phantom{\Big]}\\
  	&  & +\ q_2\ \LLabmr\phantom{\Big]} & +\ \tilde q_2\ \LLbcmb\Big]\\
  \end{array}\label{equ:dQ2}\\
\cline{1-2}
\begin{array}{rrrr}
  \ud q_1 \,=\, \e\, \Big[ &2q_1\ \LLabpr &\hskip 12pt +\ (q_1-g) \LLcp\phantom{\Big]} & \phantom{\hskip 6pt +\ (Q_2-\tilde q_2) \LLcp\Big]}\\
   	&   &+\ (g+Z) \LLcm \phantom{\Big]} & \phantom{\hskip 6pt +\ (Q_2-\tilde q_2) \LLcp\Big]} \\
  	&   &-\ Z\ \LLabcb\Big] & \phantom{\hskip 6pt +\ (Q_2-\tilde q_2) \LLcp\Big]}
  \end{array}\label{equ:dq1}\\
\begin{array}{rrrr}
  \ud q_2 \,=\, \e\, \Big[ &2q_2\ \LLabmr & \hskip 9pt+\ (q_2-Z) \LLcp\phantom{\Big]} & \phantom{\hskip 6pt +\ (Q_2-\tilde q_2) \LLcp\Big]}\\
  	&   &+\ Z\ \LLabcb\Big] & \phantom{\hskip 6pt +\ (Q_2-\tilde q_2) \LLcp\Big]}
  \end{array}\label{equ:dq2}\\
\begin{array}{rrrr}
  \ud q_3 \,=\, \e\, \Big[ &\hskip 5pt q_3\ \LLbpm &\hskip 6pt +\ (q_3+q_2)\, \LLam\phantom{\Big]} &\hskip 14pt +\ (q_3-\tilde g) \LLcp\phantom{\Big]}\\
  	&  & -\ q_2\ \LLabmr\phantom{\Big]} & +\ \tilde g\ \LLbcpb\Big]
  \end{array}\label{equ:dq3x}\\
\begin{array}{rrrr}
  \ud f \,=\, \e\, \Big[&\hskip 7pt f\ \LLam &\hskip 12pt +\ f \LLcm &\hskip 12pt +\ (f-g-\tilde g-Z) \LLbpp\phantom{\Big]}\\
  	&  & & +\ g\ \LLabpo\phantom{\Big]}\\
  	& & & +\ \tilde g\ \LLbcpo\phantom{\Big]}\\
  	& & & +\ Z\ \LLabco \Big]
  \end{array}\label{equ:f}\\
  \cline{1-2}
\begin{array}{rrrr}
  \ud g \,=\, \e\, \Big[ &\hskip 2pt 2g\ \LLabpr & \hskip 10pt+\ (g+Z)\ \LLcm\phantom{\Big]} & \phantom{\hskip 6pt +\ (Q_2-\tilde q_2) \LLcp\Big]}\\
  	&   &-\ Z\ \LLabcb\Big] & \phantom{\hskip 6pt +\ (Q_2-\tilde q_2) \LLcp\Big]}
  \end{array}\label{equ:dX-gp}\\
 \begin{array}{rrrr}
  \ud Z \,=  &\hskip 7pt 3\e\, Z\ \raisebox{-1.5pt}{\includegraphics[scale=1]{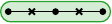}} & \phantom{\hskip 10pt+\ (g-Z)\ \LLcm\phantom{\Big]}} & \phantom{\hskip 6pt +\ (Q_2-\tilde q_2) \LLcp\Big]}
  \end{array} \label{equ:ZZZ}
\end{align}
We didn't show explicit expressions for the differentials of  $\tilde F$, $Q_3$, $\tilde q_{1,2,3}$, $\tilde g$, because these are related by symmetry to those of $F$, $Q_1$, $q_{1,2,3}$, $g$, respectively. 

\vskip 4pt
Although the above equations look rather complicated, they contain a hidden pattern that we will now reveal. The pattern is universal and will allow us to predict---by hand---the differential equations for arbitrary tree graphs.

\subsection{Kinematic Flow}
\label{ssec:rules}

Our goal is to predict the form of the connection matrix
\beq
A = \sum_i \alpha_i \,\ud \log \Phi_i(Z)\, , \label{equ:connection}
\eeq
where $\Phi_i(Z)$ are the letters of the differential equation and $\alpha_i$ are constant matrices.
 In this section, we present a set of graphical rules that will allow us to predict  \eqref{equ:connection} for all tree graphs. 
 The rules are local and can therefore be stated without committing to a specific graph. For concreteness, however, we will first illustrate the rules for the case of the three-site chain. 
 In Appendix~\ref{app:flow-examples}, we will show that the rules correctly capture more complicated examples.

\paragraph{Letters}
 To predict the alphabet of the differential equation, we start by drawing the tree graph of interest and marking all its internal lines with crosses. The dlog forms in (\ref{equ:connection}) are then given by all ``connected tubings" of the marked graph.
For the two- and three-site chains, these tubings were drawn in~\eqref{equ:Letters-2Site} and \eqref{equ:3Site-Letters}.
In this way, it is straightforward to generate the letters for arbitrary graphs. 
The size of the alphabet depends on the geometry of a graph, and grows polynomially with the number of vertices. For example, the number of letters for the $n$-site chain is given by $2n^2-2n+1$, i.e.~$5, 13, 25,\,\cdots$ for $n=2,3,4,\, \cdots$\,.
 
  \paragraph{Basis functions} Our remaining task is then to predict the matrices $\alpha_i$ in (\ref{equ:connection}). 
To do this, we draw all ``complete tubings" of the marked graph. Each distinct such tubing is then associated to a unique basis function. Unlike the tubings describing letters, the tubings representing basis functions can be disconnected.

\vskip 4pt
For the three-site chain, the set of complete tubings and associated basis functions are 
\beq
\begin{aligned}
\psi\,\   \raisebox{-3pt}{\includegraphics[scale=1]{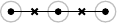}}  
\quad & \phantom{Q_1}F\,\   \raisebox{-3pt}{\includegraphics[scale=1]{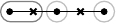}}   
\quad && \phantom{q_1} f\,\ \raisebox{-3pt}{\includegraphics[scale=1]{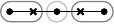}}  
 &&& \phantom{Z}g\,\ \raisebox{-3pt}{\includegraphics[scale=1]{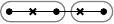}} 
\\
& \phantom{Q_1}\tilde F\,\  \raisebox{-3pt}{\includegraphics[scale=1]{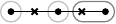}}  
&& \phantom{f} q_1\,\ \raisebox{-3pt}{\includegraphics[scale=1]{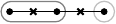}} 
&&& \phantom{Z}\tilde g\,\ \raisebox{-3pt}{\includegraphics[scale=1]{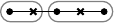}} 
 \\
& \phantom{F} Q_{1}\,\ \raisebox{-3pt}{\includegraphics[scale=1]{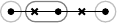}}  
&& \phantom{f}q_2\,\ \raisebox{-3pt}{\includegraphics[scale=1]{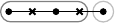}} 
&&&\phantom{g}Z\,\ \raisebox{-3pt}{\includegraphics[scale=1]{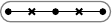}} 
\\
& \phantom{F} Q_{2} \,\ \raisebox{-3pt}{\includegraphics[scale=1]{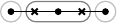}}  
&& \phantom{f}q_3\,\ \raisebox{-3pt}{\includegraphics[scale=1]{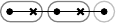}} 
\\
& \phantom{F} Q_3\,\ \raisebox{-3pt}{\includegraphics[scale=1]{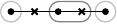}}  
&& \phantom{f}\tilde q_1\,\  \raisebox{-3pt}{\includegraphics[scale=1]{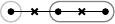}} 
\\
& && \phantom{f}\tilde q_2\,\ \raisebox{-3pt}{\includegraphics[scale=1]{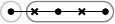}}  
\\
& && \phantom{f}\tilde q_3\,\ \raisebox{-3pt}{\includegraphics[scale=1]{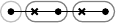}} 
\label{equ:SourceFunctions}
\end{aligned}
\eeq
These tubings are similar to those in \eqref{equ:splitting}, except that we have added extra circles without crosses (in {\color{lightgray}light gray}) to create complete tubings. The total number of basis functions is\footnote{Note that this is less than the number of bounded regions created by the singularities of the integrand in \eqref{equ:ThreeSite}, which is 25.  The latter predicts the size of the basis for generic twisted integrals, while our integrals are non-generic (only the coordinate lines are twisted), so that the differential equations for the wavefunction require only a smaller number of basis functions.} 
\beq
1+5+7+3=16\,,
\eeq
while the alternating sum of the functions appearing at each level vanishes, $1-5+7-3=0$.

\paragraph{Differential equations} In the following, we present a set of rules that will allow us to predict the differential equations for each basis function, such as those shown for the three-site chain in~\eqref{equ:SourceFunctions}. The rules are remarkably simple and universal.

\vskip 4pt
We start with the graph tubing associated to a ``parent function" of interest and then
generate a ``family tree" of its ``descendants"  according to the following rules:
\begin{enumerate}
\item {\bf Activation}\ \ We first move through the graph and ``activate" the tube enclosing each vertex.  Each activation forms a branch of the tree. These activated tubes become letters in the differential equation.

\vskip 4pt
 As an illustrative example, consider the function $Q_1$ in \eqref{equ:SourceFunctions}. Activation leads to the following three letters:
 \beq
 \begin{tikzpicture}[baseline=(current  bounding  box.center)]
\node[inner sep=0pt] at (0,0)
    {\includegraphics[scale=1]{Figures/Tubings/three/tree/threeQ1}}; 
\draw [color=gray,thick,-stealth] (1.3,0) -- (2.5,0);
\draw [color=gray,thick,-stealth] (1.3,0.15) -- (2.5,0.6);
\draw [color=gray,thick,-stealth] (1.3,-0.15) -- (2.5,-0.6);
\node[inner sep=0pt] at (3.8,0.75)
    {\includegraphics[scale=1]{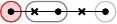}};
\node[inner sep=0pt] at (3.8,0)
    {\includegraphics[scale=1]{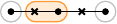}}; 
\node[inner sep=0pt] at (3.8,-0.75)
    {\includegraphics[scale=1]{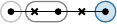}};   
\end{tikzpicture}
\eeq

\vskip 4pt
\noindent
Each graph with an activated tube can then produce further descendants.

\item {\bf Growth and merger}\ \ Activated tubes without any cross inside them can ``grow" to enclose one or more  adjacent crosses.
All possible ways in which this can happen produce new branches of the tree. If the grown tube intersects another tube, they ``merge" (i.e.~their union becomes activated).

\vskip 4pt
Returning to the example of the function $Q_1$, we have growth and merger lead to the following descendants:
 \beq
 \begin{tikzpicture}[baseline=(current  bounding  box.center)]
\node[inner sep=0pt] at (0,0)
    {\includegraphics[scale=1]{Figures/Tubings/three/tree/threeQ1}}; 
\draw [color=gray,thick,-stealth] (1.3,0) -- (2.5,0);
\draw [color=gray,thick,-stealth] (1.3,0.15) -- (2.5,0.6);
\draw [color=gray,thick,-stealth] (1.3,-0.15) -- (2.5,-0.6);
\node[inner sep=0pt] at (3.8,0.75)
    {\includegraphics[scale=1]{Figures/Tubings/three/tree/threeQ1a}};
\node[inner sep=0pt] at (3.8,0)
    {\includegraphics[scale=1]{Figures/Tubings/three/tree/threeQ1b}}; 
\node[inner sep=0pt] at (3.8,-0.75)
    {\includegraphics[scale=1]{Figures/Tubings/three/tree/threeQ1c}};   
\draw [color=gray,thick,-stealth] (5.1,0.75) -- (6.1,0.75);
\draw [color=gray,thick,-stealth] (5.1,-0.75) -- (6.1,-0.75);
\node[inner sep=0pt] at (7.3,0.75)
    {\includegraphics[scale=1]{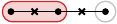}};
\node[inner sep=0pt] at (7.3,-0.75)
    {\includegraphics[scale=1]{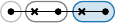}};
\end{tikzpicture}
\eeq

\vskip 4pt
\noindent
In the bottom branch, we see the growing tube encircling its neighboring cross, while in the top branch, the growing tube has merged with its neighbor: {\raisebox{-3pt}{\includegraphics[scale=1]{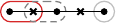}} $\equiv$ \raisebox{-3pt}{\includegraphics[scale=1]{Figures/Tubings/three/tree/threeq1ab}} .} 
\item {\bf Absorption}\ \ If an activated tube is adjacent to another tube with a cross it further merges with this tube (``absorbs" it). This absorption is directional and only occurs if the cross in the activated tube points in the direction of the other tube.

\vskip 4pt
In the tree of the function $Q_1$, there is one instance of the absorption phenomenon:
 \beq
 \begin{tikzpicture}[baseline=(current  bounding  box.center)]
\node[inner sep=0pt] at (0,0)
    {\includegraphics[scale=1]{Figures/Tubings/three/tree/threeQ1}}; 
\draw [color=gray,thick,-stealth] (1.3,0) -- (2.5,0);
\draw [color=gray,thick,-stealth] (1.3,0.15) -- (2.5,0.6);
\draw [color=gray,thick,-stealth] (1.3,-0.15) -- (2.5,-0.6);
\node[inner sep=0pt] at (3.8,0.75)
    {\includegraphics[scale=1]{Figures/Tubings/three/tree/threeQ1a}};
\node[inner sep=0pt] at (3.8,0)
    {\includegraphics[scale=1]{Figures/Tubings/three/tree/threeQ1b}}; 
\node[inner sep=0pt] at (3.8,-0.75)
    {\includegraphics[scale=1]{Figures/Tubings/three/tree/threeQ1c}};   
\draw [color=gray,thick,-stealth] (5.1,0.75) -- (6.1,0.75);
\draw [color=gray,thick,-stealth] (5.1,-0.75) -- (6.1,-0.75);
\node[inner sep=0pt] at (7.3,0.75)
    {\includegraphics[scale=1]{Figures/Tubings/three/tree/threeq1ab}};
\node[inner sep=0pt] at (7.3,-0.75)
    {\includegraphics[scale=1]{Figures/Tubings/three/tree/threeqt3c}};
\draw [color=gray,thick,-stealth] (8.5,-0.75) -- (9.5,-0.75);
\node[inner sep=0pt] at (10.7,-0.75)
    {\includegraphics[scale=1]{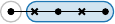}};
\end{tikzpicture}\label{equ:Q1tree}
\eeq
We see that in the bottom branch the neighboring tube is absorbed (but not in the middle branch, because there the second non-active tube does not contain a cross).
\end{enumerate}
Remarkably, this is it! These simple rules are enough to predict all differential equations by hand.

\newpage
The following procedure then lets us write down the complete set of differential equations directly from the tree structures like in (\ref{equ:Q1tree}):
\begin{enumerate}

\item To each graph in the tree, we assign the function corresponding to the graph tubing (ignoring whether tubes are active or not). The sign of the function is $(-1)^{N_{a}}$, where $N_\text{a}$ is the number of absorptions along the path from the original parent graph. If the activated tube at first step of the tree contains $n_{v}$ vertices, we include a factor of~$n_{v}$.

For the tree of the function $Q_1$, the assignments are
 \beq
 \begin{tikzpicture}[baseline=(current  bounding  box.center)]
\node[inner sep=0pt] at (0,0)
    {\includegraphics[scale=1]{Figures/Tubings/three/tree/threeQ1}}; 
\draw [color=gray,thick,-stealth] (1.3,0) -- (2.5,0);
\draw [color=gray,thick,-stealth] (1.3,0.15) -- (2.5,0.6);
\draw [color=gray,thick,-stealth] (1.3,-0.15) -- (2.5,-0.6);
\node[inner sep=0pt] at (3.8,0.75)
    {\includegraphics[scale=1]{Figures/Tubings/three/tree/threeQ1a}};
\node[inner sep=0pt] at (3.8,0)
    {\includegraphics[scale=1]{Figures/Tubings/three/tree/threeQ1b}}; 
\node[inner sep=0pt] at (3.8,-0.75)
    {\includegraphics[scale=1]{Figures/Tubings/three/tree/threeQ1c}};   
     \draw [color=gray,thick,line width=0.5pt, dashed] (1.95+0.8,-1) -- (1.95+0.8,1) --  (4.05+0.8,1) -- (4.05+0.8,-1) -- (1.95+0.8,-1); 
      \node[below] at (3.8,-1)  {$Q_1$};
\draw [color=gray,thick,-stealth] (5.1,0.75) -- (6.1,0.75);
\draw [color=gray,thick,-stealth] (5.1,-0.75) -- (6.1,-0.75);
\node[inner sep=0pt] at (7.3,0.75)
    {\includegraphics[scale=1]{Figures/Tubings/three/tree/threeq1ab}};
          \node[above] at (7.3,0.95)  {$q_1$}; 
\node[inner sep=0pt] at (7.3,-0.75)
    {\includegraphics[scale=1]{Figures/Tubings/three/tree/threeqt3c}};
      \node[below] at (7.3,-0.95)  {$\tilde q_3$}; 
\draw [color=gray,thick,-stealth] (8.5,-0.75) -- (9.5,-0.75);
\node[inner sep=0pt] at (10.7,-0.75)
    {\includegraphics[scale=1]{Figures/Tubings/three/tree/threeqt2bc}};
         \node[below] at (10.7,-0.95)  {$(-\tilde q_2)$}; 
\end{tikzpicture}
\label{equ:tree-Q1-X}
\eeq
where the minus sign on $(-\tilde q_2)$ reflects the correct grading of a descendant function coming from a single absorption.
\item For each graph in the tree, we write the letter associated to the active tube.
\item We multiply this letter by the function associated to the graph in step 1 {\it minus} the functions associated to all of its immediate descendant graphs, with an overall constant factor $\e$. 
\end{enumerate}
The sum of all letters (dlog forms) and associated coefficients is then the differential of the parent function. 
Since all descendant functions appear in pairs with opposite signs, they cancel in the sum of all coefficients, which therefore gives $n\e$ times the parent function, where $n$ is the number of vertices. 
This also means that all the $\ud\log$ forms in off-diagonal elements of the connection matrix appear in projective-invariant combinations.

\vskip 4pt
Given the tree in (\ref{equ:tree-Q1-X}),
the differential for the function $Q_1$ is
\beq
\boxed{
\begin{aligned}
\ud Q_1
&\ =  \   \e\, \Big[\,
(Q_1 -q_1)
  \raisebox{3pt}{
 \begin{tikzpicture}[baseline=(current  bounding  box.center)]
 \draw[color=Red, fill=Red!20, line width=0.6pt] (0, 0) circle (1.75mm);  
\draw[fill] (0, 0) circle (.5mm);
\draw[fill] (0.8, 0) circle (.5mm);
\draw[fill] (1.6, 0) circle (.5mm);
\draw[thick] (0, 0) -- (1.6, 0);
\node at (0.4,0)  {\Cross};
\node at (1.2,0)  {\Cross};
\end{tikzpicture}} 
\ + \ 
Q_1
 \raisebox{3pt}{
 \begin{tikzpicture}[baseline=(current  bounding  box.center)]
\node [
        draw, color=Orange, fill=Orange!20, line width=0.6pt,
        rounded rectangle,
        minimum height = 0.9em,
        minimum width = 2.5em,
        rounded rectangle arc length = 180,
    ] at (0.6,0)
    {};
\draw[fill] (0, 0) circle (.5mm);
\draw[fill] (0.8, 0) circle (.5mm);
\draw[fill] (1.6, 0) circle (.5mm);
\draw[thick] (0, 0) -- (1.6, 0);
\node at (0.4,0)  {\Cross};
\node at (1.2,0)  {\Cross};
\end{tikzpicture}}
\  + \ 
(Q_1 - \tilde q_{3})  \raisebox{3pt}{
 \begin{tikzpicture}[baseline=(current  bounding  box.center)]
 \draw[color=Blue, fill=Blue!20, line width=0.6pt] (1.6, 0) circle (1.75mm);
\draw[fill] (0, 0) circle (.5mm);
\draw[fill] (0.8, 0) circle (.5mm);
\draw[fill] (1.6, 0) circle (.5mm);
\draw[thick] (0, 0) -- (1.6, 0);
\node at (0.4,0)  {\Cross};
\node at (1.2,0)  {\Cross};
\end{tikzpicture}} 
\\
&\hspace{1.7cm} \ \ + \   q_1
 \raisebox{3pt}{
 \begin{tikzpicture}[baseline=(current  bounding  box.center)]
\node [
        draw, color=Red, fill=Red!20, line width=0.6pt,
        rounded rectangle,
        minimum height = 0.9em,
        minimum width = 3.5em,
        rounded rectangle arc length = 180,
    ] at (0.4,0)
    {};
\draw[fill] (0, 0) circle (.5mm);
\draw[fill] (0.8, 0) circle (.5mm);
\draw[fill] (1.6, 0) circle (.5mm);
\draw[thick] (0, 0) -- (1.6, 0);
\node at (0.4,0)  {\Cross};
\node at (1.2,0)  {\Cross};
\end{tikzpicture}}
\hspace{3cm} \ \ + \ \, (\tilde q_3  + \tilde q_2)
\raisebox{3pt}{
 \begin{tikzpicture}[baseline=(current  bounding  box.center)]
\node [
        draw, color=Blue, fill=Blue!20, line width=0.6pt,
        rounded rectangle,
        minimum height = 0.9em,
        minimum width = 2.5em,
        rounded rectangle arc length = 180,
    ] at (1.4,0)
    {};
\draw[fill] (0, 0) circle (.5mm);
\draw[fill] (0.8, 0) circle (.5mm);
\draw[fill] (1.6, 0) circle (.5mm);
\draw[thick] (0, 0) -- (1.6, 0);
\node at (0.4,0)  {\Cross};
\node at (1.2,0)  {\Cross};
\end{tikzpicture}} \\
&\hspace{9.4cm}  -  \tilde q_2
 \raisebox{3pt}{
 \begin{tikzpicture}[baseline=(current  bounding  box.center)]
\node [
        draw, color=Blue, fill=Blue!20, line width=0.6pt,
        rounded rectangle,
        minimum height = 0.9em,
        minimum width = 4.5em,
        rounded rectangle arc length = 180,
    ] at (1,0)
    {};
\draw[fill] (0, 0) circle (.5mm);
\draw[fill] (0.8, 0) circle (.5mm);
\draw[fill] (1.6, 0) circle (.5mm);
\draw[thick] (0, 0) -- (1.6, 0);
\node at (0.4,0)  {\Cross};
\node at (1.2,0)  {\Cross};
\end{tikzpicture}} \ \Big]
\end{aligned}}
\label{equ:dQ1}
\eeq
The letters in the first line come from the first step in the tree. Each of these letters is multiplied by the parent function $Q_1$. In the coefficients of the first and third term, we have subtracted the descendants $q_1$ and $\tilde q_3$, respectively. The letters in the second line come from the second step in the tree and are multiplied by the relevant functions $q_1$ and $\tilde q_3$. In second term, we have subtracted the descendant function $(-\tilde q_2)$. The letter in the last line arises in the last step of the tree  and is multiplied by $(-\tilde q_2)$. 

\paragraph{Integrability}  We have shown how to construct the differentials for all basis functions. In order for these equations to be self-consistent, the $\ud^2$ of any function must vanish, so that the equations are integrable. This is related to the specific rules for assigning coefficient functions to all the letters in the differential equations. In particular, the fact that each coefficient function contains a subtraction of all immediate descendants and that each function generated by absorption is graded by a minus sign are both consequences of $\ud^2 =0$. We discuss this further in Appendix~\ref{app:locality}.

\subsection{Some Examples}
\label{ssec:examples}

To illustrate the remarkable simplicity and universality of these rules, we now present some examples.
The results for the $n$-site chain, with $n=2,3,5$, are presented in this section, while results for the $4$-site chain and $4$-site star can be found in Appendix~\ref{app:flow-examples}.
All results are derived explicitly in  {\sc Mathematica} notebooks that can be downloaded at \href{https://github.com/haydenhylee/kinematic-flow}{\faGithub}.  

\subsubsection{Two-Site Chain}

We start by revisiting the two-site chain. 
As we will see, the full set of equations---displayed graphically in Section~\ref{ssec:2site-graphical}---follows straightforwardly from our rules. 

\begin{itemize}
\item The differential of the wavefunction is determined by the following tree of graph tubings:
 \beq
 \begin{tikzpicture}[baseline=(current  bounding  box.center)]
  \node at (-1.2,0)  {$\psi$\,:};
\node[inner sep=0pt] at (0,0)
   {\includegraphics[scale=1]{Figures/Tubings/two/psi/twopsi}}; 
\draw [color=gray,thick,-stealth] (0.8,0.15) -- (1.8,0.6);
\draw [color=gray,thick,-stealth] (0.8,-0.15) -- (1.8,-0.6);
\node[inner sep=0pt] at (2.65,0.7)
    {\includegraphics[scale=1]{Figures/Tubings/two/psi/twopsia}};
\node[inner sep=0pt] at (2.65,-0.7)
    {\includegraphics[scale=1]{Figures/Tubings/two/psi/twopsib}};   
    \draw [color=gray,thick,-stealth] (3.6,0.7) -- (4.6,0.7);
    \draw [color=gray,thick,-stealth] (3.6,-0.7) -- (4.6,-0.7);
    \node[inner sep=0pt] at (5.5,0.7)
    {\includegraphics[scale=1]{Figures/Tubings/two/psi/twoFa}};
    \node at (6.4,0.7)  {$F$};
       \node[inner sep=0pt] at (5.5,-0.7)
    {\includegraphics[scale=1]{Figures/Tubings/two/psi/twoFtb}};
    \node at (6.4,-0.7)  {$\tilde F$};
 \draw [color=gray,thick,line width=0.5pt, dashed] (1.95,-0.95) -- (1.95,0.95) --  (3.35,0.95) -- (3.35,-0.95) -- (1.95,-0.95); 
  \node[below] at (2.65,-.95)  {$\psi$};
\end{tikzpicture}
\label{equ:2site-psi-tree}
\eeq
In the first step, two letters are activated around each vertex. In the second step, the activated tubes grow to enclose the neighboring cross, producing the descendant functions $F$ and $\tilde F$, respectively.
The differential of the wavefunction, $\ud \psi$, is then the sum of all letters corresponding to the activated tubes in (\ref{equ:2site-psi-tree}), with coefficients determined by the corresponding graph tubing minus the associated descendant functions. 
This explains the structure of equation (\ref{equ:TwoSite-dPsi}):
\beq
\boxed{\ud \psi
\ =  \  \e\, \Big[(\psi-F)
\  \Lap
 \ + \   F
\  \Lam
 \ + \   (\psi- \tilde F)
\  \Lbp
\ \ + \ \  \tilde F
\   \Lbm \Big]}
\eeq
Note that the source functions cancel in pairs, so that the sum over all coefficients is $2\e \hs \psi$.

\item Next, we determine the differential of the function $F$. Acting on the graph tubing associated to $F$ produces the following tree structure:
 \beq
 \begin{tikzpicture}[baseline=(current  bounding  box.center)]
  \node at (-1.2,0)  {$F$\,:};
\node[inner sep=0pt] at (0,0)
   {\includegraphics[scale=1]{Figures/Tubings/two/F/twoF}}; 
\draw [color=gray,thick,-stealth] (0.8,0.15) -- (1.8,0.6);
\draw [color=gray,thick,-stealth] (0.8,-0.15) -- (1.8,-0.6);
\node[inner sep=0pt] at (2.65,0.7)
    {\includegraphics[scale=1]{Figures/Tubings/two/F/twoFa}};
\node[inner sep=0pt] at (2.65,-0.7)
    {\includegraphics[scale=1]{Figures/Tubings/two/F/twoFb}};   
    \draw [color=gray,thick,-stealth] (3.6,-0.7) -- (4.6,-0.7);
       \node[inner sep=0pt] at (5.5,-0.7)
    {\includegraphics[scale=1]{Figures/Tubings/two/F/twoZb}};
    \node at (6.4,-0.7)  {$Z$};
 \draw [color=gray,thick,line width=0.5pt, dashed] (1.95,-0.95) -- (1.95,0.95) --  (3.35,0.95) -- (3.35,-0.95) -- (1.95,-0.95); 
  \node[below] at (2.65,-.95)  {$F$};
\end{tikzpicture}
\label{equ:2site-F-tree}
\eeq
Again, in the first step, each individual tube gets activated producing two branches in the tree. In the bottom branch, the activated tube then grows and merges with its neighbor, producing the descendant function $Z$. Following our rules for assigning coefficients to each activated letter, we then simply write down equation (\ref{equ:TwoSite-dF}): 
\beq
\boxed{
\ud F 
\ =  \   \e\, \Big[ F 
\  \Lam 
 \ +  \  (F-Z) \,\Lbp 
\ + \ 
Z  \Labb \Big] }
\eeq
\item The differential of the function $\tilde F$ is related by symmetry to that of $F$; cf.~(\ref{equ:TwoSite-dtF}).
\item
Finally, we consider the differential of the function $Z$.
In this case, the graph is enclosed by a single tubing which simply gets activated:
 \beq\begin{tikzpicture}[baseline=(current  bounding  box.center)]
  \node at (-1.2,0)  {$Z$\,:};
\node[inner sep=0pt] at (0,0)
   {\includegraphics[scale=1]{Figures/Tubings/two/Z/twoZ}}; 
\draw [color=gray,thick,-stealth] (0.8,0) -- (1.8,0);
\node[inner sep=0pt] at (2.65,0.0)
    {\includegraphics[scale=1]{Figures/Tubings/two/Z/twoZgr}};
    \node at (3.7,0)  {$2Z$};
\end{tikzpicture}
\label{equ:2site-Z-tree}
\eeq
The factor of two accounts for the fact that two vertices are enclosed in the activated tube.  Hence, we get the result~\eqref{equ:TwoSite-dZ}.

\end{itemize}

\vspace{0.1cm}
\subsubsection{Three-Site Chain}

As a slightly more complicated example, we consider the case of the three-site chain. The relevant graph tubings for the letters and basis functions were shown in (\ref{equ:3Site-Letters}) and (\ref{equ:SourceFunctions}), respectively. In the following, we will derive the complete set of equations for these basis functions using the graphical rules presented above.

\paragraph{Level 1:} 
The differential of the wavefunction is given by the following evolutionary tree:
 \beq
 \begin{tikzpicture}[baseline=(current  bounding  box.center)]
  \node at (-1.9,0)  {$\psi$\,:};
\node[inner sep=0pt] at (-0.4,0)
   {\includegraphics[scale=1]{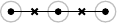}}; 
   \draw [color=gray,thick,-stealth] (0.8,0) -- (1.8,0);
\draw [color=gray,thick,-stealth] (0.8,0.15) -- (1.8,1.3);
\draw [color=gray,thick,-stealth] (0.8,-0.15) -- (1.8,-1.3);
\node[inner sep=0pt] at (3,1.4)
    {\includegraphics[scale=1]{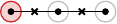}};
\node[inner sep=0pt] at (3,-1.4)
    {\includegraphics[scale=1]{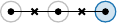}};  
    \node[inner sep=0pt] at (3,0)
    {\includegraphics[scale=1]{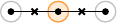}}; 
 \draw [color=gray,thick,line width=0.5pt, dashed] (1.95,-1.65) -- (1.95,1.65) --  (4.05,1.65) -- (4.05,-1.65) -- (1.95,-1.65); 
  \node[below] at (3,-1.75)  {$\psi$};
  
       \draw [color=gray,thick,-stealth] (4.2,0.15) -- (5.05,0.5);
      \draw [color=gray,thick,-stealth] (4.2,0) -- (5.05,0);
         \draw [color=gray,thick,-stealth] (4.2,-0.15) -- (5.05,-0.5);
         \node[inner sep=0pt] at (6.15,0.5)
    {\includegraphics[scale=1]{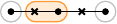}}; 
          \node[inner sep=0pt] at (6.15,0)
    {\includegraphics[scale=1]{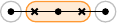}}; 
             \node[inner sep=0pt] at (6.15,-0.5)
    {\includegraphics[scale=1]{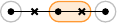}}; 
    
      \draw [color=gray,thick,-stealth] (4.2,1.4) -- (5.05,1.4);
    \draw [color=gray,thick,-stealth] (4.2,-1.4) -- (5.05,-1.4);
    \node[inner sep=0pt] at (6.15,1.4)
    {\includegraphics[scale=1]{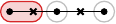}};
       \node[inner sep=0pt] at (6.15,-1.4)
    {\includegraphics[scale=1]{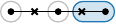}};
    
      \node[right] at (7.15,0.5)  {$Q_1$};
        \node[right] at (7.15,-0.5)  {$Q_3$};
           \node[right] at (7.15,0)  {$Q_2$};
           
            \node[right] at (7.15,1.4)  {$F$};     
                 \node[right] at (7.15,-1.4)  {$\tilde F$}; 
\end{tikzpicture}
\label{equ:3site-psi-tree}
\eeq
As in \eqref{equ:2site-psi-tree}, this describes the activation and subsequent growth of tubes around each vertex.
The only novelty is that the middle tube can grow in three different ways: i) enclosing the left cross, ii) enclosing the right cross and iii) enclosing both crosses.  Following the standard rules for assigning coefficient functions to each activated letter in the tree, we write down the result~\eqref{equ:dPsi-3S}:
\begin{equation}
\boxed{
\begin{aligned}
\ud \psi
&\ =  \ \e\, \Big[\, (\psi - F)
 \raisebox{3pt}{
 \begin{tikzpicture}[baseline=(current  bounding  box.center)]
 \draw[color=Red, fill=Red!20, line width=0.6pt] (0, 0) circle (1.75mm);
\draw[fill] (0, 0) circle (.5mm);
\draw[fill] (0.8, 0) circle (.5mm);
\draw[fill] (1.6, 0) circle (.5mm);
\draw[thick] (0, 0) -- (1.6, 0);
\node at (0.4,0)  {\Cross};
\node at (1.2,0)  {\Cross};
\end{tikzpicture}}
 \ + \ 
  (\psi -\tilde F)
 \raisebox{3pt}{
 \begin{tikzpicture}[baseline=(current  bounding  box.center)]
 \draw[color=Blue, fill=Blue!20, line width=0.6pt] (1.6, 0) circle (1.75mm);
\draw[fill] (0, 0) circle (.5mm);
\draw[fill] (0.8, 0) circle (.5mm);
\draw[fill] (1.6, 0) circle (.5mm);
\draw[thick] (0, 0) -- (1.6, 0);
\node at (0.4,0)  {\Cross};
\node at (1.2,0)  {\Cross};
\end{tikzpicture}}
 \ + \ 
\big(\psi - {\textstyle \sum}Q_i \big)  \raisebox{3pt}{
 \begin{tikzpicture}[baseline=(current  bounding  box.center)]
 \draw[color=Orange, fill=Orange!20, line width=0.6pt] (0.8, 0) circle (1.75mm);
\draw[fill] (0, 0) circle (.5mm);
\draw[fill] (0.8, 0) circle (.5mm);
\draw[fill] (1.6, 0) circle (.5mm);
\draw[thick] (0, 0) -- (1.6, 0);
\node at (0.4,0)  {\Cross};
\node at (1.2,0)  {\Cross};
\end{tikzpicture}}
 \\
&\hspace{1.1cm}\quad \ \  +  \  F
 \raisebox{3pt}{
 \begin{tikzpicture}[baseline=(current  bounding  box.center)]
\node [
        draw, color=Red, fill=Red!20, line width=0.6pt,
        rounded rectangle,
        minimum height = 0.9em,
        minimum width = 2.5em,
        rounded rectangle arc length = 180,
    ] at (0.2,0)
    {};
\draw[fill] (0, 0) circle (.5mm);
\draw[fill] (0.8, 0) circle (.5mm);
\draw[fill] (1.6, 0) circle (.5mm);
\draw[thick] (0, 0) -- (1.6, 0);
\node at (0.4,0)  {\Cross};
\node at (1.2,0)  {\Cross};
\end{tikzpicture}} 
 \hspace{1.0cm}  \ + \  \tilde F
  \raisebox{3pt}{
 \begin{tikzpicture}[baseline=(current  bounding  box.center)]
\node [
        draw, color=Blue, fill=Blue!20, line width=0.6pt,
        rounded rectangle,
        minimum height = 0.9em,
        minimum width = 2.5em,
        rounded rectangle arc length = 180,
    ] at (1.4,0)
    {};
\draw[fill] (0, 0) circle (.5mm);
\draw[fill] (0.8, 0) circle (.5mm);
\draw[fill] (1.6, 0) circle (.5mm);
\draw[thick] (0, 0) -- (1.6, 0);
\node at (0.4,0)  {\Cross};
\node at (1.2,0)  {\Cross};
\end{tikzpicture}}  
\hspace{1.5cm}\ + \   Q_{1}
\raisebox{3pt}{
 \begin{tikzpicture}[baseline=(current  bounding  box.center)]
\node [
        draw, color=Orange, fill=Orange!20, line width=0.6pt,
        rounded rectangle,
        minimum height = 0.9em,
        minimum width = 2.5em,
        rounded rectangle arc length = 180,
    ] at (0.6,0)
    {};
\draw[fill] (0, 0) circle (.5mm);
\draw[fill] (0.8, 0) circle (.5mm);
\draw[fill] (1.6, 0) circle (.5mm);
\draw[thick] (0, 0) -- (1.6, 0);
\node at (0.4,0)  {\Cross};
\node at (1.2,0)  {\Cross};
\end{tikzpicture}} 
\\[3pt]
&\hspace{9.95cm} \ \ + \   Q_{2}
\raisebox{3pt}{
 \begin{tikzpicture}[baseline=(current  bounding  box.center)]
\node [
        draw, color=Orange, fill=Orange!20, line width=0.6pt,
        rounded rectangle,
        minimum height = 0.9em,
        minimum width = 3.5em,
        rounded rectangle arc length = 180,
    ] at (0.8,0)
    {};
\draw[fill] (0, 0) circle (.5mm);
\draw[fill] (0.8, 0) circle (.5mm);
\draw[fill] (1.6, 0) circle (.5mm);
\draw[thick] (0, 0) -- (1.6, 0);
\node at (0.4,0)  {\Cross};
\node at (1.2,0)  {\Cross};
\end{tikzpicture}}
\\
&\hspace{9.95cm} \ \ + \    Q_3
\raisebox{3pt}{
 \begin{tikzpicture}[baseline=(current  bounding  box.center)]
\node [
        draw, color=Orange, fill=Orange!20, line width=0.6pt,
        rounded rectangle,
        minimum height = 0.9em,
        minimum width = 2.5em,
        rounded rectangle arc length = 180,
    ] at (1.,0)
    {};
\draw[fill] (0, 0) circle (.5mm);
\draw[fill] (0.8, 0) circle (.5mm);
\draw[fill] (1.6, 0) circle (.5mm);
\draw[thick] (0, 0) -- (1.6, 0);
\node at (0.4,0)  {\Cross};
\node at (1.2,0)  {\Cross};
\end{tikzpicture}}\ \Big] 
\end{aligned}}
\label{equ:dPsi-3S-X}
\end{equation}
As expected, the sum over all coefficients is $3\e \hs \psi$.

\paragraph{Level 2:} The differential of the wavefunction involved five source functions, associated to the following graph tubings:
\beq
\begin{aligned}
\quad & F\,\   \raisebox{-1pt}{\includegraphics[scale=1]{Figures/Tubings/three/tree/threeF}}   
\quad && Q_1\,\ \raisebox{-1pt}{\includegraphics[scale=1]{Figures/Tubings/three/tree/threeQ1}}  
\\[-1pt]
& \tilde F\,\  \raisebox{-1pt}{\includegraphics[scale=1]{Figures/Tubings/three/tree/threeFt}}  
&& Q_2\,\ \raisebox{-1pt}{\includegraphics[scale=1]{Figures/Tubings/three/tree/threeQ2}} 
\\
& 
&& Q_3\,\  \raisebox{-1pt}{\includegraphics[scale=1]{Figures/Tubings/three/tree/threeQ3}} 
\end{aligned}
\nonumber
\eeq
We will now predict the differentials of these functions.

\begin{itemize}
\item The evolutionary tree for the function $F$ is  
 \beq
 \begin{tikzpicture}[baseline=(current  bounding  box.center)]
  \node at (-1.9,0)  {$F$\,:};
\node[inner sep=0pt] at (-0.4,0)
   {\includegraphics[scale=1]{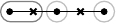}}; 
   \draw [color=gray,thick,-stealth] (0.8,0) -- (1.8,0);
\draw [color=gray,thick,-stealth] (0.8,0.15) -- (1.8,1.3);
\draw [color=gray,thick,-stealth] (0.8,-0.15) -- (1.8,-1.3);
\node[inner sep=0pt] at (3,1.4)
    {\includegraphics[scale=1]{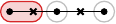}};
\node[inner sep=0pt] at (3,-1.4)
    {\includegraphics[scale=1]{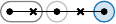}};  
    \node[inner sep=0pt] at (3,0)
    {\includegraphics[scale=1]{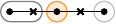}}; 
 \draw [color=gray,thick,line width=0.5pt, dashed] (1.95,-1.65) -- (1.95,1.65) --  (4.05,1.65) -- (4.05,-1.65) -- (1.95,-1.65); 
  \node[below] at (3,-1.75)  {$F$};
  
       \draw [color=gray,thick,-stealth] (4.2,0.15) -- (5.05,0.5);
      \draw [color=gray,thick,-stealth] (4.2,0) -- (5.05,0);
         \draw [color=gray,thick,-stealth] (4.2,-0.15) -- (5.05,-0.5);
         \node[inner sep=0pt] at (6.15,0.5)
    {\includegraphics[scale=1]{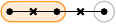}}; 
          \node[inner sep=0pt] at (6.15,0)
    {\includegraphics[scale=1]{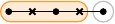}}; 
             \node[inner sep=0pt] at (6.15,-0.5)
    {\includegraphics[scale=1]{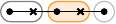}};

    \draw [color=gray,thick,-stealth] (4.2,-1.4) -- (5.05,-1.4);
       \node[inner sep=0pt] at (6.15,-1.4)
    {\includegraphics[scale=1]{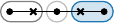}};
    
      \node[right] at (7.15,0.5)  {$q_1$};
        \node[right] at (7.15,-0.5)  {$q_3$};
           \node[right] at (7.15,0)  {$q_2$};
              
                 \node[right] at (7.15,-1.4)  {$f$}; 
\end{tikzpicture}
\label{equ:3site-F-tree}
\eeq
Here, we see two mergers to produce the tubings associated to the functions $q_1$ and $q_2$.
The tree in (\ref{equ:3site-F-tree}) then directly implies the result (\ref{equ:FR-1}):
\beq
\boxed{
\begin{aligned}
 \ud F = \e \Big[\,F\ \LLam \ +\ (F-f) &\LLcp\phantom{\Big]}  +\ (F-{\textstyle \sum}q_i) \LLbpp\phantom{\Big]}\\
 +\ f &\LLcm\phantom{\Big]} \hspace{1.32cm} +\ q_1\ \LLabpo\phantom{\Big]}\\
 &\hspace{3.55cm} +\, q_2\ \LLabmo\phantom{\Big]}\\
&\hspace{3.55cm} +\,q_3\ \LLbpm \Big]
  \end{aligned}}
  \eeq

\item The tree for the function $Q_1$ was already shown when we introduced our graphical rules in the previous section. The result for the function $Q_3$ is related to this by symmetry and therefore won't be shown explicitly. 

\item The tree for the function $Q_2$ is
 \beq
 \begin{tikzpicture}[baseline=(current  bounding  box.center)]
  \node at (-1.9,0)  {$Q_2$\,:};
\node[inner sep=0pt] at (-0.4,0)
   {\includegraphics[scale=1]{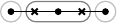}}; 
   \draw [color=gray,thick,-stealth] (0.8,0) -- (1.8,0);
\draw [color=gray,thick,-stealth] (0.8,0.15) -- (1.8,1.3);
\draw [color=gray,thick,-stealth] (0.8,-0.15) -- (1.8,-1.3);
\node[inner sep=0pt] at (3,1.4)
    {\includegraphics[scale=1]{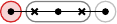}};
\node[inner sep=0pt] at (3,-1.4)
    {\includegraphics[scale=1]{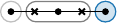}};  
    \node[inner sep=0pt] at (3,0)
    {\includegraphics[scale=1]{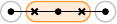}}; 
 \draw [color=gray,thick,line width=0.5pt, dashed] (1.95,-1.65) -- (1.95,1.65) --  (4.05,1.65) -- (4.05,-1.65) -- (1.95,-1.65); 
  \node[below] at (3,-1.75)  {$Q_2$};
    \draw [color=gray,thick,-stealth] (4.2,-1.4) -- (5.05,-1.4);
       \node[inner sep=0pt] at (6.15,-1.4)
    {\includegraphics[scale=1]{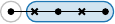}};

    \draw [color=gray,thick,-stealth] (4.2,1.4) -- (5.05,1.4);
       \node[inner sep=0pt] at (6.15,1.4)
    {\includegraphics[scale=1]{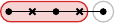}};

                 \node[right] at (7.15,1.4)  {$q_2$};                       
                 \node[right] at (7.15,-1.4)  {$\tilde q_2$}; 
\end{tikzpicture}
\label{equ:3site-Q2-tree}
\eeq
We see one pure activation in the middle branch and then the standard activation, growth and merger phenomena in the top and bottom branches.
The tree then implies directly the result (\ref{equ:dQ2}):
\beq
\boxed{
\begin{aligned}
 \ud Q_2 = \e \Big[ Q_2\ \LLbmm  +\ (Q_2-q_2)\ &\LLap\phantom{\Big]}  +\ (Q_2-\tilde q_2) \LLcp\phantom{\Big]}\\
  	 +\ q_2\ &\LLabmr\phantom{\Big]} \hspace{1.15cm} +\ \tilde q_2\ \LLbcmb\Big]
	\end{aligned}}
	\eeq
\end{itemize}

\paragraph{Level 3:}  The differentials at Level 2 have produced 7 source functions corresponding to the following graph tubings:
\begin{align}
\begin{split}
f\,\  \raisebox{-3pt}{\includegraphics[scale=1]{Figures/Tubings/three/tree/threeff}}    \hspace{1.05cm}
&q_1\,\  \raisebox{-3pt}{\includegraphics[scale=1]{Figures/Tubings/three/tree/threeqq1}}  \hspace{1cm} 
\tilde q_1\,\ \raisebox{-3pt}{\includegraphics[scale=1]{Figures/Tubings/three/tree/threeqqt1}} \\
&q_2\,\ \raisebox{-3pt}{\includegraphics[scale=1]{Figures/Tubings/three/tree/threeqq2}} \hspace{1cm} 
\tilde q_2\,\ \raisebox{-3pt}{\includegraphics[scale=1]{Figures/Tubings/three/tree/threeqqt2}} \\
&q_3\,\ \raisebox{-3pt}{\includegraphics[scale=1]{Figures/Tubings/three/tree/threeqq3}}  \hspace{1.cm}
 \tilde q_3\,\ \raisebox{-3pt}{\includegraphics[scale=1]{Figures/Tubings/three/tree/threeqqt3}}
\end{split}
\label{equ:SourceFunctions-Level2}
\end{align}
The fact that some of these tubings have disconnected tubes with crosses (in {\color{darkgray}dark gray}) will have interesting consequences.
Let's look at the differentials of these functions.

\begin{itemize}
\item The tree for the function $f$ is
 \beq
 \begin{tikzpicture}[baseline=(current  bounding  box.center)]
  \node at (-1.9,0)  {$f$\,:};
\node[inner sep=0pt] at (-0.4,0)
   {\includegraphics[scale=1]{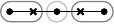}}; 
   \draw [color=gray,thick,-stealth] (0.8,0) -- (1.8,0);
\draw [color=gray,thick,-stealth] (0.8,0.15) -- (1.8,1.3);
\draw [color=gray,thick,-stealth] (0.8,-0.15) -- (1.8,-1.3);
\node[inner sep=0pt] at (3,1.4)
    {\includegraphics[scale=1]{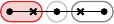}};
\node[inner sep=0pt] at (3,-1.4)
    {\includegraphics[scale=1]{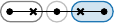}};  
    \node[inner sep=0pt] at (3,0)
    {\includegraphics[scale=1]{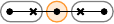}}; 
 \draw [color=gray,thick,line width=0.5pt, dashed] (1.95,-1.65) -- (1.95,1.65) --  (4.05,1.65) -- (4.05,-1.65) -- (1.95,-1.65); 
  \node[below] at (3,-1.75)  {$f$};
  
       \draw [color=gray,thick,-stealth] (4.2,0.15) -- (5.05,0.5);
      \draw [color=gray,thick,-stealth] (4.2,0) -- (5.05,0);
         \draw [color=gray,thick,-stealth] (4.2,-0.15) -- (5.05,-0.5);
         \node[inner sep=0pt] at (6.15,0.5)
    {\includegraphics[scale=1]{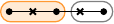}}; 
          \node[inner sep=0pt] at (6.15,-0.5)
    {\includegraphics[scale=1]{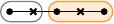}}; 
             \node[inner sep=0pt] at (6.15,0)
    {\includegraphics[scale=1]{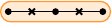}}; 
    
          \node[right] at (7.15,0.5)  {$g$};
        \node[right] at (7.15,0)  {$Z$};
           \node[right] at (7.15,-0.5)  {$\tilde g$};
         
\end{tikzpicture}
\label{equ:3site-ff-tree}
\eeq
The descendant functions $g$, $\tilde g$, $Z$ are all created by the mergers of the growing tubes with the original tubes of the function $f$.  In one case, the merger comes from the overlap of three tubes:  \raisebox{-3pt}{
\begin{tikzpicture}
\node [
        draw, color=gray, line width=0.6pt, densely dashed,
        rounded rectangle,
        minimum height = 0.9em,
        minimum width = 2.5em,
        rounded rectangle arc length = 180,
    ] at (4.2,-0.9)
    {};
    \node [
        draw, color=gray, line width=0.6pt, densely dashed,
        rounded rectangle,
        minimum height = 0.9em,
        minimum width = 2.5em,
        rounded rectangle arc length = 180,
    ] at (5.4,-0.9)
    {};
      \node [
        draw, color=Orange, line width=0.6pt,
        rounded rectangle,
        minimum height = 0.9em,
        minimum width = 3.5em,
        rounded rectangle arc length = 180,
    ] at (4.8,-0.9)
    {};
\draw[fill] (4,-0.9) circle (.5mm);
\draw[fill] (4.8,-0.9) circle (.5mm);
\draw[fill] (5.6,-0.9) circle (.5mm);
\draw[thick] (4,-0.9) -- (5.6,-0.9);
\node at (4.4,-0.9)  {\Cross};
\node at (5.2,-0.9)  {\Cross};
\node[right] at (5.8,-0.9)  {$\equiv$};
    \node [
        draw, color=Orange, fill=Orange!20, line width=0.6pt,
        rounded rectangle,
        minimum height = 0.9em,
        minimum width = 5.5em,
        rounded rectangle arc length = 180,
    ] at (7.5,-0.9)
    {};

\draw[fill] (6.7, -0.9) circle (.5mm);
\draw[fill] (7.5, -0.9) circle (.5mm);
\draw[fill] (8.3, -0.9) circle (.5mm);
\draw[thick] (6.7, -0.9) -- (8.3,-0.9);
\node at (7.1,-0.9)  {\Cross};
\node at (7.9,-0.9)  {\Cross};
\end{tikzpicture}} .
Assigning the correct coefficients to each letter in the tree then straightforwardly reproduces the result \eqref{equ:f}: 
\beq
\boxed{
\begin{aligned}
 \ud f = \e \Big[ f\ \LLam \ +\ f \LLcm  +\ (f-g-\tilde g-Z) &\LLbpp\phantom{\Big]}\\
  	+\ g\ &\LLabpo\phantom{\Big]}\\
  	 +\ \tilde g\ &\LLbcpo\phantom{\Big]}\\
   +\ Z\ &\LLabco \Big]
\end{aligned}}
\eeq	
\item The tree for the function $q_1$ is
 \beq
 \begin{tikzpicture}[baseline=(current  bounding  box.center)]
  \node at (-1.9,0)  {$q_1$\,:};
\node[inner sep=0pt] at (-0.4,0)
   {\includegraphics[scale=1]{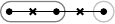}}; 
\draw [color=gray,thick,-stealth] (0.8,0.15) -- (1.8,0.6);
\draw [color=gray,thick,-stealth] (0.8,-0.15) -- (1.8,-0.6);
\node[inner sep=0pt] at (3,0.7)
    {\includegraphics[scale=1]{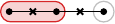}};
\node[inner sep=0pt] at (3,-0.7)
    {\includegraphics[scale=1]{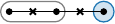}};  
 \draw [color=gray,thick,line width=0.5pt, dashed] (1.95,-0.95) -- (1.95,0.95) --  (4.05,0.95) -- (4.05,-0.95) -- (1.95,-0.95); 
  \node[below] at (3,-1.05)  {$q_1$};
  \node[above] at (3,1.05)  {$2q_1$};  
  
    \draw [color=gray,thick,-stealth] (4.2,-0.7) -- (5.0,-0.7);
       \node[inner sep=0pt] at (6.1,-0.7)
    {\includegraphics[scale=1]{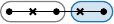}};                       
                 \node[right] at (7.1,-0.7)  {$g$}; 
                  \draw [color=gray,thick,-stealth] (7.55,-0.7) -- (8.3,-0.7); 
                       \node[inner sep=0pt] at (9.4,-0.7)
                      {\includegraphics[scale=1]{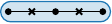}};        
                        \node[right] at (10.3,-0.7)  {$(-Z)$};  
\end{tikzpicture}
\label{equ:3site-q1-tree}
\eeq
Note that the activated letter in the top branch is multiplied by $2 q_1$ because in contains two enclosed vertices.
In the bottom branch, we first have the standard growth of the activated letter (to produce the function $g$) and then the absorption of the neighboring tube (to give the function $Z$). Importantly, the coefficient for the last tubing comes with an extra minus sign. Taking this into account, the tree in (\ref{equ:3site-q1-tree}) leads to (\ref{equ:dq1}): 
\beq
\boxed{
\begin{aligned}
  \ud q_1 = \e \Big[\ 2q_1\ \LLabpr \ +\ (q_1-g) &\LLcp\phantom{\Big]} \\
   	+\ (g+Z) &\LLcm \phantom{\Big]} \\
 -\ Z\ &\LLabcb\ \Big] 
\end{aligned}}
\eeq

\item The tree for the function $q_2$ is
 \beq
 \begin{tikzpicture}[baseline=(current  bounding  box.center)]
  \node at (-1.9,0)  {$q_2$\,:};
\node[inner sep=0pt] at (-0.4,0)
   {\includegraphics[scale=1]{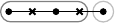}}; 
\draw [color=gray,thick,-stealth] (0.8,0.15) -- (1.8,0.6);
\draw [color=gray,thick,-stealth] (0.8,-0.15) -- (1.8,-0.6);
\node[inner sep=0pt] at (3,0.7)
    {\includegraphics[scale=1]{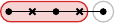}};
\node[inner sep=0pt] at (3,-0.7)
    {\includegraphics[scale=1]{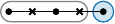}};  
 \draw [color=gray,thick,line width=0.5pt, dashed] (1.95,-0.95) -- (1.95,0.95) --  (4.05,0.95) -- (4.05,-0.95) -- (1.95,-0.95); 
  \node[below] at (3,-1.05)  {$q_2$};
  \node[above] at (3,1.05)  {$2q_2$};  
  
    \draw [color=gray,thick,-stealth] (4.2,-0.7) -- (5.0,-0.7);
       \node[inner sep=0pt] at (6.1,-0.7)
    {\includegraphics[scale=1]{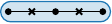}};                       
                 \node[right] at (7.1,-0.7)  {$Z$}; 
\end{tikzpicture}
\label{equ:3site-q2-tree}
\eeq
Here, we just see the standard activation, growth and merger phenomena in action.
The tree in \eqref{equ:3site-q2-tree} then correctly leads to (\ref{equ:dq2}):
\beq
\boxed{
\begin{aligned}
 \ud q_2 = \e \Big[\,2q_2\ \LLabmr \ +\ (q_2-Z) &\LLcp\phantom{\Big]} \\
  	+\ Z\ &\LLabcb\,\Big] 
\end{aligned}}
\eeq

\item The tree for the function $q_3$ is
 \beq
 \begin{tikzpicture}[baseline=(current  bounding  box.center)]
  \node at (-1.9,0)  {$q_3$\,:};
\node[inner sep=0pt] at (-0.4,0)
   {\includegraphics[scale=1]{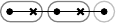}}; 
   \draw [color=gray,thick,-stealth] (0.8,0) -- (1.8,0);
\draw [color=gray,thick,-stealth] (0.8,0.15) -- (1.8,1.3);
\draw [color=gray,thick,-stealth] (0.8,-0.15) -- (1.8,-1.3);
\node[inner sep=0pt] at (3,1.4)
    {\includegraphics[scale=1]{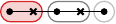}};
\node[inner sep=0pt] at (3,-1.4)
    {\includegraphics[scale=1]{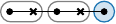}};  
    \node[inner sep=0pt] at (3,0)
    {\includegraphics[scale=1]{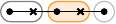}}; 
 \draw [color=gray,thick,line width=0.5pt, dashed] (1.95,-1.65) -- (1.95,1.65) --  (4.05,1.65) -- (4.05,-1.65) -- (1.95,-1.65); 
  \node[below] at (3,-1.75)  {$q_3$};

    \draw [color=gray,thick,-stealth] (4.2,-1.4) -- (5.05,-1.4);
       \node[inner sep=0pt] at (6.15,-1.4)
    {\includegraphics[scale=1]{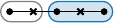}};

    \draw [color=gray,thick,-stealth] (4.2,1.4) -- (5.05,1.4);
       \node[inner sep=0pt] at (6.15,1.4)
    {\includegraphics[scale=1]{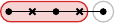}};

                 \node[right] at (7.15,1.4)  {$(-q_2)$};                       
                 \node[right] at (7.15,-1.4)  {$\tilde g$}; 
\end{tikzpicture}
\label{equ:3site-q3-tree}
\eeq
In the top branch, we first have the activation of the tube around vertex $1$ and then the absorption of the neighboring tube (to produce the tubing of the function $q_2$).   In the middle branch, we have the activation of the tube around vertex $2$, but {\it no} absorption, demonstrating the directional character of the absorption rule.
Finally, in the bottom branch, we have the standard activation, growth and merger of tubes.
The tree in \eqref{equ:3site-q3-tree} then gives (\ref{equ:dq3x}): \beq
\boxed{
\begin{aligned}
  \ud q_3 = \e \Big[ \, q_3\, \LLbpm \ +\ (q_3+q_2)\, &\LLam\phantom{\Big]} \ +\ (q_3-\tilde g) \LLcp\phantom{\Big]}\\
  	 -\ q_2\ &\LLabmr\phantom{\Big]} \hspace{1.05cm}\ +\ \tilde g\ \LLbcpb\, \Big]
\end{aligned}}
\eeq

\item The trees for the functions $\tilde q_{1,2,3}$ (and the associated differentials) are related by symmetry to those of $q_{1,2,3}$ and therefore are not reproduced here.
\end{itemize}

\paragraph{Level 4:}  At Level $3$, we obtained three new source functions corresponding to the following graph tubings: 
\begin{align}
\begin{split}
&g\, \raisebox{3pt}{
 \begin{tikzpicture}[baseline=(current  bounding  box.center)]
 \node [
        draw, color=gray, line width=0.6pt,
        rounded rectangle,
        minimum height = 0.9em,
        minimum width = 3.5em,
        rounded rectangle arc length = 180,
    ] at (0.4,0)
    {};
 \node [
        draw, color=gray, line width=0.6pt,
        rounded rectangle,
        minimum height = 0.9em,
        minimum width = 2.5em,
        rounded rectangle arc length = 180,
    ] at (1.4,0)
    {};
\draw[fill] (0, 0) circle (.5mm);
\draw[fill] (0.8, 0) circle (.5mm);
\draw[fill] (1.6, 0) circle (.5mm);
\draw[thick] (0, 0) -- (1.6, 0);
\node at (0.4,0)  {\Cross};
\node at (1.2,0)  {\Cross};
\end{tikzpicture}}\hspace{1cm}
\tilde g\, \raisebox{3pt}{
 \begin{tikzpicture}[baseline=(current  bounding  box.center)]
\node [
        draw, color=gray, line width=0.6pt,
        rounded rectangle,
        minimum height = 0.9em,
        minimum width = 3.5em,
        rounded rectangle arc length = 180,
    ] at (1.2,0)
    {};
     \node [
        draw, color=gray, line width=0.6pt,
        rounded rectangle,
        minimum height = 0.9em,
        minimum width = 2.5em,
        rounded rectangle arc length = 180,
    ] at (0.2,0)
    {};
\draw[fill] (0, 0) circle (.5mm);
\draw[fill] (0.8, 0) circle (.5mm);
\draw[fill] (1.6, 0) circle (.5mm);
\draw[thick] (0, 0) -- (1.6, 0);
\node at (0.4,0)  {\Cross};
\node at (1.2,0)  {\Cross};
\end{tikzpicture}} 
 \hspace{1cm}
Z\, \raisebox{3pt}{
 \begin{tikzpicture}[baseline=(current  bounding  box.center)]
\node [
        draw, color=gray, line width=0.6pt,
        rounded rectangle,
        minimum height = 0.9em,
        minimum width = 5.5em,
        rounded rectangle arc length = 180,
    ] at (0.8,0)
    {};
\draw[fill] (0, 0) circle (.5mm);
\draw[fill] (0.8, 0) circle (.5mm);
\draw[fill] (1.6, 0) circle (.5mm);
\draw[thick] (0, 0) -- (1.6, 0);
\node at (0.4,0)  {\Cross};
\node at (1.2,0)  {\Cross};
\end{tikzpicture}}
\\
\end{split}
\label{equ:SourceFunctions-Level3}
\end{align}
To complete our analysis, we consider their differentials.

\begin{itemize}
\item The tree for the function $g$ is
 \beq
 \begin{tikzpicture}[baseline=(current  bounding  box.center)]
  \node at (-1.9,0)  {$g$\,:};
\node[inner sep=0pt] at (-0.4,0)
   {\includegraphics[scale=1]{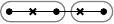}}; 
\draw [color=gray,thick,-stealth] (0.8,0.15) -- (1.8,0.6);
\draw [color=gray,thick,-stealth] (0.8,-0.15) -- (1.8,-0.6);
\node[inner sep=0pt] at (3,0.7)
    {\includegraphics[scale=1]{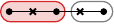}};
\node[inner sep=0pt] at (3,-0.7)
    {\includegraphics[scale=1]{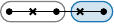}};  
 \draw [color=gray,thick,line width=0.5pt, dashed] (1.95,-0.95) -- (1.95,0.95) --  (4.05,0.95) -- (4.05,-0.95) -- (1.95,-0.95); 
  \node[below] at (3,-1.05)  {$g$};
  \node[above] at (3,1.05)  {$2g$};  
  
    \draw [color=gray,thick,-stealth] (4.2,-0.7) -- (5.0,-0.7);
       \node[inner sep=0pt] at (6.1,-0.7)
    {\includegraphics[scale=1]{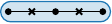}};                       
                 \node[right] at (7.1,-0.7)  {$(-Z)$}; 
\end{tikzpicture}
\label{equ:3site-g-tree}
\eeq
which is similar to the tree in (\ref{equ:3site-q2-tree}) for $q_2$,
expect that the function $Z$ is created by absorption (rather than merger) and therefore comes with a minus sign.
We then read off correctly the result (\ref{equ:dX-gp}): 
\beq
\boxed{
\begin{aligned}
\ud g = \e \Big[ \, 2g\ \LLabpr \ +\ (g+Z)\, &\LLcm\phantom{\Big]} \\
  	-\ Z\ &\LLabcb\Big] 
\end{aligned}}
\eeq

\item Finally, the tree for the function $Z$ is simply
 \beq
 \begin{tikzpicture}[baseline=(current  bounding  box.center)]
  \node at (-1.9,0)  {$Z$\,:};
\node[inner sep=0pt] at (-0.4,0)
   {\includegraphics[scale=1]{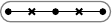}}; 
   \draw [color=gray,thick,-stealth] (0.8,0) -- (1.8,0);
    \node[inner sep=0pt] at (3,0)
    {\includegraphics[scale=1]{Figures/Tubings/three/Z/threeZgr}}; 
 \node at (4.4,0)  {$3Z$}; 
\end{tikzpicture}
\label{equ:3site-Z-tree}
\eeq
which leads to (\ref{equ:ZZZ}) for $\ud Z$. 
\end{itemize}

\vspace{0.1cm}
\subsubsection{Five-Site Chain}

Above we have used our graphical method to derive the complete set of differential equations for the two- and three-site chains.  Readers who would like to experience the full power of the kinematic flow are encouraged to look at Appendix~\ref{app:flow-examples} for further nontrivial applications of the method.  Here, we restrict ourselves to giving an interesting example of the absorption phenomenon that arises for the case of the five-site chain.

\vskip 4pt
We leave it as an exercise to the reader to draw all complete tubings of the marked five-site chain to determine the set of basis functions.
The number of functions at each level are $1$, $11$, $46$, $90$, $81$, $27$, so that
\beq
\begin{aligned}
1 \ + \ 11 \ + \ 46 \ + \ 90 \ + \ 81\ + \ 27 &\ = \ 256\,,\\
1 \ - \ 11 \ + \ 46 \ - \ 90 \ + \ 81\ - \ 27 &\ = \ 0\,.\phantom{56}
\end{aligned}
\eeq
The total number of functions again satisfies $4^e$, with $e = 4$ being the number of edges of the graph.

\vskip 4pt
Naturally, we won't present the differentials for all basis functions (although the reader could derive them using the {\sc Mathematica} notebooks provided at \href{https://github.com/haydenhylee/kinematic-flow}{\faGithub}). 
Instead, we will just highlight one particularly interesting example associated to the following graph tubing 
\beq
   \raisebox{-3pt}{\includegraphics[scale=1]{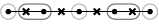}}
\eeq

\vskip 4pt
Let us construct the evolutionary tree for this function.
The first two steps are
 \beq
 \begin{tikzpicture}[baseline=(current  bounding  box.center)]
   \node at (-1.9,0)  {$F$\,:};
\node[inner sep=0pt] at (-0.1,0)
    {\includegraphics[scale=1]{Figures/Tubings/five/fiveF}}; 
\draw [color=gray,thick,-stealth] (1.3,0) -- (2.3,0);
\draw [color=gray,thick,-stealth] (1.3,0.1) -- (2.3,0.6);
\draw [color=gray,thick,-stealth] (1.3,-0.1) -- (2.3,-0.6);
\draw [color=gray,thick,-stealth] (1.3,0.25) -- (2.3,1.3);
\draw [color=gray,thick,-stealth] (1.3,-0.25) -- (2.3,-1.3);
\node[inner sep=0pt] at (3.8,1.4)
    {\includegraphics[scale=1]{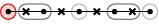}};
\node[inner sep=0pt] at (3.8,0.7)
    {\includegraphics[scale=1]{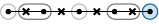}};
\node[inner sep=0pt] at (3.8,0)
    {\includegraphics[scale=1]{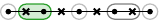}}; 
\node[inner sep=0pt] at (3.8,-0.7)
    {\includegraphics[scale=1]{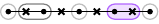}}; 
    \node[inner sep=0pt] at (3.8,-1.4)
    {\includegraphics[scale=1]{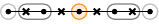}}; 
    \draw [color=gray,thick,-stealth] (5.3,1.4) -- (6.1,1.4);
    \draw [color=gray,thick,-stealth] (5.3,0.7) -- (6.1,0.7);  
        \draw [color=gray,thick,-stealth] (5.3,-1.4) -- (6.1,-1.4); 
        \draw [color=gray,thick,-stealth] (5.3,-1.35) -- (6.1,-0.95); 
         \draw [color=gray,thick,-stealth] (5.3,-1.45) -- (6.1,-1.85);           
        \node[inner sep=0pt] at (7.6,1.4)
    {\includegraphics[scale=1]{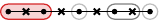}};
            \node[inner sep=0pt] at (7.6,0.7)
    {\includegraphics[scale=1]{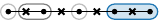}};
       \node[right] at (8.9,1.4)  {$H_1$};
              \node[right] at (8.9,0.7)  {$H_2$};
                              \node[inner sep=0pt] at (7.6,-0.9)
    {\includegraphics[scale=1]{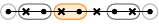}};   
                  \node[inner sep=0pt] at (7.6,-1.4)
    {\includegraphics[scale=1]{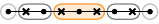}};   
                    \node[inner sep=0pt] at (7.6,-1.9)
    {\includegraphics[scale=1]{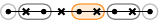}};  
           \node[right] at (8.9,-0.9)  {$G_1$};   
       \node[right] at (8.9,-1.4)  {$G_2$}; 
              \node[right] at (8.9,-1.9)  {$G_3$};       
\end{tikzpicture}
\label{equ:5site-tree}
\eeq
In the top two branches, we have the standard activation, growth and merger of tubes. This leads to the descendant functions $H_1$ and $H_2$.  This explains part of the differential of $F$:
\beq
\boxed{
\begin{aligned}
\ \ud F \,\supset \, 
\big(F-H_1 \big) &\raisebox{3pt}{
 \begin{tikzpicture}[baseline=(current  bounding  box.center)]
  \draw[color=Red, fill=Red!20, line width=0.6pt] (0, 0) circle (1.3mm);
 \draw[fill] (0, 0) circle (.4mm);
\draw[fill] (0.6, 0) circle (.4mm);
\draw[fill] (1.2, 0) circle (.4mm);
\draw[fill] (1.8, 0) circle (.4mm);
\draw[fill] (2.4, 0) circle (.4mm);
\draw[thick] (0, 0) -- (2.4, 0);
\node at (0.3,0)  {\Cross};
\node at (0.9,0)  {\Cross};
\node at (1.5,0)  {\Cross};
\node at (2.1,0)  {\Cross};
\end{tikzpicture}} 
\hskip 6.5pt  + \ H_1 \raisebox{3pt}{
 \begin{tikzpicture}[baseline=(current  bounding  box.center)]
    \node [
        draw, color=Red, fill=Red!20, line width=0.6pt,
        rounded rectangle,
        minimum height = .4mm,
        minimum width = 2.1em,
        rounded rectangle arc length = 180,
    ] at (0.15,0)
    {};
\draw[fill] (0, 0) circle (.4mm);
\draw[fill] (0.6, 0) circle (.4mm);
\draw[fill] (1.2, 0) circle (.4mm);
\draw[fill] (1.8, 0) circle (.4mm);
\draw[fill] (2.4, 0) circle (.4mm);
\draw[thick] (0, 0) -- (2.4, 0);
\node at (0.3,0)  {\Cross};
\node at (0.9,0)  {\Cross};
\node at (1.5,0)  {\Cross};
\node at (2.1,0)  {\Cross};
\end{tikzpicture}}
\\[6pt]
\big(F-H_2 \big) &\, \raisebox{3pt}{
 \begin{tikzpicture}[baseline=(current  bounding  box.center)]
  \draw[color=Blue, fill=Blue!20, line width=0.6pt] (2.4, 0) circle (1.3mm);
 \draw[fill] (0, 0) circle (.4mm);
\draw[fill] (0.6, 0) circle (.4mm);
\draw[fill] (1.2, 0) circle (.4mm);
\draw[fill] (1.8, 0) circle (.4mm);
\draw[fill] (2.4, 0) circle (.4mm);
\draw[thick] (0, 0) -- (2.4, 0);
\node at (0.3,0)  {\Cross};
\node at (0.9,0)  {\Cross};
\node at (1.5,0)  {\Cross};
\node at (2.1,0)  {\Cross};
\end{tikzpicture}} 
\hskip 3.6pt + \ H_2\hskip 2.45pt \raisebox{3pt}{
 \begin{tikzpicture}[baseline=(current  bounding  box.center)]
    \node [
        draw, color=Blue, fill=Blue!20, line width=0.6pt,
        rounded rectangle,
        minimum height = .4mm,
        minimum width = 2.1em,
        rounded rectangle arc length = 180,
    ] at (2.25,0)
    {};
\draw[fill] (0, 0) circle (.4mm);
\draw[fill] (0.6, 0) circle (.4mm);
\draw[fill] (1.2, 0) circle (.4mm);
\draw[fill] (1.8, 0) circle (.4mm);
\draw[fill] (2.4, 0) circle (.4mm);
\draw[thick] (0, 0) -- (2.4, 0);
\node at (0.3,0)  {\Cross};
\node at (0.9,0)  {\Cross};
\node at (1.5,0)  {\Cross};
\node at (2.1,0)  {\Cross};
\end{tikzpicture}}
\\[6pt]
\ + \ F\ &\,\raisebox{3pt}{
 \begin{tikzpicture}[baseline=(current  bounding  box.center)]
    \node [
        draw, color=Green, fill=Green!20, line width=0.6pt,
        rounded rectangle,
        minimum height = .4mm,
        minimum width = 2.1em,
        rounded rectangle arc length = 180,
    ] at (0.45,0)
    {};
\draw[fill] (0, 0) circle (.4mm);
\draw[fill] (0.6, 0) circle (.4mm);
\draw[fill] (1.2, 0) circle (.4mm);
\draw[fill] (1.8, 0) circle (.4mm);
\draw[fill] (2.4, 0) circle (.4mm);
\draw[thick] (0, 0) -- (2.4, 0);
\node at (0.3,0)  {\Cross};
\node at (0.9,0)  {\Cross};
\node at (1.5,0)  {\Cross};
\node at (2.1,0)  {\Cross};
\end{tikzpicture}} 
\\[6pt]
\ + \ F\ &\,\raisebox{3pt}{
 \begin{tikzpicture}[baseline=(current  bounding  box.center)]
    \node [
        draw, color=Purple, fill=Purple!20, line width=0.6pt,
        rounded rectangle,
        minimum height = .4mm,
        minimum width = 2.1em,
        rounded rectangle arc length = 180,
    ] at (1.95,0)
    {};
\draw[fill] (0, 0) circle (.4mm);
\draw[fill] (0.6, 0) circle (.4mm);
\draw[fill] (1.2, 0) circle (.4mm);
\draw[fill] (1.8, 0) circle (.4mm);
\draw[fill] (2.4, 0) circle (.4mm);
\draw[thick] (0, 0) -- (2.4, 0);
\node at (0.3,0)  {\Cross};
\node at (0.9,0)  {\Cross};
\node at (1.5,0)  {\Cross};
\node at (2.1,0)  {\Cross};
\end{tikzpicture}} 
\end{aligned}}
\eeq
In the bottom branch of (\ref{equ:5site-tree}), the middle tube is activated and then grows in three different ways, producing the functions $G_1$, $G_2$, $G_3$. 
The latter tubings then grow further through two stages of absorption:
 \beq
 \begin{tikzpicture}[baseline=(current  bounding  box.center)]
 
\node[inner sep=0pt] at (3.8,0)
    {\includegraphics[scale=1]{Figures/Tubings/five/fiveG1}}; 
    \node[inner sep=0pt] at (3.8,-1.4)
    {\includegraphics[scale=1]{Figures/Tubings/five/fiveG2}}; 
    \node[inner sep=0pt] at (3.8,-2.8)
    {\includegraphics[scale=1]{Figures/Tubings/five/fiveG3}}; 
               \node[right] at (1.75,0)  {$G_1$};   
       \node[right] at (1.75,-1.4)  {$G_2$}; 
              \node[right] at (1.75,-2.8)  {$G_3$};   
      \draw [color=gray,thick,-stealth] (5.3,0) -- (6.1,0); 
      \draw [color=gray,thick,-stealth] (5.3,-2.8) -- (6.1,-2.8);       
     \draw [color=gray,thick,-stealth] (5.3,-1.4) -- (6.1,-1.4); 
        \draw [color=gray,thick,-stealth] (5.3,-1.35) -- (6.1,-0.95); 
         \draw [color=gray,thick,-stealth] (5.3,-1.45) -- (6.1,-1.85);  
                              \node[inner sep=0pt] at (7.6,-0.9)
    {\includegraphics[scale=1]{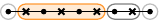}};   
                  \node[inner sep=0pt] at (7.6,-1.4)
    {\includegraphics[scale=1]{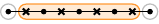}};   
                    \node[inner sep=0pt] at (7.6,-1.9)
    {\includegraphics[scale=1]{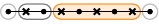}};  
           \node[right] at (8.9,-0.9)  {$(-J_1)$};   
       \node[right] at (8.9,-1.4)  {$(-K)$}; 
              \node[right] at (8.9,-1.9)  {$(-J_2)$}; 
     \node[inner sep=0pt] at (7.6,0)
    {\includegraphics[scale=1]{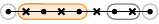}};   
                    \node[inner sep=0pt] at (7.6,-2.8)
    {\includegraphics[scale=1]{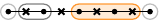}};  
            \node[right] at (8.9,0)  {$(-L_1)$}; 
                    \node[right] at (8.9,-2.8)  {$(-L_2)$}; 
          \draw [color=gray,thick,-stealth] (10.1,-0.9) -- (10.9,-0.9);   
                  \draw [color=gray,thick,-stealth] (10.1,-1.9) -- (10.9,-1.9);
                             \node[inner sep=0pt] at (12.4,-0.9)
    {\includegraphics[scale=1]{Figures/Tubings/five/fiveK}};   
                  \node[inner sep=0pt] at (12.4,-1.9)
    {\includegraphics[scale=1]{Figures/Tubings/five/fiveK}};  
           \node[right] at (13.75,-0.9)  {$K$};     
                      \node[right] at (13.75,-1.9)  {$K$};            
\end{tikzpicture}
\label{equ:5site-tree2}
\eeq
In the top and bottom branches, absorption produces the tubings associated to the functions $-L_1$ and $-L_2$, respectively. In the middle branch, absorption first occurs in three different ways: i) absorbing the tube to the left, ii) absorbing the tube to the right, and iii) absorbing the tubes to both sides simultaneously. This produces the functions $-J_1$, $-J_2$ and $-K$, respectively. The tubings associated to $J_1$ and $J_2$ then grow further through a second absorption, creating the function $+K$.

\vskip 4pt
The rest of the differential of $F$ then is
\beq
\boxed{
\begin{aligned}
\ \ud F \,\supset \, 
 \big(F- {\textstyle \sum} G_i \big) \raisebox{3pt}{
 \begin{tikzpicture}[baseline=(current  bounding  box.center)]
  \draw[color=Orange, fill=Orange!20, line width=0.6pt] (1.2, 0) circle (1.3mm);
 \draw[fill] (0, 0) circle (.4mm);
\draw[fill] (0.6, 0) circle (.4mm);
\draw[fill] (1.2, 0) circle (.4mm);
\draw[fill] (1.8, 0) circle (.4mm);
\draw[fill] (2.4, 0) circle (.4mm);
\draw[thick] (0, 0) -- (2.4, 0);
\node at (0.3,0)  {\Cross};
\node at (0.9,0)  {\Cross};
\node at (1.5,0)  {\Cross};
\node at (2.1,0)  {\Cross};
\end{tikzpicture}}\ + \,(G_3 + L_2) &\raisebox{3pt}{
 \begin{tikzpicture}[baseline=(current  bounding  box.center)]
    \node [
        draw, color=Orange, fill=Orange!20, line width=0.6pt,
        rounded rectangle,
        minimum height = .4mm,
        minimum width = 2.1em,
        rounded rectangle arc length = 180,
    ] at (1.35,0)
    {};
\draw[fill] (0, 0) circle (.4mm);
\draw[fill] (0.6, 0) circle (.4mm);
\draw[fill] (1.2, 0) circle (.4mm);
\draw[fill] (1.8, 0) circle (.4mm);
\draw[fill] (2.4, 0) circle (.4mm);
\draw[thick] (0, 0) -- (2.4, 0);
\node at (0.3,0)  {\Cross};
\node at (0.9,0)  {\Cross};
\node at (1.5,0)  {\Cross};
\node at (2.1,0)  {\Cross};
\end{tikzpicture}}\\
-\hskip 2.5pt L_2\hskip 4pt &\raisebox{3pt}{
 \begin{tikzpicture}[baseline=(current  bounding  box.center)]
  \node [
        draw, color=Orange, fill=Orange!20, line width=0.6pt,
        rounded rectangle,
        minimum height = .4mm,
        minimum width = 3.6em,
        rounded rectangle arc length = 180,
    ] at (1.65,0)
    {};
\draw[fill] (0, 0) circle (.4mm);
\draw[fill] (0.6, 0) circle (.4mm);
\draw[fill] (1.2, 0) circle (.4mm);
\draw[fill] (1.8, 0) circle (.4mm);
\draw[fill] (2.4, 0) circle (.4mm);
\draw[thick] (0, 0) -- (2.4, 0);
\node at (0.3,0)  {\Cross};
\node at (0.9,0)  {\Cross};
\node at (1.5,0)  {\Cross};
\node at (2.1,0)  {\Cross};
\end{tikzpicture}}\\
+ \ (G_1 + L_1) &\raisebox{3pt}{
 \begin{tikzpicture}[baseline=(current  bounding  box.center)]
   \node [
        draw, color=Orange, fill=Orange!20, line width=0.6pt,
        rounded rectangle,
        minimum height = .4mm,
        minimum width = 2.1em,
        rounded rectangle arc length = 180,
    ] at (1.05,0)
    {};
\draw[fill] (0, 0) circle (.4mm);
\draw[fill] (0.6, 0) circle (.4mm);
\draw[fill] (1.2, 0) circle (.4mm);
\draw[fill] (1.8, 0) circle (.4mm);
\draw[fill] (2.4, 0) circle (.4mm);
\draw[thick] (0, 0) -- (2.4, 0);
\node at (0.3,0)  {\Cross};
\node at (0.9,0)  {\Cross};
\node at (1.5,0)  {\Cross};
\node at (2.1,0)  {\Cross};
\end{tikzpicture}}\\
-\hskip 2.5pt L_1\hskip 4pt &\raisebox{3pt}{
 \begin{tikzpicture}[baseline=(current  bounding  box.center)]
  \node [
        draw, color=Orange, fill=Orange!20, line width=0.6pt,
        rounded rectangle,
        minimum height = .4mm,
        minimum width = 3.6em,
        rounded rectangle arc length = 180,
    ] at (0.75,0)
    {};
\draw[fill] (0, 0) circle (.4mm);
\draw[fill] (0.6, 0) circle (.4mm);
\draw[fill] (1.2, 0) circle (.4mm);
\draw[fill] (1.8, 0) circle (.4mm);
\draw[fill] (2.4, 0) circle (.4mm);
\draw[thick] (0, 0) -- (2.4, 0);
\node at (0.3,0)  {\Cross};
\node at (0.9,0)  {\Cross};
\node at (1.5,0)  {\Cross};
\node at (2.1,0)  {\Cross};
\end{tikzpicture}}\\[4pt]
+ \ \big(G_2 + J_1 + J_2 + K \big)  &\raisebox{3pt}{
 \begin{tikzpicture}[baseline=(current  bounding  box.center)]
  \node [
        draw, color=Orange, fill=Orange!20, line width=0.6pt,
        rounded rectangle,
        minimum height = .4mm,
        minimum width = 3em,
        rounded rectangle arc length = 180,
    ] at (1.2,0)
    {};
\draw[fill] (0, 0) circle (.4mm);
\draw[fill] (0.6, 0) circle (.4mm);
\draw[fill] (1.2, 0) circle (.4mm);
\draw[fill] (1.8, 0) circle (.4mm);
\draw[fill] (2.4, 0) circle (.4mm);
\draw[thick] (0, 0) -- (2.4, 0);
\node at (0.3,0)  {\Cross};
\node at (0.9,0)  {\Cross};
\node at (1.5,0)  {\Cross};
\node at (2.1,0)  {\Cross};
\end{tikzpicture}}\\
- \ \big(J_1  + K \big)  &\raisebox{3pt}{
 \begin{tikzpicture}[baseline=(current  bounding  box.center)]
  \node [
        draw, color=Orange, fill=Orange!20, line width=0.6pt,
        rounded rectangle,
        minimum height = .4mm,
        minimum width = 4.4em,
        rounded rectangle arc length = 180,
    ] at (1.5,0)
    {};
\draw[fill] (0, 0) circle (.4mm);
\draw[fill] (0.6, 0) circle (.4mm);
\draw[fill] (1.2, 0) circle (.4mm);
\draw[fill] (1.8, 0) circle (.4mm);
\draw[fill] (2.4, 0) circle (.4mm);
\draw[thick] (0, 0) -- (2.4, 0);
\node at (0.3,0)  {\Cross};
\node at (0.9,0)  {\Cross};
\node at (1.5,0)  {\Cross};
\node at (2.1,0)  {\Cross};
\end{tikzpicture}}\\
- \ \big(J_2 + K \big)  &\raisebox{3pt}{
 \begin{tikzpicture}[baseline=(current  bounding  box.center)]
  \node [
        draw, color=Orange, fill=Orange!20, line width=0.6pt,
        rounded rectangle,
        minimum height = .4mm,
        minimum width = 4.4em,
        rounded rectangle arc length = 180,
    ] at (0.9,0)
    {};
\draw[fill] (0, 0) circle (.4mm);
\draw[fill] (0.6, 0) circle (.4mm);
\draw[fill] (1.2, 0) circle (.4mm);
\draw[fill] (1.8, 0) circle (.4mm);
\draw[fill] (2.4, 0) circle (.4mm);
\draw[thick] (0, 0) -- (2.4, 0);
\node at (0.3,0)  {\Cross};
\node at (0.9,0)  {\Cross};
\node at (1.5,0)  {\Cross};
\node at (2.1,0)  {\Cross};
\end{tikzpicture}}\\
+\hskip 2.5pt K\hskip 5pt  &\raisebox{3pt}{
 \begin{tikzpicture}[baseline=(current  bounding  box.center)]
  \node [
        draw, color=Orange, fill=Orange!20, line width=0.6pt,
        rounded rectangle,
        minimum height = .4mm,
        minimum width = 6em,
        rounded rectangle arc length = 180,
    ] at (1.2,0)
    {};
\draw[fill] (0, 0) circle (.4mm);
\draw[fill] (0.6, 0) circle (.4mm);
\draw[fill] (1.2, 0) circle (.4mm);
\draw[fill] (1.8, 0) circle (.4mm);
\draw[fill] (2.4, 0) circle (.4mm);
\draw[thick] (0, 0) -- (2.4, 0);
\node at (0.3,0)  {\Cross};
\node at (0.9,0)  {\Cross};
\node at (1.5,0)  {\Cross};
\node at (2.1,0)  {\Cross};
\end{tikzpicture}\ }
\end{aligned}}
\eeq
Note that the coefficient of the letter in the last line comes from the three occurrences of the function $K$ in \eqref{equ:5site-tree2}: $-K + K+K = K$.

\vspace{0.25cm}
\begin{center}
***
\end{center}
The examples above (and those in Appendix~\ref{app:flow-examples}) capture all phenomena that can occur in the kinematic flow.   All the complexity of the equations for larger tree graphs
 arises from the repeated application of a few simple rules. Do these rules have a life of their own? Can they be derived from a deeper mathematical structure? We leave these intriguing questions to future work.

\newpage
\section{Beyond Single Graphs}
\label{sec:Channels}

So far, we have studied the differential equations for individual Feynman graphs. In the context of scattering amplitudes, the most nontrivial structures arise when we combine together the different channels contributing to a given process. (Indeed, in many cases individual Feynman diagrams are not even physically meaningful.) We therefore expect additional simplifications to occur when we consider the differential equations for the analogous objects in cosmology. We will now discuss this sum over graphs. One intriguing feature 
is that the equations for the different channels share common functions, with a geometric structure underpinning the commonalities. There is again a graphical algorithm that predicts the structure of the differential equations and we will illustrate it in a number of examples.

\vskip 4pt
 The simplest theory where individual graphs combine together irreducibly is so-called ${\rm tr}\,\phi^3$ theory. This is the theory of a matrix-valued scalar field $\phi^{a}_b $ that has 
 indices  transforming in the fundamental and the anti-fundamental of SU$(N)$. The minimal interaction in this theory is a cubic coupling, so that the action takes the form
\beq
S =  \int \ud^4 x \sqrt{-g}\left(-\frac{1}{2}\partial_\mu \phi^a_b \partial^\mu \phi^b_a - \frac{g}{3!} \phi^a_b  \phi^b_c  \phi^c_a \right) .
\label{eq:BAP3}
\eeq
The wavefunction (and amplitudes) in this model can be written in terms of flavor-ordered partial wavefunctions multiplied by flavor factors. For a fixed ordering of external particles, this flavor-ordering restricts the possible processes that contribute to the amplitude/wavefunction (see e.g.~\cite{Arkani-Hamed:2023lbd}). In particular, at tree level, the flavor-ordered wavefunction will only get  
contributions from planar diagrams that respect the given cyclic ordering.\footnote{More explicitly, we can write the full wavefunction coefficient as a sum over partial wavefunctions weighted by flavor factors
\be
\psi_n = \sum_{\Gamma}\,F(\Gamma)\, \psi_n^{(\Gamma)}\,,
\ee
where $\Gamma$ is a doubled cubic graph obtained by applying the Feynman rules of~\eqref{eq:BAP3} and $F(\Gamma)$ is the contraction of external flavor indices for each particle. At tree level, two graphs will lead to the same flavor factor if they share the same ordering of external particles. (More generally, two graphs will have the same color factor if they have the same topology, thought of as marked surfaces~\cite{Arkani-Hamed:2023lbd}.) It is worth noting that these partial amplitudes/wavefunctions are the same as those of bi-adjoint $\phi^3$ theory when both orderings of external indices in the bi-adjoint theory are taken to be the same as in the ${\rm tr}\,\phi^3$ theory.}

\subsection{Triangulations of Polygons}
\label{sec:triangulations}

The various exchange diagrams contributing to a given flavor-ordered wavefunction are each naturally associated to triangulations of polygons. Recall that an $n$-point function depends on $n$ (spatial) momentum vectors $\vec k_a$. As a consequence of translation invariance, these momenta must sum to zero, and therefore define a closed $n$-gon. 
A basis of independent kinematic variables is given by the side lengths and diagonals of the polygon.
The compatible exchange channels that contribute to a given flavor-ordered process are then associated to the possible triangulations of this kinematic polygon, and the flat-space wavefunction can naturally be  read off from these triangulations. From this point of view, we can think of the tree graphs studied earlier as corresponding to a particular choice of triangulation.

\paragraph{Graphs and triangulations:}
The simplest exchange graph is the two-site chain, which in $\phi^3$ theory corresponds to a four-point function. The external momenta $\vec{k}_1, \vec{k}_2, \vec{k}_3, \vec{k}_4$ form a closed quadrilateral (which, for simplicity, we will draw as a square).
There are two possible triangulations of this quadrilateral, corresponding to the $s$ and $t$-channel exchanges:\footnote{Note that, for planar graphs, $u$-channel exchange is not compatible with our chosen cyclic ordering, so it does not contribute to this flavor-ordered object.}
\begin{equation*}
 \raisebox{5pt}{
 \begin{tikzpicture}[baseline=(current  bounding  box.center)]
 \draw[lightgray, line width=1.pt] (0,0) -- (-0.25,0.75);
\draw[lightgray, line width=1.pt] (0,0) -- (0.25,0.75);
\draw[lightgray, line width=1.pt] (1.4,0) -- (1.15,0.75);
\draw[lightgray, line width=1.pt] (1.4,0) -- (1.65,0.75);
\draw[lightgray, line width=2.pt] (-0.5,0.75) -- (1.9,0.75);
\node[above] at (-0.25,0.75) {\scriptsize $1$};
\node[above] at (0.25,0.75) {\scriptsize $2$};
\node[above] at (1.15,0.75) {\scriptsize $3$};
\node[above] at (1.65,0.75) {\scriptsize $4$};
\node[below] at (.7,-.15) {\small $s$-channel};
\draw[thick] (0, 0) -- (1.4, 0);
\draw[fill,color=Red] (0, 0) circle (.5mm);
\draw[fill,color=Blue] (1.4, 0) circle (.5mm);
   \draw[black, line width=0.5pt] (3,-0.5) -- (4.25,0.75);
 \draw[Red, line width=1.pt] (3,-0.5) -- (3,0.75) -- (4.25,0.75);
  \draw[Blue, line width=1.pt] (3,-0.5) -- (4.25,-0.5) -- (4.25,0.75);
  \node[left,color=Red] at (3,0.125) {\scriptsize $k_1$};
  \node[above,color=Red] at (3.625,0.75) {\scriptsize $k_2$};
    \node[right,color=Blue] at (4.25,0.125) {\scriptsize $k_3$};
  \node[below,color=Blue] at (3.625,-0.5) {\scriptsize $k_4$};
\end{tikzpicture}} 
\hspace{1.6cm}
 \raisebox{5pt}{
 \begin{tikzpicture}[baseline=(current  bounding  box.center)]
 \draw[lightgray, line width=1.pt] (0,0) -- (-0.25,0.75);
\draw[lightgray, line width=1.pt] (0,0) -- (0.25,0.75);
\draw[lightgray, line width=1.pt] (1.4,0) -- (1.15,0.75);
\draw[lightgray, line width=1.pt] (1.4,0) -- (1.65,0.75);
\draw[lightgray, line width=2.pt] (-0.5,0.75) -- (1.9,0.75);
\node[above] at (-0.25,0.75) {\scriptsize $1$};
\node[above] at (0.25,0.75) {\scriptsize $4$};
\node[above] at (1.15,0.75) {\scriptsize $2$};
\node[above] at (1.65,0.75) {\scriptsize $3$};
\node[below] at (.7,-.15) {\small $t$-channel};
\draw[thick] (0, 0) -- (1.4, 0);
\draw[fill,color=Red] (0, 0) circle (.5mm);
\draw[fill,color=Blue] (1.4, 0) circle (.5mm);
   \draw[black, line width=0.5pt] (3,0.75) -- (4.25,-0.5);
 \draw[Red, line width=1.pt] (3,0.75)  -- (3,-0.5) -- (4.25,-0.5);
  \draw[Blue, line width=1.pt] (3,0.75) -- (4.25,0.75) -- (4.25,-0.5);
  \node[left,color=Red] at (3,0.125) {\scriptsize $k_1$};
  \node[above,color=Blue] at (3.625,0.75) {\scriptsize $k_2$};
    \node[right,color=Blue] at (4.25,0.125) {\scriptsize $k_3$};
  \node[below,color=Red] at (3.625,-0.5) {\scriptsize $k_4$};
\end{tikzpicture}} 
\end{equation*}
At the graph level, these two triangulations reflect different permutations of the external lines. 

\vskip 4pt
Things get more interesting at five points. In $\phi^3$ theory, the five-point function arises from a three-site graph, with a particular ordering of the external lines corresponding to a specific triangulation of the kinematic pentagon:
\beq
\hspace{-0.4cm}
 \raisebox{-12pt}{
\begin{tikzpicture}[line width=1. pt, scale=1.5]
\draw[fill=black] (0,0) -- (2,0);
\draw[lightgray, line width=1.pt] (0,0) -- (-0.25,0.75);
\draw[lightgray, line width=1.pt] (0,0) -- (0.25,0.75);
\draw[lightgray, line width=1.pt] (1,0) -- (1,0.75);
\draw[lightgray, line width=1.pt] (2,0) -- (1.75,0.75);
\draw[lightgray, line width=1.pt] (2,0) -- (2.25,0.75);
\draw[lightgray, line width=2.pt] (-0.5,0.75) -- (2.5,0.75);
\draw[Red,fill=Red] (0,0) circle (.03cm);
\draw[Orange,fill=Orange] (1,0) circle (.03cm);
\draw[Blue,fill=Blue] (2,0) circle (.03cm);
\node[above] at (-0.25,0.75) {\small $1$};
\node[above] at (0.25,0.75) {\small $2$};
\node[above] at (1,0.75) {\small $3$};
\node[above] at (1.75,0.75) {\small $4$};
\node[above] at (2.25,0.75) {\small $5$};
\end{tikzpicture}
} 
\hspace{0.85cm} \raisebox{-40pt}{
\begin{tikzpicture}[line width=1. pt, scale=1.2]
\coordinate (1) at (0,1);
\coordinate (2) at (0.951,0.309);
\coordinate (3) at (0.588,-0.809);
\coordinate (4) at (-0.588,-0.809);
\coordinate (5) at (-0.951,0.309);
\draw[black, line width=0.5pt] (4) -- (1);
\draw[black, line width=0.5pt] (4) -- (2);
\draw[Red, line width=1.pt] (4) -- (5) -- (1);
\draw[Orange, line width=1.pt] (1) -- (2);
\draw[Blue, line width=1.pt] (2) -- (3) -- (4);
\node[color=Red,scale=1] at (-0.94,-0.26) {\small $k_1$};
\node[color=Red,scale=1] at (-0.62,0.83) {\small $k_2$};
\node[color=Orange,scale=1] at (0.62,0.8) {\small $k_3$};
\node[color=Blue,scale=1] at (1,-0.26) {\small $k_4$};
\node[color=Blue,scale=1,below] at (0,-0.82) {\small $k_5$};
\end{tikzpicture}}
\nonumber
\eeq
Note that each of the diagonals present in this triangulation correspond to the pairs of momenta that combine into the internal lines. There are five triangulations of a pentagon, and each corresponds to an inequivalent exchange channel
\beq 
\begin{tikzpicture}[line width=1.pt, scale=.85]
\coordinate (1) at (0,1);
\coordinate (2) at (0.951,0.309);
\coordinate (3) at (0.588,-0.809);
\coordinate (4) at (-0.588,-0.809);
\coordinate (5) at (-0.951,0.309);
\draw[gray, line width=0.5pt] (4) -- (1);
\draw[gray, line width=0.5pt] (4) -- (2);
\draw[black, line width=1.pt] (1) -- (2) -- (3) -- (4) -- (5) -- (1);
\node[below] at (0,-0.9) {I};
\end{tikzpicture}
\qquad
\begin{tikzpicture}[line width=1.pt, scale=.85]
\coordinate (1) at (0,1);
\coordinate (2) at (0.951,0.309);
\coordinate (3) at (0.588,-0.809);
\coordinate (4) at (-0.588,-0.809);
\coordinate (5) at (-0.951,0.309);
\draw[gray, line width=0.5pt] (5) -- (2);
\draw[gray, line width=0.5pt] (5) -- (3);
\draw[black, line width=1.pt] (1) -- (2) -- (3) -- (4) -- (5) -- (1);
\node[below] at (0,-0.9) {II};
\end{tikzpicture}
\qquad
\begin{tikzpicture}[line width=1.pt, scale=.85]
\coordinate (1) at (0,1);
\coordinate (2) at (0.951,0.309);
\coordinate (3) at (0.588,-0.809);
\coordinate (4) at (-0.588,-0.809);
\coordinate (5) at (-0.951,0.309);
\draw[gray, line width=0.5pt] (4) -- (1);
\draw[gray, line width=0.5pt] (3) -- (1);
\draw[black, line width=1.pt] (1) -- (2) -- (3) -- (4) -- (5) -- (1);
\node[below] at (0,-0.9) {III};
\end{tikzpicture}
\qquad
\begin{tikzpicture}[line width=1.pt, scale=.85]
\coordinate (1) at (0,1);
\coordinate (2) at (0.951,0.309);
\coordinate (3) at (0.588,-0.809);
\coordinate (4) at (-0.588,-0.809);
\coordinate (5) at (-0.951,0.309);
\draw[gray, line width=0.5pt] (5) -- (2);
\draw[gray, line width=0.5pt] (4) -- (2);
\draw[black, line width=1.pt] (1) -- (2) -- (3) -- (4) -- (5) -- (1);
\node[below] at (0,-0.9) {IV};
\end{tikzpicture}
\qquad
\begin{tikzpicture}[line width=1.pt, scale=.85]
\coordinate (1) at (0,1);
\coordinate (2) at (0.951,0.309);
\coordinate (3) at (0.588,-0.809);
\coordinate (4) at (-0.588,-0.809);
\coordinate (5) at (-0.951,0.309);
\draw[gray, line width=0.5pt] (3) -- (1);
\draw[gray, line width=0.5pt] (3) -- (5);
\draw[black, line width=1.pt] (1) -- (2) -- (3) -- (4) -- (5) -- (1);
\node[below] at (0,-0.9) {V};
\end{tikzpicture}
\nonumber
\eeq
For more general polygons, the number of possible triangulations (and hence exchange channels) is counted by the Catalan numbers. The number of triangulations of an $n$-gon is the Catalan number, $C_{n-2}$.
The first few Catalan numbers are $C_{n-2}=2,5,14, 42, \cdots$ for $n=4,5,6,7,\cdots$\,.

\noindent
\paragraph{From polygons to the wavefunction:} 
The flat-space wavefunction has a beautifully simple characterization in terms of these triangulations of kinematic polygons. The wavefunction coefficient for the exchange in a given channel is encoded geometrically in terms of the perimeters of the sub-polygons that appear in a given triangulation. The full wavefunction for the sum over graphs then arises from summing over all triangulations. We can therefore write an abstract formula for the flat-space wavefunction as:\footnote{The analogous formula for the flat-space scattering amplitude is to sum over the (squared) diagonal lengths of a given triangulation, rather than the perimeters of polygons, together with a sum over all triangulations~\cite{Arkani-Hamed:2017mur}.}
\beq
\psi_{\rm flat} = \sum_{\cal T} \sum_{\cal P} \prod_a \frac{1}{P_a}\,,
\label{eq:polygonsandWF}
\eeq
where ${\cal T}$ is the set of distinct triangulations, ${\cal P}$ are the sets of maximal non-overlapping sub-polygons of a given triangulation 
 and $P_a$ are the perimeters of the (sub)polygons. This combinatorial definition is the generalization to the ${\rm tr}\,\phi^3$ theory of the rule presented in Section~\ref{ssec:FlatSpace} to derive the wavefunction for individual graphs in terms of non-overlapping graph tubings.

\vskip4pt
It is simplest to demonstrate this via examples. First consider the four-point function and look at the triangulation corresponding to $s$-channel exchange: 
\begin{equation*}
 \raisebox{5pt}{
 \begin{tikzpicture}[baseline=(current  bounding  box.center),scale=.65]
 \draw[black, line width=1.5pt] (-1,-1) -- (-1,1) -- (1,1) -- (1,-1) -- (-1,-1);
 \draw[Red, line width=1.5pt] (-0.94,-0.9) -- (-0.94,0.94) -- (0.9,0.94) -- (-0.94,-0.9);
  \draw[Blue, line width=1.5pt] (-0.9,-0.94) -- (0.94,-0.94) -- (0.94,0.9) -- (-0.9,-0.94);
\node at (-1.2,-1.2) {\small  1};
\node at (-1.2,1.2) {\small 2};
\node at (1.2,1.2) {\small 3};
\node at (1.2,-1.2) {\small  4};
\end{tikzpicture}} 
\end{equation*}
This triangulation contains three non-overlapping sub-polygons---the two triangles and the full square. Taking the inverse of the product of their perimeters, we obtain the $s$-channel contribution to the flat-space wavefunction:  
\be
\begin{aligned}
\hat\psi^{(s)}_4 &= \frac{1}{P_{1234}{\color{Red} P_{123}}{\color{Blue} P_{134}}}\\
&=  \frac{1}{k_{1234} \,{\color{Red} (k_{12}+s)}{\color{Blue} (k_{34}+s)}}\,,
\end{aligned}
\label{eq:squareWF}
\ee
where $s \equiv |\vec{k_1}+\vec{k}_2|$.
This matches the result~\eqref{eq:flatWF} derived from graph tubings in Section~\ref{ssec:FlatSpace}. 
Summing over the two possible triangulations, we recover the full flavor-ordered wavefunction arising from the sum of $s$ and $t$-channel exchanges.

\vskip4pt
A more interesting example is the five-point function, where the various channels correspond to triangulations of a pentagon. The novelty in this case is that there are two inequivalent choices of maximal non-overlapping polygons for each triangulation, and we are instructed to sum over both of them. For example, for Channel ${\rm I}$, we have
\beq 
\begin{aligned}
\hat\psi^{({\rm I})}_5 \ \ &= \  
 \raisebox{-28pt}{\begin{tikzpicture}[line width=1.pt, scale=0.85]

\coordinate (3) at (0,1);
\coordinate (4) at (0.951,0.309);
\coordinate (5) at (0.588,-0.809);
\coordinate (1) at (-0.588,-0.809);
\coordinate (2) at (-0.951,0.309);

\draw[gray, line width=0.5pt] (4) -- (1);
\draw[gray, line width=0.5pt] (1) -- (3);
\draw[black, line width=1.pt] (1) -- (2) -- (3) -- (4) -- (5) -- (1);
\draw[Purple, line width=1.pt] (1) -- (2) -- (3) -- (4) -- (1);
\node[below] at (-0.62,-0.78) {\footnotesize  1};
\node[below] at (0.62,-0.78) {\footnotesize  5};
\node at (-1.15,0.35) {\footnotesize  2};
\node at (1.15,0.35) {\footnotesize  4};
\node[above] at (0,0.95) {\footnotesize 3};
\end{tikzpicture}}
 \ \ \,+ \ \ 
 \raisebox{-28pt}{\begin{tikzpicture}[line width=1.pt, scale=0.85]

\coordinate (3) at (0,1);
\coordinate (4) at (0.951,0.309);
\coordinate (5) at (0.588,-0.809);
\coordinate (1) at (-0.588,-0.809);
\coordinate (2) at (-0.951,0.309);

\draw[gray, line width=0.5pt] (4) -- (1);
\draw[gray, line width=0.5pt] (1) -- (3);
\draw[black, line width=1.pt] (1) -- (2) -- (3) -- (4) -- (5) -- (1);
\draw[brown, line width=1.pt] (1) -- (3) -- (4) -- (5) -- (1);
\node[below] at (-0.62,-0.78) {\footnotesize  1};
\node[below] at (0.62,-0.78) {\footnotesize  5};
\node at (-1.15,0.35) {\footnotesize  2};
\node at (1.15,0.35) {\footnotesize  4};
\node[above] at (0,0.95) {\footnotesize 3};
\end{tikzpicture}}
 \\[8pt]
 &= \ \frac{1}{P_{12345}\, P_{123} \,P_{134}\, P_{145}} \left[ {\color{Purple}\frac{1}{P_{1234}}} + {\color{brown}\frac{1}{P_{1345}}}  \right] \\
 &= \ \frac{1}{k_{12345}\,(k_{12}+y_{12})\,(k_3+y_{12}+y_{45})\,(k_{45}+y_{45})} \left[ {\color{Purple}\frac{1}{k_{123}+y_{45}}} + {\color{brown}\frac{1}{k_{345}+y_{12}}}  \right] .
 \end{aligned}
 \label{eq:5ptWFpent}
\eeq
The purple and brown quadrilaterals are incompatible with each other (they overlap), so we must make a choice of one of them to include in our set of perimeters. We then sum over both choices to obtain the contribution to the wavefunction from Channel ${\rm I}$, matching the result of graph tubings~\eqref{equ:3pt-tubing}. The full flavor-ordered wavefunction is the sum over all triangulations.

\paragraph{Recursion:}
The algorithm we just presented is essentially recursive, and we can exploit this structure to efficiently generate wavefunction coefficients with large multiplicity. In particular, we can express an $n$-point wavefunction in terms of products of lower-point wavefunctions. Given a kinematic $n$-gon, its wavefunction coefficient can be obtained by splitting it into pairs of polygons by drawing diagonals. 
For each splitting of the polygon, we multiply the wavefunctions associated to these smaller polygons together, and then sum over all possible ways of splitting the polygon by picking a diagonal. The wavefunction of the original polygon is then given by this sum, multiplied by one over the perimeter of the original $n$-gon. (From the bulk perspective, this recursive formula is obtained by applying the time translation operator as in~\cite{Arkani-Hamed:2017fdk}.) If we restrict to a single triangulation---by restricting the possible diagonals that we can use to split the polygon---this recursion reduces exactly to the graphical recursion formula of~\cite{Arkani-Hamed:2017fdk}. We can see this recursive structure in the examples presented above. The expression~\eqref{eq:squareWF} is directly a product of the wavefunctions associated to the triangles in its triangulation, divided by the total perimeter. The five-point example~\eqref{eq:5ptWFpent} is a bit more nontrivial: the two terms can be written as the products of the wavefunctions associated to the two possible quadrilaterals and their complementary triangles, again divided by the perimeter.

\subsection{Letters and Functions}
\label{ssec:letterssec5}
In Section~\ref{sec:GraphicalRules}, we saw that the letters and functions appearing in the system of differential equations for the cosmological wavefunction are naturally associated to tubings of marked graphs. When we sum over graphs, these get replaced with shaded sub-polygons of the kinematic polygon. 

\paragraph{Letters:} Recall that for a single graph, letters were associated with the connected tubings of a marked graph. Here, the letters appearing in the differential equation are given by the perimeters of ``shaded sub-polygons" of the kinematic polygon. When we select a sub-polygon, if it is not the full polygon, we have to draw an internal line, and the energy corresponding to this internal line can appear with a minus sign. This is equivalent to the tubings that enclose one of the markings of the graph. We will use dashed lines to indicate that the sign of the internal energy has been flipped. This is essentially just a re-packaging of information present in the marked graph, and so we can freely translate back and forth between the two pictures. Some examples for the five-point function in Channel ${\rm I}$ are 
\begin{align}
\begin{split}
\raisebox{7pt}{
 \begin{tikzpicture}[baseline=(current  bounding  box.center)]
 \draw[lightgray, line width=1.pt] (0,0) -- (-0.25,0.75);
\draw[lightgray, line width=1.pt] (0,0) -- (0.25,0.75);
\draw[lightgray, line width=1.pt] (.8,0) -- (.8,0.75);
\draw[lightgray, line width=1.pt] (1.6,0) -- (1.35,0.75);
\draw[lightgray, line width=1.pt] (1.6,0) -- (1.85,0.75);
\draw[lightgray, line width=2.pt] (-0.5,0.75) -- (2.1,0.75);
\node[above] at (-0.25,0.75) {\scriptsize $\color{gray}{1}$};
\node[above] at (0.25,0.75) {\scriptsize $\color{gray}{2}$};
\node[above] at (.8,0.75) {\scriptsize $\color{gray}{3}$};
\node[above] at (1.35,0.75) {\scriptsize $\color{gray}{4}$};
\node[above] at (1.85,0.75) {\scriptsize $\color{gray}{5}$};
 \draw[color=Red, fill=Red!20, line width=0.6pt] (0, 0) circle (1.75mm);
\draw[fill] (0, 0) circle (.5mm);
\draw[fill] (0.8, 0) circle (.5mm);
\draw[fill] (1.6, 0) circle (.5mm);
\draw[thick] (0, 0) -- (1.6, 0);
\node at (0.4,0)  {\Cross};
\node at (1.2,0)  {\Cross};
\end{tikzpicture}} &\qquad \Leftrightarrow \qquad
 \raisebox{-15pt}{\begin{tikzpicture}[line width=1.pt, scale=0.85]
\coordinate (3) at (0,1);
\coordinate (4) at (0.951,0.309);
\coordinate (5) at (0.588,-0.809);
\coordinate (1) at (-0.588,-0.809);
\coordinate (2) at (-0.951,0.309);
\fill[color=Red!20] (1) -- (2) -- (3) -- (1);
\draw[Red, line width=1.pt] (1)  -- node[] {} (3);
\draw[black, line width=1.pt] (1) -- (2) -- (3) -- (4) -- (5) -- (1);
\end{tikzpicture}}
\ \,\equiv \, \ud \log(k_{12}+y_{12}) \\[4pt]
 \raisebox{7pt}{
 \begin{tikzpicture}[baseline=(current  bounding  box.center)]
 \draw[lightgray, line width=1.pt] (0,0) -- (-0.25,0.75);
\draw[lightgray, line width=1.pt] (0,0) -- (0.25,0.75);
\draw[lightgray, line width=1.pt] (.8,0) -- (.8,0.75);
\draw[lightgray, line width=1.pt] (1.6,0) -- (1.35,0.75);
\draw[lightgray, line width=1.pt] (1.6,0) -- (1.85,0.75);
\draw[lightgray, line width=2.pt] (-0.5,0.75) -- (2.1,0.75);
\node[above] at (-0.25,0.75) {\scriptsize $\color{gray}{1}$};
\node[above] at (0.25,0.75) {\scriptsize $\color{gray}{2}$};
\node[above] at (.8,0.75) {\scriptsize $\color{gray}{3}$};
\node[above] at (1.35,0.75) {\scriptsize $\color{gray}{4}$};
\node[above] at (1.85,0.75) {\scriptsize $\color{gray}{5}$};
\node [
        draw, color=Red, fill=Red!20, line width=0.6pt,
        rounded rectangle,
        minimum height = 0.9em,
        minimum width = 2.5em,
        rounded rectangle arc length = 180,
    ] at (0.2,0)
    {};
\draw[fill] (0, 0) circle (.5mm);
\draw[fill] (0.8, 0) circle (.5mm);
\draw[fill] (1.6, 0) circle (.5mm);
\draw[thick] (0, 0) -- (1.6, 0);
\node at (0.4,0)  {\Cross};
\node at (1.2,0)  {\Cross};
\end{tikzpicture}} &\qquad \Leftrightarrow \qquad
 \raisebox{-15pt}{\begin{tikzpicture}[line width=1.pt, scale=0.85]
\coordinate (3) at (0,1);
\coordinate (4) at (0.951,0.309);
\coordinate (5) at (0.588,-0.809);
\coordinate (1) at (-0.588,-0.809);
\coordinate (2) at (-0.951,0.309);
\fill[color=Red!20] (1) -- (2) -- (3) -- (1);
\draw[Red, line width=1.pt, dashed] (1) --  (3);
\draw[black, line width=1.pt] (1) -- (2) -- (3) -- (4) -- (5) -- (1);
\end{tikzpicture}}
\ \,\equiv \, \ud \log(k_{12}-y_{12})
\\[4pt]
 \raisebox{7pt}{
 \begin{tikzpicture}[baseline=(current  bounding  box.center)]
 \draw[lightgray, line width=1.pt] (0,0) -- (-0.25,0.75);
\draw[lightgray, line width=1.pt] (0,0) -- (0.25,0.75);
\draw[lightgray, line width=1.pt] (.8,0) -- (.8,0.75);
\draw[lightgray, line width=1.pt] (1.6,0) -- (1.35,0.75);
\draw[lightgray, line width=1.pt] (1.6,0) -- (1.85,0.75);
\draw[lightgray, line width=2.pt] (-0.5,0.75) -- (2.1,0.75);
\node[above] at (-0.25,0.75) {\scriptsize $\color{gray}{1}$};
\node[above] at (0.25,0.75) {\scriptsize $\color{gray}{2}$};
\node[above] at (.8,0.75) {\scriptsize $\color{gray}{3}$};
\node[above] at (1.35,0.75) {\scriptsize $\color{gray}{4}$};
\node[above] at (1.85,0.75) {\scriptsize $\color{gray}{5}$};
\node [
        draw, color=Orange, fill=Orange!20, line width=0.6pt,
        rounded rectangle,
        minimum height = 0.9em,
        minimum width = 2.5em,
        rounded rectangle arc length = 180,
    ] at (0.6,0)
    {};
\draw[fill] (0, 0) circle (.5mm);
\draw[fill] (0.8, 0) circle (.5mm);
\draw[fill] (1.6, 0) circle (.5mm);
\draw[thick] (0, 0) -- (1.6, 0);
\node at (0.4,0)  {\Cross};
\node at (1.2,0)  {\Cross};
\end{tikzpicture}} &\qquad \Leftrightarrow \qquad
 \raisebox{-15pt}{\begin{tikzpicture}[line width=1.pt, scale=0.85]
\coordinate (3) at (0,1);
\coordinate (4) at (0.951,0.309);
\coordinate (5) at (0.588,-0.809);
\coordinate (1) at (-0.588,-0.809);
\coordinate (2) at (-0.951,0.309);
\fill[color=Orange!20] (1) -- (3) -- (4) -- (1);
\draw[Orange, line width=1.pt, dashed] (1) --  (3);
\draw[Orange, line width=1.pt] (1) --  (4);
\draw[black, line width=1.pt] (1) -- (2) -- (3) -- (4) -- (5) -- (1);
\end{tikzpicture}}
\ \, \equiv\,  \ud \log(k_3-y_{12}+y_{45}) \end{split} 
\end{align}
We can therefore enumerate all letters that will appear in the differential equations by listing all possible sub-polygons, allowing for internal lines to either be solid (so that the corresponding energy comes with a $+$ sign) or dashed (so that the internal energy comes with a $-$ sign).  In some sense the assignment of letters to shaded sub-polygons is more natural than the graph picture, because we can directly think of the letters as cataloguing the possible interesting kinematic limits where correlations could in principle have singularities. Conceptually, there are two interesting classes of limits. In the first, we analytically continue the lengths of the momenta, so that the length of the perimeter of some sub-polygon (which can be the full polygon) vanishes. These kinematic limits correspond to the shadings with solid internal lines. The second class of limits are those where we collapse some external lines to a diagonal, so that their momentum vectors become collinear. These limits correspond to the shadings with dashed internal lines.

\vskip4pt
Already at this level, we see a small compression of information: many sub-polygons appear in more than one triangulation, and so these letters are repeated in the differential equations when we split the full wavefunction into a sum over graphs. A simple example is provided at five points, where the letter/polygon
\begin{equation*}
\raisebox{-15pt}{\begin{tikzpicture}[line width=1.pt, scale=0.75]
\coordinate (3) at (0,1);
\coordinate (4) at (0.951,0.309);
\coordinate (5) at (0.588,-0.809);
\coordinate (1) at (-0.588,-0.809);
\coordinate (2) at (-0.951,0.309);
\fill[color=Red!20] (1) -- (2) -- (3) -- (4) -- (1);
\draw[Red, line width=1.pt,dashed] (1) --  (4);
\draw[black, line width=1.pt] (1) -- (2) -- (3) -- (4) -- (5) -- (1);
\end{tikzpicture}} \ \ \,
\equiv\,  \ud \log(k_{123}-y_{45}) 
\end{equation*}
appears in both Channel ${\rm I}$ and ${\rm IV}$, and is therefore associated to {\it both} of the following tubings in the graph picture
\begin{equation*}
 \raisebox{5pt}{
 \begin{tikzpicture}[baseline=(current  bounding  box.center)]
 \draw[lightgray, line width=1.pt] (0,0) -- (-0.25,0.75);
\draw[lightgray, line width=1.pt] (0,0) -- (0.25,0.75);
\draw[lightgray, line width=1.pt] (.8,0) -- (.8,0.75);
\draw[lightgray, line width=1.pt] (1.6,0) -- (1.35,0.75);
\draw[lightgray, line width=1.pt] (1.6,0) -- (1.85,0.75);
\draw[lightgray, line width=2.pt] (-0.5,0.75) -- (2.1,0.75);
\node[above] at (-0.25,0.75) {\scriptsize \color{gray}{$1$}};
\node[above] at (0.25,0.75) {\scriptsize \color{gray}{$2$}};
\node[above] at (.8,0.75) {\scriptsize \color{gray}{$3$}};
\node[above] at (1.35,0.75) {\scriptsize \color{gray}{$4$}};
\node[above] at (1.85,0.75) {\scriptsize \color{gray}{$5$}};
\node[below] at (.8,-.25) {\small Channel I};
 \node [
        draw, color=Red,fill=Red!20, line width=0.6pt,
        rounded rectangle,
        minimum height = 0.9em,
        minimum width = 4.5em,
        rounded rectangle arc length = 180,
    ] at (.6,0)
    {};
\draw[fill] (0, 0) circle (.5mm);
\draw[fill] (0.8, 0) circle (.5mm);
\draw[fill] (1.6, 0) circle (.5mm);
\draw[thick] (0, 0) -- (1.6, 0);
\node at (0.4,0)  {\Cross};
\node at (1.2,0)  {\Cross};
\end{tikzpicture}} 
\hspace{1.6cm}
 \raisebox{5pt}{
 \begin{tikzpicture}[baseline=(current  bounding  box.center)]
 \draw[lightgray, line width=1.pt] (0,0) -- (-0.25,0.75);
\draw[lightgray, line width=1.pt] (0,0) -- (0.25,0.75);
\draw[lightgray, line width=1.pt] (.8,0) -- (.8,0.75);
\draw[lightgray, line width=1.pt] (1.6,0) -- (1.35,0.75);
\draw[lightgray, line width=1.pt] (1.6,0) -- (1.85,0.75);
\draw[lightgray, line width=2.pt] (-0.5,0.75) -- (2.1,0.75);
\node[above] at (-0.25,0.75) {\scriptsize \color{gray}{$3$}};
\node[above] at (0.25,0.75) {\scriptsize \color{gray}{$2$}};
\node[above] at (.8,0.75) {\scriptsize \color{gray}{$1$}};
\node[above] at (1.35,0.75) {\scriptsize \color{gray}{$4$}};
\node[above] at (1.85,0.75) {\scriptsize \color{gray}{$5$}};
\node[below] at (.8,-.25) {\small Channel IV};
 \node [
        draw, color=Red,fill=Red!20, line width=0.6pt,
        rounded rectangle,
        minimum height = 0.9em,
        minimum width = 4.5em,
        rounded rectangle arc length = 180,
    ] at (.6,0)
    {};
\draw[fill] (0, 0) circle (.5mm);
\draw[fill] (0.8, 0) circle (.5mm);
\draw[fill] (1.6, 0) circle (.5mm);
\draw[thick] (0, 0) -- (1.6, 0);
\node at (0.4,0)  {\Cross};
\node at (1.2,0)  {\Cross};
\end{tikzpicture}} 
\end{equation*}
where the external lines are permuted. We see that the polygon picture is therefore somewhat more economical than the split into graphs. In particular, the total number of letters is smaller than the naive count obtained by summing up the letters for each individual channel.

\paragraph{Functions:} The source functions appearing in the equations also have a natural geometric interpretation: they are associated to (possibly disconnected) shaded sub-polygons, where all shaded triangles have at least one marked (dashed) internal edge. There is one exception to this general assignment, which is that a complete triangulation of the polygon, with no shadings or dashed internal edges is also naturally associated to a function (we will see that the function is the piece of the wavefunction that arises from the corresponding exchange channel). 
This enumeration is the natural translation of the association of marked graph tubings to functions in Section~\ref{sec:GraphicalRules}. We can therefore list the set of functions graphically. Much in the same way that only one tube can contain the cross between two vertices, if two sub-polygons are disjoint but share a border, only one of them can involve a negative energy (dashed line). In cases where disjoint polygons are adjacent, we will therefore draw two-sided borders, to make it clear to which polygon the dashed line belongs.

\vskip 4pt
As before, it is simplest to illustrate this with some examples. For the four-point function, there are $7$ source functions that appear when we differentiate the wavefunction, associated to the following ``shaded triangulations":
\begin{equation*}
 \begin{tikzpicture}[baseline=(current  bounding  box.center)]
          \draw[gray, line width=1.pt] (-2.75,3.5) -- (-1.75,4.5); 
 \draw[black, line width=1pt] (-2.75,3.5) -- (-2.75,4.5) -- (-1.75,4.5) -- (-1.75,3.5) -- (-2.75,3.5); 
\node[below] at (-2.25,3.5) {$\psi^{(s)}$};
     \draw[gray, line width=1.pt] (2.75,3.5) -- (1.75,4.5); 
 \draw[black, line width=1pt] (2.75,3.5) -- (2.75,4.5) -- (1.75,4.5) -- (1.75,3.5) -- (2.75,3.5); 
\node[below] at (2.25,3.5) {$\psi^{(t)}$};
    \fill[gray!20] (-3.5,1.5) -- (-3.5,2.5) -- (-2.5,2.5) -- (-3.5,1.5);
    \draw[gray, line width=1.pt, dashed] (-2.5,2.5) -- (-3.5,1.5); 
  \draw[black, line width=1pt] (-3.5,1.5) -- (-3.5,2.5) -- (-2.5,2.5) -- (-2.5,1.5) -- (-3.5,1.5); 
  \node[below] at (-3,1.5) {$F^{(s)}$};
    \fill[gray!20] (3.5,1.5) -- (3.5,2.5) -- (2.5,2.5) -- (3.5,1.5);
    \draw[gray, line width=1.pt, dashed] (2.5,2.5) -- (3.5,1.5); 
  \draw[black, line width=1pt] (3.5,1.5) -- (3.5,2.5) -- (2.5,2.5) -- (2.5,1.5) -- (3.5,1.5); 
  \node[below] at (3,1.5) {$\tilde F^{(t)}$};
    \fill[gray!20] (-2,1.5) -- (-1,1.5) -- (-1,2.5) -- (-2,1.5);
    \draw[gray, line width=1.pt, dashed] (-1,2.5) -- (-2,1.5); 
  \draw[black, line width=1pt] (-2,1.5) -- (-2,2.5) -- (-1,2.5) -- (-1,1.5) -- (-2,1.5); 
  \node[below] at (-1.5,1.5) {$\tilde F^{(s)}$};
    \fill[gray!20] (2,1.5) -- (1,1.5) -- (1,2.5) -- (2,1.5);
    \draw[gray, line width=1.pt, dashed] (1,2.5) -- (2,1.5); 
  \draw[black, line width=1pt] (2,1.5) -- (2,2.5) -- (1,2.5) -- (1,1.5) -- (2,1.5); 
  \node[below] at (1.5,1.5) {$F^{(t)}$};
\draw[black, fill=gray!20, line width=1pt] (-0.5,-0.5) -- (-0.5,0.5) -- (0.5,0.5) -- (0.5,-0.5) -- (-0.5,-0.5); 
\node[below] at (0,-0.5) {$Z$};
\end{tikzpicture}
\end{equation*}
Importantly, the source functions are associated to shaded sub-polygons, rather than channels per se. In this case, we see that the function $Z$ is ``shared" between the $s$ and $t$-channel contributions, in the sense that it would appear as a source in both sets of equations if we were to split the full answer into channels. This is more interesting than the fact that the same letters are shared between channels. When we consider the derivation of differential equations from the growth of polygons, we will see that this implies that the functions that appear depend only on the shaded sub-polygon, and not upon the path of shadings that we take to arrive there.  These shared functions are a nontrivial compatibility condition between the various channels.

\vskip4pt
At five points, there are more allowed triangulations, and so things are more interesting.
The allowed subdivisions of the kinematic pentagon that contribute to Channel I are:
\begin{equation*}
 \begin{tikzpicture}[baseline=(current  bounding  box.center)]
  \node[inner sep=0pt] at (-7,-2.3)
    {\includegraphics[scale=0.8]{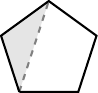}}; 
  \node[inner sep=0pt] at (-5,-2.3)
    {\includegraphics[scale=0.8]{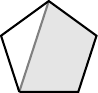}}; 
   \node[inner sep=0pt] at (-3,-2.3)
    {\includegraphics[scale=0.8]{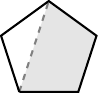}}; 
   \node[inner sep=0pt] at (-1,-2.3)
    {\includegraphics[scale=0.8]{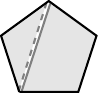}}; 
      \node[inner sep=0pt] at (1,-2.3)
    {\includegraphics[scale=0.8]{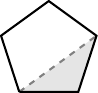}}; 
      \node[inner sep=0pt] at (3,-2.3)
    {\includegraphics[scale=0.8]{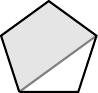}}; 
      \node[inner sep=0pt] at (5,-2.3)
    {\includegraphics[scale=0.8]{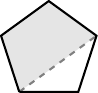}}; 
      \node[inner sep=0pt] at (7,-2.3)
    {\includegraphics[scale=0.8]{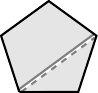}}; 
   \node[inner sep=0pt] at (-7,-3.3)
    {$F^{(\rm I,III)}$}; 
  \node[inner sep=0pt] at (-5,-3.3)
      {$\tilde q_1^{(\rm I,III)}$};
   \node[inner sep=0pt] at (-3,-3.3)
      {$\tilde q_2^{(\rm I,III)}$};
   \node[inner sep=0pt] at (-1,-3.3)
    {$\tilde g^{(\rm I,III)}$};
      \node[inner sep=0pt] at (1,-3.3)
    {$\tilde F^{(\rm I,IV)}$};
      \node[inner sep=0pt] at (3,-3.3)
      {$q_1^{(\rm I,IV)}$};
      \node[inner sep=0pt] at (5,-3.3)
      {$q_2^{(\rm I,IV)}$};
      \node[inner sep=0pt] at (7,-3.3)
    {$g^{(\rm I,IV)}$};
  \node[inner sep=0pt] at (-7,0)
    {\includegraphics[scale=0.8]{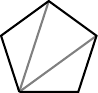}}; 
  \node[inner sep=0pt] at (-5,0)
    {\includegraphics[scale=0.8]{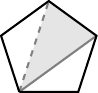}}; 
   \node[inner sep=0pt] at (-3,0)
    {\includegraphics[scale=0.8]{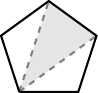}}; 
   \node[inner sep=0pt] at (-1,0)
    {\includegraphics[scale=0.8]{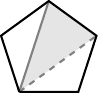}}; 
      \node[inner sep=0pt] at (1,0)
    {\includegraphics[scale=0.8]{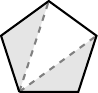}}; 
      \node[inner sep=0pt] at (3,0)
    {\includegraphics[scale=0.8]{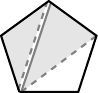}}; 
      \node[inner sep=0pt] at (5,0)
    {\includegraphics[scale=0.8]{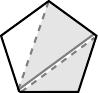}}; 
      \node[inner sep=0pt] at (7,0)
    {\includegraphics[scale=0.8]{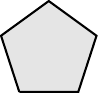}}; 
   \node[inner sep=0pt] at (-7,-1)
    {$\psi^{(\rm I)}$}; 
  \node[inner sep=0pt] at (-5,-1)
      {$Q_1^{(\rm I)}$};
   \node[inner sep=0pt] at (-3,-1)
      {$Q_2^{(\rm I)}$};
   \node[inner sep=0pt] at (-1,-1)
    {$Q_3^{(\rm I)}$};
      \node[inner sep=0pt] at (1,-1)
    {$f^{(\rm I)}$};
      \node[inner sep=0pt] at (3,-1)
      {$q_3^{(\rm I)}$};
      \node[inner sep=0pt] at (5,-1)
      {$\tilde q_3^{(\rm I)}$};
      \node[inner sep=0pt] at (7,-1)
    {$Z^{(\text{I--V})}$};
\end{tikzpicture}
\end{equation*}

\vskip 4pt
\noindent
There are two somewhat conceptually distinct classes of functions: those whose collections of sub-polygons are sensitive to the full triangulation, and those that only know about a partial triangulation, and are therefore shared between different channels. The sub-polygons appearing on the first line (except $Z$) are shaded pieces of the full triangulation I, and their corresponding functions therefore only appear in Channel I. Conversely, the functions in the second line are only sensitive to a partial triangulation, and therefore are shared between the Channels I and~III (or IV). Finally, the function $Z$ is shared between all five channels.
In the four-point case, only the function $Z$ was shared between the two channels, but here there is a richer structure of shared functions. In Section~\ref{sec:associahedron}, we will see how this can be understood in terms of a ``dressed" associahedron structure underlying the kinematic polygon.

\subsection{Differential Equations}
\label{ssec:polygondifferential}
Now that we understand how to enumerate the letters and functions that will appear in the system of equations, we would like to derive the differential equations themselves. As in Section~\ref{sec:GraphicalRules}, the differential equations have a relatively simple diagrammatic interpretation, but now in terms of the growth of sub-polygons inside the kinematic polygon, rather than the growth of tubings on graphs. The rules are the obvious translation of the graph-by-graph rules articulated in Section~\ref{sec:GraphicalRules}. For convenience, we summarize them in the following inset before discussing examples.

\vskip10pt
\begin{eBox}
\vskip1pt
\noindent
{\small {\bf Kinematic flow:} 
We begin with the shaded polygon associated to a parent function of interest and perform the following sequence of operations to generate a tree of descendants: 
\begin{enumerate}
\item First, we {\bf activate} all shaded sub-polygons of the parent function. We further activate all triangles that don't overlap with the shaded sub-polygons of the parent function. Each activation produces a branch of the tree.

As an example, we consider the tree for the function $Q_1^{\rm (I)}$:  
\beq
 \begin{tikzpicture}[baseline=(current  bounding  box.center)]
\node[inner sep=0pt] at (-0.1,0)
    {\includegraphics[scale=0.6]{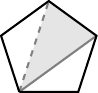}}; 
    \draw [color=gray,thick,-stealth] (0.6,0.3) -- (1.45,1.1);
     \draw [color=gray,thick,-stealth] (0.6,-0.3) -- (1.45,-1.1);
\draw [color=gray,thick,-stealth] (0.6,0) -- (1.45,0);
\node[inner sep=0pt] at (2.15,1.3)
    {\includegraphics[scale=0.6]{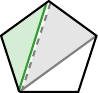}}; 
\node[inner sep=0pt] at (2.15,0)
    {\includegraphics[scale=0.6]{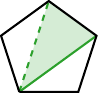}}; 
  \node[inner sep=0pt] at (2.15,-1.3)
    {\includegraphics[scale=0.6]{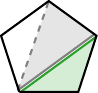}};   
\end{tikzpicture}
\eeq
Activation has created three branches: one from the activation of the gray-shaded triangle and two from that of the two triangles outside this shading.
\item Activated triangles with no dashed lines can then ``{\bf grow}" by changing internal edges from solid to dashed in all possible ways. If a dashed edge is adjacent to a dashed edge of an un-activated gray-shaded polygon, they further {\bf merge} (producing a new descendant) and the union of polygons becomes active.

For the function $Q_1^{\rm (I)}$, this growth and merger leads to
\beq
 \begin{tikzpicture}[baseline=(current  bounding  box.center)]
\node[inner sep=0pt] at (-0.1,0)
    {\includegraphics[scale=0.6]{Figures/Shadings/Pentagon/Tree/pentQ}}; 
    \draw [color=gray,thick,-stealth] (0.6,0.3) -- (1.45,1.1);
     \draw [color=gray,thick,-stealth] (0.6,-0.3) -- (1.45,-1.1);
\draw [color=gray,thick,-stealth] (0.6,0) -- (1.45,0);
\node[inner sep=0pt] at (2.15,1.3)
    {\includegraphics[scale=0.6]{Figures/Shadings/Pentagon/Tree/pentL11}}; 
\node[inner sep=0pt] at (2.15,0)
    {\includegraphics[scale=0.6]{Figures/Shadings/Pentagon/Tree/pentL3a}}; 
  \node[inner sep=0pt] at (2.15,-1.3)
    {\includegraphics[scale=0.6]{Figures/Shadings/Pentagon/Tree/pentQ44}};  
    \draw [color=gray,thick,-stealth] (2.85,1.3) -- (3.7,1.3);
      \node[inner sep=0pt] at (4.35,1.3)
    {\includegraphics[scale=0.6]{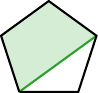}}; 
        \draw [color=gray,thick,-stealth] (2.85,-1.3) -- (3.7,-1.3);
  \node[inner sep=0pt] at (4.35,-1.3)
    {\includegraphics[scale=0.6]{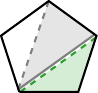}};  
\end{tikzpicture}
\eeq
In the top and bottom branches, the solid internal lines on the green triangles turned to dashed lines. In the top branch, this created two adjacent dashed lines, so the two triangles merged into a larger quadrilateral.

\item If a polygon with a dashed internal edge is adjacent to an un-activated gray-shaded polygon with a different dashed internal edge, the active polygon can {\bf absorb} the un-activated polygon, producing a new descendant and activating the union of the two polygons.

For the function $Q_1^{\rm (I)}$, we have one instance of an absorption phenomenon:
 \beq
 \begin{tikzpicture}[baseline=(current  bounding  box.center)]
\node[inner sep=0pt] at (-0.1,0)
    {\includegraphics[scale=0.6]{Figures/Shadings/Pentagon/Tree/pentQ}}; 
    \draw [color=gray,thick,-stealth] (0.6,0.3) -- (1.45,1.1);
     \draw [color=gray,thick,-stealth] (0.6,-0.3) -- (1.45,-1.1);
\draw [color=gray,thick,-stealth] (0.6,0) -- (1.45,0);
\node[inner sep=0pt] at (2.15,1.3)
    {\includegraphics[scale=0.6]{Figures/Shadings/Pentagon/Tree/pentL11}}; 
\node[inner sep=0pt] at (2.15,0)
    {\includegraphics[scale=0.6]{Figures/Shadings/Pentagon/Tree/pentL3a}}; 
  \node[inner sep=0pt] at (2.15,-1.3)
    {\includegraphics[scale=0.6]{Figures/Shadings/Pentagon/Tree/pentQ44}};  
    \draw [color=gray,thick,-stealth] (2.85,1.3) -- (3.7,1.3);
      \node[inner sep=0pt] at (4.35,1.3)
    {\includegraphics[scale=0.6]{Figures/Shadings/Pentagon/Tree/pentF5}}; 
        \draw [color=gray,thick,-stealth] (2.85,-1.3) -- (3.7,-1.3);
  \node[inner sep=0pt] at (4.35,-1.3)
    {\includegraphics[scale=0.6]{Figures/Shadings/Pentagon/Tree/pentQ4}};  
            \draw [color=gray,thick,-stealth] (5,-1.3) -- (5.85,-1.3);         
            \node[inner sep=0pt] at (6.5,-1.3)
    {\includegraphics[scale=0.6]{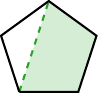}}; 
\end{tikzpicture}
\eeq

\end{enumerate}
Once we have this tree of descendants, we read off the differential equation following a similar procedure to before:
\begin{enumerate}
\item First, we assign a function to each leaf of the tree, which is just the function associated to the shaded sub-polygons present with a prefactor which is $(-1)^{N_a}$, where $N_a$ is the number of absorptions between the original shaded polygon and the leaf of interest. For immediate descendants of the original polygon, also multiply by $N_\bigtriangleup$, which is the number of triangles present in a triangulation of the activated polygon.

\item To generate the differential of the original function, we consider each node of the directed tree structure, and for the shaded polygon associated to each node,
we write the letter associated to its activated sub-polygon.

\item We then multiply this letter by the function associated to the shaded polygon minus the function associated to its immediate descendants, with an overall constant factor $\e$.
\end{enumerate}
}
\vskip1pt
\end{eBox}

\vskip 4pt
\paragraph{Four-point function:} 
As before, it is useful to consider some examples to build intuition. We first consider the four-point function.
The system of differential equations for the four-point function can be understood from successive shadings of a square. 
The tree associated to the differential of the wavefunction is
 \beq
 \begin{tikzpicture}[baseline=(current  bounding  box.center)]
\node[inner sep=0pt] at (0,0)
    {\includegraphics[scale=0.8]{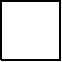}}; 
    \draw [color=gray,thick,-stealth] (0.6,0.5) -- (1.45,1.65);
     \draw [color=gray,thick,-stealth] (0.6,-0.5) -- (1.45,-1.65);
\draw [color=gray,thick,-stealth] (0.6,0.2) -- (1.45,0.65);
\draw [color=gray,thick,-stealth] (0.6,-0.2) -- (1.45,-0.65);
 \draw [color=gray,thick,line width=0.5pt, dashed] (1.6,0.2) -- (1.6,2.4) --  (2.6,2.4) -- (2.6,0.2) -- (1.6,0.2); 
   \node[above] at (2.1,2.45) {$\psi^{(s)}$};
    \draw [color=gray,thick,line width=0.5pt, dashed] (1.6,-0.2) -- (1.6,-2.4) --  (2.6,-2.4) -- (2.6,-0.2) -- (1.6,-0.2); 
   \node[below] at (2.1,-2.45) {$\psi^{(t)}$};
\node[inner sep=0pt] at (2.1,1.9)
    {\includegraphics[scale=0.8]{Figures/Shadings/Square/sqpsisr}};
\node[inner sep=0pt] at (2.1,0.7)
    {\includegraphics[scale=0.8]{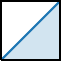}};
\node[inner sep=0pt] at (2.1,-0.7)
    {\includegraphics[scale=0.8]{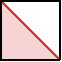}};
    \node[inner sep=0pt] at (2.1,-1.9)
    {\includegraphics[scale=0.8]{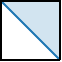}};
   \draw [color=gray,thick,-stealth] (2.8,1.9) -- (3.7,1.9); 
   \draw [color=gray,thick,-stealth] (2.8,0.7) -- (3.7,0.7); 
   \draw [color=gray,thick,-stealth] (2.8,-0.7) -- (3.7,-0.7);
   \draw [color=gray,thick,-stealth] (2.8,-1.9) -- (3.7,-1.9); 
\node[inner sep=0pt] at (4.3,1.9)
    {\includegraphics[scale=0.8]{Figures/Shadings/Square/sqFsr}};
\node[inner sep=0pt] at (4.3,0.7)
    {\includegraphics[scale=0.8]{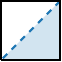}};
\node[inner sep=0pt] at (4.3,-0.7)
    {\includegraphics[scale=0.8]{Figures/Shadings/Square/sqFtr}};
    \node[inner sep=0pt] at (4.3,-1.9)
    {\includegraphics[scale=0.8]{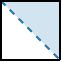}};  
   \node[right] at (4.7,1.9+.05) {$F^{(s)}$};  
   \node[right] at (4.7,0.7+.05) {$\tilde F^{(s)}$};  
      \node[right] at (4.7,-1.9+.05) {$\tilde F^{(t)}$};  
   \node[right] at (4.7,-0.7+.05) {$F^{(t)}$};         
\end{tikzpicture}
\label{eq:fullpsitree}
\eeq
We started with the blank kinematic square and then generated a first level of descendants from all possible shaded triangles that have an external edge.  Each of these activated triangles can develop a dashed internal line (which is the analogue of the growing tube in the single-graph case). This produces the descendants $F^{(s,t)}$ and $\tilde F^{(s,t)}$. In this case, the evolution stops at the second level. We can then read off the differential by writing down the letters corresponding to all active (shaded) triangles. Each letter is multiplied by the coefficient function for that shaded sub-polygon minus the coefficient function associated to the descendant one level down the chain (with an overall factor of $\e$).
This procedure leads to 
\be
\begin{aligned}
\ud\psi = \e\, \bigg[~~
&\raisebox{-8pt}{
\includegraphics[scale=0.8]{Figures/Shadings/Square/sqpsisr}
   }
   \big(\psi^{(s)}-F^{(s)}\big)\, 
   +
   \raisebox{-8pt}{
\includegraphics[scale=0.8]{Figures/Shadings/Square/sqFsr}
   }
   F^{(s)}
   \\[4pt]
+
& \raisebox{-8pt}{
\includegraphics[scale=0.8]{Figures/Shadings/Square/sqpsisb}
}
   \big(\psi^{(s)}-\tilde F^{(s)}\big)\,
   +
   \raisebox{-8pt}{
\includegraphics[scale=0.8]{Figures/Shadings/Square/sqFtsb}
   }
\tilde F^{(s)}
   \\[4pt]
   +
& \raisebox{-8pt}{
\includegraphics[scale=0.8]{Figures/Shadings/Square/sqpsitr} 
}  %
   \big(\psi^{(t)}-F^{(t)}\big)\
   +
   \raisebox{-8pt}{
\includegraphics[scale=0.8]{Figures/Shadings/Square/sqFtr} 
}
F^{(t)}\\[4pt]
+
& \raisebox{-8pt}{
\includegraphics[scale=0.8]{Figures/Shadings/Square/sqpsitb} 
}
   \big(\psi^{(t)}-\tilde F^{(t)}\big)\ 
   +
   \raisebox{-8pt}{
\includegraphics[scale=0.8]{Figures/Shadings/Square/sqFttb} 
}
\tilde F^{(t)}\bigg]\,.
\end{aligned}
\label{eq:psisquareeq}
\ee
\vskip 4pt
\noindent
Notice that the wavefunction coefficients corresponding  $s$ and $t$-channel exchanges appear as sources in~\eqref{eq:psisquareeq}. Taking their differentials just reproduces the equations for single graphs, e.g.~\eqref{equ:TwoSite-dPsi} for $\psi^{(s)}$. In this example, the differential of the sum of graphs is just the sum of the differentials for each graph.

\vskip4pt
We can then continue to construct the differential of the source functions $F^{(s,t)}$ and $\tilde F^{(s,t)}$. For concreteness, we will consider $F^{(s,t)}$, but the other functions $\tilde F^{(s,t)}$ satisfy very similar equations, which can be obtained by permutation. The differentials of $F^{(s,t)}$ can be computed from the following tree:
 \beq
 \begin{tikzpicture}[baseline=(current  bounding  box.center)]
\node[inner sep=0pt] at (0.1,1.4)
    {\includegraphics[scale=0.8]{Figures/Shadings/Square/sqFs}}; 
    \node[inner sep=0pt] at (0.1,-1.4)
    {\includegraphics[scale=0.8]{Figures/Shadings/Square/sqFt}}; 
    \draw [color=gray,thick,-stealth] (0.6,1.5) -- (1.45,1.9);
     \draw [color=gray,thick,-stealth] (0.6,1.3) -- (1.45,0.9);
     \draw [color=gray,thick,-stealth] (0.6,-1.5) -- (1.45,-1.9);
     \draw [color=gray,thick,-stealth] (0.6,-1.3) -- (1.45,-0.9);
 \draw [color=gray,thick,line width=0.5pt, dashed] (1.6,0.3) -- (1.6,2.5) --  (2.6,2.5) -- (2.6,0.3) -- (1.6,0.3); 
   \node[above] at (2.1,2.55) {$F^{(s)}$};
    \draw [color=gray,thick,line width=0.5pt, dashed] (1.6,-0.3) -- (1.6,-2.5) --  (2.6,-2.5) -- (2.6,-0.3) -- (1.6,-0.3); 
   \node[below] at (2.1,-2.55) {$F^{(t)}$};
\node[inner sep=0pt] at (2.1,2)
    {\includegraphics[scale=0.8]{Figures/Shadings/Square/sqFsr}};
\node[inner sep=0pt] at (2.1,0.8)
    {\includegraphics[scale=0.8]{Figures/Shadings/Square/sqFsZ}};
\node[inner sep=0pt] at (2.1,-2)
    {\includegraphics[scale=0.8]{Figures/Shadings/Square/sqFtr}};
    \node[inner sep=0pt] at (2.1,-0.8)
    {\includegraphics[scale=0.8,angle=90]{Figures/Shadings/Square/sqFsZ}};
   \draw [color=gray,thick,-stealth] (2.8,0.7) -- (3.7,0.2); 
   \draw [color=gray,thick,-stealth] (2.8,-0.7) -- (3.7,-0.2);
\node[inner sep=0pt] at (4.2,0)
    {\includegraphics[scale=0.8]{Figures/Shadings/Square/sqZb}}; 
   \node[right] at (4.65,0) {$Z$};    
\end{tikzpicture}
\eeq
We see that in both the $s$ and $t$-channel contributions, the same function $Z$ is created by growth of an activated triangle (blue) and its merger with the un-activated triangle (gray). 
The differential equations for $F^{(s)}$ and $F^{(t)}$ therefore are
\begin{align}
\ud F^{(s)} &= \e\,\bigg[\    
\raisebox{-8pt}{\includegraphics[scale=0.8]{Figures/Shadings/Square/sqFsr} }F^{(s)} 
 \  + \
\raisebox{-8pt}{\includegraphics[scale=0.8]{Figures/Shadings/Square/sqFsZ} } (F^{(s)}  - Z)
  \ + \
\raisebox{-8pt}{\includegraphics[scale=0.8]{Figures/Shadings/Square/sqZb} }Z\bigg]\,,
\label{equ:dFs} \\
\ud F^{(t)} &= \e\,\bigg[\    
\raisebox{-8pt}{\includegraphics[scale=0.8]{Figures/Shadings/Square/sqFtr} }F^{(t)} 
 \  + \ \hskip 1pt
\raisebox{-8pt}{\includegraphics[scale=0.8,angle=90]{Figures/Shadings/Square/sqFsZ} } (F^{(t)}  - Z)
  \ + \ \hskip 1pt
\raisebox{-8pt}{\includegraphics[scale=0.8]{Figures/Shadings/Square/sqZb} }Z\bigg]\,.
\label{equ:dFt}
\end{align}
Crucially, the function $Z$ is the same in (\ref{equ:dFs}) and (\ref{equ:dFt}), as well as in their tilded counterparts.
In this particular example, there is only one shared function because the two triangulations do not share any sub-polygons (aside from the full square), so the structure is not that impressive, but we will see that many more functions are shared in other examples. Broadly speaking, it is this structure that makes splitting the wavefunction into channels slightly artificial.

\vskip 4pt
Finally, taking a differential of the function $Z$, we obtain the equation
\be
\ud Z = 
\raisebox{3pt}{
\begin{tikzpicture}[baseline=(current  bounding  box.center),scale=0.8]
     \fill[Green!20] (-2.75,3.5) -- (-2.75,4.5) -- (-1.75,4.5) -- (-1.75,3.5) -- (-2.75,3.5);
 \draw[black, line width=1pt] (-2.75,3.5) -- (-2.75,4.5) -- (-1.75,4.5) -- (-1.75,3.5) -- (-2.75,3.5); 
\end{tikzpicture}
} 
 2\e  Z\,,
\ee
which is the same as~\eqref{equ:TwoSite-dZ}, and the system closes.

\paragraph{Five-point function:} The five-point function has a richer structure that reveals more of the universal features of the system of differential equations. In this case, there are 41 letters and 56 independent functions. Although it is possible to enumerate all of these and derive the differential equations that they all satisfy, this is quite a large number of equations and does not lead to a huge amount of insight. Instead, we will show some representative examples that illustrate some of the structure.

\vskip4pt
Recall that the five-point function is related to triangulations of a pentagon and that there are five different exchange channels that contribute to the full wavefunction. When we differentiate the full wavefunction, the pieces that comprise these channels get shuffled together in a nontrivial way. We first consider the differential of the wavefunction. The full tree of descendants is fairly large, so we will only show a representative piece of it: 
\begin{align}
 \begin{tikzpicture}[baseline=(current  bounding  box.center)]
\node[inner sep=0pt] at (-0.1,0)
    {\includegraphics[scale=0.6]{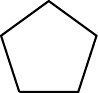}}; 
    \draw [color=gray,thick,-stealth] (0.6,0.3) -- (1.45,1.1);
     \draw [color=gray,thick,-stealth] (0.6,-0.3) -- (1.45,-1.1);
	\draw [color=gray,thick,-stealth] (0.6,0) -- (1.45,0);
\node[inner sep=0pt] at (2.15,1.3)
    {\includegraphics[scale=0.6]{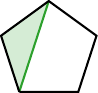}}; 
\node[inner sep=0pt] at (2.1,0)
    {\includegraphics[angle=72,scale=0.6]{Figures/Shadings/Pentagon/Tree/pentL1}}; 
  \node[inner sep=0pt] at (2.15,-1.3)
    {\includegraphics[scale=0.6]{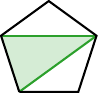}};  
     \node[inner sep=0pt] at (2.15,-1.3-1.)
     {\vdots};
    \draw [color=gray,thick,-stealth] (2.85,1.6-1.3) -- (3.7,2.4-1.3);
    \draw [color=gray,thick,-stealth] (2.85,1.6) -- (3.7,2.4);
      \node[inner sep=0pt] at (4.35,2.65)
    {\includegraphics[scale=0.6]{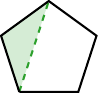}}; 
          \node[inner sep=0pt] at (4.35,2.65-1.3)
    {\includegraphics[angle=72, scale=0.6]{Figures/Shadings/Pentagon/Tree/pentL1a}}; 
   \draw [color=gray,thick,-stealth] (0.6+2.25,0.3-1.3) -- (1.45+2.25,1.1-1.3);
     \draw [color=gray,thick,-stealth] (0.6+2.25,-0.3-1.3) -- (1.45+2.25,-1.1-1.3);
	\draw [color=gray,thick,-stealth] (0.6+2.25,0-1.3) -- (1.45+2.25,0-1.3);
     \node[inner sep=0pt] at (2.15+2.25,1.3-1.3)
    {\includegraphics[scale=0.6]{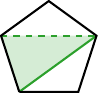}}; 
	\node[inner sep=0pt] at (2.1+2.25,0-1.3)
    {\includegraphics[scale=0.6]{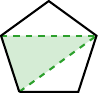}}; 
	  \node[inner sep=0pt] at (2.15+2.25,-1.3-1.3)
    {\includegraphics[scale=0.6]{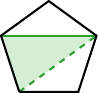}};  
\end{tikzpicture}
\end{align}
At the first level, we generate all possible triangles, which then ``grow" by turning a solid internal line into a dashed one (in all possible ways). There are four possible triangles with only one internal side, and five triangles with two internal sides. We have shown the part of the differential relevant for the 
$\partial_{k_1}$-derivative of the wavefunction. Notice that the shaded triangles with only one internal side (and two external sides) are naturally associated to the sum of two kinematic channels, while the  triangles with two internal lines belong to a single channel.

\vskip4pt
\newpage
We can similarly derive the differential equations that each of the source functions satisfy. Here, we just wish to highlight one interesting feature. The tree of descendants associated to \raisebox{-2pt}{\includegraphics[scale=0.25]{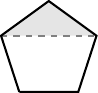}} and \raisebox{-2pt}{\includegraphics[scale=0.25]{Figures/Shadings/Pentagon/Tree/pentQ}}, which correspond to the functions $F^{\rm  (II,IV)}$ and $Q_1^{\rm (I)}$ in {\it different} channels, takes the schematic form
 \beq
 \begin{tikzpicture}[baseline=(current  bounding  box.center)]
\node[inner sep=0pt] at (-0.1,4)
    {\includegraphics[scale=0.6]{Figures/Shadings/Pentagon/Tree/pentFc2}}; 
    \draw [color=gray,thick,-stealth] (0.6,4) -- (1.45,4);
        \draw [color=gray,thick,-stealth] (0.6,3.6) -- (1.45,2.6);
             \node[inner sep=0pt] at (2.15,5.2) {\vdots};
             \node[inner sep=0pt] at (2.15,3) {\vdots};
            \draw [color=gray,thick,-stealth] (0.6,3.8) -- (1.45,3.2);
           \draw [color=gray,thick,-stealth] (0.6,4.2) -- (1.45,4.8);
                    \draw [color=gray,thick,-stealth] (0.6,4.4) -- (1.45,5.4);
    \draw [color=gray,thick,-stealth] (2.85,3.7) -- (3.7,2.9);
      \draw [color=gray,thick,-stealth] (2.85,4) -- (3.7,4);
        \draw [color=gray,thick,-stealth] (2.85,4.3) -- (3.7,5.1);
    \node[inner sep=0pt] at (2.15,4)
    {\includegraphics[scale=0.6]{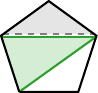}}; 
       \node[inner sep=0pt] at (4.35,5.35)
    {\includegraphics[scale=0.6]{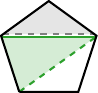}}; 
           \node[inner sep=0pt] at (4.35,4)
    {\includegraphics[scale=0.6]{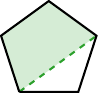}}; 
\node[inner sep=0pt] at (-0.1,0)
    {\includegraphics[scale=0.6]{Figures/Shadings/Pentagon/Tree/pentQ}}; 
    \draw [color=gray,thick,-stealth] (0.6,0.3) -- (1.45,1.1);
     \draw [color=gray,thick,-stealth] (0.6,-0.3) -- (1.45,-1.1);
\draw [color=gray,thick,-stealth] (0.6,0) -- (1.45,0);
\node[inner sep=0pt] at (2.15,1.3)
    {\includegraphics[scale=0.6]{Figures/Shadings/Pentagon/Tree/pentL11}}; 
\node[inner sep=0pt] at (2.15,0)
    {\includegraphics[scale=0.6]{Figures/Shadings/Pentagon/Tree/pentL3a}}; 
  \node[inner sep=0pt] at (2.15,-1.3)
    {\includegraphics[scale=0.6]{Figures/Shadings/Pentagon/Tree/pentQ44}};  
    \draw [color=gray,thick,-stealth] (2.85,1.6) -- (3.7,2.4);
      \node[inner sep=0pt] at (4.35,2.65)
    {\includegraphics[scale=0.6]{Figures/Shadings/Pentagon/Tree/pentF5}}; 
        \draw [color=gray,thick,-stealth] (2.85,-1.3) -- (3.7,-1.3);
  \node[inner sep=0pt] at (4.35,-1.3)
    {\includegraphics[scale=0.6]{Figures/Shadings/Pentagon/Tree/pentQ4}};  
            \draw [color=gray,thick,-stealth] (5,-1.3) -- (5.85,-1.3);         
            \node[inner sep=0pt] at (6.5,-1.3)
    {\includegraphics[scale=0.6]{Figures/Shadings/Pentagon/Tree/pentF7}}; 
\end{tikzpicture}
\eeq

\vskip 4pt
\noindent
An interesting feature is that \raisebox{-2pt}{\includegraphics[scale=0.25]{Figures/Shadings/Pentagon/Tree/pentF5}} appears in two branches (i.e.~it arises from two different sequences of shadings of triangles). Nevertheless, it is associated in both cases to the {\it same} function. This is an illustration of the fact that the functions that appear in the differential system are associated to shaded sub-polygons, rather than to channels, and that they are independent of the path of shadings taken to shade a given sub-polygon.

\vskip4pt
We see from these examples that certain functions  are associated to specific channels, while others appear in the equations for multiple channels. In the following section, we will see that there is an intriguing geometric structure underlying the way that functions are shared between channels, and this will allow us to easily count the total number of functions and letters of the full system.

\subsection{Dressed Associahedra}
\label{sec:associahedron}

Both the letters and source functions that will appear in the differential equations are naturally associated to sub-polygons of the kinematic polygon. Many of these substructures appear in multiple full triangulations of the polygon, and therefore would appear in multiple channels if we were to split apart the final wavefunction coefficient. Assuring the compatibility of these different channel decompositions is fairly nontrivial; one aspect of this is counting the number of independent letters and functions that appear in the full system. In the context of scattering amplitudes, the compatibility of different  channels is guaranteed by the fact that they can be assigned to the vertices of a polytope---the {\it associahedron}~\cite{Arkani-Hamed:2017mur}. We will see that the objects appearing in cosmology can also be associated to a sort of ``dressed" associahedron, which captures the compatibility relations between various channels.\footnote{Of course, an important difference is that, in the amplitudes case, the relevant associahedron can be thought of as living directly in the space of kinematic variables, and the canonical function of this positive geometry is the corresponding scattering amplitude~\cite{Arkani-Hamed:2017mur}. Finding the analogous object for the cosmological wavefunction would be extremely interesting.}

\paragraph{Associahedron:} The associahedron 
relevant for an $n$-point function 
is an $(n-3)$-dimensional polytope, $A_n$, which has $C_{n-2}$ vertices. These vertices can be placed in one-to-one correspondence with the complete triangulations of a regular $n$-gon. Moreover, the codimension-$k$ boundaries are in one-to-one correspondence with the partial triangulations of an $n$-gon obtained by removing $k$ internal lines. The compatibility condition between boundaries is that a boundary of codimension-$k$ is adjacent to a boundary of codimension $m > k$ if and only if the partial triangulation associated to the latter boundary can be obtained from the former by the addition of $m-k$ diagonals. The simplest associahedron, $A_3$, corresponding to a kinematic triangle (or a three-point function), is just a point. The associahedron for the four-point function is a line segment, and the one for a five-point function is a pentagon.

\paragraph{Letters, sources and dressing:} The associahedron naturally captures the relations between different triangulations (and hence channels), and so makes it easy to visualize how letters and functions are shared between channels. 

\begin{figure}[t]
    \centering
 \begin{tikzpicture}[baseline=(current  bounding  box.center)]
  \draw[Blue, line width=0.5pt] (-2.35,-0.35) -- (-1.65,0.35);
    \draw[Blue, line width=0.5pt] (2.35,-0.35)  -- (1.65,0.35);
    \draw[Blue, line width=0.5pt] (-2.35,-0.35) -- (-2.35,0.35) -- (-1.65,0.35) -- (-1.65,-0.35) -- (-2.35,-0.35);
    \draw[Blue, line width=0.5pt] (2.35,-0.35) -- (2.35,0.35) -- (1.65,0.35) -- (1.65,-0.35) -- (2.35,-0.35);
        \draw[Green, line width=0.5pt] (-0.35,0.2) -- (-0.35,0.9) -- (0.35,0.9) -- (0.35,0.2) -- (-0.35,0.2) ;
 \draw[Green, line width=1.5pt] (-1.35,0) -- (1.35,0) ;
 \draw[fill, Blue] (-1.35, 0) circle (.75mm);
  \draw[fill, Blue] (1.35, 0) circle (.75mm);
   \node[Blue, above] at (2,0.37) {\small $3$};
   \node[Blue, above] at (-2,0.37) {\small $3$};
   \node[Green,above] at (0,0.9) {\small $1$};
   \node at (-2.75,0) {$=$};
      \node at (2.75,0) {$=$};
   \draw [thick,decorate,
    decoration = {calligraphic brace}] (-3.2,0.4) --  (-3.2,-0.4);
   \node[inner sep=0pt] at (-3.7-1.8,0)
    {\includegraphics[scale=0.55]{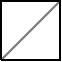}}; 
   \node[inner sep=0pt] at (-3.7-.9,0)
    {\includegraphics[scale=0.55]{Figures/Shadings/Square/sqFs}}; 
\node[inner sep=0pt] at (-3.7,0)
    {\includegraphics[scale=0.55]{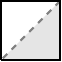}}; 
       \draw [thick,decorate,
    decoration = {calligraphic brace}] (-5.5-.5,-0.4) --  (-5.5-.5,0.4);
     \draw [thick,decorate,
    decoration = {calligraphic brace}] (3.2,-0.4) --  (3.2,0.4);
   \node[inner sep=0pt] at (3.7+1.8,0)
    {\includegraphics[scale=0.55]{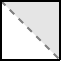}}; 
   \node[inner sep=0pt] at (3.7+.9,0)
    {\includegraphics[scale=0.55]{Figures/Shadings/Square/sqFt}}; 
\node[inner sep=0pt] at (3.7,0)
    {\includegraphics[scale=0.55]{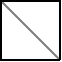}}; 
       \draw [thick,decorate,
    decoration = {calligraphic brace}] (5.5+.5,0.4) --  (5.5+.5,-0.4);
\end{tikzpicture}
\caption{Illustration of the dressed associahedron for the four-point function. The left and right vertices correspond to the triangulations associated to the $s$ and $t$-channel exchanges, respectively, each of which is dressed with $3$ source functions $\{\psi,F,\tilde F\}$ for the given channel. The edge corresponds to the function $Z$, which is associated to the un-triangulated square and is shared between the two channels.}
\label{equ:4pt-assoc}
\end{figure}

\vskip4pt
It is simplest to begin by considering the associahedron for the four-point function (see Fig.\,\ref{equ:4pt-assoc}).
The two vertices correspond to the triangulations associated to the $s$ and $t$-channel exchanges, while the edge connecting them corresponds to the un-triangulated square.
Each vertex is dressed by three functions---$\{\psi^{(s)}, F^{(s)}, \tilde F^{(s)}\}$ and $\{\psi^{(t)}, F^{(t)}, \tilde F^{(t)}\}$---while the edge contains the shared function~$Z$.
Similarly, one letter, $k_{1234}$, is shared between the two channels and four letters---$\{k_{12} \pm s,  k_{34}\pm s\}$ and $\{k_{14} \pm t,  k_{23}\pm t\}$---are associated uniquely to each channel.

 \begin{figure}[t!]
\centering
\includegraphics[scale=.9]{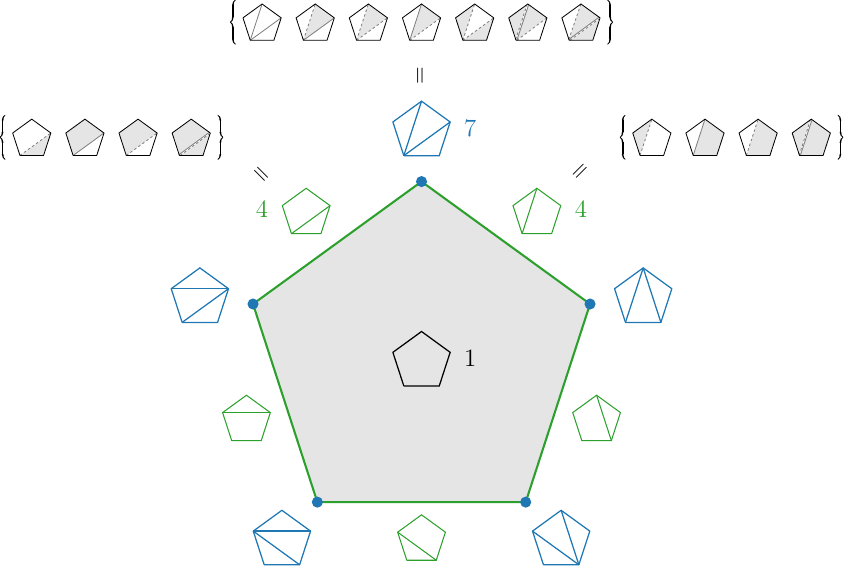}
\caption{Illustration of the five-point associahedron and the shared functions between the various channels. Each vertex corresponds to a complete triangulation, and there are $7$ source functions specific to a given channel. The edges connecting the various vertices correspond to partial triangulations with a single internal line, and there are $4$ functions associated to each edge that are shared between the two channels corresponding to the connected vertices. The center of the pentagon corresponds to the un-triangulated pentagon (and associated source function), which is shared between all channels.}
\label{fig:shared}
\end{figure}

\vskip4pt
A more nontrivial example is provided by the five-point function, where both the kinematic polygon and its associahedron are pentagons. The associahedron and the relevant (partial) triangulations are visualized in Figure~\ref{fig:shared}. Each vertex (exchange channel) is connected via edges to two other vertices. The partial triangulations associated to these edges correspond to letters and functions that are shared between the two channels.
All vertices are codimension-two boundaries of the bulk of the pentagon, which corresponds to the un-triangulated pentagon. This central pentagon is associated to the letter $k_{12345}$ and the source function $Z$, which are both shared between all five channels.

\vskip4pt
Since each boundary of the associahedron corresponds to multiple letters/functions, we can think of decorating the associahedron by numbers that count the multiplicity of these objects. At each vertex of the five-point associahedron, there are $4$ letters and $7$ source functions that are sensitive to the full triangulation. At each edge, there are $4$ letters and also $4$ functions that are shared between adjacent channels. Finally, in the middle of the pentagon there is a single letter and a single function that is shared between all the channels. Notice that this dressing makes it straightforward to count the total number of functions/letters, but also the number in a given channel. To count the letters and functions in a given channel, we select a given vertex and add up the numbers associated to that vertex, the edges connected to it, and the bulk of the polytope. This tells us that each channel will have $4+4+4+1 = 13$ letters and $7+4+4+1 =16$ functions (matching our earlier counts). In order to obtain the full count, we just add up the numbers associated to the bulk and the boundaries of all codimensions. In this case, we find that the sum of graphs will have $41$ letters and $56$ functions. (Note that this is substantially fewer than the naive count of $5\times13 = 65$ letters and $5\times 16 = 80$ functions, which doesn't take the sharing of functions into account.)

\begin{figure}[t!]
\centering
\includegraphics[scale=.9]{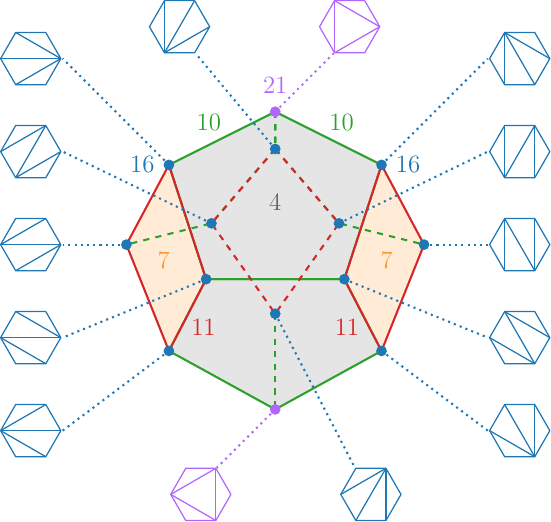}
\caption{Associahedron for the six-point function, with vertices labeled with their corresponding triangulation. The two different colorings for the vertices correspond to different graph topologies---purple for the star and blue for the chain. Source functions and letters that are sensitive to the full triangulation are naturally associated to the vertices, while shared functions are associated to edges and faces. Shown is the number of source functions associated to each vertex/edge/face. The bulk of the polytope corresponds to the one function shared by all channels. }
\label{fig:shared2}
\end{figure}

\vskip4pt
As a final example, we can consider the six-point function, which has three internal lines. An interesting novelty of this case is that there are two different graph topologies---the four-site chain and the four-site star---that both contribute in different exchange channels. In terms of the associahedron (which is now a three-dimensional polytope, visualized in Figure~\ref{fig:shared2}) this means that there are now two types of vertices, which are connected to different edges and faces. As before, we can associate a certain number of functions and letters to each vertex, which depend on the full triangulation, and other functions and letters to the edges and faces that are shared between the vertices. 
Note that each exchange channel has the same number of letters and functions, including channels with different graph topologies. Since there are two different kinds of vertices, which are connected to a different facet structure, it is nontrivial that the numbers add up to give the same total. In this case, it corresponds to the identity 
\beq
21 + 3\times10 + 3\times4 + 1 =  16 + 2\times11 + 10 + 7 + 2\times4 + 1\,.
\eeq
Finally, the total number of independent functions is given by
\beq
 \underbrace{12\times 16 + 2\times 21}_{\rm vertices}\ +\ \underbrace{12\times 11+9\times 10}_{\rm edges}\ +\ \underbrace{6\times 4+3\times 7}_{\rm faces}\ +\ \underbrace{1\times 1}_{\rm volume} = 502\,,
\eeq
as can be read off from the dressed associahedron shown in Figure~\ref{fig:shared2}. 
Notice that this is significantly less than the naive counting of $64\times 14=896$ functions, where 64 is the size of the basis for a single four-site graph and 14 is the number of channels for the six-point function. 

\vspace{0.25cm}
\begin{center}
***
\end{center}
So far, the dressed associahedra are just bookkeeping devices, which organize the information about shared functions in the differential equations for the individual channels.
This is to be contrasted with the case for scattering amplitudes where associahedra have a life of their own~\cite{Arkani-Hamed:2017mur}. In particular, the canonical function of the associahedron directly gives the corresponding flavor-ordered amplitude. Moreover, each facet of the geometry describes a particular factorization channel of the scattering process. It remains an interesting challenge for the future to breathe similar physics life into the dressed associahedron underlying the cosmological wavefunction.

\newpage
\section{Outlook and Speculations}
\label{sec:Outlook}

How does the strength of correlations change as we smoothly vary kinematic parameters?
This is one of the most natural questions to ask of cosmological correlation functions.
In this paper, we showed that the cosmological wavefunction is part of a finite-dimensional vector space whose basis members satisfy a system of first-order differential equations in the spatial momenta.  We saw that kinematic derivatives generate an evolution through the space of functions with a given structure of singularities. Remarkably, this ``kinematic flow" can be derived completely abstractly in terms of the local evolution of tubes of a marked graph, subject to relatively simple rules, which replaces the picture of bulk time evolution in standard cosmological perturbation theory.  This unexpected pattern in the equations inspires belief in the existence of some autonomously defined mathematical structure that directly computes cosmological correlations, and we hope that this work serves as an invitation to find such a structure.

\vskip4pt
For scattering amplitudes, irreducible structures arise in the sum over the Feynman diagrams, while the individual diagrams are often not physically meaningful objects.  One of the simplest incarnations of this phenomenon arises in ${\rm tr}\, \phi^3$ theory, which  is therefore also a natural place to look for similar structures in the cosmological setting.  By considering the differential equations satisfied by the full flavor-ordered wavefunction, we have seen the first indications of such a larger structure. The nontrivial feature of the differential system is that certain functions and symbol letters are shared between would-be different kinematic channels in a relatively intricate way. This pattern of shared letters/functions can naturally be understood in terms of a ``dressed associahedron," but the full geometric picture still remains to be explored.

\vskip4pt
It seems clear that we are just enjoying the first vista along a long journey, but there is an intriguing glimpse of a structure in the distance.  There are clear paths forward, in both physical and mathematical directions, and we hope that by combining these investigations this larger structure will emerge.

\subsection*{Physics} 

Although we have considered a somewhat special physical setup of conformally coupled scalars in a power-law cosmology, similar techniques can be employed in more general situations. 

\begin{itemize}
\item {\it Massive particles:} An important extension is to consider particles with generic masses. 
While we have treated the cosmology dependence as the twist deformation parameter, it is equally possible to work in exact de Sitter space and treat the masses of particles as twist deformations, providing a different way to compute these correlation functions, some of which are of phenomenological interest~\cite{Arkani-Hamed:2015bza}. 

\item {\it Spin:}  So far, we have only studied scalar theories.  Theories with spinning particles---especially gauge theories and gravity---are particularly interesting because the most drastic simplifications are then expected in the sum over graphs. This remains an important avenue of further investigation.

\item {\it Loops:} An important challenge is to understand how to perform loop integrals in cosmology. In this case, there are two types of integrations: those over the energies of the external states, which turn flat-space correlators into cosmological correlators, and integrals over the loop momenta. We expect that techniques similar to those developed in this paper can be used for both types of integrals. Very little is known about the function space of loop correlators, and understanding the structure of their differential equations and the associated space of functions will provide valuable insights.

\item {\it Analytic structure:}  Our understanding of the analytic structure of cosmological correlators is still rather limited, although interesting progress was recently made in the case of the flat-space wavefunction \cite{Salcedo:2022aal} (see also~\cite{Arkani-Hamed:2017fdk, Arkani-Hamed:2018bjr, Jazayeri:2021fvk, Baumann:2021fxj}). 
In this work, we discovered that their is a simple pattern in the singularities of the differential equations satisfied by cosmological correlators.  It would be interesting to relate these insights more directly to the analytic structure of the wavefunction. 

\item {\it Space of functions:}  What classes of correlation functions can arise from consistent cosmological evolution? Surprisingly little is known about this in a systematic fashion. In this paper, we have started to answer this question for the case of conformally coupled scalars in a power-law FRW background. One-site (contact) graphs lead to simple power-law solutions, while two-site (single-exchange) graphs give power laws and hypergeometric functions. Three-site (double-exchange) graphs contain power laws, hypergeometric functions and integrals of hypergeometric functions. A similar hierarchical structure is expected in more general examples, but this remains to be further explored. 

\item {\it Recovering time:}  It would be important to understand how to systematically ``integrate in" time from this boundary point of view. Given the differential equations satisfied by the cosmological wavefunction, how do we see explicitly that their solutions can be written in terms of integrals over an auxiliary ``time" variable, without knowing this a priori? This is a concrete and natural step to better understand the encoding of time in these static observables.
\end{itemize}

\subsection*{Mathematics}
Our investigation has revealed several interesting mathematical structures which deserve to be understood more fully in their own right. It is natural to expect that a more complete mathematical understanding will also lead to further physical insights.

\begin{itemize}
\item {\it Twists:} The twisted integrals describing cosmological correlations are highly non-generic. In particular, the physical requirement of bulk locality imposes important restrictions on which hypersurfaces (in kinematic space) can appear as twist factors. We are therefore led to a rather special subset of twisted integrals as being of physical interest. It may be illuminating to study further which classes of twisted integrals can reflect consistent physical processes and which don't. This may provide an autonomous mathematical perspective to fundamental physical principles. In other words, maybe the principles of locality and causality arise as the answers to specific mathematical questions.

\item {\it Tubings:} We introduced tubings of marked graphs to define a specific set of basis functions. It is in this particular basis that the equations have the structures that we described. It may be instructive to understand further what is special about this choice of basis in the eyes of a mathematician, along with the more general exploration of the natural notion of compatibility of tubings in the cosmology context.

\item {\it Flow:} We discovered simple rules underlying the ``evolution" of the letters in the differential equations. Despite being easy to apply, the rules have some subtle features (like the directionality of the absorption rule; see Section~\ref{sec:GraphicalRules}), which are necessary for $\ud^2 = 0$.  It would be nice to find the deeper mathematical raison d'\^etre of these rules. 

\item {\it Explicit solutions:} We have derived systems of equations satisfied by cosmological wavefunctions, but have only solved these equations in some simple examples. An important challenge is to understand how to solve these equations in the most efficient way. There is an interesting hierarchical structure to the equations (see Appendix~\ref{app:functions}) that it would be important to explore further, along with developing a systematic understanding of the expansion of the solutions around all singularities.

\item {\it Geometry:} The equations for the sum over graphs contain shared functions that are naturally associated with the geometry of a ``dressed associahedron" (see Section~\ref{sec:Channels}).  The key difference between this associahedron structure and the one that appears in the context of scattering amplitudes~\cite{Arkani-Hamed:2017mur}, is that the canonical function of the ordinary associahedron directly gives the corresponding flavor-ordered amplitude. It would be extremely interesting to find the geometry that directly computes the wavefunction in ${\rm tr}\,\phi^3$ theory.
\end{itemize}

In the Introduction, we originally raised the speculative question of emergent time by suggesting that the breakdown of the classical spacetime near the Big Bang singularity requires a more ``timeless" description of our cosmological origin. The differential equations discussed in this paper are indeed formulated purely in terms of the kinematic variables of degrees of freedom living on the boundary of the spacetime, with no explicit reference to cosmological time evolution. 
 To view this as an emergence of time, however, requires that these equations take on a life of their own, arising from a different set of physical and mathematical principles. The graphical rules that we have presented can be thought of as a first step in this direction, but they still rely on some auxiliary structures that we must introduce, like an underlying marked graph or kinematic polygon. A more radical reformulation would explain the necessity of these objects as well.
 This would be an important step towards a truly timeless theory of cosmology.

\newpage
\paragraph{Acknowledgments} We are grateful to Ana Ach\'ucarro, Paolo Benincasa, Jan de Boer, Alessandra Caraceni, J.\,J.\,Carrasco, Xingang Chen, Claude Duhr, Alex Edison, Carolina Figueiredo, Dan Green, Thomas Grimm, Song He, Johannes Henn, Arno Hoefnagels,  Yu-tin Huang, Michael Jones, Manki Kim, Barak Kol, Chia-Kai Kuo, Daniel Longenecker, Manuel Loparco, Scott Melville, Sebastian Mizera, Matteo Parisi, Julio Parra-Martinez, Nic Pavao, Andrzej Pokraka, Oliver Schlotterer, Leonardo Senatore, Chia-Hsien Shen, Melissa Sherman-Bennett, John Stout,  Bernd Sturmfels, Simon Telen, Jaroslav Trnka, Kamran Salehi Vaziri, Dong-Gang Wang and Alexander Zhiboedov for insightful discussions.  Thanks to Harry Goodhew and Kamran Salehi Vaziri for corrections to a previous version of the manuscript.

\vskip 4pt
This project was initiated at the  online workshop ``Cosmological Correlators"~\cite{Workshop}, and we thank all participants for providing an exciting atmosphere during the workshop. GLP presented some of these results at various conferences and workshops---QCD meets Gravity 2021, Bethe Forum ``Inflation," DITP Holography Workshop at KU Leuven, Holography Workshop at Trinity College Dublin, Amplitudes 2022, MPI Group Retreat, Iberian Strings, ``DarkCosmoGrav" in Pisa, ``Scattering Amplitudes and Cosmology" at ICTP, ``Quantum de Sitter Universe" at DAMTP, ``Recent Trends in and out of the Swampland" at IBS, ``Gravitational Waves in the Southern Hemisphere" at ICTP-SAIFR---and he is grateful for many interesting discussions and feedback from the participants.  DB and AJ thank the participants of the Simons Symposium ``Amplitudes Meet Cosmology"~\cite{Simons} for helpful discussions. GLP thanks the Universities of Leiden and Amsterdam for their hospitality and support. GLP also thanks the Weizmann Institute of Science---in particular, Ofer Aharony and Kfir Blum---for hosting an extended visit, and all the discussions with the groups at Weizmann, Hebrew University, Technion, and Tel Aviv University.  NAH and DB thank the participants of the workshop ``Current Topics in Fundamental Physics" (organized by Johannes Henn) for interesting discussions.

\vskip 4pt
NAH is supported by the US Department of Energy (DOE) under contract DE–SC0009988.
DB is supported by a Yushan Professorship at National Taiwan University funded by the Ministry of Education (MOE) NTU-112V2004-1. AH is supported by DOE (HEP) Award DE-SC0011632 and by the Walter Burke Institute for Theoretical Physics. AJ is supported in part by DOE (HEP) Award DE-SC0009924. HL is supported by the Kavli Institute for Cosmological Physics at the University of Chicago through an endowment from the Kavli Foundation and its founder Fred Kavli. GLP is supported by Scuola Normale, by a Rita-Levi Montalcini Fellowship from the Italian Ministry of Universities and Research (MUR), and by INFN (IS GSS-Pi). 

\vskip 4pt
The research of NAH and DB is funded by the European Union (ERC, UNIVERSE PLUS, 101118787). Views and opinions expressed are however those of the author(s) only and do not necessarily reflect those of the European Union or the European Research Council Executive Agency. Neither the European Union nor the granting authority can be held responsible for them.

\appendix
\newpage
\addtocontents{toc}{\protect\setcounter{tocdepth}{1}}
\section{Some Mathematics}
\label{app:Maths}

There are a number of mathematical ideas that underpin the analysis in the main text. To provide a self-contained introduction to this circle of ideas, we summarize some of them in this appendix. In particular, ideas from projective geometry, positive geometry, twisted cohomology, and hyperplane arrangements appear in various places in the text, and formalizing these concepts sheds some light on the arenas in which they appear.

\subsection{Projective Geometry}

Many of the geometric constructions considered in the main text are most naturally considered in projective space.\footnote{For general introductions to projective geometry, see e.g.~\cite{hilbert1952geometry,hartshorne1967foundations,richter2011perspectives}.} {\it Projective $n$-space}, ${\mathbb P}^n$, is the space of all lines through the origin in $n+1$ dimensions.\footnote{This can be any vector space, but we will mainly consider real projective space, which is the space of lines through the origin in ${\mathbb R}^{n+1}$.} Broadly speaking, projective space is useful when thinking about any question that does not explicitly involve distances, volumes, or other dimensionful measures. In particular, questions of incidence---like whether a given point is on a given line, or determining the intersection point between two lines---are naturally formulated and answered projectively.

\paragraph{Projective space:}
As an example, ${\mathbb P}^1$ is the space of all lines through the origin in ${\mathbb R}^2$. We can specify a point on the projective line by the pair
\be
\left(
\begin{array}{c}
y\\
x
\end{array}
\right)
\sim
t
\left(
\begin{array}{c}
y\\
x
\end{array}
\right) ,
\label{eq:P1coords}
\ee
where $\sim$ indicates that we should identify vectors that differ by an overall rescaling. A convenient way to label the points is by their intersection with some reference line:
\begin{equation*}
\includegraphics[width=0.3 \textwidth]{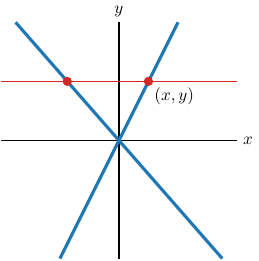}
\end{equation*}
For concreteness, we have taken the line $y = 1$. This is equivalent to using the rescaling freedom to fix the top component of the lines in~\eqref{eq:P1coords} to be $1$; rescalings are equivalent to sliding the red line up and down. Notice that this coordinatization obviously misses the $x$-axis, which does not intersect the red line. This ``line at infinity" can be covered by choosing a different reference line.

\vskip4pt
Two-dimensional projective space ${\mathbb P}^2$ has more interesting features. This is the space of lines through the origin in three-dimensions, which we can write as
\be
X^I = \Bigg(
\begin{array}{c}
z\\[-4pt]
x\\[-4pt]
y
\end{array}
\Bigg)\,, \quad {\rm with} \quad X^I \sim t X^I\,.
\label{eq:projective2spacecoords}
\ee
As before, we can coordinatize ${\mathbb P}^2$ by labeling lines by their intersection points with some reference plane. It is convenient to take the plane $z=1$. This  projective space is therefore similar to the ordinary Euclidean plane, with the addition of a ``line at infinity" that is missed by any particular choice of coordinates. What the coordinatization~\eqref{eq:projective2spacecoords} makes manifest is the SL$(3)$ invariance of projective space, which maps lines in the projective plane into other lines. The ${\mathbb R^2}\rtimes$SL$(2)$ subgroup is the obvious set of affine linear transformations, which maps a plane vector as $\vec x \mapsto {\mathbb L}\,\vec x +\vec t$, with ${\mathbb L}$ an SL$(2)$ matrix and $\vec t$ a vector. The SL$(3)$ group then acts as 
\be
\Bigg(
\begin{array}{c}
z\\[-4pt]
x\\[-4pt]
y
\end{array}
\Bigg)= \
\left(
\begin{array}{cc}
c ~\,&\,~ \vec b\\[4pt]
\vec t ~\,&\,~ {\mathbb L}
\end{array}
\right)
\Bigg(
\begin{array}{c}
z\\[-4pt]
x\\[-4pt]
y
\end{array}
\Bigg) \,,
\label{eq:sl3matrix}
\ee
with $\vec b$ a vector and $c$ a constant.
Using our rescaling freedom to set $z=1$, performing an SL$(3)$ transformation~\eqref{eq:sl3matrix} and then re-fixing the gauge $z=1$, we find that the SL$(3)$ acts nonlinearly on a vector in the projective plane:
\be
\vec x\mapsto \frac{{\mathbb L}\,\vec x+\vec t}{\vec b\cdot\vec x+c}\,.
\ee
We see that the choice of a line at infinity causes the SL$(3)$ to be realized on the coordinates nonlinearly because it does not leave the line at infinity invariant.\footnote{The affine group ${\mathbb R^2}\rtimes$SL$(2)$ is precisely the subgroup that maps points on the line at infinity to other points on the same line. One can require further structures to be left invariant, and this will further reduce the symmetry group (for example, fixing two points on the line at infinity reduces the symmetries to the $(1+1)d$ Poincar\'e group). Klein's Erlangen program was essentially to proceed in this way and classify geometries by the subset of projective transformations that they preserve.}

\vskip4pt
Projective space is well adapted to ask questions of incidence, so let's see how this works. Recall that {\it points} in ${\mathbb P}^2$ can be represented as $X^I$, with a raised index, as in~\eqref{eq:projective2spacecoords}. We can therefore represent a line in projective space as
\be
L_I X^I \equiv  Ax+By+Cz = 0\,.
\ee
Using the rescaling freedom to set $z=1$ reproduces the usual formula for a line, $Ax+By+C = 0$. It is convenient to associate lines to the coefficients $L_I$, with a lowered index.
The only tensor that is invariant under the SL$(3)$ symmetries of projective space is the Levi--Civita symbol $\epsilon_{IJK}$. Given two points $X_1^I$ and $X_2^J$, we can then construct the line that connects them as
\be
L_I^{(12)} = \epsilon_{IJK}X_1^J X_2^K\,.
\ee
The antisymmetry of the Levi--Civita symbol makes it clear that $L_I^{(12)} X_1^I = L_I^{(12)} X_2^I = 0$, so that both the points $X_1$ and $X_2$ are on the line, as desired. It is straightforward to check whether a third point $Y^I$ is on the line by checking whether
\be
L_I^{(12)} Y^I =  \epsilon_{IJK}Y^IX_1^J X_2^K\overset{?}{=} 0\,.
\ee
Similarly, given two lines $L^1_I$ and $L^2_J$, their intersection point $X^I_{(12)}$ is
\be
X^I_{(12)} = \epsilon^{IJK}L^1_J L^2_K\,.
\ee
Note that, in all of these examples, the lines and points are the only things that we can construct from the objects at hand with the right index structure. For example, three points all lie on a line if
\be
\epsilon_{IJK}X_1^IX_2^JX_3^K \equiv \langle X_1 X_2 X_3\rangle = 0\,,
\ee
and three lines meet at a point if
\be
\epsilon^{IJK}L^1_IL^2_JL^3_K \equiv \langle L_1 L_2 L_3\rangle = 0\,.
\ee

\vskip4pt
Another nice feature of projective space is that all pairs of lines intersect somewhere, even parallel ones. From the perspective of the projective plane, parallel lines meet on the line at infinity.\footnote{We can see this explicitly by considering the two lines $L^1_I X^I = Ax+By+C = 0$ and $L^2_I X^I = Ax+By+C' = 0$. Their intersection point is then
\be
X_{(12)}^I = \epsilon^{IJK}L^1_JL^2_k = 
\left(
\begin{array}{c}
0\\
A(C-C')\\
B(C-C')
\end{array}
\right) .
\ee
Recall that points with $z=0$ do not intersect the plane $z = 1$, and so lie at infinity. More generally, we can think of the gauge where we set $z=1$ as covering the parts of ${\mathbb P}^2$ for which $X^IL_I^\infty \neq 0$, with
\be
L_I^\infty = 
\left(
\begin{array}{c}
1\\
0\\
0
\end{array}
\right) .
\label{eq:P2Linf}
\ee
This coordinate chart therefore covers all points that do not lie on the line at infinity.
}

\vskip4pt
We can also consider more complicated structures in projective space. For example, we can express a conic $Ax^2+Bxy+Cy^2+Dx+Ey+F = 0$ as $C_{IJ}X^IX^J= 0$. Note that $C_{IJ}$ has $5$ independent components after modding out by the overall rescaling freedom, so any five points determine a conic that passes through them. It is then possible to use similar techniques to construct various tangents to and points on the conic. For example, one can prove {\it Pascal's Theorem}, which states that if $6$ points happen to lie on a conic, if we look at the inscribed hexagon obtained by connecting these points pairwise, then the three intersection points of opposing sides will lie on a single line.

\vskip4pt
An interesting feature of projective space is so-called {\it projective duality}. In the context of ${\mathbb P}^2$, it is particularly easy to see: we have assigned points as $X^I$ with raised indices and lines as $L_I$ with lowered indices, but clearly nothing changes if we swap every raised index into a lowered one, and vice versa. This implies that any projective statement involving points and lines continues to be true if we everywhere interchange the words points and lines. As an example, the projective dual of Pascal's theorem is {\it Brianchon's theorem}, which states that if six lines inscribe a conic, then the lines formed by connecting opposite vertices of the hexagon will meet at a point.

\vskip4pt
We can also consider higher-dimensional projective spaces, where planes are now associated to objects with additional indices. For example, in ${\mathbb P}^3$, points are labeled by $X^I$, planes are labeled by $H_I$ (or equivalently antisymmetric tensors $H^{IJK}$) and lines correspond to antisymmetric tensors~$L^{IJ}$. The invariant tensor is now $\epsilon_{IJKL}$, and we can use a similar logic to the ${\mathbb P}^2$ case to see that just like two points determine a line, two planes intersect along a line. In ${\mathbb P}^n$, a $(p-1)$-plane corresponds to an antisymmetric tensor $H^{I_1\cdots I_p}$, or equivalently $H_{J_1\cdots J_{n+1-p}}$ if we use the invariant tensor $\epsilon_{I_1\cdots I_{n+1}}$ to lower the indices. In general dimensions, projective duality interchanges $p$-planes and $(n-p-1)$-planes.

\paragraph{Volumes:} Although there is no notion of volume in projective space, it is nevertheless often useful to think about volumes projectively. This requires introducing some additional structure in the form of a line/plane at infinity $L^\infty_{I}$. We can then construct volumes in a projectively invariant way. 

\vskip4pt
For example, in ${\mathbb P}^1$, we define volumes by choosing a point at infinity $Y^I =(1,0)$. It is then natural to describe the points~\eqref{eq:P1coords} in the chart that sets $y=1$. We can then write the length of the interval between the points $X_a^I$ and $X_b^J$ as
\be
L = \frac{\langle X_a X_b\rangle}{\langle Y X_a\rangle\langle YX_b\rangle}\,,
\label{eq:lineinterval}
\ee
where $\langle X_aX_b\rangle\equiv \epsilon_{IJ} X_a^I X_b^J$. Notice that despite the fact that the choice of a point at infinity breaks the symmetries of projective space down to those of the affine line, it is still possible to write the length of an interval in a way that respects the scaling invariance of projective space. If we re-scale all the coordinates $X$ by the same amount, the expression~\eqref{eq:lineinterval} is invariant.

\vskip4pt
We can also talk about volumes in ${\mathbb P}^2$. In this case the most relevant example will be the area of a triangle. Given three points $X_1^I, X_2^J, X_3^K$, the area of the triangle defined by these points can be written as
\be
A = \frac{\langle X_1X_2X_3\rangle}{(L^\infty\cdot X_1)(L^\infty\cdot X_2)(L^\infty\cdot X_3)}\,,
\ee
which involves a choice of line at infinity, which it is often convenient to choose to be~\eqref{eq:P2Linf}. We can also define the area of a triangle in terms of the lines that define it, rather than the points that are its vertices. Imagine that a triangle is bounded by the lines $L^a_I, L^b_J, L^c_K$. The area of this triangle then is
\be
A  = \frac{\langle L^a L^b L^c\rangle^2}{\langle L^\infty L^a L^b\rangle\langle L^\infty L^b L^c\rangle\langle L^\infty L^c L^a\rangle}\,,
\ee
which is the only way to build an invariant out of these objects with the correct scaling weight.

\vskip4pt
In higher dimensions, it is also easy to write expressions for the volumes of simplices. Given a plane at infinity $H_I^\infty$, the volume of a simplex in ${\mathbb P}^N$ with vertices $X_1,\cdots X_{N+1}$ is
\be
V = \frac{\langle X_1\cdots X_{N+1}\rangle}{(H^\infty\cdot X_1)\cdots (H^\infty\cdot X_{N+1})}\,.
\ee
Similarly, if we define a simplex via the $N+1$ hyperplanes $H_I^{a}$ that bound it, we can write its volume as
\be
V = \frac{\langle 1\cdots (n+1)\rangle^N}{ \langle \infty\, 1\cdots n\rangle\cdots \langle \infty\, 2\cdots (n+1)\rangle}\,,
\ee
where we have used the shorthand of replacing a plane with its label inside the brackets.

\paragraph{Differential forms:} In the main text, we study families of integrals of differential forms that live on ${\mathbb R}^n$ minus some singular divisors corresponding to a hyperplane arrangement. It is often useful to describe these differential forms projectively, so we now want to consider how to write differential (top) forms on projective space.

\vskip4pt
We first consider one-dimensional projective space ${\mathbb P}^1$. Naively, any form built from $X$ and $\ud X$ would be acceptable as long as it is invariant under overall re-scalings of $X$. Hoewever, in order for differential forms to be well-defined, they must actually be invariant under {\it local} GL$(1)$ rescalings that send $X^I \mapsto t(X)X^I$. This means that a well-defined differential form on ${\mathbb P}^1$ will be writeable as $\langle X\ud X\rangle f(X)$, where $f$ is a function of weight $-2$ under local rescalings. As a concrete example, consider two points $X_a$ and $X_b$. It is possible to construct a well-defined projectively invariant differential form as
\be
\Omega = \frac{\langle ab\rangle\langle X\ud X\rangle}{\langle X a\rangle\langle X b\rangle}\,.
\label{eq:P1form-X}
\ee
We can write this form more explicitly by choosing a coordinate chart where
\be
X^I = 
\left(
\begin{array}{c}
1\\
x
\end{array}
\right),
\quad 
\ud X^I = 
\left(
\begin{array}{c}
0\\
\ud x
\end{array}
\right) ,
\quad 
X^I_a = 
\left(
\begin{array}{c}
1\\
a
\end{array}
\right) ,
\quad 
X^I_b = 
\left(
\begin{array}{c}
1\\
b
\end{array}
\right) ,
\ee
so that~\eqref{eq:P1form-X} is
\be
\Omega_1 = \frac{(b-a)\ud x}{(x-a)(x-b)}\,.
\label{eq:P1form}
\ee

The construction in a general projective space ${\mathbb P}^n$ is quite similar: forms can be written as $\Omega = \langle X\ud X\cdots \ud X\rangle f(X)$, which involves $n$ different $\ud X$s and where $f(X)$ has scaling weight $-(n+1)$. For example, given three lines in ${\mathbb P}^2$ forming a triangle, the natural differential form is
\be
\Omega_2 = \frac{\langle X\ud X\ud X\rangle \langle L_1L_2L_3\rangle}{(L_1\cdot X)(L_2\cdot X)(L_3\cdot X)}\,.
\label{eq:P2form}
\ee
More generally, given a simplex in ${\mathbb P}^n$ (built from $n+1$ hyperplanes), the following top form can be constructed\footnote{Note that in the main text---in particular in Section~\ref{sec:MoreComplex}---we continue to denote hyperplanes by $L_{a,I}$, like lines, and write projective coordinates as $P^I$ in order to not cause confusion with the kinematic variables $X_i$.} 
\be
\Omega_{n+1} = \frac{\langle X\ud X\cdots \ud X\rangle \langle H_1H_2\cdots H_{n+1}\rangle}{(H_1\cdot X)(H_2\cdot X)\cdots (H_{n+1}\cdot X)}\,.
\label{eq:Pnform}
\ee
We will now see that these differential forms have the interpretation of {\it canonical forms} for the corresponding simplicial positive geometries.

\subsection{Canonical Forms}
\label{app:canforms}

The concept of canonical forms has played an important role in defining our basis of integrals in the main text.
In this section, we will give a pedagogical introduction to canonical forms for readers who haven't come across them before.

\vskip 4pt
Canonical forms are naturally associated to {\it positive geometries}~\cite{Arkani-Hamed:2017tmz}. Roughly speaking, a positive geometry is a convex region with boundaries of all codimension (which themselves are lower-dimensional positive geometries), with a distinguished top-dimensional form called its {\it canonical form}. For our purposes, it will suffice to consider a region whose boundaries are hyperplanes (meaning, linear functions of the coordinates).\footnote{The canonical form exists for more complicated positive regions, depending on the genus of the curve parametrizing the boundary. For example, in two dimensions, the boundary may be a conic section cut by lines. We do not use canonical forms in this broader sense throughout the text, but see \cite{Arkani-Hamed:2017tmz} for more details.} The  canonical form associated with this region is the {\it unique} form with logarithmic singularities along all its boundaries (and nowhere else), whose residue on the boundaries is the canonical form of the lower-dimensional positive geometry. Throughout this paper, we choose the residues at the vertices of the bounded regions to be $\pm 1$, which fixes the canonical form up to an overall sign (or orientation).

\vskip 4pt
It is best to illustrate the meaning of canonical forms by looking at examples:
\begin{itemize}
\item {\bf Interval:} Consider the following bounded interval in one dimension:
\begin{center}
\begin{tikzpicture}

\draw[black,->,line width=0.75pt] (-4.5,0) -- (1.75,0);
\draw[black] (1.75,0) node[right] {$x$};

\draw[Blue!40,line width=3pt] (-2.5,0) -- (0,0);

\filldraw[Red] (0,0) circle (3pt);
\node[Red,below] at (0,-0.05) {$-b$};

\filldraw[Blue] (-2.5,0) circle (3pt);
\node[Blue,below] at (-2.5,-0.07) {$-a$};
\end{tikzpicture}
\end{center} 

\vspace{-0.3cm}
We would like to define a one-form on this interval with logarithmic singularities only at the endpoints $x=-a$ and $x=-b$. Clearly, this form must be a linear combination of $\ud x/(x+a)$ and $\ud x/(x+b)$. It seems that we can set the residues to be whatever we wish. However, requiring that there are no singularities anywhere else imposes a restriction, as both of these forms, on their own, also have a singularity at infinity. In order to cancel such a singularity (also logarithmic), we must subtract the two terms; i.e.~the residues at $-a$ and $-b$ must be equal in magnitude, but have opposite signs. Taking the residues to have unit magnitudes fixes the canonical form on the interval up to an overall sign:
\beq
\Omega =\frac{ \ud x}{x+a}-\frac{\ud x}{x+b}=\frac{(b-a) \,\ud x}{(x+a)(x+b)} =\ud \log \left(\frac{x+a}{x+b}\right) .
\label{equ:interval-form}
\eeq
We used this result in Section~\ref{ssec:dS}. Notice also that this canonical form can be written in a projectively invariant way as in~\eqref{eq:P1form}.

\item {\bf Triangle:} Next, we move to bounded regions in two dimensions. We start with the example of a triangle. For simplicity, we take two of its lines to coincide with the coordinate axes, $L_1=x_1$ and $L_2=x_2$, and the third line, $L_3=a_1 x_1+a_2 x_2+a_3$, to be generic: 
\begin{center}
\begin{tikzpicture}

\draw[black,->,line width=0.75pt] (-3,0) -- (1.75,0);
\draw[black] (1.75,0) node[right] {$x_1$};

\draw[black,->,line width=0.75pt] (-2.5,-0.5) -- (-2.5,4);
\draw[black] (-2.5,4) node[above] {$x_2$};

\filldraw[Blue!20] (-2.5, 3) -- (0.5,0) -- (-2.5,0) -- (-2.5, 3);

\draw[Blue!50,line width=1.5pt] (-2.5,-0.5) -- (-2.5,3.75);
\draw[Red!50,line width=1.5pt] (-3,0) -- (1.5,0);
\draw[Green!50,line width=1.5pt] (-2.75, 3.25) -- (0.75,-0.25);

\draw[Blue,line width=2.5pt] (-2.5,0) -- (-2.5,3);
\draw[Red,line width=2.5pt] (-2.5,0) -- (0.5,0);
\draw[Green,line width=2.5pt] (-2.5,3) -- (0.5,0);

\draw[Red] (-1,0.) node[below] {$L_2$};
\draw[Blue] (-2.5,1.5) node[left] {$L_1$};
\draw[Green] (-0.15,1.5) node[left] {$L_3$};

\filldraw (0.5,0) circle (3pt);
\filldraw (-2.5,0) circle (3pt);
\filldraw (-2.5,3) circle (3pt);

\end{tikzpicture}
\end{center}
A natural guess for the canonical form on this triangle is 
\beq
\Omega = \frac{\ud x_1 \wedge \ud x_2}{L_1 L_2 L_3}\,,
\label{equ:triangle-form}
\eeq
which is correct up to normalization.
A nice feature is that the properly-normalized canonical form associated to a triangle can be expressed as the projectively invariant form~\eqref{eq:P2form}, which reduces to~\eqref{equ:triangle-form} up to normalization after choosing a coordinate chart. 

\vskip 4pt
Besides having the correct singularities on the codimension-one boundaries, the form~\eqref{equ:triangle-form} does a bit more, which goes at the heart of the definition of a canonical form. Consider taking the limit $L_1=x_1\to 0$, i.e.~going to the boundary defined by $L_1$. The residue of (\ref{equ:triangle-form}) at this locus is
\beq
{\rm Res}[\Omega]\Big|_{L_1=0} = \frac{1}{a_2} \frac{\ud x_2}{ x_2 (x_2+a_3/a_2) }\,,
\eeq
which, up to an overall factor, is the canonical form on the interval $[-a_3/a_2,0]$; cf.~(\ref{equ:interval-form}).
Sending $\Omega \to a_3 \Omega$, would then also fix the correct normalization of the form. (For a generic non-right triangle, the precise normalization factor will be different.)
We have just displayed an important general feature of canonical forms: the residue of a canonical form on any of its boundaries is the canonical form of the boundary.

The result (\ref{equ:triangle-form}) can also be written as 
\beq
\Omega = \ud \log \left(\frac{L_1}{L_2}\right) \wedge \ud \log \left(\frac{L_2}{L_3}\right) ,
\eeq
which is independent of the choice of repeated line (despite the apparent asymmetry among the lines).

\item {\bf Quadrilateral:}
As a more nontrivial example, 
consider a quadrilateral bounded by four lines $\{ L_i =0\}$. It is tempting to guess that the canonical form is 
\beq
\Omega \stackrel{?}{=} \frac{\ud x_1 \wedge \ud x_2}{L_1 L_2 L_3 L_4}\,.
\label{equ:quad-form}
\eeq 
This time, however, this guess is wrong. It is easy to understand this by looking at the drawing: 
\begin{center}
\begin{tikzpicture}

\draw[black,->,line width=0.75pt] (-3,0) -- (2.15,0);
\draw[black] (2.15,0) node[right] {$x_1$};

\draw[black,->,line width=0.75pt] (-2.5,-0.5) -- (-2.5,4.25);
\draw[black] (-2.5,4.25) node[above] {$x_2$};

\filldraw[Blue!20] (-2,0.5) -- (-1.,2.5) -- (0.1,1.5) -- (0.5,0.5) -- (-2,0.5);

\draw[Red!50,line width=1.5pt] (-2.5,0.5) -- (1.75,0.5);
\draw[Blue!50,line width=1.5pt] (-2.25,0.) -- (-0.375,3.75);
\draw[Orange!50,line width=1.5pt] (-0.9,4) -- (0.7,0.);
\draw[Green!50,line width=1.5pt] (1.75,0) -- (-2.5,3.864);

\draw[Red] (1.75,0.5) node[right] {\small $L_1$};
\draw[Green] (-2.25,3.75) node[right] {\small $L_4$};
\draw[Blue] (-0.5,3.95) node[right] {\small $L_2$};
\draw[Orange] (-0.9,4) node[above] {\small $L_3$};
\draw[black] (0.8,1.2) node[right] {\small $L_N$};

\draw[line width=0.75pt] (-0.92,3.8) -- (1.52,0);

\draw[Red,line width=2.5pt] (-2,0.5) -- (0.5,0.5);
\draw[Blue,line width=2.5pt] (-2,0.5) -- (-1.,2.5);
\draw[Green,line width=2.5pt] (0.1,1.5) -- (-1.,2.5);
\draw[Orange,line width=2.5pt] (0.1,1.5) -- (0.5,0.5);

\filldraw (-2,0.5) circle (3pt);
\filldraw (0.5,0.5) circle (3pt);
\filldraw (-1.,2.5) circle (3pt);
\filldraw (0.1,1.5) circle (3pt);

\draw[fill=white,line width=0.75pt] (1.2,0.5) circle (2.75pt);
\draw[fill=white,line width=0.75pt] (-0.6,3.3) circle (2.75pt);

\end{tikzpicture}
\end{center}
In the limit  $L_1\to 0$, the residue of the canonical form should give the canonical form on the bounded interval contained by the line $L_1$. However, the ansatz in (\ref{equ:quad-form}) also contains a spurious pole at $L_1 =0\cap L_4=0$, which lies outside of our bounded region!  A similar reasoning implies that there is another spurious pole when $L_2\to0 \cap L_3 \to 0$. To fix this, we must add something to the numerator of (\ref{equ:quad-form}) which offsets these unwanted poles.   The minimal way of removing both poles is to include a numerator that is linear in the coordinates. It is the line $L_N$ connecting the white dots in the figure above. Thus, up to an overall normalization (which we can include in the definition of $L_N$), 
we obtain
\beq
\Omega =\frac{L_N\,\ud x_1 \wedge \ud x_2}{L_1 L_2 L_3 L_4}\, .
\eeq 
Below we will discuss a property of canonical forms---triangulation independence---that explains why the numerator is a line in this example (and not a more complicated curve).
\item {\bf Pentagon:} A pentagon is determined by five lines $\{L_i=0\}$, which intersect at 10 different points:
\begin{center}
\begin{tikzpicture}

\coordinate (A) at (-2.05,1.22);
\coordinate (B) at (0.34,4.07);
\coordinate (C) at (-1.56,3.26);
\coordinate (D) at (1.9,3.05);
\coordinate (E) at (0.76,0.78);

\coordinate (1) at (-0.9,2.59);
\coordinate (2) at (-0.4,3.18);
\coordinate (3) at (0.47,3.12);
\coordinate (4) at (0.54,2.42);
\coordinate (5) at (-0.37,2.);

\begin{scope}[rotate=30]
\draw (1,2) ellipse (2.2cm and 1.7cm);
\end{scope}

\draw[black,->,line width=0.75pt] (-3,0) -- (2.6,0);
\draw[black] (2.6,0) node[right] {$x_1$};

\draw[black,->,line width=0.75pt] (-2.5,-0.5) -- (-2.5,4.6);
\draw[black] (-2.5,4.6) node[above] {$x_2$};

\filldraw[Blue!20] (1) -- (2) -- (3) -- (4) -- (5) -- (1);

\draw[Red!50,line width=1.5pt] (0.83,0.) -- (0.3,4.48);
\draw[Blue!50,line width=1.5pt] (-2.5,0.7) -- (0.7,4.46);
\draw[Orange!50,line width=1.5pt] (-2.2,3.3) -- (2.35,3.);
\draw[Green!50,line width=1.5pt] (1.5,0) -- (-1.95,3.7);
\draw[Purple!50,line width=1.5pt] (-2.5,1.02) --  (2.35,3.25);

\draw[fill=white,line width=0.75pt] (A) circle (2.75pt);
\draw[fill=white,line width=0.75pt] (B) circle (2.75pt);
\draw[fill=white,line width=0.75pt] (C) circle (2.75pt);
\draw[fill=white,line width=0.75pt] (D) circle (2.75pt);
\draw[fill=white,line width=0.75pt] (E) circle (2.75pt);

\draw[Blue,line width=2.5pt] (1) -- (2);
\draw[Orange,line width=2.5pt] (2) -- (3);
\draw[Red,line width=2.5pt] (3) -- (4);
\draw[Purple,line width=2.5pt] (4) -- (5);
\draw[Green,line width=2.5pt] (1) -- (5);

\filldraw (1) circle (3pt);
\filldraw (2) circle (3pt);
\filldraw (3) circle (3pt);
\filldraw (4) circle (3pt);
\filldraw (5) circle (3pt);

\draw[Red] (0.3,4.48) node[above] {\small $L_1$};
\draw[Blue] (0.8,4.4) node[above] {\small $L_2$};
\draw[Orange] (2.6,2.6) node[above] {\small $L_3$};
\draw[Green] (-2.05,3.7) node[above] {\small $L_4$};
\draw[Purple] (2.6,3.05) node[above] {\small $L_5$};

\draw[black] (1.58,1.6) node[right] {\small $C_N$};
\end{tikzpicture}
\end{center}
 Five of these points are the vertices of the pentagon and the five other points lie outside of it.
As in the previous example, the absence of spurious poles at the intersections outside the pentagon fixes the numerator of the canonical form associated to the pentagon.  There exists a unique conic~$C_N$ passing through any given five points on the plane. We pick the conic section going through all spurious poles to be the numerator of the canonical form: 
\beq
\Omega=\frac{C_N\, \ud x_1 \wedge \ud x_2}{L_1 L_2 L_3 L_4 L_5}\, .
\eeq 
\item {\bf Polygons:} For general polygons, it is often simpler to determine the canonical forms by triangulation. The canonical form of a polygon is the sum of canonical forms of the triangles in a triangulation~\cite{Arkani-Hamed:2017tmz}.\footnote{The generalization to higher dimensions is the obvious one: that the canonical form of a polytope is the sum of those of the simplices in a triangulation.}
 The uniqueness property of canonical forms guarantees that any triangulation will give the same answer. The only important step is to cancel residues from the lines used to glue the triangles together, as these correspond to spurious singularities. This also explains why the degree of the numerator increases with the number of sides of the polygon. As we need to glue more triangles together, when putting everything under a common denominator, we form a numerator of higher and higher degree.
\item {\bf Simplices:} The generalization of a triangle in higher dimensions is a simplex. It consists of the intersection of $d+1$ hyperplanes in $d$ dimensions. If we parametrize the equations of the hyperplanes as $L_i=H_i\cdot X$, with $X$ being the projective coordinates on ${\mathbb R}^d$, then the canonical form of a simplex is~\eqref{eq:Pnform}. Upon choosing the section $\{X_i=x_i, X_{d+1}=1\}$, this can be written as
\beq
\Omega = \frac{ |H_1 \cdots H_{d+1}|}{L_1 \cdots L_{d+1}} \, \ud x_1 \wedge \cdots \wedge \ud x_d\, ,
\eeq
where the numerator is the determinant of the matrix built out of the coefficients of all hyperplanes, which fixes the normalization for the simplex to have unit residues at the corners. For the case of the triangle discussed above, picking two of the lines to be parallel to the coordinate axes gives $|H_1 H_2 H_3|=a_3$.
\end{itemize}
In higher dimensions, and for a larger number of hyperplanes, the bounded regions
become harder to visualize and it is more difficult to build the canonical form of a given polytope in a systematic way. With the examples listed above, one should have enough intuition to determine the canonical forms of general hyperplane arrangements by either triangulation or educated guesswork.
A deeper dive into the subject and some of its applications can be found in~\cite{Arkani-Hamed:2017tmz,Arkani-Hamed:2017mur}.

\subsection{Twisted Cohomology}
\label{app:twistedcohomology}

In this section,   we give a brief account of twisted de Rham cohomology, focusing on  results needed in the main text. More comprehensive expositions can be found in the classic books~\cite{aomoto2011theory,yoshida2013hypergeometric} and the reviews~\cite{Mizera:2019ose,Cacciatori:2021nli,Weinzierl:2022eaz,Abreu:2022mfk}. Recent applications to the study of various aspects of Feynman integrals include~\cite{brown2015feynman,Mastrolia:2018uzb,Abreu:2019wzk,Caron-Huot:2021xqj}.

\paragraph{Twisted integrals:} 
Let $\mathbb{C}^n$ be $n$-dimensional complex space. We are interested in integrals of forms defined on this space, which we will allow to be singular on a set of hypersurfaces defined by the polynomial equations $f_i(z_1,\ldots,z_n)=0$, for $i=1,\ldots, m$.
The union of these hypersurfaces is the divisor $D$. 
In order to understand the space of possible integrals, we will therefore be interested in the cohomology of differential forms on the space $M \equiv \mathbb{C}^n \setminus D$.

\vskip 4pt
Consider differential $n$-forms $\Omega$, which are holomorphic on $M$, but may be singular on the divisor, and so take the form
\beq
\Omega = \frac{N(z)}{f_1^{p_1} \cdots  f_m^{p_m} } \,\ud z_1 \wedge \cdots \wedge \ud z_n \,, \quad 
\label{eq:holomorphicforms}
\eeq
where $N(z)$ is a polynomial in $z=(z_1,\ldots, z_n)$ and $p_i$ are integers. 
In general, we are then interested in {\it twisted integrals} of the form
\be
I \equiv \int_{\gamma} \Omega\,\prod_{i=1}^m f_i(z)^{\varepsilon_i}\,,
\label{eq:twistedintdef}
\ee
with $\gamma$ some $n$-dimensional integration domain. The twist parameters $\varepsilon_i$ appearing in~\eqref{eq:twistedintdef} are {\it not} integers ($\varepsilon_i\in {\mathbb C}\setminus{\mathbb Z}$), so the integrand is in general multivalued. We want to understand  how many independent integrals there are of the form~\eqref{eq:twistedintdef}.  We begin with a reminder of the concept of ordinary (de Rham) cohomology of differential forms and then explain how it generalizes to twisted cohomology.

\paragraph{de Rham complex:} 
When all the $\varepsilon_i=0$ (and on a suitable integration domain without a boundary), the answer to this question is provided by de Rham cohomology, which counts equivalence classes of integrands. 
In particular, since $\partial \gamma =0$, Stokes' theorem implies that $I$ will not change if we shift $\Omega \to \Omega +\ud\xi$, with $\xi$  an arbitrary $(n-1)$-form. 
Two $n$-forms are therefore equivalent upon integration if they only differ by the exterior derivative of a $(n-1)$-form:  
 \beq
 \Omega \sim \Omega + \ud \xi\,.
\eeq
The vector space of  such equivalence classes is precisely the de Rham cohomology group $H^n(M)$, and so we see that the independent integrals of the form $I$ are in one-to-one correspondence with nontrivial cocycles in the de Rham cohomology. 

\paragraph{Twisted de Rham complex:} 
In this paper, we are instead interested in  integrals of the form~(\ref{eq:twistedintdef}), with nontrivial twist factors
\beq
u \equiv \prod_{i=1}^m f_i(z)^{\e_i}\,.
\eeq
The integration domain $\gamma$ now may have a boundary contained in the divisor $D$, as long as the twist factor $u$ vanishes sufficiently quickly on $\partial \gamma$. Stokes' theorem then implies that 
\beq
\int_\gamma \ud(u \xi) = \int_{\partial \gamma} u \xi = 0\,.
\eeq
This can also be written as
\beq
\int_\gamma \ud(u \xi) = \int_\gamma (\ud u \wedge \xi + u\hskip 1pt \ud \xi) = \int_\gamma u\, (\ud + \ud \log u \hskip 1pt \wedge) \xi \equiv   \int_\gamma u \nabla_\omega \hskip 1pt \xi =0\,,
\label{eq:stokes}
\eeq
where we have introduced the {\it covariant derivative}
\beq
\nabla_\omega \equiv \ud + \omega \, \wedge\,,
\label{equ:nab}
\eeq
defined in terms of the closed, holomomorphic $1$-form 
\be
\omega \equiv \ud\log u = \sum_{i=1}^m\varepsilon_i \frac{\ud f_i}{f_i}\,.
\ee
Two rational forms then yield equivalent integrals if they differ by the covariant derivative of a $(n-1)$-form:  
 \beq
 \Omega \sim \Omega + \nabla_\omega \hskip 1pt \xi\,.
\eeq
The set of equivalence classes of forms $\Omega$ with respect to the twisted differential (\ref{equ:nab}) then defines the twisted (de Rham) cohomology group $H^n_\omega(M, \nabla_\omega) = {\rm Ker}\nabla_\omega/{\rm Im}\nabla_\omega$.
This cohomology provides a basis of independent rational integrands $\Omega$ appearing in~(\ref{eq:twistedintdef}) and hence defines the space of independent master integrals.

\paragraph{A simple example:} It is worthwhile to illustrate the above considerations with a simple example. One important lesson is that twisting an integral typically reduces the dimensionality of the cohomology.

\vskip4pt
Consider the space of forms defined on $M = \mathbb{C}\setminus \{0,-1,\infty\}$ (that is, the complex line minus three points). In ordinary cohomology, the space of holomorphic $1$-forms is spanned by
\beq
\Omega_1 \equiv \frac{\ud z}{z}\,,\quad\Omega_2 \equiv \frac{\ud z}{z+1}\,. 
\eeq
It is easy to check that all other holomorphic $1$-forms in this space can be reduced to a linear combination of those two forms, plus total derivatives. For example,
\beq
\frac{ \ud z}{z^2(z+1)^3} = 3 \Omega_2-3\Omega_1 -\ud\left(\frac{2+9z+6z^2}{2z(1+z^2)} \right).
\eeq
We therefore see that $\Omega_1$ and $\Omega_2$ act as a basis for the de Rham cohomology $H^1(M)$.

\vskip4pt
Next, we consider the twisted cohomology defined by the twist function $u = z^\e$. 
It is easy to see then that, inside the twisted integral, the one-form $\Omega_1$ becomes a total derivative
\beq
\int u \hskip 1pt \Omega_1 = \int  z^\e \frac{\ud z}{z} 
 =\frac{1}{\e} \int \ud (z^{\e})\, ,
 \label{eq:extderiv}
\eeq 
or, equivalently, we see that %
\be
\Omega_1 = \nabla_\omega\left(\frac{1}{\e}\right) ,
\ee
where $\omega = \e\hskip 1pt \ud z/z$. 
As a result, the basis is now one-dimensional. In principle, we could have
introduced more general twists,
 but for this one-dimensional problem, we always end up removing a single member of the basis (a specific linear combination of the two members of the ordinary de Rham basis, determined by our choice of twist function).

\paragraph{A vanishing theorem:} As we consider higher-dimensional examples, the space of integrals becomes seemingly much more complex, and the structure of twisted cohomology would seem forbiddingly complicated. However, Aomoto~\cite{aomoto1975vanishing} proved a powerful statement that organizes our thinking about the problem. It turns out that for generic twists, only the $n$-dimensional twisted cohomology is nonzero in an $n$-dimensional space, i.e.~$H^k_\omega(M,\nabla_\omega) = 0$, for $n\neq k$. This means that only $n$-forms in $M$ are nontrivial.\footnote{In a complex space, the nontrivial cohomology is in holomorphic $n$-forms, so they have half the total dimension, while in a real space, these are top-dimensional real $n$-forms.} 
An important consequence is that the dimension of the twisted cohomology is closely related to the Euler character of $M$. In general, the Euler character is given by the sum of (signed) dimensions of the de Rham cohomology, but since only one of them is non-vanishing, we have~\cite{aomoto2011theory} 
\be
{\rm dim} \,H^n_\omega(M,\nabla_\omega) = (-1)^n \chi(M)\,.
\label{eq:twistedcohm}
\ee
This result is particularly useful because it allows us to determine geometrically the dimension of the twisted cohomology of 
top forms ($n$-forms), and consequently the number of master integrals required to span the vector space of integrals. It turns out that the spaces $M$ of interest in this paper have sufficient structure that $\chi(M)$ can be computed by simple counting.

\subsection{Hyperplane Arrangements}
\label{app:hyperplanes}

The singularities of the integrals that arise in the computation of the FRW wavefunction have a particularly simple structure---compared to generic twisted integrals in~\eqref{eq:twistedintdef}---in terms of hyperplane arrangements.  
In the following, we provide a bit of mathematical background on the theory of hyperplane arrangements (see~\cite{zaslavsky1975facing,stanley2004introduction,orlik1992arrangements} for more comprehensive treatments). 

\paragraph{Definitions:}
Recall that a {\it hyperplane} $H$ in an $n$-dimensional vector space $V$ (for our purposes, typically ${\mathbb C}^n$ or ${\mathbb R}^n$) is an affine subspace of dimension $(n-1)$
defined by 
\be
c_1 z_1 +\cdots + c_n z_n = a\,.
\ee
A finite set of such hyperplanes is the {\it hyperplane arrangement} ${\cal A}$.\footnote{In general, both hyperplanes and hyperplane arrangements can be defined with respect to an arbitrary vector space over an arbitrary field.} The {\it rank} of an arrangement
is the dimension of the space spanned by the vectors normal to the hyperplanes, and an arrangement is called {\it essential} if it has maximal rank (i.e.~the dimension of the ambient vector space).

\paragraph{Intersection, deletion, and restriction:} There are several natural operations one can perform on a hyperplane arrangement. The first is to consider the {\it intersections} of the hyperplanes that comprise an arrangement ${\cal A}$. We denote the set of all such 
 intersections by $L({\cal A})$.  This includes the ambient vector space $V$ itself, thought of as an intersection with the empty set. 
Note that 
$L({\cal A})$ has a natural partial order by reverse inclusion, so that $X\leq Y$ if $X\supseteq Y$, for $X,Y\in L({\cal A})$.
 The collection of a subset of the planes, ${\cal B} \subseteq  {\cal A}$, is called a {\it subarrangement}.
 
\vskip4pt 
An important subarrangement of ${\cal A}$ is obtained by {\it deletion}, wherein we choose a distinguished hyperplane $H_0$ and remove it from the arrangement, producing a new arrangement.

\vskip4pt
Another natural operation is {\it restriction}. As in deletion, we choose a distinguished hyperplane, but now envision it as the ambient vector space of a hyperplane arrangement, and think of the intersections of all the other planes that intersect it as hyperplanes in this ambient space. In order to define this operation formally, we first have to develop some notation.
 Given an element $X\in L({\cal A})$, 
 we first define the subarrangement of planes that contain $X$ as
\be
{\cal A}_X\equiv \{H\in {\cal A} \,|\, X \subseteq H\}\,.
\ee
The restriction of ${\cal A}$ to $X$ is then given by 
\be
{\cal A}^X\equiv \{X \cap H \neq \varnothing \,|\, H\in {\cal A}/{\cal A}_X \}\,.
\ee
The latter describes the set of hyperplanes in the space $X$ that arise from the intersection of the planes not in ${\cal A}_X$ with $X$. 

\begin{figure}[t!]
\centering
\includegraphics[width=0.95 \textwidth]{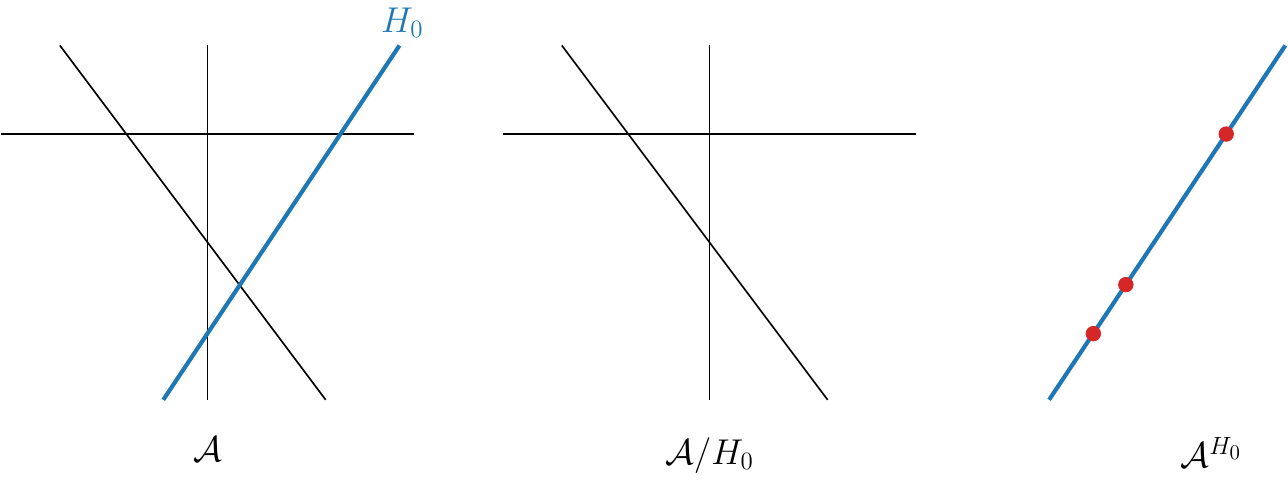}
\caption{Illustration of a hyperplane arrangement ${\cal A}$ with a distinguished hyperplane $H_0$ ({\it left}). 
Deleting $H_0$ gives the arrangement ${\cal A}/H_0$ ({\it middle}), while the restriction to $H_0$ is ${\cal A}^{H_0}$ ({\it right}). Note that the numbers of (bounded) regions in this triple of arrangements satisfy the relations~\eqref{eq:unboundcount} and~\eqref{eq:boundcount}. Concretely, we see that there are $11=7+4$ total regions in the arrangement ${\cal A}$, and $3=1+2$ bounded regions.}
\label{fig:HyperplaneArrangement}
\end{figure}

\vskip4pt
Given a hyperplane arrangement ${\cal A}$ and a choice of {\it distinguished hyperplane} $H_0\in {\cal A}$, we can form a {\it triple}
\be
({\cal A}, {\cal A}/H_0, {\cal A}^{H_0})\,,
\label{eq:triple}
\ee
which consists of the arrangement itself, the arrangement with $H_0$ deleted, and the arrangement obtained by restriction to $H_0$ (see Fig.\:\ref{fig:HyperplaneArrangement}). This procedure of {\it deletion-restriction} plays an important role in the combinatorics of hyperplane arrangements.

\paragraph{Regions:}
One important feature of hyperplane arrangements is that they partition the vector space $V$ into various {\it regions}. We denote the number of regions by ${\cal R}({\cal A})$. Some of the regions can be {\it bounded} and we denote their number by ${\cal B}({\cal A})$. These bounded regions will play an important role in connecting hyperplane arrangements to twisted cohomology.
The number of regions satisfy nice relations under deletion-restriction: Given a triple, the number of regions and bounded regions are related by\footnote{There is a small technical requirement on the bounded region relation, which is that the deletion operation should not change the rank of the hyperplane arrangement. In the case that it does, ${\cal B}({\cal A})$ is instead zero~\cite{stanley2004introduction}.}
\begin{align}
\label{eq:unboundcount}
{\cal R}({\cal A}) &= {\cal R}({\cal A}/H_0) +{\cal R}({\cal A}^{H_0})\,,\\
{\cal B}({\cal A}) &= {\cal B}({\cal A}/H_0) +{\cal B}({\cal A}^{H_0})\,.
\label{eq:boundcount}
\end{align}
One can then use these formulas directly to compute the number of (bounded) regions recursively. 
In Figure~\ref{fig:HyperplaneArrangement}, the number of (bounded) regions can simply be counted and the above relations are easily verified.

\paragraph{Poincar\'e polynomial:} In higher-dimensional examples, it typically becomes extremely hard
 to visualize the hyperplane arrangements, so that simply counting the (bounded) regions is not an option.
Fortunately, there is a systematic way to proceed using the {\it Poincar\'e polynomial}, $\pi({\cal A},t)$. Beyond this, the connection between the number or regions and the Poincar\'e polynomials serves as an important bridge between the combinatorics of hyperplane arrangements, topology, and twisted cohomology.

\vskip 4pt
The Poincar\'e polynomial can be given a formal definition,\footnote{Explicitly, the Poincar\'e polynomial is defined by
\be
\pi({\cal A},t) = \sum_{X\,\in\, L} \mu(V,X) (-t)^{r(X)}\,,
\label{eq:poincaredef}
\ee
where the sum runs over all elements of the poset of intersections $L({\cal A})$. Here, $r(X)$ is the rank of the element $X$ and $ \mu(V,X)$ is the {\it M\"obius function}, which satisfies
\begin{align}
\mu(X, X) &= 1\,,\\
\sum_{X\leq Z\leq Y} \mu(Z,Y) &= 0\,,\quad \text{if}~X<Y\,,
\end{align}
for elements $X,Y,Z\in L({\cal A})$. The definition is somewhat unwieldy when it comes to actual calculations and it is typically easier to use the formula~\eqref{eq:poincarerecurs}.
}
but, for our purposes, it is sufficient to note how it behaves under the operation of deletion-restriction. Given a triple of arrangements~\eqref{eq:triple}, we have~\cite{brylawski1972decomposition,zaslavsky1975facing}
\be
\pi({\cal A},t)  = \pi({\cal A}/H_0,t) +t \,\pi({\cal A}^{H_0},t) \,.
\label{eq:poincarerecurs}
\ee
Starting from the Poincar\'e polynomial of the empty arrangement, $\pi( \varnothing,t) =1$, we can use this relation to recursively compute the polynomial of more complicated arrangements. 

\vskip4pt
A remarkable feature of the Poincar\'e polynomial is that, for real arrangements, it counts both the total number of regions that the hyperplanes split $V$ into and the number of bounded regions (for an essential arrangement)~\cite{zaslavsky1975facing,orlik1992arrangements}:\footnote{The proof of this result is actually not too difficult. The basic idea is to note that both the number of regions and the Poincar\'e polynomial satisfy the same relation under the deletion-restriction operation. Since the formula~\eqref{eq:regioncount} is true for the empty arrangement, it is then true generally.}
\begin{align}
{\cal R}({\cal A}) &= \pi({\cal A},1)\,,  \label{eq:regioncount} \\
 {\cal B}({\cal A}) &=(-1)^n \pi({\cal A},-1)\,.
\end{align}
A priori, it may not obvious why this fact is of interest to us. However, the Poincar\'e polynomial is also a generating functional for Betti numbers, so that the Euler character of the hyperplane arrangement is
\be
\chi({\cal A}) = \pi({\cal A},-1)\,.
\ee
Comparing this to~\eqref{eq:twistedcohm}, we see that the number of bounded regions {\it also} computes the dimension of the twisted cohomology on the space $M$ where the hyperplane arrangement ${\cal A}$ is the singular divisor! In many cases, we can therefore read off the number of master integrals geometrically by counting the number of bounded regions, or equivalently by computing $\pi({\cal A},-1)$.

\paragraph{Counting bounded regions:} 
As a practical matter, it is often difficult to actually compute the Poincar\'e polynomial. However, there is a fascinating way to proceed. We instead consider a hyperplane arrangement over the finite field ${\mathbb F}_p$. The cardinality of the complement of the hyperplane arrangement, $\lvert \bar{\cal A}\rvert_p$, is then given by~\cite{orlik1992arrangements}\footnote{The proof again proceeds by noting that these quantities satisfy the same deletion-restriction relations.}
\be
\lvert \bar{\cal A}\rvert_p = p^{n}\pi({\cal A},-p^{-1})\,.
\ee
We can therefore extract $\pi({\cal A},t)$ by performing the following computation. We consider a hyperplane arrangement over the finite field ${\mathbb F}_p$ corresponding to the lattice $({\mathbb Z}/p)^n$, where $p$ is a prime number. We then remove from this lattice all points that are intersected by the hyperplanes. 
The resulting number of points---the size of the complement---is a function of the prime $p$. It turns out that $\lvert \bar{\cal A}\rvert_p$ is a polynomial of degree $n$, so determining its general form is just a matter of empirically evaluating and counting $\lvert \bar{\cal A}\rvert_p$ for a small set of primes. Given this result, we then have
\beq
{\cal B}({\cal A}) =\lvert \bar{\cal A}\rvert_{p=1}\, .
\eeq
For the single exchange integral discussed in Section~\ref{sec:TwoSite}, for example, this procedure gives $\lvert \bar{\cal A}\rvert_p =p^2-5p+8$, and so
\beq
\pi({\cal A},t) = 1+5t+8t^2\,.
\eeq
As we can see from Figure~\ref{fig:regions}, there are indeed $\pi({\cal A},1)= 14$ total regions and $\pi({\cal A},-1)= 4$ bounded regions. Since $\pi$ is a generating function for Betti numbers, its coefficients count the dimensions of de Rham cohomology.  In this case, we have $\{1,5,8\}$-dimensional spaces of $\{0,1,2\}$-forms.

\subsection{The Hypergeometric Function} 
We can illustrate all of these mathematical constructions in a simple example: the hypergeometric function.
This is also a toy example of the cosmological integrals considered in the main text.

\vskip4pt
Imagine that we are interested in computing the following integral 
\beq
I(X) = \int_0^\infty \ud x\, \frac{x^{\e_1}}{(x+X)^{\e_2} (x+1)^{\e_3} }\, \frac{1}{x(x+X)} \, ,
\label{equ:I}
\eeq
where $X$ is a real constant and $\e_i$ are independent twist parameters. We will now show that this integral is part of a finite-dimensional vector space of integrals, and therefore satisfies a differential equation.

\paragraph{Master integrals:}
The first step is to enlarge the problem and think of~\eqref{equ:I} as an element of the following family of integrals 
\beq
I_{n_1 n_2 n_3}(X) \equiv \int  \frac{{\color{Red}x}^{\e_1}}{{\color{Blue}(x+X)}^{\e_2} {\color{Green}(x+1)}^{\e_3} }
 \, \Omega\,, \qquad \Omega \equiv \frac{ {\rm d} x}{\colornucleus{Red}{x}^{n_1} \colornucleus{Blue}{(x+X)}^{n_2} \colornucleus{Green}{(x+1)}^{n_3} }\,,
 \label{eq:genintshyperg}
\eeq
where $n_i$ are positive integers, and we require $\e_1-n_1 > -1$, and $n_2+\e_2+n_3+\e_3+n_1-\e_1 > 1$, in order for the integrals to converge. The integral in (\ref{equ:I}) corresponds to $I_{110}$.

\vskip4pt
The integrals~\eqref{eq:genintshyperg} can be evaluated explicitly for particular values of the parameters; for example, they evaluate to power laws when $\e_2 = \e_3 = 0$. We are, however, interested in the case where the twists $\e_i$ are generic. In that case, we can think of the integrands in~\eqref{eq:genintshyperg} as rational functions twisted by factors with the same singularities. The locations of these singularities define a hyperplane arrangement, which in this case corresponds to points on the line:
\begin{equation*}
\includegraphics[scale=1]{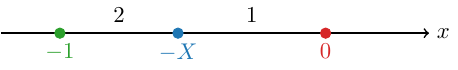}
\end{equation*}
As we have described above, the twisted integrals~\eqref{eq:genintshyperg} form a finite-dimensional vector space whose size is given by the number of bounded regions enclosed by the singular loci, $x = \{0,-X,-1\}$. We see from the hyperplane arrangement that there are two bounded intervals, and hence the basis for the family of integrals  in~\eqref{eq:genintshyperg} is two-dimensional.

\vskip 4pt
A convenient choice of basis is given by the integrals arising from the twists times the canonical forms for the bounded intervals.  As shown in Appendix~\ref{app:canforms}, the canonical forms associated with the intervals  $[-1,-X]$ and $[-X,0]$  are
\beq
\begin{aligned}
\Omega_1 &\equiv \Omega_{[-X,0]} =\frac{X}{\color{Red}{x}\color{Blue}{(x+X)}}\,\ud x\,,\\
\Omega_2 & \equiv \Omega_{[-1,-X]} =\frac{(1-X)}{\color{Blue}{(x+X)}\color{Green}{(x+1)}}\,\ud x\,,
\end{aligned}
\label{eq:intervalcanforms}
\eeq
and the canonical basis of integrals is therefore
\beq
\vec{I} \equiv \left[ \begin{array}{c} I_1 \\ I_2 \end{array} \right]  =\int_0^\infty \frac{x^{\e_1}}{(x+X)^{\e_2}(x+1)^{\e_3}} \left[ \begin{array}{c} \Omega_1 \\ \Omega_2 \end{array} \right] .
\eeq
Our original integral (\ref{equ:I}) is $I_1$, but any integral of the form~\eqref{eq:genintshyperg} can be written as a linear combination of $I_1$ and $I_2$ with $X$-dependent coefficients.

\paragraph{Differential equation:}
It is now mechanical to derive a first-order matrix differential equation satisfied by these basis integrals: if we differentiate $I_1$ or $I_2$ with respect to the parameter $X$, we have to be able to write the result back in terms of $I_1$ and $I_2$. This procedure is completely systematic, but in general can be somewhat tedious. It is at this point that our choice of using canonical forms to define the basis integrals pays off. The derivatives of canonical forms with respect to $X$ can be written as derivatives with respect to the integration variable $x$. Concretely, we have
\begin{align}
\label{eq:aderivs1}
\partial_X\Omega_1 &= -\partial_x\left(\frac{1}{X} {\color{Red}{x}}\,\Omega_1\right) = -\partial_x\left(\frac{1}{1-X} {\color{Green}{(x+1)}}\,\Omega_2\right) ,\\
\partial_X\Omega_2 &= \partial_x\left(\frac{1}{X} {\color{Red}{x}}\,\Omega_1\right) = \partial_x\left(\frac{1}{1-X} {\color{Green}{(x+1)}}\,\Omega_2\right) .
\label{eq:bderivs2}
\end{align}
These formulas are the un-integrated analogues of~\eqref{equ:Formula}.
Using these relations, it is straightforward to use integration by parts to derive the matrix differential equation for ${\cal I}$. For example, consider the derivative of the first basis integral
\beq
\partial_X I_1 = \int_0^\infty  \frac{x^{\e_1}}{(x+X)^{\e_2}(x+1)^{\e_3}}\left(\partial_X\Omega_1-\frac{\e_2}{x+X}\,\Omega_1\right) .
\label{eq:derivhyper1}
\eeq
Using~\eqref{eq:aderivs1}, we can write the first term as a derivative with respect to $x$ and then integrate by parts. This leads to 
\beq
\hspace{-0.2cm}\int_0^\infty \frac{x^{\e_1}\, \partial_X\Omega_1}{(x+X)^{\e_2}(x+1)^{\e_3}} = \int_0^\infty \frac{x^{\e_1}}{(x+X)^{\e_2}(x+1)^{\e_3}} \left(\frac{\e_1}{X}\,\Omega_1-\frac{\e_3}{1-X}\,\Omega_2-\frac{\e_2}{X}\frac{x}{x+X}\,\Omega_1\right) .
\label{eq:derivhyper2}
\eeq
The first two terms are already proportional to the basis elements $I_1$ and $I_2$. The  last term in~\eqref{eq:derivhyper2} combines with the last term in~\eqref{eq:derivhyper1} as
\beq
-\frac{\e_2}{x+X}\,\Omega_1-\frac{\e_2}{X}\frac{x}{x+X}\,\Omega_1 = -\frac{\varepsilon_2}{X}\,\Omega_1\,.
\eeq
Putting everything together, we get
\beq
\partial_X I_1 = \frac{\e_1-\e_2}{X} I_1-\frac{\e_3}{1-X} I_2\,.
\label{eq:psi2solve}
\eeq
Notice that the right-hand side is proportional to the twist parameters. This is a consequence of the relations~\eqref{eq:aderivs1} and \eqref{eq:bderivs2}, which guarantee that we can always treat the derivatives on the canonical forms by integrating by parts, so that they act on the twisted points. Following the same recipe,  the derivative of the second basis integral becomes
\beq
\partial_X I_2 = -\frac{\e_1}{X}I_1+\frac{\e_2+\e_3}{1-X}I_2\,.
\label{eq:aderivpsi1}
\eeq
Defining $\ud \equiv \ud X\,\partial_X $, we can combine (\ref{eq:psi2solve}) and (\ref{eq:aderivpsi1}) into a single matrix equation 
\beq
\ud  \,\vec{I} = A \,\vec{I}\, ,
\label{eq:1storderhyper}
\eeq
where the connection matrix $A$ is
\beq
A= \left(\left[\begin{matrix} 1 &0 \\-1& 0 \end{matrix}\right]\e_1+ \left[\begin{matrix} -1 &0 \\0 & 0 \end{matrix}\right]\e_2\right) \ud \log X 
+\left(\left[\begin{matrix} 0 &0 \\0 & -1 \end{matrix}\right]\e_2+ \left[\begin{matrix} 0&1\\0 & -1 \end{matrix}\right]\e_3\right) \ud \log(1-X)\, .
\eeq
We see that $A$ is a sum of dlog forms, $A = \sum_i \alpha_i(\e)\, \ud \log \Phi_i(X)$, and hence satisfies $\ud A=0$, i.e.~it is a flat connection.  Acting with $\ud$ on (\ref{eq:1storderhyper}) gives $\ud A +A\wedge A=0$,
so that $A\wedge A=0$, which is a nontrivial constraint on the entries of $A$.  Finally, note that all nonzero entries of $A$ are $\pm 1$. This is a consequence of the basis of canonical forms having unit residues.

\paragraph{Hypergeometric solution:} The solutions to (\ref{eq:1storderhyper}) can be written in terms of the Gauss hypergeometric function ${}_2F_1$. To see this, we combine \eqref{eq:psi2solve} and \eqref{eq:aderivpsi1} into a single second-order equation for $I_1$ (which is the original integral~\eqref{equ:I} that we want to compute).
Defining $I_1(X) \equiv X \hat I_1(X)$ and $X\equiv 1-z$, we find
\beq
\partial_z^2 \hat I_1 + \frac{c-z(1 + a + b)}{z(1-z)} \hskip 1pt \partial_z\hat I_1 -\frac{ab}{z(1-z)} \hskip 1pt \hat I_1=0\,,
\label{eq:hypergeomeq1}
\eeq
with the parameters
\beq
\begin{aligned}
a &\equiv 1+\e_2\,, \\
 b &\equiv 1+\e_2+\e_3-\e_1 \,, \\
  c &\equiv 1+\e_2+\e_3\,.
  \end{aligned}
\eeq
We recognize (\ref{eq:hypergeomeq1}) as the hypergeometric differential equation whose solution is 
\beq
\hat I_1(z) \propto\,{}_2F_1\left[\begin{array}{c}
a,\,b\\[-3pt]
c
\end{array}\Big\rvert \,z\,\right]\,.
\eeq
Evaluating $I_1$ for a specific value of $X$ fixes the overall normalization and we eventually get 
\beq
I_1(X) = \frac{\Gamma[1+\e_2+\e_3-\e_1]\Gamma[\e_1]}{\Gamma[1+\e_2+\e_3]}\, X\,{}_2F_1\left[\begin{array}{c}
1+\e_2\ ,\,1+\e_2+\e_3-\e_1\\[-2pt]
1+\e_2+\e_3
\end{array}\Big\rvert \,1-X\,\right] .
\label{equ:I1-Sol}
\eeq
Of course, in this case, we could also have found this solution by identifying the original integral~\eqref{equ:I} with the integral representation of the hypergeometric function. 
The second basis integral $I_2$ is a linearly independent hypergeometric function with slightly different parameters. We won't write its explicit form, but any other member of the hypergeometric family can be written as a linear combination of these two functions.

\paragraph{Gauss contiguous relations:}
Since the vector space of functions is two-dimensional, any three elements of the integral family~\eqref{eq:genintshyperg} must be linearly dependent. The resulting linear relations are the so-called contiguous relations of the hypergeometric function. They relate different ${}_2F_1$ functions with  values of $\e_{1,2,3}$ shifted by units.\footnote{These are the hypergeometric analogue of the cosmology shift relations discussed in Section~\ref{sec:cosmoshift}.} That these relations exist becomes a trivial consequence of the finite vector space structure of the space of integrals. In order to make a more direct connection to the Gauss contiguous relations, it is helpful to choose a different normalization of the integrands, and consider instead
\beq
\begin{aligned}
F(A,B) &\equiv \int_0^\infty \ud x \frac{ x^{\e_1}}{\colornucleus{Blue}{(x+X)}^{\e_2+A} \colornucleus{Green}{(x+1)}^{\e_3+B}} \, ,\\
G(A,B) &\equiv \int_0^\infty \frac{\ud x}{\color{Red}{x}} \frac{ x^{\e_1}}{\colornucleus{Blue}{(x+X)}^{\e_2+A} \colornucleus{Green}{(x+1)}^{\e_3+B}}\, ,
\end{aligned}
\label{equ:FG}
\eeq
so that $I_1 = G(1,0)/X$ and $I_2= F(1,1)/(1-X)$.
It must be possible to choose any two representatives from $F$ and $G$ and transform all others into this basis. This implies relations between the functions with shifted values of $A$ and $B$. For example, we can use the identity
\beq
\frac{1}{x} - \frac{1}{x+X} - \frac{X}{x(x+X)}=0\,,
\eeq
to derive such relations by integrating. We know that
\beq
 \int_0^\infty \ud x \frac{ x^{\e_1}}{(x+X)^{\e_2+A}(x+1)^{\e_3+B}}\left[\frac{1}{x} - \frac{1}{x+X} - \frac{X}{x(x+X)}\right] = 0\,.
\eeq
Translating this into a statement about the integrals~\eqref{equ:FG}, we find
\beq
G(A,B) -F(A+1,B) - XG(A+1,B) = 0\,.
\label{eq:shift1}
\eeq
Employing similar partial fraction identities, we can also derive the shift relations
\beq
\begin{aligned}
G(A,B) -F(A,B+1) - G(A,B+1) &= 0\,,\\
F(A+1,B) -F(A,B+1) -(1-X) F(A+1,B+1) &= 0\,.
\end{aligned}
\label{eq:shift2}
\eeq
Finally, we can use the total derivative relation\footnote{The total derivative of the integrand of $G$  diverges at $x=0$ and so does not lead to an interesting relation.}
\beq
\int_0^\infty \ud\left(\frac{ x^{\e_1}}{(x+X)^{\e_2+A}(x+1)^{\e_3+B}}\right) = 0\,,
\eeq
to infer the relation
\beq
\e_1G(A,B)-(\e_2+A)F(A+1,B)-(\e_3+B)F(A,B+1) = 0\,.
\label{eq:shift3}
\eeq
One can prove that the shift relations~\eqref{eq:shift1},~\eqref{eq:shift2} and~\eqref{eq:shift3} are sufficient to transform any element of the integral family into a chosen basis, and so all other shift relations can be obtained from these identities. These relations can be translated into a more familiar form by using the relation between the functions $F$ and $G$ and the hypergeometric function. Note that these shift relations can be used to derive the differential equation~\eqref{eq:1storderhyper} in a more brute-force way: one just differentiates the basis elements, and then uses the shift relations to rewrite the result back in the original basis. Since the vector space is finite dimensional, this always has to be possible and leads to a differential equation for the basis functions.

\newpage
\section{Kinematic Flow: Further Examples}
\label{app:flow-examples}

In Section~\ref{sec:GraphicalRules}, we introduced graphical rules to obtain the differential equations for arbitrary tree graphs. We illustrated these rules using two, three and, five-site chains.  In this appendix, we present further examples.

\subsection{Four-Site Chain}

Consider the case of the four-site chain; cf.~(\ref{equ:psi4c}).
The basis functions are associated with the following complete tubings of the marked graph:
\begin{align}
\begin{split}
\psi \raisebox{3pt}{
}
\end{split} 
\end{align}
As before, we have organized the tubings by the number of vertices enclosed by tubes that also contain a cross. This separates the functions into the sources that appear at each level of the differential equations. To label the large number of functions, we are forced to introduce a slightly cumbersome index notation. The functions in the second column each only have one tube that contains crosses. Their first subscript is the vertex associated to this tube.
The labels $1$ and $4$ are assigned to the tubings around the end vertices, whereas the notation $i_r$ for $r\in\{1,2,3\}$ indicate the three different ways we can draw tubings encircling an inner vertex $i\in\{2,3\}$.  The multiple subscript labels in the other columns then mean a union of these ``elementary" tubes. This composition is independent of the ordering, 
so that e.g.~$F_{41} = F_{14}$.

\vskip 4pt
The total number of functions in this example is
\beq
1 \ + \ 8 \ + \ 22 \ + \ 24 \ + \ 9\ = \ 64\, ,
\eeq
so the basis is 64-dimensional. This number is consistent with the general formula $4^e$, where $e=3$ is the number of edges of the graph.
Adding the number of functions with alternating signs, we again find $1  -  8 +  22  -  24  +  9 = 0$.
We will present the differential equations for a representative subset of these functions.

\paragraph{Level 1:}
The differential of $\psi$ is
\begin{align}
\begin{split}
\ud \psi
\, =  \, \e\, \Big[ (\psi - F_1)
 &\raisebox{3pt}{
} \Big]\,,
\label{equ:psi4-eqn}
\end{split}
\end{align}
which has the same structure as for the three-site chain. The sum of all coefficients is $4\e\hs\psi$, as expected.

\paragraph{Level 2:} The differentials for the new source functions work in exactly the same way as for the three-site chain and are not all repeated here.

\vskip 4pt
The most nontrivial example is the differential of $F_{2_1}$: 
\begin{align}
\begin{split}
\ud F_{2_1} \, =\,\,\e\, \Big[ &F_{2_1}
\raisebox{3pt}{
 %
}\, \Big] \, .\\
\end{split}
\end{align}
This is very similar to the case of $\ud Q_1$ in (\ref{equ:dQ1}): this time, we get two letters from the absorption of adjacent tubes:
\beq
\begin{aligned}
& %
 
\end{aligned}
\eeq
 The rules by which these letters arise and the corresponding assignment of coefficients, however, is the same as before. The sum of all coefficients is $4\e\hs F_{2_1}$, as expected.

\paragraph{Level 3:} At Level 3, three tubes can now merge into a single letter, like what happened for the function $f$ in~\eqref{equ:3site-ff-tree}.
As an example of this phenomenon, consider the function associated to the graph  
\!\!\raisebox{2.5pt}{
%
},
which we denote by $F_{1 3_2}$. Its differential is
\begin{align}
\begin{split}
\ud F_{1 3_2} \, =\, \,\e\,\Big[ &F_{13_2}\,\big(
\raisebox{3pt}{
 %
} \, \Big]\, .
\end{split}
\end{align}
The term in the last line arises from the merger of a new tube around vertex $2$ with the two pre-existing tubes around vertices $1$ and $3$: \!\!\raisebox{2.5pt}{
%
}

\vskip 4pt
Another interesting example is the differential of~$F_{2_1 4}$, which is the function associated to the graph
\!\!\raisebox{2.5pt}{
 %
}  \, \Big]\, .
\end{split}
\end{align}
In this case, there is an instance of a tube created by merger further absorbing an adjacent tube: 
\beq
 \raisebox{2.5pt}{%
}
\eeq
The graphical rules for this and the function assignments are the same as before.

\paragraph{Level 4:} Next, we consider the most nontrivial example at Level 4, which is the function associated to the graph
\!\!\raisebox{2.5pt}{
 %
}\,, which we denote by $F_{2_1 3_1 4} $. Its differential is
\begin{align}
\begin{split} 
\ud F_{2_1 3_1 4} \,=\, \e\,\Big[\, (F_{2_1 3_1 4}- F_{1 2_1 3_1 4})
\raisebox{3pt}{
 %
} \, \Big]\,.
\end{split}
\end{align}
Note that the tubing around vertex 3 gets activated to produce the letter in the first line. 
It then absorbs the tube on its left (but not that to its right), as expected from the directionality of the absorption rule:
\beq
 %
 
\eeq
In the last three lines, we see that the activated tube grows {\it twice} by absorbing the neighboring tube: 
\beq
 %
 }
\label{equ:B11}
\eeq
The first absorption produces the function $- F_{2_1 3_2 4}$, while the second gives $(-1)^2 F_{2_2 3_2 4} = F_{2_2 3_2 4}$.
The signs are consistent with $(-1)^{N_a}$, where $N_a$ is the number of absorptions along the path from the original parent graph.
Taking this into account, 
the coefficients assigned to each letter conform to the expected pattern.

\paragraph{Level 5:}  Finally, a nontrivial example at Level 5 is given by the function $F_{12_1 3_1 4}$, which is associated to the graph tubing
\!\!\raisebox{2.5pt}{
 %
} \,\Big]\, .
\end{split}
\end{align}
Again, we have an instance of an activated tube growing twice by absorbing a neighboring tube, 
 but it works just as in (\ref{equ:B11}) and is completely predicted by the graphical rules.

\newpage
\subsection{Four-Site Star}

Another interesting test case is the four-site star; cf.~(\ref{equ:4Site-Star}). In this case, there is not enough room for complicated merger and absorption phenomena, so the equations will actually be simpler than those for the four-site chain.

\vskip 4pt
We leave it as an exercise to the reader to draw the set of complete tubings of the marked four-site star (see \eqref{starreplacemets} for some examples).
There are now 10 tubings encircling a single vertex in tubes that also contain a cross, $24$ tubings with $2$ encircled vertices, $23$ tubings with $3$ encircled  vertices and finally $7$ tubings with $4$ encircled vertices.
The total number of basis functions is
\beq
1 \ + \ 10 \ + \ 24 \ + \ 22 \ + \ 7\ = \ 64\,,
\eeq
and their alternating sum again vanishes 
\beq
1 \ - \ 10 \ + \ 24 \ - \ 22 \ + \ 7\ = \ 0\,.\phantom{1}
\eeq
In the following, we present the differentials for a small subset of representative basis functions.

\vspace{0.3cm}
\paragraph{Level 1:}
The differential of the wavefunction $\psi$ is
\begin{align}
\begin{split}
\ud \psi
\, =\, \, &\e\, \left[(\psi - F_1) 
\raisebox{3pt}{
} 
\ \right] ,
\end{split}
\end{align}

\vspace{0.2cm}
\noindent
which has
exactly the expected pattern of activated tubes on each vertex.

\newpage
\paragraph{Level 2:} Next, we look at the differentials for some of the source functions. Nothing really special will happen, and we  just get the usual picture of activations, growth and mergers.

\vskip 4pt
The most nontrivial example is associated to the graph tubing 
\raisebox{3pt}{
 %
}\,,
which has differential
\begin{align}
\begin{split}
\ud F_{4_1} \,=\ \e\, &\left[F_{4_1}
\raisebox{3pt}{
 %
}  \ \ \right] .
 \end{split}
 \label{equ:XXX}
\end{align}
In the first line, we see the activation of the original tube around vertex $4$, as well as the growth of the tube around vertex $1$ and its merger with the tube around vertex $4$. The terms in the second and third lines, are the activation and growth of tubes around vertices $2$ and $3$.
In the last line, we have the usual absorption phenomenon.

\paragraph{Level 3:} Nothing surprising happens  at Level 3 either.
Here is an example for the graph
\!\!\raisebox{3pt}{
 %
} \ \ \right] .
 \end{split}
\end{align}
This is very similar to (\ref{equ:XXX}), except that we don't get the first term in the second line. Instead, we get
 a single activated tube around vertex $2$ (see the first term in the third line).

\paragraph{Levels 4 and 5:} Predicting the equations at Levels 4 and 5 is left as an exercise to the reader. It is a straightforward application of our graphical rules, without any surprises.

\newpage
\section{Integrability from Locality}
\label{app:locality}

An important (and nontrivial) feature of the graphical rules presented in Section~\ref{sec:GraphicalRules} is that they imply that $\ud^2 = 0$. This is required for consistency/integrability of the differential equations, and motivates this graphical procedure from an abstract perspective. Given a cochain complex, with a differential defined in some way, it is natural to study the cohomology of this complex and understand its features. In principle, this would provide a completely orthogonal starting point that would lead us to the study of the same set of equations.

\vskip4pt
In this appendix, we will to describe how the graphical rules guarantee that $\ud^2 = 0$, and to explore both the rigidity of the construction, and its relation to spacetime locality. 

\subsection{Preliminaries}

The nilpotency of $\ud$ is related to two essential features:
\begin{itemize}
\item The letters that appear in the differential equations satisfy {\it three-letter relations} that encode the linear dependencies of these quantities. These relations imply certain identities satisfied by the $\ud\log$ forms that appear when we take $\ud^2$ of a function, which cause it to vanish.
\vskip4pt
At their core, the letter relations can be understood as a consequence of the way that the marked graph encodes kinematics, because of the locality of this encoding (which is a reflection of spacetime locality), the letters associated to certain tubings have to be related. 
Essentially, the letter associated to a given tubing is related to the sum of letters for all possible ways of splitting the tubing into smaller tubes.

\vskip 4pt
Recall that the letters appearing in the equations for the two-site chain are
\beq
\begin{aligned} 
\hat B_1 &\equiv X_1 + Y\,, &\quad \hat B_2 &\equiv X_2+Y\,, &\quad  \hat B_3 &\equiv X_1+X_2\,, \\
 \hat B_4 &\equiv  X_1-Y\,, &\quad \hat B_5 &\equiv X_2-Y\,. & &
\end{aligned}
\eeq
By inspection, it is easy to see that these satisfy\footnote{Unlike in the rest of the paper, the tubings here represent the letters themselves and {\it not} the corresponding dlog forms.}
\beq
\begin{aligned}
\hat B_3 &=\hat B_2 + \hat B_4    \quad \Leftrightarrow \quad
\raisebox{3pt}{
 \begin{tikzpicture}[baseline=(current  bounding  box.center)]
    \node [
        draw, color=gray, fill=gray!20, line width=0.6pt,
        rounded rectangle,
        minimum height = .4mm,
        minimum width = 3em,
        rounded rectangle arc length = 180,
    ] at (0.3,0)
    {};
\draw[fill] (0, 0) circle (.4mm);
\draw[fill] (0.6, 0) circle (.4mm);
\draw[thick] (0, 0) -- (0.6, 0);
\node at (0.3,0)  {\Cross};
\end{tikzpicture}} 
\ = \
 \raisebox{3pt}{
 \begin{tikzpicture}[baseline=(current  bounding  box.center)]
    \draw[color=gray, fill=gray!20, line width=0.6pt] (0.6, 0) circle (1.3mm);
\draw[fill] (0, 0) circle (.4mm);
\draw[fill] (0.6, 0) circle (.4mm);
\draw[thick] (0, 0) -- (0.6, 0);
\node at (0.3,0)  {\Cross};
\end{tikzpicture}} \ + 
\raisebox{3pt}{
 \begin{tikzpicture}[baseline=(current  bounding  box.center)]
    \node [
        draw, color=gray, fill=gray!20, line width=0.6pt,
        rounded rectangle,
        minimum height = .4mm,
        minimum width = 2.1em,
        rounded rectangle arc length = 180,
    ] at (0.15,0)
    {};
\draw[fill] (0, 0) circle (.4mm);
\draw[fill] (0.6, 0) circle (.4mm);
\draw[thick] (0, 0) -- (0.6, 0);
\node at (0.3,0)  {\Cross};
\end{tikzpicture}}\,,\\
\hat B_3 &= \hat B_1 + \hat B_5  \quad \Leftrightarrow \quad
\raisebox{3pt}{
 \begin{tikzpicture}[baseline=(current  bounding  box.center)]
    \node [
        draw, color=gray, fill=gray!20, line width=0.6pt,
        rounded rectangle,
        minimum height = .4mm,
        minimum width = 3em,
        rounded rectangle arc length = 180,
    ] at (0.3,0)
    {};
\draw[fill] (0, 0) circle (.4mm);
\draw[fill] (0.6, 0) circle (.4mm);
\draw[thick] (0, 0) -- (0.6, 0);
\node at (0.3,0)  {\Cross};
\end{tikzpicture}}
\ = \
 \raisebox{3pt}{
 \begin{tikzpicture}[baseline=(current  bounding  box.center)]
    \draw[color=gray, fill=gray!20, line width=0.6pt] (0, 0) circle (1.3mm);
\draw[fill] (0, 0) circle (.4mm);
\draw[fill] (0.6, 0) circle (.4mm);
\draw[thick] (0, 0) -- (0.6, 0);
\node at (0.3,0)  {\Cross};
\end{tikzpicture}} \ + 
\raisebox{3pt}{
 \begin{tikzpicture}[baseline=(current  bounding  box.center)]
    \node [
        draw, color=gray, fill=gray!20, line width=0.6pt,
        rounded rectangle,
        minimum height = .4mm,
        minimum width = 2.1em,
        rounded rectangle arc length = 180,
    ] at (0.45,0)
    {};
\draw[fill] (0, 0) circle (.4mm);
\draw[fill] (0.6, 0) circle (.4mm);
\draw[thick] (0, 0) -- (0.6, 0);
\node at (0.3,0)  {\Cross};
\end{tikzpicture}} \,,
\end{aligned} 
\label{equ:3letters}
\eeq
which encode the two ways of splitting the tubing of the full graph into smaller tubes.
Versions of these relations exist for all graphs and all ways of splitting tubes into smaller pieces.

\vskip4pt
We can understand the consequences of these relations geometrically. In the space of kinematic variables, each letter can be thought of as defining a hyperplane, the three-letter relations imply the incidence of three of these hyperplanes along a codimension-two surface. This degeneration implies that the canonical form (written in $\ud\log$ form) vanishes, implying relations between wedge products of $\ud\log$ forms. As we will see, $\ud^2$ of a function is always proportional to these identically zero combinations.
\item The other nontrivial feature is the precise assignment of functions and signs to the graphical rules. To illustrate this, recall the tree structure~\eqref{equ:Q1tree} of the function $Q_1$ for the three-site chain:
 \beq
 \begin{tikzpicture}[baseline=(current  bounding  box.center)]
\node[inner sep=0pt] at (0,0)
    {\includegraphics[scale=1]{Figures/Tubings/three/tree/threeQ1}}; 
\draw [color=gray,thick,-stealth] (1.3,0) -- (2.5,0);
\draw [color=gray,thick,-stealth] (1.3,0.15) -- (2.5,0.6);
\draw [color=gray,thick,-stealth] (1.3,-0.15) -- (2.5,-0.6);
\node[inner sep=0pt] at (3.8,0.75)
    {\includegraphics[scale=1]{Figures/Tubings/three/tree/threeQ1a}};
\node[inner sep=0pt] at (3.8,0)
    {\includegraphics[scale=1]{Figures/Tubings/three/tree/threeQ1b}}; 
\node[inner sep=0pt] at (3.8,-0.75)
    {\includegraphics[scale=1]{Figures/Tubings/three/tree/threeQ1c}};   
     \draw [color=gray,thick,line width=0.5pt, dashed] (1.95+0.8,-1) -- (1.95+0.8,1) --  (4.05+0.8,1) -- (4.05+0.8,-1) -- (1.95+0.8,-1); 
      \node[below] at (3.8,-1)  {$Q_1$};
\draw [color=gray,thick,-stealth] (5.1,0.75) -- (6.1,0.75);
\draw [color=gray,thick,-stealth] (5.1,-0.75) -- (6.1,-0.75);
\node[inner sep=0pt] at (7.3,0.75)
    {\includegraphics[scale=1]{Figures/Tubings/three/tree/threeq1ab}};
          \node[above] at (7.3,0.95)  {$q_1$}; 
\node[inner sep=0pt] at (7.3,-0.75)
    {\includegraphics[scale=1]{Figures/Tubings/three/tree/threeqt3c}};
      \node[below] at (7.3,-0.95)  {$\tilde q_3$}; 
\draw [color=gray,thick,-stealth] (8.5,-0.75) -- (9.5,-0.75);
\node[inner sep=0pt] at (10.7,-0.75)
    {\includegraphics[scale=1]{Figures/Tubings/three/tree/threeqt2bc}};
         \node[below] at (10.7,-0.95)  {$(-\tilde q_2)$}; 
\end{tikzpicture}
\label{equ:tree-Q1}
\eeq
In addition to the functions being assigned to the corresponding graph tubings in the tree, there were two noteworthy aspects of the rules: {\it i}) the {\it subtraction} of each immediate descendant from its parent function and {\it ii}) the {\it sign} of the descendant functions given by $(-1)^{N_a}$, with $N_a$ being the number of absorptions along the path from the original graph. 
To elucidate the necessity of these features, we will consider a modified version of the graphical rules that leads to the following differential equation for $Q_1$:
\beq
\begin{aligned}
  \ud Q_1 = \e\, \Big[  Q_1 \LLbmp  \, +\, (Q_1- {\color{regal}c_S}\hs q_1)\, &\LLap \,+\, \ \  (Q_1-{\color{regal}c_S}\hs \tilde q_3) \LLcp\phantom{\Big]}\\
   +\ q_1\, &\LLabpr  \,+ \, (\tilde q_3+{\color{Orange}c_A}{\color{regal} c_S}\hs \tilde q_2) \LLcm\phantom{\Big]}\\
&\hspace{3.5cm} -\ {\color{Orange}c_A}\hs\tilde q_2\ \LLbcmb\Big]
  \end{aligned}
  \label{eq:modifiedgraph}
  \eeq
where the coefficient ${\color{regal}c_S}$ parameterizes the subtraction of descendants and ${\color{Orange}c_A}$ is an extra grading for functions created by absorption.  
In what follows, we illustrate through examples that ${c_S=c_A=1}$ is a consequence of the integrability of our differential equations.
\end{itemize}
We will now show how these two features guarantee that $\ud^2 = 0$ in a nontrivial way.

\newpage
\subsection{Two-Site Chain}

We first consider the example of the two-site chain.
In that case, the three-letter relations shown in (\ref{equ:3letters}) imply the following identities between triplets of dlog forms:
\begin{align}
\begin{split}
{\color{Red} R_1} &\equiv \partial[\hat B_2 \hat B_3  \hat B_4] = - [\hat B_3  \hat B_4] + [\hat B_2   \hat B_4] - [\hat B_2 \hat B_3]   = 0\,, \\
{\color{Blue} R_2}&\equiv  \partial[\hat B_1 \hat B_3  \hat B_5] = - [\hat B_3  \hat B_5] + [\hat B_1   \hat B_5] - [\hat B_1 \hat B_3]   = 0\,,
\end{split}
\label{equ:identities}
\end{align}
where $[\hat B_i \hat B_j] \equiv \ud \log \hat B_i \wedge \ud \log \hat B_j$.

\vskip 4pt
The modified graphical rules~\eqref{eq:modifiedgraph} give the following differential equations for the two-site chain:
\beq
\begin{aligned}
\ud \psi &= \e \big[ (\psi-{\color{regal}c_S} F) \hs \ud \log \hat B_1 + F \hs \ud \log \hat B_4 +  (\psi-{\color{regal}c_S} \tilde F)\hs \ud \log \hat B_2 +  \tilde F\hs \ud \log \hat B_5 \big] \,,\\
\ud F &= \e \big[ F \hs \ud \log \hat B_4 + (F-{\color{regal}c_S} Z) \hs \ud \log \hat B_2 +  Z \hs \ud \log \hat B_3 \big]\,,\\
\ud \tilde F &= \e \big[ \tilde F \hs \ud \log \hat B_5 + (\tilde F-{\color{regal}c_S} Z) \hs \ud \log \hat B_1 + Z \hs \ud \log \hat B_3 \big]\,,\\
\ud Z &= 2\e \hs Z\hs \ud \log \hat B_3\, .
\end{aligned}
\label{equ:2site-eqns}
\eeq
Taking a further differential of these equations, we get
\begin{align}
\begin{split}
\ud^2 \psi &= \e^2 \Big[Z({\color{Red} R_1}  + {\color{Blue} R_2}) + (1-{\color{regal}c_S})\Big(Z\big[ \ud\log( \hat B_3/ \hat B_5)\wedge\ud\log \hat B_1-\ud\log( \hat B_3/ \hat B_4)\wedge\ud\log \hat B_2 \big] \\
&\hspace{3cm} + F\,\ud\log\hat B_4\wedge\ud\log\hat B_1+ \tilde F\,\ud\log\hat B_5\wedge\ud\log\hat B_2\Big)\Big]\,,
\end{split}\\[4pt]
\ud^2 F &= \e^2Z \Big[{\color{Red} R_1} +(1-{\color{regal}c_S}) \,\ud\log(2\hat B_3/\hat B_4) \wedge\ud\log\hat B_2  \Big] \,,\\
\ud^2 \tilde F &=\e^2Z\Big[ {\color{Blue} R_2} +(1-{\color{regal}c_S})\, \ud\log(2\hat B_3/\hat B_5) \wedge\ud\log\hat B_1  \Big]\,,\\[4pt]
\ud^2 Z &= 0\, .
\end{align}
Note that there are non-vanishing combinations of $\ud\log$ forms proportional to $1-c_S$. Integrability of these equations then forces $c_S=1$. (Since this system does not have any absorptions, we get no constraint on $c_A$, as expected.)
The assignment of source functions in (\ref{equ:2site-eqns})---including their signs---is then precisely such that the $\ud^2$ of all functions is proportional to the vanishing three-letter identities in (\ref{equ:identities}). Note that $\ud^2 \psi$ contains $R_1+R_2$ because there are two ways of reaching the function~$Z$.
Graphically, we can represent this as
\beq
 \begin{tikzpicture}[baseline=(current  bounding  box.center)]
 \node at (0,1.25)  {$\psi$};
 \node[gray] at (-1,0)  {$F$};
 \node[gray] at (1,0)  {$\tilde F$};
 \node at (-0.03,-1.3)  {$Z$};
  \node at (-0.75,-0.7)  {\small \color{Red}$1$};
    \node at (0.7,-0.7)  {\small \color{Blue}$2$};
 \draw[thick,color=gray] (-0.12,1.1) --  (-0.84,0.2);
 \draw[thick,color=gray] (0.12,1.1) --  (0.88,0.15);
  \draw[thick,color=Red] (-0.12,-1.1) --  (-0.88,-0.15);
   \draw[thick,color=Blue] (0.12,-1.1) --  (0.84,-0.2);
 \end{tikzpicture}
 \nonumber
 \eeq
 As we will see, this tree structure will arise locally in more complicated examples.

\vspace{0.25cm}
\newpage
\subsection{Three-Site Chain}

Next, we consider the case of the three-site chain, which displays the full complexity of the most general case. As before, it is useful to enumerate the letters appearing in the equations:
\beq
\begin{aligned}
\hat B_1 &\equiv X_1 + Y & \quad \hat B_2 &\equiv  X_2+Y+Y'  &\quad  \hat B_3 &\equiv X_3 + Y'  &\quad \hat B_4 &\equiv X_1 + X_2+X_3 \\
\hat B_7 &\equiv X_1-Y  & \quad  \hat B_8 &\equiv X_2-Y+Y'  & \quad \hat B_{11} &\equiv X_3-Y'  & &\\
& & \quad  \hat B_9 &\equiv X_2+Y-Y' & & & \hat B_5 &\equiv X_1+X_2+Y'   \\
& & \quad \hat B_{10} &\equiv X_2-Y-Y'  & & & \hat B_{12} &\equiv X_1+X_2-Y'  \\
& & & & & & \hat B_6 &\equiv X_2+X_3+Y  \\
& & & & & & \hat B_{13} &\equiv X_2+X_3-Y \,,
\end{aligned}
\eeq
which satisfy the following 
{\it three-letter relations}:
\beq
\begin{aligned}
1:& & \hat B_5 + \hat B_{11} - \hat B_4  &= 0  \quad \Leftrightarrow \quad
 \raisebox{3pt}{
 \begin{tikzpicture}[baseline=(current  bounding  box.center)]
 \node [
        draw, color=gray, fill=gray!20,  line width=0.6pt,
        rounded rectangle,
        minimum height = .4mm,
        minimum width = 3em,
        rounded rectangle arc length = 180,
    ] at (0.3,0)
    {};
\draw[fill] (0, 0) circle (.4mm);
\draw[fill] (0.6, 0) circle (.4mm);
\draw[fill] (1.2, 0) circle (.4mm);
\draw[thick] (0, 0) -- (1.2, 0);
\node at (0.3,0)  {\Cross};
\node at (0.9,0)  {\Cross};
\end{tikzpicture}} \ + 
\raisebox{3pt}{
 \begin{tikzpicture}[baseline=(current  bounding  box.center)]
    \node [
        draw, color=gray, fill=gray!20, line width=0.6pt,
        rounded rectangle,
        minimum height = .4mm,
        minimum width = 2.1em,
        rounded rectangle arc length = 180,
    ] at (1.05,0)
    {};
\draw[fill] (0, 0) circle (.4mm);
\draw[fill] (0.6, 0) circle (.4mm);
\draw[fill] (1.2, 0) circle (.4mm);
\draw[thick] (0, 0) -- (1.2, 0);
\node at (0.3,0)  {\Cross};
\node at (0.9,0)  {\Cross};
\end{tikzpicture}}\ - 
\raisebox{3pt}{
 \begin{tikzpicture}[baseline=(current  bounding  box.center)]
    \node [
        draw, color=gray, fill=gray!20, line width=0.6pt,
        rounded rectangle,
        minimum height = .4mm,
        minimum width = 4.4em,
        rounded rectangle arc length = 180,
    ] at (0.6,0)
    {};
\draw[fill] (0, 0) circle (.4mm);
\draw[fill] (0.6, 0) circle (.4mm);
\draw[fill] (1.2, 0) circle (.4mm);
\draw[thick] (0, 0) -- (1.2, 0);
\node at (0.3,0)  {\Cross};
\node at (0.9,0)  {\Cross};
\end{tikzpicture}} \ = \ 0
\\
2:& & \hat B_{12} + \hat B_{3} - \hat B_4  &= 0  \quad \Leftrightarrow \quad
 \raisebox{3pt}{
 \begin{tikzpicture}[baseline=(current  bounding  box.center)]
 \node [
        draw, color=gray, fill=gray!20,  line width=0.6pt,
        rounded rectangle,
        minimum height = .4mm,
        minimum width = 3.6em,
        rounded rectangle arc length = 180,
    ] at (0.45,0)
    {};
\draw[fill] (0, 0) circle (.4mm);
\draw[fill] (0.6, 0) circle (.4mm);
\draw[fill] (1.2, 0) circle (.4mm);
\draw[thick] (0, 0) -- (1.2, 0);
\node at (0.3,0)  {\Cross};
\node at (0.9,0)  {\Cross};
\end{tikzpicture}} \ + 
\raisebox{3pt}{
 \begin{tikzpicture}[baseline=(current  bounding  box.center)]
   \draw[color=gray, fill=gray!20, line width=0.6pt] (1.2, 0) circle (1.3mm);
\draw[fill] (0, 0) circle (.4mm);
\draw[fill] (0.6, 0) circle (.4mm);
\draw[fill] (1.2, 0) circle (.4mm);
\draw[thick] (0, 0) -- (1.2, 0);
\node at (0.3,0)  {\Cross};
\node at (0.9,0)  {\Cross};
\end{tikzpicture}}\ - 
\raisebox{3pt}{
 \begin{tikzpicture}[baseline=(current  bounding  box.center)]
    \node [
        draw, color=gray, fill=gray!20, line width=0.6pt,
        rounded rectangle,
        minimum height = .4mm,
        minimum width = 4.4em,
        rounded rectangle arc length = 180,
    ] at (0.6,0)
    {};
\draw[fill] (0, 0) circle (.4mm);
\draw[fill] (0.6, 0) circle (.4mm);
\draw[fill] (1.2, 0) circle (.4mm);
\draw[thick] (0, 0) -- (1.2, 0);
\node at (0.3,0)  {\Cross};
\node at (0.9,0)  {\Cross};
\end{tikzpicture}} \ = \ 0 \\
3:& & \hat B_{6} + \hat B_{7} - \hat B_4  &= 0  \quad \Leftrightarrow \quad
 \raisebox{3pt}{
 \begin{tikzpicture}[baseline=(current  bounding  box.center)]
 \node [
        draw, color=gray, fill=gray!20,  line width=0.6pt,
        rounded rectangle,
        minimum height = .4mm,
        minimum width = 3em,
        rounded rectangle arc length = 180,
    ] at (0.9,0)
    {};
\draw[fill] (0, 0) circle (.4mm);
\draw[fill] (0.6, 0) circle (.4mm);
\draw[fill] (1.2, 0) circle (.4mm);
\draw[thick] (0, 0) -- (1.2, 0);
\node at (0.3,0)  {\Cross};
\node at (0.9,0)  {\Cross};
\end{tikzpicture}} \ + 
\raisebox{3pt}{
 \begin{tikzpicture}[baseline=(current  bounding  box.center)]
    \node [
        draw, color=gray, fill=gray!20, line width=0.6pt,
        rounded rectangle,
        minimum height = .4mm,
        minimum width = 2.1em,
        rounded rectangle arc length = 180,
    ] at (0.15,0)
    {};
\draw[fill] (0, 0) circle (.4mm);
\draw[fill] (0.6, 0) circle (.4mm);
\draw[fill] (1.2, 0) circle (.4mm);
\draw[thick] (0, 0) -- (1.2, 0);
\node at (0.3,0)  {\Cross};
\node at (0.9,0)  {\Cross};
\end{tikzpicture}}\ - 
\raisebox{3pt}{
 \begin{tikzpicture}[baseline=(current  bounding  box.center)]
    \node [
        draw, color=gray, fill=gray!20, line width=0.6pt,
        rounded rectangle,
        minimum height = .4mm,
        minimum width = 4.4em,
        rounded rectangle arc length = 180,
    ] at (0.6,0)
    {};
\draw[fill] (0, 0) circle (.4mm);
\draw[fill] (0.6, 0) circle (.4mm);
\draw[fill] (1.2, 0) circle (.4mm);
\draw[thick] (0, 0) -- (1.2, 0);
\node at (0.3,0)  {\Cross};
\node at (0.9,0)  {\Cross};
\end{tikzpicture}} \ = \ 0 \\
4:& & \hat B_{13} + \hat B_{1} - \hat B_4  &= 0  \quad \Leftrightarrow \quad
 \raisebox{3pt}{
 \begin{tikzpicture}[baseline=(current  bounding  box.center)]
 \node [
        draw, color=gray, fill=gray!20,  line width=0.6pt,
        rounded rectangle,
        minimum height = .4mm,
        minimum width = 3.6em,
        rounded rectangle arc length = 180,
    ] at (0.75,0)
    {};
\draw[fill] (0, 0) circle (.4mm);
\draw[fill] (0.6, 0) circle (.4mm);
\draw[fill] (1.2, 0) circle (.4mm);
\draw[thick] (0, 0) -- (1.2, 0);
\node at (0.3,0)  {\Cross};
\node at (0.9,0)  {\Cross};
\end{tikzpicture}} \ + 
\raisebox{3pt}{
 \begin{tikzpicture}[baseline=(current  bounding  box.center)]
   \draw[color=gray, fill=gray!20, line width=0.6pt] (0, 0) circle (1.3mm);
\draw[fill] (0, 0) circle (.4mm);
\draw[fill] (0.6, 0) circle (.4mm);
\draw[fill] (1.2, 0) circle (.4mm);
\draw[thick] (0, 0) -- (1.2, 0);
\node at (0.3,0)  {\Cross};
\node at (0.9,0)  {\Cross};
\end{tikzpicture}}\ - 
\raisebox{3pt}{
 \begin{tikzpicture}[baseline=(current  bounding  box.center)]
    \node [
        draw, color=gray, fill=gray!20, line width=0.6pt,
        rounded rectangle,
        minimum height = .4mm,
        minimum width = 4.4em,
        rounded rectangle arc length = 180,
    ] at (0.6,0)
    {};
\draw[fill] (0, 0) circle (.4mm);
\draw[fill] (0.6, 0) circle (.4mm);
\draw[fill] (1.2, 0) circle (.4mm);
\draw[thick] (0, 0) -- (1.2, 0);
\node at (0.3,0)  {\Cross};
\node at (0.9,0)  {\Cross};
\end{tikzpicture}} \ = \ 0\\[-6pt]
  \cline{1-5}
5:& & \hat B_{2} + \hat B_{7} - \hat B_5  &= 0  \quad \Leftrightarrow \quad
 \raisebox{3pt}{
 \begin{tikzpicture}[baseline=(current  bounding  box.center)]
    \draw[color=gray, fill=gray!20, line width=0.6pt] (0.6, 0) circle (1.3mm);
\draw[fill] (0, 0) circle (.4mm);
\draw[fill] (0.6, 0) circle (.4mm);
\draw[fill] (1.2, 0) circle (.4mm);
\draw[thick] (0, 0) -- (1.2, 0);
\node at (0.3,0)  {\Cross};
\node at (0.9,0)  {\Cross};
\end{tikzpicture}} \ \, + 
\raisebox{3pt}{
 \begin{tikzpicture}[baseline=(current  bounding  box.center)]
    \node [
        draw, color=gray, fill=gray!20, line width=0.6pt,
        rounded rectangle,
        minimum height = .4mm,
        minimum width = 2.1em,
        rounded rectangle arc length = 180,
    ] at (0.15,0)
    {};
\draw[fill] (0, 0) circle (.4mm);
\draw[fill] (0.6, 0) circle (.4mm);
\draw[fill] (1.2, 0) circle (.4mm);
\draw[thick] (0, 0) -- (1.2, 0);
\node at (0.3,0)  {\Cross};
\node at (0.9,0)  {\Cross};
\end{tikzpicture}}\ - 
 \raisebox{3pt}{
 \begin{tikzpicture}[baseline=(current  bounding  box.center)]
 \node [
        draw, color=gray, fill=gray!20,  line width=0.6pt,
        rounded rectangle,
        minimum height = .4mm,
        minimum width = 3em,
        rounded rectangle arc length = 180,
    ] at (0.3,0)
    {};
\draw[fill] (0, 0) circle (.4mm);
\draw[fill] (0.6, 0) circle (.4mm);
\draw[fill] (1.2, 0) circle (.4mm);
\draw[thick] (0, 0) -- (1.2, 0);
\node at (0.3,0)  {\Cross};
\node at (0.9,0)  {\Cross};
\end{tikzpicture}} \,\ = \ 0 \\
6:& & \hat B_{1} + \hat B_{8} - \hat B_5  &= 0  \quad \Leftrightarrow \quad
 \raisebox{3pt}{
 \begin{tikzpicture}[baseline=(current  bounding  box.center)]
    \draw[color=gray, fill=gray!20, line width=0.6pt] (0, 0) circle (1.3mm);
\draw[fill] (0, 0) circle (.4mm);
\draw[fill] (0.6, 0) circle (.4mm);
\draw[fill] (1.2, 0) circle (.4mm);
\draw[thick] (0, 0) -- (1.2, 0);
\node at (0.3,0)  {\Cross};
\node at (0.9,0)  {\Cross};
\end{tikzpicture}} \ \, + 
\raisebox{3pt}{
 \begin{tikzpicture}[baseline=(current  bounding  box.center)]
    \node [
        draw, color=gray, fill=gray!20, line width=0.6pt,
        rounded rectangle,
        minimum height = .4mm,
        minimum width = 2.1em,
        rounded rectangle arc length = 180,
    ] at (0.45,0)
    {};
\draw[fill] (0, 0) circle (.4mm);
\draw[fill] (0.6, 0) circle (.4mm);
\draw[fill] (1.2, 0) circle (.4mm);
\draw[thick] (0, 0) -- (1.2, 0);
\node at (0.3,0)  {\Cross};
\node at (0.9,0)  {\Cross};
\end{tikzpicture}}\ - 
 \raisebox{3pt}{
 \begin{tikzpicture}[baseline=(current  bounding  box.center)]
 \node [
        draw, color=gray, fill=gray!20,  line width=0.6pt,
        rounded rectangle,
        minimum height = .4mm,
        minimum width = 3em,
        rounded rectangle arc length = 180,
    ] at (0.3,0)
    {};
\draw[fill] (0, 0) circle (.4mm);
\draw[fill] (0.6, 0) circle (.4mm);
\draw[fill] (1.2, 0) circle (.4mm);
\draw[thick] (0, 0) -- (1.2, 0);
\node at (0.3,0)  {\Cross};
\node at (0.9,0)  {\Cross};
\end{tikzpicture}} \,\ = \ 0\\[-6pt]
  \cline{1-5}
7:& & \hat B_7 + \hat B_{9} - \hat B_{12}  &= 0  \quad \Leftrightarrow \quad
 \raisebox{3pt}{
 \begin{tikzpicture}[baseline=(current  bounding  box.center)]
    \node [
        draw, color=gray, fill=gray!20, line width=0.6pt,
        rounded rectangle,
        minimum height = .4mm,
        minimum width = 2.1em,
        rounded rectangle arc length = 180,
    ] at (0.15,0)
    {};
\draw[fill] (0, 0) circle (.4mm);
\draw[fill] (0.6, 0) circle (.4mm);
\draw[fill] (1.2, 0) circle (.4mm);
\draw[thick] (0, 0) -- (1.2, 0);
\node at (0.3,0)  {\Cross};
\node at (0.9,0)  {\Cross};
\end{tikzpicture}} \ \, + 
\raisebox{3pt}{
 \begin{tikzpicture}[baseline=(current  bounding  box.center)]
    \node [
        draw, color=gray, fill=gray!20, line width=0.6pt,
        rounded rectangle,
        minimum height = .4mm,
        minimum width = 2.1em,
        rounded rectangle arc length = 180,
    ] at (0.75,0)
    {};
\draw[fill] (0, 0) circle (.4mm);
\draw[fill] (0.6, 0) circle (.4mm);
\draw[fill] (1.2, 0) circle (.4mm);
\draw[thick] (0, 0) -- (1.2, 0);
\node at (0.3,0)  {\Cross};
\node at (0.9,0)  {\Cross};
\end{tikzpicture}}\ - 
 \raisebox{3pt}{
 \begin{tikzpicture}[baseline=(current  bounding  box.center)]
 \node [
        draw, color=gray, fill=gray!20,  line width=0.6pt,
        rounded rectangle,
        minimum height = .4mm,
        minimum width = 3.6em,
        rounded rectangle arc length = 180,
    ] at (0.45,0)
    {};
\draw[fill] (0, 0) circle (.4mm);
\draw[fill] (0.6, 0) circle (.4mm);
\draw[fill] (1.2, 0) circle (.4mm);
\draw[thick] (0, 0) -- (1.2, 0);
\node at (0.3,0)  {\Cross};
\node at (0.9,0)  {\Cross};
\end{tikzpicture}} \,\ = \ 0\\
8:& & \hat B_1 + \hat B_{10} - \hat B_{12}  &= 0\quad \Leftrightarrow \quad
 \raisebox{3pt}{
 \begin{tikzpicture}[baseline=(current  bounding  box.center)]
        \draw[color=gray, fill=gray!20, line width=0.6pt] (0, 0) circle (1.3mm);
\draw[fill] (0, 0) circle (.4mm);
\draw[fill] (0.6, 0) circle (.4mm);
\draw[fill] (1.2, 0) circle (.4mm);
\draw[thick] (0, 0) -- (1.2, 0);
\node at (0.3,0)  {\Cross};
\node at (0.9,0)  {\Cross};
\end{tikzpicture}} \ \, + 
\raisebox{3pt}{
 \begin{tikzpicture}[baseline=(current  bounding  box.center)]
    \node [
        draw, color=gray, fill=gray!20, line width=0.6pt,
        rounded rectangle,
        minimum height = .4mm,
        minimum width = 3em,
        rounded rectangle arc length = 180,
    ] at (0.6,0)
    {};
\draw[fill] (0, 0) circle (.4mm);
\draw[fill] (0.6, 0) circle (.4mm);
\draw[fill] (1.2, 0) circle (.4mm);
\draw[thick] (0, 0) -- (1.2, 0);
\node at (0.3,0)  {\Cross};
\node at (0.9,0)  {\Cross};
\end{tikzpicture}}\ - 
 \raisebox{3pt}{
 \begin{tikzpicture}[baseline=(current  bounding  box.center)]
 \node [
        draw, color=gray, fill=gray!20,  line width=0.6pt,
        rounded rectangle,
        minimum height = .4mm,
        minimum width = 3.6em,
        rounded rectangle arc length = 180,
    ] at (0.45,0)
    {};
\draw[fill] (0, 0) circle (.4mm);
\draw[fill] (0.6, 0) circle (.4mm);
\draw[fill] (1.2, 0) circle (.4mm);
\draw[thick] (0, 0) -- (1.2, 0);
\node at (0.3,0)  {\Cross};
\node at (0.9,0)  {\Cross};
\end{tikzpicture}} \,\ = \ 0\\[-6pt]
  \cline{1-5}
9:& & \hat B_{2} + \hat B_{11} - \hat B_{6}  &= 0 \quad \Leftrightarrow \quad
 \raisebox{3pt}{
 \begin{tikzpicture}[baseline=(current  bounding  box.center)]
    \draw[color=gray, fill=gray!20, line width=0.6pt] (0.6, 0) circle (1.3mm);
\draw[fill] (0, 0) circle (.4mm);
\draw[fill] (0.6, 0) circle (.4mm);
\draw[fill] (1.2, 0) circle (.4mm);
\draw[thick] (0, 0) -- (1.2, 0);
\node at (0.3,0)  {\Cross};
\node at (0.9,0)  {\Cross};
\end{tikzpicture}} \ \, + 
\raisebox{3pt}{
 \begin{tikzpicture}[baseline=(current  bounding  box.center)]
    \node [
        draw, color=gray, fill=gray!20, line width=0.6pt,
        rounded rectangle,
        minimum height = .4mm,
        minimum width = 2.1em,
        rounded rectangle arc length = 180,
    ] at (1.05,0)
    {};
\draw[fill] (0, 0) circle (.4mm);
\draw[fill] (0.6, 0) circle (.4mm);
\draw[fill] (1.2, 0) circle (.4mm);
\draw[thick] (0, 0) -- (1.2, 0);
\node at (0.3,0)  {\Cross};
\node at (0.9,0)  {\Cross};
\end{tikzpicture}}\ - 
 \raisebox{3pt}{
 \begin{tikzpicture}[baseline=(current  bounding  box.center)]
 \node [
        draw, color=gray, fill=gray!20,  line width=0.6pt,
        rounded rectangle,
        minimum height = .4mm,
        minimum width = 3em,
        rounded rectangle arc length = 180,
    ] at (0.9,0)
    {};
\draw[fill] (0, 0) circle (.4mm);
\draw[fill] (0.6, 0) circle (.4mm);
\draw[fill] (1.2, 0) circle (.4mm);
\draw[thick] (0, 0) -- (1.2, 0);
\node at (0.3,0)  {\Cross};
\node at (0.9,0)  {\Cross};
\end{tikzpicture}} \,\ = \ 0\\
10:& & \hat B_{3} + \hat B_{9} - \hat B_{6}  &= 0\quad \Leftrightarrow \quad
 \raisebox{3pt}{
 \begin{tikzpicture}[baseline=(current  bounding  box.center)]
    \draw[color=gray, fill=gray!20, line width=0.6pt] (1.2, 0) circle (1.3mm);
\draw[fill] (0, 0) circle (.4mm);
\draw[fill] (0.6, 0) circle (.4mm);
\draw[fill] (1.2, 0) circle (.4mm);
\draw[thick] (0, 0) -- (1.2, 0);
\node at (0.3,0)  {\Cross};
\node at (0.9,0)  {\Cross};
\end{tikzpicture}} \ \, + 
\raisebox{3pt}{
 \begin{tikzpicture}[baseline=(current  bounding  box.center)]
    \node [
        draw, color=gray, fill=gray!20, line width=0.6pt,
        rounded rectangle,
        minimum height = .4mm,
        minimum width = 2.1em,
        rounded rectangle arc length = 180,
    ] at (0.75,0)
    {};
\draw[fill] (0, 0) circle (.4mm);
\draw[fill] (0.6, 0) circle (.4mm);
\draw[fill] (1.2, 0) circle (.4mm);
\draw[thick] (0, 0) -- (1.2, 0);
\node at (0.3,0)  {\Cross};
\node at (0.9,0)  {\Cross};
\end{tikzpicture}}\ - 
 \raisebox{3pt}{
 \begin{tikzpicture}[baseline=(current  bounding  box.center)]
 \node [
        draw, color=gray, fill=gray!20,  line width=0.6pt,
        rounded rectangle,
        minimum height = .4mm,
        minimum width = 3em,
        rounded rectangle arc length = 180,
    ] at (0.9,0)
    {};
\draw[fill] (0, 0) circle (.4mm);
\draw[fill] (0.6, 0) circle (.4mm);
\draw[fill] (1.2, 0) circle (.4mm);
\draw[thick] (0, 0) -- (1.2, 0);
\node at (0.3,0)  {\Cross};
\node at (0.9,0)  {\Cross};
\end{tikzpicture}} \,\ = \ 0\\[-6pt]
  \cline{1-5}
11:& & \hat B_{8} + \hat B_{11} - \hat B_{13}  &= 0 \quad \Leftrightarrow \quad
 \raisebox{3pt}{
 \begin{tikzpicture}[baseline=(current  bounding  box.center)]
    \node [
        draw, color=gray, fill=gray!20, line width=0.6pt,
        rounded rectangle,
        minimum height = .4mm,
        minimum width = 2.1em,
        rounded rectangle arc length = 180,
    ] at (0.45,0)
    {};
\draw[fill] (0, 0) circle (.4mm);
\draw[fill] (0.6, 0) circle (.4mm);
\draw[fill] (1.2, 0) circle (.4mm);
\draw[thick] (0, 0) -- (1.2, 0);
\node at (0.3,0)  {\Cross};
\node at (0.9,0)  {\Cross};
\end{tikzpicture}} \ \, + 
\raisebox{3pt}{
 \begin{tikzpicture}[baseline=(current  bounding  box.center)]
    \node [
        draw, color=gray, fill=gray!20, line width=0.6pt,
        rounded rectangle,
        minimum height = .4mm,
        minimum width = 2.1em,
        rounded rectangle arc length = 180,
    ] at (1.05,0)
    {};
\draw[fill] (0, 0) circle (.4mm);
\draw[fill] (0.6, 0) circle (.4mm);
\draw[fill] (1.2, 0) circle (.4mm);
\draw[thick] (0, 0) -- (1.2, 0);
\node at (0.3,0)  {\Cross};
\node at (0.9,0)  {\Cross};
\end{tikzpicture}}\ - 
 \raisebox{3pt}{
 \begin{tikzpicture}[baseline=(current  bounding  box.center)]
 \node [
        draw, color=gray, fill=gray!20,  line width=0.6pt,
        rounded rectangle,
        minimum height = .4mm,
        minimum width = 3.6em,
        rounded rectangle arc length = 180,
    ] at (0.75,0)
    {};
\draw[fill] (0, 0) circle (.4mm);
\draw[fill] (0.6, 0) circle (.4mm);
\draw[fill] (1.2, 0) circle (.4mm);
\draw[thick] (0, 0) -- (1.2, 0);
\node at (0.3,0)  {\Cross};
\node at (0.9,0)  {\Cross};
\end{tikzpicture}} \,\ = \ 0\\
12:& & \hat B_{3} + \hat B_{10} - \hat B_{13}  &= 0\quad \Leftrightarrow \quad
 \raisebox{3pt}{
 \begin{tikzpicture}[baseline=(current  bounding  box.center)]
        \draw[color=gray, fill=gray!20, line width=0.6pt] (1.2, 0) circle (1.3mm);
\draw[fill] (0, 0) circle (.4mm);
\draw[fill] (0.6, 0) circle (.4mm);
\draw[fill] (1.2, 0) circle (.4mm);
\draw[thick] (0, 0) -- (1.2, 0);
\node at (0.3,0)  {\Cross};
\node at (0.9,0)  {\Cross};
\end{tikzpicture}} \ \, + 
\raisebox{3pt}{
 \begin{tikzpicture}[baseline=(current  bounding  box.center)]
    \node [
        draw, color=gray, fill=gray!20, line width=0.6pt,
        rounded rectangle,
        minimum height = .4mm,
        minimum width = 3em,
        rounded rectangle arc length = 180,
    ] at (0.6,0)
    {};
\draw[fill] (0, 0) circle (.4mm);
\draw[fill] (0.6, 0) circle (.4mm);
\draw[fill] (1.2, 0) circle (.4mm);
\draw[thick] (0, 0) -- (1.2, 0);
\node at (0.3,0)  {\Cross};
\node at (0.9,0)  {\Cross};
\end{tikzpicture}}\ - 
 \raisebox{3pt}{
 \begin{tikzpicture}[baseline=(current  bounding  box.center)]
 \node [
        draw, color=gray, fill=gray!20,  line width=0.6pt,
        rounded rectangle,
        minimum height = .4mm,
        minimum width = 3.6em,
        rounded rectangle arc length = 180,
    ] at (0.75,0)
    {};
\draw[fill] (0, 0) circle (.4mm);
\draw[fill] (0.6, 0) circle (.4mm);
\draw[fill] (1.2, 0) circle (.4mm);
\draw[thick] (0, 0) -- (1.2, 0);
\node at (0.3,0)  {\Cross};
\node at (0.9,0)  {\Cross};
\end{tikzpicture}} \,\ = \ 0
\end{aligned}
\label{equ:3letters2}
\eeq
These relations again imply identities between triplets of dlog forms:
\beq
\begin{aligned}
R_1 &\equiv \partial[\hat B_4 \hat B_5  \hat B_{11}]   = 0\,,& R_2 &\equiv \partial[\hat B_3 \hat B_4  \hat B_{12}] = 0\,, & 
R_3 &\equiv \partial[\hat B_4 \hat B_6  \hat B_{7}] =  0\,, \\
R_4 &\equiv \partial[\hat B_1 \hat B_4  \hat B_{13}] = 0\,, &
R_5 &\equiv  \partial[\hat B_2 \hat B_5  \hat B_{7}] = 0\,, &
R_6 &\equiv  \partial[\hat B_1 \hat B_5  \hat B_{8}]  = 0\,,  \\ 
R_7 &\equiv \partial[\hat B_7 \hat B_9  \hat B_{12}] = 0\,, &
R_8 &\equiv  \partial[\hat B_1 \hat B_{10}  \hat B_{12}] = 0\,, &
R_9 &\equiv  \partial[\hat B_2 \hat B_6  \hat B_{11}]  = 0\,,  \\ 
R_{10} &\equiv \partial[\hat B_3 \hat B_6  \hat B_{9}] = 0\,, &
R_{11} &\equiv  \partial[\hat B_8 \hat B_{11}  \hat B_{13}] = 0\,, &
R_{12} &\equiv  \partial[\hat B_3 \hat B_{10}  \hat B_{13}]  = 0\,,  \\ 
\end{aligned}
\label{equ:identitiesX}
\eeq
where $[\hat B_i \hat B_j \hat B_k] \equiv \ud \log \hat B_i \wedge \ud \log \hat B_j \wedge \ud \log \hat B_k$.

\vskip 10pt
Computing $\ud^2 \psi$ with the modified graphical rules~\eqref{eq:modifiedgraph}, we find 
\beq
\begin{aligned}
\ud^2 \psi &= \e^2\Big[q_1 ({\color{Purple}R_5}+{\color{Green}R_6}) + q_2({\color{Red}R_7}-{\color{Blue}R_8}) - \tilde q_2 ({\color{brown}R_{11}}+{\color{Orange}R_{12}}) + \tilde q_1 ({\color{regal}R_9}+{\color{teal}R_{10}})\Big] +\cdots\,,
\end{aligned}
\eeq
where the ellipsis denotes non-vanishing combinations of dlog forms that are proportional to $1-c_S$, $1-c_A$ and $1-c_Ac_S$. Integrability then forces $c_S=c_A=1$.

\vskip4pt
After imposing these conditions, we see that integrability is related to the identities in~(\ref{equ:identitiesX}). Graphically, we express this as
\beq
 \begin{tikzpicture}[baseline=(current  bounding  box.center)]
 \node at (0,1.5)  {$\psi$};
  \node[gray] at (0,0)  {$Q_2$};
 \node[gray] at (-1.8,0)  {$F$};
 \node[gray] at (1.8,0)  {$\tilde F$};
  \node[gray] at (-3.6,0)  {$Q_1$};
   \node[gray] at (3.6,0)  {$Q_3$};
    \node at (-2.7,-1.5)  {$q_1$};
 \node at (-0.9,-1.5)  {$q_2$};
  \node at (0.9,-1.5)  {$\tilde q_2$};
      \node at (2.7,-1.5)  {$\tilde q_1$};
  \node at (-1.5,-0.9)  {\footnotesize \color{Red}$7$};
    \node at (-0.45,-0.9)  {\footnotesize \color{Blue}$8$};
      \node at (1.5,-0.9)  {\footnotesize \color{Orange}$12$};
          \node at (0.3,-0.9)  {\footnotesize \color{brown}$11$};
          
     \node at (-3.3,-0.9)  {\footnotesize \color{Green}$6$};
    \node at (-2.25,-0.9)  {\footnotesize \color{Purple}$5$};       
         
   \node at (3.2,-0.9)  {\footnotesize \color{teal}$10$};
    \node at (2.15,-0.9)  {\footnotesize \color{regal}$9$};  
    
 \draw[gray, thick] (-0.13,1.3) --  (-3.54,0.2); 
  \draw[gray, thick] (0.13,1.3) --  (3.54,0.23); 
 \draw[gray, thick] (-0.12,1.25) --  (-1.74,0.2);
  \draw[gray, thick] (0.12,1.25) --  (1.74,0.2);
  \draw[thick,color=Red] (-1.,-1.3) --  (-1.74,-0.15);
    \draw[thick,color=Blue] (-0.9,-1.3) --  (-0.15,-0.15);   
      \draw[thick,color=Orange] (1.,-1.3) --  (1.7,-0.2);
       \draw[thick,color=brown] (0.8,-1.3) --  (0.05,-0.15); 
       
         \draw[thick,color=Green] (-2.8,-1.3) --  (-3.54,-0.15);
    \draw[thick,color=Purple] (-2.7,-1.3) --  (-1.95,-0.15);  
          \draw[thick,color=teal] (2.7,-1.3) --  (3.44,-0.15);
    \draw[thick,color=regal] (2.6,-1.3) --  (1.85,-0.15);  
  \draw[gray, thick] (0,1.25) --  (0,0.25);  
 \end{tikzpicture}
 \nonumber
 \eeq
 Inspection of (\ref{equ:3letters2}) naturally explains that appearance to the relevant three-letter relations. Each identity corresponds to a splitting of the letters associated to the source functions $q_{1,2}$ and~$\tilde q_{1,2}$.
 
 \vskip 4pt
 The $\ud^2$ of the functions $F$ and $\tilde F$ is
\beq
 \begin{aligned}
 \ud^2 F &= \e^2 \Big[q_1\hs {\color{Purple}R_5} + q_2\hs {\color{Red}R_7} + \tilde g({\color{regal}R_9}+{\color{teal}R_{10}}) + Z ({\color{darkblue2}R_2}-{\color{reddish}R_1})\Big] = 0\,,\\
 \ud^2 \tilde F &= \e^2 \Big[\tilde q_1\hs {\color{regal}R_9} + \tilde q_2\hs {\color{brown}R_{11}} + g({\color{Purple}R_5}+{\color{Green}R_{6}}) + Z ({\color{magenta}R_4}-{\color{SkyBlue}R_3})\Big] = 0\,.
 \end{aligned}
 \label{equ:d2F}
 \eeq
Graphically, the result in (\ref{equ:d2F}) corresponds to
  \beq
 \begin{tikzpicture}[baseline=(current  bounding  box.center)]
  \node at (-3.3,1.32)  {$F$};
    \node at (3.35,1.32)  {$\tilde F$};
 \node at (-1.2,0)  {$q_2$};
   \node at (-2.6,0)  {$q_1$};
\node[gray] at (-4,0)  {$q_3$};
\node[gray] at (-5.4,0)  {$f$};
 \node at (1.2,0)  {$\tilde q_2$};
   \node at (2.6,0)  {$\tilde q_1$};
   \node[gray] at (4,0)  {$\tilde q_3$};
\node[gray] at (5.4,0)  {$f$};
 \node at (-0.03,-1.3)  {$Z$};
  \node at (-4.7,-1.3)  {$\tilde g$};
   \node at (4.7,-1.3)  {$g$};
  \node at (-0.78,-0.7)  {\footnotesize \color{darkblue2}$2$};
    \node at (0.78,-0.7)  {\footnotesize \color{magenta}$4$};

    \node at (-2.0,-0.7)  {\footnotesize \color{reddish}$1$};
    \node at (1.9,-0.7)  {\footnotesize \color{SkyBlue}$3$};    
              
    \node at (-1.75,0.7)  {\footnotesize \color{Red}$7$};
        \node at (-3.2,0.5)  {\footnotesize \color{Purple}$5$};
          \node at (1.75,0.7)  {\footnotesize \color{brown}$11$};
               \node at (3.1,0.5)  {\footnotesize \color{regal}$9$};
  \draw[thick,color=darkblue2] (-0.12,-1.1) --  (-1.,-0.2);
   \draw[thick,color=magenta] (0.12,-1.1) --  (1.,-0.2);
   
    \draw[thick,color=Green] (4.6,-1.1) --  (4.1,-0.2);
        \node at (4.2,-0.7)  {\footnotesize \color{Green}$6$};
        \draw[thick,color=Purple] (4.8,-1.1) --  (5.3,-0.2);
            \node at (5.2,-0.7)  {\footnotesize \color{Purple}$5$};
            
        \draw[thick,color=teal] (-4.6,-1.1) --  (-4.1,-0.2);
        \node at (-4.15,-0.7)  {\footnotesize \color{teal}$10$};
        \draw[thick,color=regal] (-4.8,-1.1) --  (-5.3,-0.2);
            \node at (-5.2,-0.7)  {\footnotesize \color{regal}$9$};
                    
 \draw[thick,color=Red] (-3.2,1.1) --  (-1.3,0.2);
 \draw[thick,color=brown] (3.2,1.1) --  (1.3,0.2);
  
   \draw[thick,color=Purple] (-3.3,1.1) --  (-2.75,0.2);
     \draw[thick,color=gray] (-3.4,1.1) --  (-4,0.2);
   \draw[thick,color=gray] (-3.5,1.1) --  (-5.3,0.2);

   \draw[thick,color=regal] (3.3,1.1) --  (2.75,0.2);
     \draw[thick,color=gray] (3.4,1.1) --  (4,0.2);
   \draw[thick,color=gray] (3.5,1.1) --  (5.3,0.2);
        
     \draw[thick,color=reddish] (-2.6,-0.17) --  (-0.15,-1.2);
          \draw[thick,color=SkyBlue] (2.47,-0.2) --  (0.15,-1.2);
          
 \end{tikzpicture}
 \nonumber
 \eeq
 Each labelled line again corresponds to a specific splitting of the letters associated to the source functions. Somewhat remarkably, the functions whose letters do not have a possible three letter relation with the parent function (like $f$ for $F$) identically drop out of the $\ud^2$ equation, while those that do have three letter relations appear, but proportional to the identically vanishing combination. (We saw a similar phenomenon for $\psi$, where $Q_i, F, \tilde F$ did not appear in $\ud^2\psi$ either.)
 
 \vskip 4pt
The $\ud^2$ of the $Q_i$ functions is 
 \beq
 \begin{aligned}
 \ud^2 Q_1 &= \e^2 \Big[q_1 \hs {\color{Green}R_6} - \tilde q_2\hs {\color{brown}R_{11}} - Z({\color{reddish}R_1}+ {\color{magenta}R_4})\Big] = 0\, , \\
 \ud^2 Q_2 &= \e^2 \Big[-q_2\hs {\color{Blue}R_8} -\tilde q_2\hs {\color{Orange}R_{12}} + Z({\color{darkblue2}R_2}+{\color{magenta}R_4})\Big] = 0\,,\\
  \ud^2 Q_3 &= \e^2 \Big[q_2 \hs {\color{Red}R_7} + \tilde q_1\hs {\color{teal}R_{10}} - Z({\color{darkblue2}R_2}+ {\color{SkyBlue}R_3})\Big] = 0\, , 
 \end{aligned}
 \eeq
 which has the following graphical representation
 \beq
 \begin{tikzpicture}[baseline=(current  bounding  box.center)]
 \node at (0,1.3)  {$Q_2$};
 \node at (-0.97,0)  {$\tilde q_2$};
 \node at (1,0)  {$q_2$};
  \node at (-3,0)  {$q_1$};
   \node at (3.05,0)  {$\tilde q_1$};
    \node at (-2,1.3)  {$Q_1$};
        \node at (2,1.3)  {$Q_2$};
 \node at (-0.03,-1.25)  {$Z$};
  \node at (-0.7,-0.7)  {\footnotesize \color{magenta}$4$};
    \node at (0.7,-0.7)  {\footnotesize \color{darkblue2}$2$};
      \node at (-0.7,0.7)  {\footnotesize \color{Blue}$8$};
    \node at (0.7,0.7)  {\footnotesize \color{Orange}$12$};
  \node at (2.7,0.7)  {\footnotesize \color{teal}$10$};
        \node at (1.3,0.7)  {\footnotesize \color{Red}$7$};
    \node at (-2.15,-0.7)  {\footnotesize \color{reddish}$1$};
    \node at (2.15,-0.7)  {\footnotesize \color{SkyBlue}$3$};     
     
              \node at (-2.7,0.7)  {\footnotesize \color{Green}$6$};
    \node at (-1.3,0.7)  {\footnotesize \color{brown}$11$};
    
 \draw[thick,color=Blue] (-0.12,1.1) --  (-0.88,0.15);
 \draw[thick,color=Orange] (0.12,1.1) --  (0.88,0.15);
  \draw[thick,color=Red] (1.88,1.1) --  (1.12,0.15);
 \draw[thick,color=teal] (2.12,1.1) --  (2.88,0.15);
  \draw[thick,color=magenta] (-0.12,-1.1) --  (-0.88,-0.2);
   \draw[thick,color=darkblue2] (0.12,-1.1) --  (0.84,-0.2);
    \draw[thick,color=Green] (-2.12,1.1) --  (-2.88,0.15);
 \draw[thick,color=brown] (-1.88,1.1) --  (-1.12,0.15);
 
     \draw[thick,color=reddish] (-2.88,-0.15) --  (-0.15,-1.2);
          \draw[thick,color=SkyBlue] (2.88,-0.15) --  (0.15,-1.2);
 \end{tikzpicture}
 \nonumber
 \eeq
The $\ud^2$ of the remaining functions is 
 \begin{align}
 \begin{split}
 \ud^2 q_1 &= -2\e^2 Z \hs {\color{reddish}R_1} \, , \\
 \ud^2 q_2 &= 2\e^2 Z\hs {\color{darkblue2}R_2} \,, \\
 \ud^2 q_3 &= \e^2 \big[q_2\hs {\color{Red}R_7} + \tilde g\hs {\color{teal}R_{10}} - Z({\color{darkblue2}R_2}+{\color{SkyBlue}R_3})\big]\,,
 \end{split}
\begin{split}
  \ud^2 \tilde q_1 &= -2\e^2 Z \hs {\color{SkyBlue}R_3} \, , \\
   \ud^2 \tilde q_2 &= 2\e^2 Z\hs {\color{magenta}R_4}\,, \\
 \ud^2 \tilde q_3 &= \e^2 \big[-\tilde q_2\hs {\color{brown}R_{11}} + g\hs {\color{Green}R_{6}} -Z({\color{reddish}R_1}+{\color{magenta}R_4})\big]\,,
 \end{split}\\[6pt]
 \begin{split}
 \ud^2 f &=\e^2 \big[ g \hs {\color{Purple}R_5} + \tilde g\hs  {\color{regal}R_{9}} - Z\hs ({\color{reddish}R_1}+{\color{SkyBlue}R_3})\big]\,,
\end{split}\\[6pt]
\begin{split}
 \ud^2 g &= -2\e^2  Z \hs {\color{reddish}R_1} \, , \\
 \ud^2 \tilde g &= -2\e^2 Z\hs {\color{SkyBlue}R_3} \,, \\
 \ud^2 Z &= 0\,,
 \end{split}
 \end{align}
 which graphically corresponds to
  \beq
 \begin{tikzpicture}[baseline=(current  bounding  box.center)]
 \node at (0,1.3)  {$f$};
 \node at (-1,0)  {$\tilde g$};
 \node at (1,0)  {$g$};
  \node at (-3,0)  {$q_2$};
   \node at (3.05,0)  {$\tilde q_2$};
    \node at (-2,1.3)  {$q_3$};
        \node at (2,1.3)  {$\tilde q_3$};
 \node at (-0.03,-1.25)  {$Z$};
  \node at (-0.7,-0.7)  {\footnotesize \color{SkyBlue}$3$};
    \node at (0.7,-0.7)  {\footnotesize \color{reddish}$1$};
      \node at (-0.7,0.7)  {\footnotesize \color{regal}$9$};
    \node at (0.7,0.7)  {\footnotesize \color{Purple}$5$};
  \node at (2.7,0.7)  {\footnotesize \color{brown}$11$};
        \node at (1.3,0.7)  {\footnotesize \color{Green}$6$};
    \node at (-2.15,-0.7)  {\footnotesize \color{darkblue2}$2$};
    \node at (2.15,-0.7)  {\footnotesize \color{magenta}$4$};     
     
              \node at (-2.7,0.7)  {\footnotesize \color{Red}$7$};
    \node at (-1.3,0.7)  {\footnotesize \color{teal}$10$};
    
 \draw[thick,color=regal] (-0.12,1.1) --  (-0.88,0.15);
 \draw[thick,color=Purple] (0.12,1.1) --  (0.88,0.15);
  \draw[thick,color=Green] (1.88,1.1) --  (1.12,0.15);
 \draw[thick,color=brown] (2.12,1.1) --  (2.88,0.15);
  \draw[thick,color=SkyBlue] (-0.12,-1.1) --  (-0.88,-0.2);
   \draw[thick,color=reddish] (0.12,-1.1) --  (0.84,-0.2);
    \draw[thick,color=Red] (-2.12,1.1) --  (-2.88,0.15);
 \draw[thick,color=teal] (-1.88,1.1) --  (-1.12,0.15);
 
     \draw[thick,color=darkblue2] (-2.88,-0.15) --  (-0.15,-1.2);
          \draw[thick,color=magenta] (2.88,-0.15) --  (0.15,-1.2);
 \end{tikzpicture}
 \nonumber
 \eeq
This again displays the same pattern.

\subsection{Further Examples}

We have demonstrated, for the case of the two- and three-site chains, that  $\ud^2$ of any function is proportional to vanishing three-letter identities. Though we have not attempted to prove this fact abstractly, the integrability of the equations for more complicated graphs is analyzed in the {\sc Mathematica} notebooks provided at \href{https://github.com/haydenhylee/kinematic-flow}{\faGithub}. In all cases, we find the same pattern.

\newpage
\section{Space of Functions}
\label{app:functions}

In this appendix, we elucidate some properties of the solutions to the kinematic flow equations. We describe the boundary conditions that fix all the master integrals, and show that the complexity of the solutions grows linearly with the size of the graph (although the size of the basis grows exponentially).  We also give a bulk interpretation of all master integrals.  It turns out that each master integral (in a suitable basis) computes a specific ``timeline of events", and that the wavefunction is given by the superposition of all these timelines.

\subsection{Boundary Conditions}

 To solve the system of differential equations, we must impose suitable boundary conditions.
One strategy---which we used in Section~\ref{sec:TwoSite}---is use the explicit form of the master integrals and evaluate them at specific kinematics.
In this section, however, we present a more physically satisfying way of fixing boundary conditions. 
First, we demand the absence of any ``folded singularities".\footnote{The folded limit corresponds to the limit where any sum of external energies becomes equal to an internal energy. It is called folded because in this limit some sides of the kinematic polygon in momentum space ``fold" onto an internal line.} This provides a nontrivial boundary condition for many master integrals. 
Second, we impose consistent factorization of the wavefunction in a {\it single} kinematic limit, which fixes it in terms of lower-point data. This constraint is strong enough to fix all remaining integration constants. Let us elaborate on these points:

\begin{itemize}
\item Folded singularities are absent in all basis integrals. This is evident from the form of the integrand, as folded combinations of kinematic variables 
never appear in denominators. Removing them in the answers fixes many coefficients at once. The precise counting is not entirely obvious, as we are dealing with functions in many variables. Sometimes removing a folded singularity in one variable leaves a whole functions worth of freedom in other variables, and allows us to set many coefficients to zero. Some folded singularities (like the ones that depend on sums of external energies, rather than on a single external energy) are automatically absent once we remove all folded singularities in each individual energy.

\item After all folded singularities are removed, one is left with fewer constants to fix than there are partial energy singularities. This means that we can fix all of them by taking various limits of partial energies. In fact, in the examples we will discuss, a single factorization limit  
fixes all remaining freedom. 
Factorization at the other partial energy singularities as well as the flat-space limit are then consequences of the structure of the equations.
\end{itemize}
\noindent
In the following, we illustrate this procedure in a few examples.

\paragraph{Two-site chain:}
In Section~\ref{sec:TwoSite}, 
we analyzed the two-site chain and showed that the answer for general $\e$ can be written in terms of hypergeometric functions. To understand the structure of the equations, it is instructive to take the flat-space limit, $\e\to-1$, where all master integrals evaluate to rational functions.\footnote{Strictly speaking, the integrals are doubly logarithmically divergent, scaling as $(\e+1)^{-2}$ in the flat-space limit, but since the differential equations are completely regular in this limit,  we can solve them and fix the normalization on physical grounds. Alternatively, one can 
 absorb this divergence in a coupling constant of the diagram.}

\vskip 4pt
The system of differential equations for $\e=-1$ takes the form
\begin{align}
&\ud Z = -2 Z\, \ud \log(X_1+X_2)\,,\nonumber\\[4pt]
\begin{split}
\ud F &= -\, \big[ F\,\ud \log(X_1-Y)  + (F-  Z)\, \ud \log(X_2+Y) + Z\, \ud \log(X_1+X_2) \big] \,,  \\[4pt]
\ud \tilde F &=  -\, \big[ \tilde F\,\ud \log(X_2-Y)  + (\tilde F-  Z)\, \ud \log(X_1+Y) + Z\, \ud \log(X_1+X_2) \big]\,,  \\[4pt]
\end{split} \label{equ:2SiteEqnsapp}\\
&\ud \psi = - \, \big[(\psi - F)\, \log(X_1+Y) + F\, \ud \log(X_1-Y) +  (\psi - \tilde F)\, \ud \log(X_2+Y) +\tilde F\,\ud \log(X_2-Y)  \big]\,.
\nonumber
\end{align}
Solving these equations starting with $Z$, we find
\beq
\begin{aligned}\label{eq:flatsol}
Z&=\frac{c_{Z}}{(X_1+X_2)^2}\,,\\
F&=\frac{c_{F}}{(X_1-Y)(X_2+Y)}-\frac{X_1-Y}{2(X_2+Y)}\,Z\,, \\
\tilde F&=\frac{c_{\tilde F}}{(X_2-Y)(X_1+Y)}-\frac{X_2-Y}{2(X_1+Y)}\,Z\,,  \\
\psi& =\frac{c_{\psi}}{(X_1+Y)(X_2+Y)}+(F+\tilde F-Z)\,,
\end{aligned}
\eeq
with four constants of integration $c_Z$, $c_F$, $c_{\tilde F}$, $c_\psi$.
Demanding the absence of folded singularities at $X_{1,2} = Y$ sets $c_{F}=c_{\tilde F}=0$.  Next, we take the factorization limit $X_1 \to -Y$ to obtain
\beq
\lim_{X_1 \to -Y}\psi = \frac{1}{X_1+Y}\left[\frac{c_{\psi}}{X_2+Y} -\frac{c_{Z}}{2(X_2-Y)} \right] .
\eeq
The coefficient in front of each term is determined by the three-point function squared. Normalizing it to unity sets $c_Z=2$. 
Choosing $c_\psi = + 1$, the solution for $\psi$ is proportional to the wavefunction coefficient (\ref{eq:flatWF}):
\beq
\psi=-\frac{2Y}{(X_1+X_2)(X_1+Y)(X_2+Y)} \, .
\eeq
If, instead, we set $c_\psi = -1$, then $\psi$ becomes the corresponding in-in correlator. In this way, we see that the wavefunction and correlator solve the same differential equations, the difference between them just being which boundary conditions we specify in
 the factorization limit.

\vskip 4pt
We can also phrase this discussion slightly more abstractly, in a way that is applicable to any~$\e$. Since the equations are first order in the $X$ variables, the solutions are completely determined by the value of the function at a given point in the kinematic space. However, it is often more convenient to inspect the homogeneous solutions to the equations and try to fix their free parameters by utilizing the singular limits.  For generic $\e$, the homogeneous solutions are
\be
\begin{aligned}
Z &=c_{Z} (X_1+X_2)^{2\e}\,,\\
F_h&=c_{F} (X_1-Y)^\e(X_2+Y)^\e\,, \\
\tilde F_h &=c_{\tilde F} (X_1+Y)^\e(X_2-Y)^\e\,,\\
\psi_h &=c_{\psi} (X_1+Y)^\e(X_2+Y)^\e\,.
\end{aligned}
\ee
We can see that the free parts of~\eqref{eq:flatsol} precisely agree with these solutions in the limit $\e\to-1$. Since the homogeneous solutions for $F$ and $\tilde F$ contain folded singularities, requiring the absence of these singularities (which otherwise would propagate to $\psi$) fixes $c_{F} = c_{\tilde F} = 0$. Factorization is more subtle, because we need to know the full solution $\psi$, and the singularities come from both the $\ud\log$-forms appearing in the equations and from the solutions themselves. 
Although the solution $Z$ itself is not singular in the partial energy limit, it appears in the $\psi$ equation as a coefficient of $\ud\log$-forms that have these singularities. We can see this explicitly in the flat-space solution~\eqref{eq:flatsol}. This shows that the parameter $c_Z$ is relevant in the factorization limit.

\vskip 4pt
As for the ``space of functions" in this flat-space case, all of the functions~\eqref{eq:flatsol} are of course rational, but they still have an organization in order of increasing complexity. In particular, the source $Z$ has only one singularity, while $F,\tilde F$ have singularities at two locations. These objects then feed into $\psi$, which itself is singular at three locations. At finite $\e$, the analogue is that both power-law functions and hypergeometric functions appear.  Power laws are the functions that appear for the single-site case, i.e.~contact diagrams. Hypergeometric functions solve a first-order differential equation with a power law as an inhomogeneous source term. This hierarchical structure persists in more complicated cases where more complicated hypergeometric-type functions solve first-order equations with hypergeometric sources, and so on.

\vskip 4pt
It is worth remarking that the de Sitter limit ($\e=0$) is somewhat special. Since the diagonal terms of the matrix $A$ vanish in that limit, boundary conditions should be fixed in a slightly different way. There are still four integration constants $c_\psi, c_F, c_{\tilde F}, c_Z$. The absence of folded singularities fixes $ c_F$ and $c_{\tilde F}$, but this time factorization only fixes $c_Z$.  The constant $c_\psi$ must be fixed by another method, as now the homogeneous solution of the $\psi$ equation is just a constant, so there is no singularity that depends on its value.
What we have been neglecting is that the wavefunction has a {\it zero} at $Y=0$. This is evident from the integral representation\footnote{From bulk perturbation theory, this zero is obvious from a cancellation of the three terms in the bulk-to-bulk propagator in the soft limit, $2k_I G\xrightarrow{\ k_I\to 0\ } \theta(\eta_1-\eta_2)+\theta(\eta_2-\eta_1)-1=0$.}
\beq
\psi = -\int_0^\infty \ud x_1\int_0^\infty \ud x_2  \, (x_1 x_2)^\e\frac{2 Y }{(X_1 + x_1 +Y)(X_2+x_2+Y)(X_1+X_2+x_1+x_2)}\,.
\eeq
Strictly speaking, we should divide by $Y$ to obtain the physical wavefunction that derives from a bulk time integral, but then the integrand ceases to be a canonical form.
For generic $\e$, the vanishing of $\psi$ at $Y\to 0$ is a consequence of nontrivial hypergeometric identities, and so follows from the other boundary conditions; in the de Sitter limit, on the other hand,  we impose this as a separate condition and it then fixes the constant $c_\psi$.

\paragraph{Three-site chain:} Let us further illustrate these considerations for the case of the three-site chain. In that case, we find (again in flat space):
\begin{itemize}
\item There are 16 integration constants,  12 of which are fixed by demanding the absence of folded singularities as $\{X_1-Y,X_2-Y-Y',X_2\mp Y\pm Y',X_3-Y', X_1+X_2-Y',X_2+X_3-Y\}\to 0$. The last two folded singularities (involving sums of external energies) are automatically absent after removing the folded singularities involving a single $X_i$. The counting is not obvious, because we can set many constants to zero with a few folded limits, since the resulting numerators are functions of the other kinematic variables.
\item A single factorization limit---in either the partial energies corresponding to the end of the chain or in the middle---fixes the remaining constants of integrations.
\item The interpretation of the 16 basis functions is similar to the case of the two-site chain, but now the source functions have higher-order poles not only in $(X_1+X_2+X_3)$, but also in $(X_1+X_2+Y')$ and $(X_2+X_3+Y)$. The precise map is given in the next section.
\end{itemize}

\vskip 4pt
At finite $\e$, despite the appearance of multiple letters in sources in the differential equations, we can explicitly see that the complexity of functions only grows linearly in the number of vertices. Namely, beyond power laws and hypergeometric functions, we find a two-variable generation of the hypergeometric function, which satisfies a first-order differential equations with a hypergeometric functions as a source. 

\paragraph{Four-site chain:}
For the four-site chain, things work essentially the same, only the numbers involved change. 
There are 64 integration constants, 56 of which are fixed by demanding the absence of folded singularities. Moreover, a single factorization limit, in any of the partial energies, fixes the remaining 8 constants of integration. 
Notice that the number of leftover constants to fix is always equal to or smaller than the number of factorization poles, excluding the total energy.

\vskip 4pt
At finite $\e$, yet another class of generalized hypergeometric functions will appear, whose differentials are given by linear combinations of the three-site solutions. It is interesting that despite the exponential growth of the number of master integrals with the number of sites, the complexity of the functions only grows linearly. 
While this is somewhat non-obvious from the differential equations themselves, it is manifest from the time integral perspective. 

\subsection{Bulk Interpretation}

Somewhat remarkably, all of the master integrals appearing in the differential system have a physical interpretation. This can be contrasted with other applications of differential equations methods, where typically one is only interested in a particular subset of master integrals, while the remaining functions are not necessarily physical objects.

\paragraph{Two-site chain:}
Let us return to the two-site case, where the master integrals were the wavefunction $\psi$ and the three auxiliary functions $F,\tilde F$ and $Z$.  
The explicit solutions~\eqref{eq:flatsol} suggest that we can perform a change of basis that gives functions with simple bulk interpretations: 
\begin{align}
	\raisebox{-10pt}{\includegraphics[scale=1]{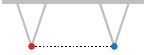}} &\equiv \psi-F-\tilde F+Z = \frac{1}{(X_1+Y)(X_2+Y)} \label{equ:psi-bulk}\,,\\[-1pt]
	\cline{1-2}\nonumber\\[-18pt]
	\raisebox{-10pt}{\includegraphics[scale=1]{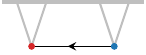}} &\equiv F-\frac{1}{2}Z = -\frac{1}{(X_2+Y)(X_1+X_2)}\,,\\
	\raisebox{-10pt}{\includegraphics[scale=1]{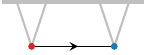}} &\equiv \tilde F-\frac{1}{2} Z=-\frac{1}{(X_1+Y)(X_1+X_2)}\,,\\[-1pt]
		\cline{1-2}\nonumber\\[-18pt]
	\raisebox{-10pt}{\includegraphics[scale=1]{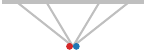}} &\equiv	Z = \frac{2}{(X_1+X_2)^2}\,, \label{equ:Z-bulk}
\end{align}
where we have also shown the resulting expressions for $\e=-1$ on the right-hand side.

\vskip 4pt
Recall that the bulk-to-bulk propagator~\eqref{eq:Gbb} has three components: the left-time-ordered piece $\theta(\eta_1-\eta_2)$, the right-time-ordered piece $\theta(\eta_1-\eta_2)$, and the non-time-ordered piece.  
In the above Feynman diagrams, the direction of the arrows indicates the flow of time, whereas the dotted line represents the absence of time ordering. 
In the rotated basis, the function (\ref{equ:psi-bulk})  corresponds to the homogeneous part of $\psi$, which comes from the non-time-ordered piece.
The interpretation of the other functions is also interesting. Both $F$ and $\tilde F$ have a piece proportional to $Z$. After removing this piece, we are left with functions that arise from the piece of the bulk-to-bulk propagator with a specific time ordering.
Finally, $Z$ does not have a totally obvious interpretation in terms of a bulk diagram due to the presence of the double pole. 
Mechanically, it comes from inserting the dimensionless factor $\sqrt{\eta_1 \eta_2}\,\delta(\eta_1-\eta_2)$ in the time integrals, which can be interpreted as ``collapsing"  the vertices, as depicted in~(\ref{equ:Z-bulk}).
Note that this is also the source function in the second-order inhomogeneous equation satisfied by the wavefunction coefficient.

\vskip 4pt
Using the integral representations, this map can also be demonstrated for general $\e$. However, even after changing basis, matching to bulk perturbation theory requires some integration by parts. For example, consider the combination of master integrals
\beq
F-\frac{1}{2}Z=\int \ud x_1 \ud x_2\, \frac{(x_1 x_2)^{\e-1}}{(x_1+x_2+X_1+X_2)}\left[\frac{x_2(X_1-Y)}{	x_2+X_2+Y}-\frac{X_1+X_2}{2}\right] .
\eeq
This should match
 the (Schwinger-parametrized) time integral for the relevant time ordering: 
\begin{align}
F_{\rm bulk}&=\int \ud \eta_1 \ud \eta_2 (\eta_1 \eta_2)^{\e-1} \theta(\eta_1-\eta_2) e^{i X_1 \eta_1}e^{i X_2 \eta_2}e^{-i Y(\eta_1-\eta_2)} \nonumber
 \\&= \int\ud x_1 \ud x_2\,  \frac{(x_1 x_2)^\e}{(x_2+X_2+Y)(x_1+x_2+X_1+X_2)} \, .
\end{align}
It is straightforward, but tedious, to show that these expressions only differ by a total derivative, 
\begin{align}
F-\frac{1}{2}Z + F_{\rm bulk}&=\int \ud x_1 \ud x_2\, \frac{\partial}{\partial x_2}\frac{(x_1 x_2)^\e}{2\e}\left[\frac{1}{x_1+x_2+X_1+X_2}-\frac{X_2+Y}{x_2+X_2+Y}  \right] \\
&\quad -\frac{\partial}{\partial x_1}\frac{(x_1 x_2)^\e}{2\e}\left[\frac{1}{x_1+x_2+X_1+X_2} + \frac{X_2+Y}{\e(x_2+X_2+Y)^2}-\frac{1}{x_2+X_2+Y} \right] .\nonumber
\end{align}

\paragraph{Three-site chain:}
For general tree graphs, we can understand the size of the basis from the bulk as follows. For each propagator, we have four contributing components: two time-ordered pieces, one non-time-ordered piece, and a delta function insertion that collapses the propagator. 
For multiple propagators, we basically take all possible combinations of these components.
This gives $4^e$ for the size of the basis, with $e$ the number of propagators, which indeed matches our counting in terms of the
number of (disconnected) tubings of a graph.

\vskip 4pt
In flat space, it is straightforward to compute the master integrals for the three-site chain explicitly and find the rotated basis. 
It is easier to work out the change of basis from the ``bottom up," starting from the master integrals with the most twist factors and work towards the original wavefunction. Explicitly, we obtain 
\begin{align*}
	\raisebox{-10pt}{\includegraphics[scale=1]{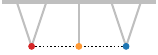}} &\equiv \psi -(F+\tilde F+{\textstyle\sum_i} Q_i)+(f+{\textstyle\sum_i} (q_i+ \tilde q_i)) - (g+\tilde g +Z)\nonumber =\frac{1}{4X_{1}^+X_{2}^+X_{3}^+}\,,\\[-1pt]
  \cline{1-2}\nonumber\\[-18pt]
	\raisebox{-10pt}{\includegraphics[scale=1]{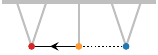}} &\equiv F - f +\frac{1}{2}g+\tilde g -\frac{1}{2}q_1-q_2-q_3+Z=-\frac{1}{4X_{12}^+X_{2}^+X_{3}^+}\,,\\
	\raisebox{-10pt}{\includegraphics[scale=1]{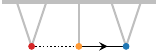}} &\equiv \tilde F-f+\frac{1}{2}\tilde g+g-\frac{1}{2}\tilde q_1-\tilde q_2-\tilde q_3+Z=-\frac{1}{4X_{1}^+X_{23}^+X_{2}^+}\,,\\
	\raisebox{-10pt}{\includegraphics[scale=1]{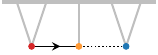}} &\equiv Q_1-\frac{1}{2}q_1-\tilde q_3+\frac{1}{2}g=-\frac{1}{4X_{1}^+X_{12}^+X_{3}^+}\,,\\
	\raisebox{-10pt}{\includegraphics[scale=1]{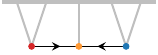}} &\equiv Q_2-\frac{1}{2}(q_2+\tilde q_2)+\frac{1}{3}Z=\frac{1}{4X_{123}X_{1}^+X_{3}^+}\,,\\
	\raisebox{-10pt}{\includegraphics[scale=1]{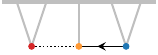}} &\equiv Q_3-\frac{1}{2}\tilde q_1-q_3+\frac{1}{2}\tilde g=-\frac{1}{4X_{1}^+X_{23}^+X_{3}^+}\,,\\[-1pt]
  \cline{1-2}\nonumber\\[-18pt]
	\raisebox{-10pt}{\includegraphics[scale=1]{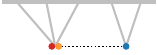}} &\equiv q_1-g=\frac{1}{2(X_{12}^+)^2X_{3}^+}\,,\\
	\raisebox{-10pt}{\includegraphics[scale=1]{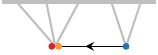}} &\equiv q_2-\frac{1}{3}Z=-\frac{1}{2X_{123}^2X_{3}^+}\,,\\
	\raisebox{-10pt}{\includegraphics[scale=1]{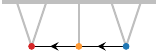}} &\equiv q_3+\frac{1}{2}(q_2-\tilde g)-\frac{1}{3}Z=\frac{1}{4X_{123}X_{23}^+X_{3}^+}\,,\\
	\raisebox{-10pt}{\includegraphics[scale=1]{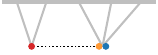}} &\equiv \tilde q_1-\tilde g=\frac{1}{2X_{1}^+(X_{23}^+)^2}\,,\\
	\raisebox{-10pt}{\includegraphics[scale=1]{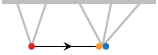}} &\equiv \tilde q_2-\frac{1}{3}Z=-\frac{1}{2X_{123}^2X_{1}^+}\,,\\
	\raisebox{-10pt}{\includegraphics[scale=1]{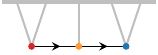}} &\equiv \tilde q_3+\frac{1}{2}(\tilde q_2- g)-\frac{1}{3}Z=\frac{1}{4X_{123}X_{1+}X_{12+}}\,,\\
	\raisebox{-10pt}{\includegraphics[scale=1]{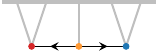}} &=f - \frac{1}{2}(g+\tilde g)-\frac{2}{3}Z=\frac{1}{4X_{123}X_{23}^+X_{2}^+}+\frac{1}{4X_{123}X_{12}^+X_{2}^+}\,,\\[-1pt]
  \cline{1-2}\nonumber\\[-18pt]
	\raisebox{-10pt}{\includegraphics[scale=1]{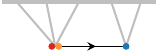}} &\equiv g+\frac{1}{3}Z=-\frac{1}{2X_{123}(X_{12}^+)^2}-\frac{1}{2X_{123}^2X_{12}^+}\,,\\
	\raisebox{-10pt}{\includegraphics[scale=1]{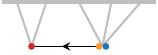}} &\equiv \tilde g+\frac{1}{3}Z=-\frac{1}{2X_{123}(X_{23}^+)^2}-\frac{1}{2X_{123}^2X_{23}^+}\,,\\
	\raisebox{-10pt}{\includegraphics[scale=1]{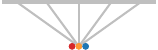}} &\equiv Z=\frac{3}{2X_{123}^3}\,,
\end{align*}
where we used the notation defined in \eqref{threesiteX} for the energy variables.
The boundary-to-bulk map above is valid at any $\e$, while the explicit results involving rational functions are for $\e=-1$.

\vskip 4pt
\newpage
It is also interesting to apply the basis rotation, $\vec{I} \to \vec{I}^{\hskip 2pt \prime}$, for generic $\e$.
The $\tilde A$-matrix then takes the form
 \be
\tilde A' =  \left[\renewcommand{\arraystretch}{0.7} {\footnotesize{\begin{array}{cccccccccccccccc}
 \OC & \phantom{0} & \phantom{0} & \phantom{0} & \phantom{0} & \phantom{0} & \phantom{0} & \phantom{0} & \GC & \GC & \phantom{0} & \phantom{0} & \phantom{0} & \phantom{0} & \phantom{0} & \phantom{0} \\
 \phantom{0} & \OC & \phantom{0} & \phantom{0} & \phantom{0} & \phantom{0} & \phantom{0} & \phantom{0} & \phantom{0} & \phantom{0} & \GC & \GC & \phantom{0} & \phantom{0} & \phantom{0} & \phantom{0} \\
 \phantom{0} & \phantom{0} & \OC & \phantom{0} & \phantom{0} & \phantom{0} & \phantom{0} & \phantom{0} & \GC & \phantom{0} & \phantom{0} & \GC & \phantom{0} & \phantom{0} & \phantom{0} & \phantom{0} \\
 \phantom{0} & \phantom{0} & \phantom{0} & \OC & \phantom{0} & \phantom{0} & \phantom{0} & \phantom{0} & \phantom{0} & \GC & \GC & \phantom{0} & \phantom{0} & \phantom{0} & \phantom{0} & \phantom{0} \\
 \phantom{0} & \phantom{0} & \phantom{0} & \phantom{0} & \GC  & \phantom{0} & \phantom{0} & \phantom{0} & \phantom{0} & \phantom{0} & \phantom{0} & \phantom{0} & \phantom{0} & \BC & \phantom{0} & \phantom{0} \\
 \phantom{0} & \phantom{0} & \phantom{0} & \phantom{0} & \phantom{0} & \GC & \phantom{0} & \phantom{0} & \phantom{0} & \phantom{0} & \phantom{0} & \phantom{0} & \phantom{0} & \phantom{0} & \BC & \phantom{0} \\
 \phantom{0} & \phantom{0} & \phantom{0} & \phantom{0} & \phantom{0} & \phantom{0} & \GC & \phantom{0} & \phantom{0} & \phantom{0} & \phantom{0} & \phantom{0} & \phantom{0} & \BC & \phantom{0} & \phantom{0} \\
 \phantom{0} & \phantom{0} & \phantom{0} & \phantom{0} & \phantom{0} & \phantom{0} & \phantom{0} & \GC & \phantom{0} & \phantom{0} & \phantom{0} & \phantom{0} & \phantom{0} & \phantom{0} & \BC & \phantom{0} \\
 \phantom{0} & \phantom{0} & \phantom{0} & \phantom{0} & \phantom{0} & \phantom{0} & \phantom{0} & \phantom{0} & \GC & \phantom{0} & \phantom{0} & \phantom{0} & \phantom{0} & \phantom{0} & \phantom{0} & \BC \\
 \phantom{0} & \phantom{0} & \phantom{0} & \phantom{0} & \phantom{0} & \phantom{0} & \phantom{0} & \phantom{0} & \phantom{0} & \GC & \phantom{0} & \phantom{0} & \phantom{0} & \phantom{0} & \phantom{0} & \BC \\
 \phantom{0} & \phantom{0} & \phantom{0} & \phantom{0} & \phantom{0} & \phantom{0} & \phantom{0} & \phantom{0} & \phantom{0} & \phantom{0} & \GC & \phantom{0} & \phantom{0} & \phantom{0} & \phantom{0} & \BC \\
 \phantom{0} & \phantom{0} & \phantom{0} & \phantom{0} & \phantom{0} & \phantom{0} & \phantom{0} & \phantom{0} & \phantom{0} & \phantom{0} & \phantom{0} & \GC & \phantom{0} & \phantom{0} & \phantom{0} & \BC \\
 \phantom{0} & \phantom{0} & \phantom{0} & \phantom{0} & \phantom{0} & \phantom{0} & \phantom{0} & \phantom{0} & \phantom{0} & \phantom{0} & \phantom{0} & \phantom{0} & \BC & \phantom{0} & \phantom{0} & \phantom{0} \\
 \phantom{0} & \phantom{0} & \phantom{0} & \phantom{0} & \phantom{0} & \phantom{0} & \phantom{0} & \phantom{0} & \phantom{0} & \phantom{0} & \phantom{0} & \phantom{0} & \phantom{0} & \BC & \phantom{0} & \phantom{0} \\
 \phantom{0} & \phantom{0} & \phantom{0} & \phantom{0} & \phantom{0} & \phantom{0} & \phantom{0} & \phantom{0} & \phantom{0} & \phantom{0} & \phantom{0} & \phantom{0} & \phantom{0} & \phantom{0} & \BC & \phantom{0} \\
 \phantom{0} & \phantom{0} & \phantom{0} & \phantom{0} & \phantom{0} & \phantom{0} & \phantom{0} & \phantom{0} & \phantom{0} & \phantom{0} & \phantom{0} & \phantom{0} & \phantom{0} & \phantom{0} & \phantom{0} & \BC \\
\end{array}}}  \begin{matrix} \phantom{\hskip 0pt} \\[110pt] \phantom{\hskip 0pt}\end{matrix} \right] ,
\ee
where the basis vector is arranged as
\begin{align}
	\vec{I}^{\hskip 2pt \prime} = ( \orange{Q_2'},\,\orange{f'},\,\orange{q_3'},\,\orange{\tilde{q}_3^{\hskip 1pt \prime}},\,\green{F'},\,\green{\tilde{F}'},\,\green{Q_1'},\,\green{Q_3'},\,\green{q_2'},\,\green{\tilde{q}_2^{\hskip 1pt \prime}},\,\green{g'},\,\green{\tilde{g}'},\,\blue{\psi'},\,\blue{q_1'},\,\blue{\tilde{q}_1^{\hskip 1pt \prime}},\,\blue{Z'} )\,.
\end{align}
We see that, in this basis, the matrix becomes very sparse and exhibits a hierarchical structure. 
These characteristics are advantageous when it comes to solving the system of equations.

\vskip 4pt
The differentials of the basis functions can be represented as
\begin{align}
  \ud Q_2' &= \e \Big[ \hskip 3pt(Q_2'-\tfrac{1}{2}q_2')\, \LLap \hskip -15pt && + (Q_2'+\tfrac{1}{2}(q_2'+\tilde q_2^{\hskip 1pt \prime})) \LLbmm \hskip -10pt && + \hskip 1pt (Q_2'-\tfrac{1}{2}\tilde q_2^{\hskip 1pt \prime}) \LLcp\Big]\nn\\
  \ud f' &= \e \Big[ \hskip 8pt(f'+\tfrac{1}{2}g')\, \LLam \hskip -15pt && + \hskip 7pt(f'-\tfrac{1}{2}(g'+\tilde g')) \LLbpp \hskip -10pt && + \hskip 6.3pt(f'+\tfrac{1}{2}\tilde g') \LLcm\Big]\nn\\
   \ud q_3' &= \e \Big[ \hskip 6.5pt(q_3'+\tfrac{1}{2}q_2')\, \LLam \hskip -15pt && +  \hskip 5pt(q_3'+\tfrac{1}{2}(\tilde g'-q_2'))\LLbpm \hskip -10pt && +\hskip 6pt(q_3'-\tfrac{1}{2}\tilde g')\LLcp\Big]\nn\\
   \cline{1-6}
   \ud F' &= \e \Big[ \hskip 5pt(F'+\tfrac{1}{2}q_1')\, \LLam \hskip -15pt && +\hskip 34pt (F'-\tfrac{1}{2}q_1') \LLbpp \hskip -10pt && + \hskip 41.3ptF' \LLcp\Big]\nn\\
   \ud Q_1' &= \e \Big[\hskip 3pt (Q_1'-\tfrac{1}{2}q_1')\, \LLap \hskip -15pt && +\hskip 32pt (Q_1'+\tfrac{1}{2}q_1') \LLbmp \hskip -10pt && + \hskip 39.3pt Q_1' \LLcp\Big]\nn\\
   \ud q_2' &= \e \Big[ (2q_2'+\tfrac{2}{3}Z')\, \LLabmr \hskip -10pt && \hskip -10pt && +\hskip 3.5pt (q_2'-\tfrac{2}{3}Z') \LLcp\Big]\nn\\
   \ud g' &= \e \Big[ (2g_2'-\tfrac{2}{3}Z')\, \LLabpr \hskip -15pt && \hskip -10pt && + \hskip 4.7pt (g'+\tfrac{2}{3}Z') \LLcm\Big]\nn\\
   \cline{1-6}
	\ud \psi' &= \e \Big[\hskip 5pt \psi'\, \LLap +\, \psi'\,\LLbpp+ \, \psi'\, \LLcp\Big] \hskip -200pt &&  \hskip -200pt && \nn\\
	\ud q_1' &= \e \Big[ 2q_1'\, \LLabpr  \hskip 84.3pt +\, q_1'\, \LLcp\Big]\hskip -200pt && \hskip -200pt && \label{equ:3ptbulkbasis}\\[2pt]
	\ud Z' &= \hskip 6pt 3\e\,Z'\,\raisebox{-1.5pt}{\includegraphics[scale=1]{Figures/Tubings/three/Z/threeZgr}}\nn
\end{align}
where the expressions for $\tilde q_3^{\hskip 1pt \prime},\tilde F',Q_3^{\prime},\tilde q_2^{\hskip 1pt \prime},\tilde g',\tilde q_1^{\hskip 1pt \prime}$ are given by symmetry.  
We see that these equations exhibit an interesting pattern, with the tubings associated to the basis functions simply become activated (but do not grow) under the differential. 
Understanding the ``flow'' in this basis is an interesting problem that we leave to future work.

\newpage

\newpage
\section{Relations to Symbology}
\label{app:symbology}

In this work, we have described how to construct the system of differential equations satisfied by the wavefunction of
 conformally coupled scalars in an FRW background.  In the de Sitter limit ($\varepsilon \to 0$), these differential equations can be used to compute what is called the {\it symbol} of the associated pure transcendental functions~\cite{goncharov2013simple,Goncharov:2010jf,Duhr:2011zq,Duhr:2019tlz}.  The symbol of such functions was first discussed in~\cite{Arkani-Hamed:2017fdk} from the point of view of the cosmological polytope, and a recursive rule for constructing the symbol from discontinuities was the subject of~\cite{Hillman:2019wgh}.  The symbol limit therefore serves as a useful cross-check of our calculations.  In this appendix, we define the symbol of a transcendental function and describe some of its properties.

\vskip 4pt
The recursive rule in particular provides a complementary way of constructing the symbol with an analogous set of associated graphical rules.  These two sets of graphical rules---symbol recursion and kinematic flow---are not obviously identical, and are built upon two complementary consistency conditions: factorization and local time evolution.  Kinematic flow is a more microscopic analogue of the kinematic differential operator which collapses the exchanged particle's propagator to a contact term, and is therefore tied to locality and consistency of bulk time evolution~\cite{Arkani-Hamed:2018bjr}. The singularities of the wavefunction, on the other hand, must reflect the consistency of factorization of processes at large spacetime separation (therefore encoding locality in a somewhat difference sense).   
In the language of the momentum polygon on the future boundary, kinematic flow tells us what happens when we change the side lengths of the polygon and the discontinuities tell us what happens when we cut it into pieces.  We have control over both descriptions in the de Sitter limit, captured cleanly by the symbol.  This limit is therefore a valuable arena in which we can probe the relationship between these two conditions. 

\subsection{Properties of the Symbol}

We begin by reviewing some elementary features of the symbol map.
A pure transcendental function of weight $n$ (denoted by $F^{(n)}$) can be defined recursively as a function whose differential is given by
\begin{equation}
	\label{equ:transcdef}
	\ud F^{(n)} = \sum\limits_i F_i^{(n-1)} \ud \log R_i \, ,
\end{equation}
where $R_i$ are rational functions of the variables associated to the differential.  This condition is equivalent to saying that $F^{(n)}$ is a function that can be written as a linear combination of $n$-fold iterated integrals of the form
\begin{equation}
\label{eq:transcintdef}
  F^{(n)} = \int \ud\log R_1 \circ \dots \circ \ud\log R_n \, ,
\end{equation}
with $R_1$ being the innermost integral and these iterated integrals are again defined recursively.\footnote{More explicitly, the composition of iterated integrals is defined recursively as in~\cite{Goncharov:2010jf}
\be
\int_a^b \ud\log R_1 \circ \dots \circ \ud\log R_n =\int_a^b \left(\int _a^t\ud\log R_1 \circ \dots \circ \ud\log R_{n-1}\right)\ud\log R_{n}(t)\,,
\ee
so that $F^{(n)}$ is obtained by repeated integration. An example are the classical polylogarithms~\eqref{eq:polylogs}.
}
 While there is no generic means by which to map iterated integrals to integrated functions, this representation carries more content than a $n$-fold integral of a random rational integrand.  

\paragraph{Definition:}
The data of iterated integrals is neatly packaged in the {\it symbol}.  The symbol of (\ref{equ:transcdef}) is defined recursively by 
\begin{equation}
	\label{equ:symboldef}
	\mathcal{S}(F^{(n)}) = \sum\limits_i\mathcal{S}(F_i^{(n-1)})\otimes R_i \, .
\end{equation}
Formally, the symbol is an element of the tensor product space of rational functions~\cite{goncharov2013simple,Goncharov:2010jf}, which motivates the use of the
notation $\otimes$ since $\mathcal{S}(F^{(n)})$ indeed possesses a multilinearity worthy of the tensor product. To understand this, one should implicitly think of each entry in the symbol (elements separated by $\otimes$) as implicitly being inside $\log$ functions. The symbol then has the expected distributive properties of products of logarithms; e.g. 
\begin{equation}
    a_1\otimes a_2-a_1\otimes b_2 = a_1 \otimes \frac{a_2}{b_2}\, .
\end{equation}
We see from (\ref{equ:transcdef}) that the symbol manifestly contains the information needed to extract the differential equations of $F^{(n)}$,  organized  in a graded, first-order form. That is, like the equations in the main text, the symbol defines an associated set of first-order inhomogeneous differential equations graded by transcendental weight. Less obvious at this stage is that, read from left to right, the symbol captures the sequences of branch points of $F^{(n)}$.  The innermost iterated integral determines the branch point on the first sheet. We will illustrate this explicitly in the case of classical polylogarithms below. 

\paragraph{Shuffle product:} A function that is the product of two transcendental functions of weights $n$ and $m$, respectively, has weight $n+m$, and the symbol of the product is the {\it shuffle product} of the two symbols of the constituent functions. Given a symbol, the ordering of entries defines a {\it word} (product of group elements), which is just the product of symbol entries. The shuffle product of two symbols is the
 sum over all ways of interlacing the symbol entries between a pair of words, while maintaining the internal order of each symbol.  Denoting the shuffle product by $\shuffle$, we have
\be
\big(a_{1}\otimes \dots \otimes a_{n} \big)\,\shuffle\, \big(a_{n+1} \otimes \dots \otimes a_m\big) = \sum_{\sigma \in \Sigma(n, m)} a_{\sigma(1)}\otimes \dots \otimes a_{\sigma(m)} \, ,
\ee
where $\Sigma(n, m)$ is the set of all shuffles on a 
word of length $n$ and a word of length $m$, i.e. the subset of the permutations $S_{n+m}$ defined by  
\be
	\Sigma(n, m) = \left \{ \sigma \in S_{n+m} \big | \sigma^{-1}(1) < \dots < \sigma^{-1}(n)\hspace{3mm} {\rm and}  \hspace{3mm}\sigma^{-1}(n+1) < \dots < \sigma^{-1}(m)\right \}\, .
\ee
An explicit example is the following. If we have two symbols of length 2, their shuffle product is
\begin{align}
	\big(a_1\otimes a_2\big)\,\shuffle\,\big(b_1\otimes b_2\big) =&~ a_1\otimes a_2 \otimes b_1\otimes b_2+a_1\otimes b_1 \otimes a_2 \otimes b_2+b_1\otimes a_1 \otimes a_2\otimes b_2\nonumber\\&+a_1\otimes b_1 \otimes b_2\otimes a_2+b_1\otimes a_1 \otimes b_2\otimes a_2+b_1\otimes b_2 \otimes a_1\otimes a_2\, ,
\end{align}
which preserves the relative ordering of the $a$'s and $b$'s, but interleaves them in all possible ways.

\vskip8pt
\noindent
{\bf Integrability:} In order for a symbol to be integrable to a parent function, it must satisfy certain compatibility conditions.
In particular, given a symbol of the form
\be
	{\cal S} = \sum\limits_{I = (i_1, \cdots, i_n)}c_{I}\, a_{i_1}\otimes \dots \otimes a_{i_n}\,,
\ee
we must have 
\be
 \sum\limits_{I = (i_1, \cdots, i_n)}c_{I} \,a_{i_1}\otimes \dots \otimes a_{i_{p-1}}\otimes a_{i_{p+2}} \otimes a_{i_n} \,\ud\log a_{i_{p}}  \wedge \ud\log a_{i_{p+1}} = 0 \,,
\ee
 for any fixed $p$. 
 This condition is nontrivial, as it is possible to write down symbols which do not obey the integrability condition and therefore do not lift to a function. 
Consider the following two examples
\begin{align}
	\label{eq:nonint}
	\mathcal{S}_{\text{non-int}} &= (1-z)\otimes w+w\otimes z\,, \\
	\label{eq:nonint}
	\mathcal{S}_{\text{int}} &= (1-z)\otimes w+w\otimes (1-z)\,.
\end{align}
The first is not integrable, while the second
integrates to $\log(1-z)\log(w)$.  Integrability is interesting from the point of view of our ab initio graphical rules for constructing the differential equations---and also from the singularity-based symbol recursion---as neither manifests the fact that the objects 
 actually have integral representations.

\paragraph{Classical polylogarithms:} It is perhaps useful to illustrate these concepts with an explicit example. The simplest class of polylogarithms are the classical polylogarithms, defined recursively via $\text{Li}_1(z) = -\log(1-z)$ and 
\begin{equation}
    \text{Li}_n(z) = \int\limits_0^z \text{Li}_{n-1}(t)\,\ud\log t \quad \Leftrightarrow \quad \ud\,\text{Li}_n(z) = \text{Li}_{n-1}(z )\,\ud\log z \, .
    \label{eq:polylogs}
\end{equation}
These functions have branch points at $z=1$. It is instructive to consider the
discontinuity across the cut at real $z > 1$.  The relevant contour to compute this discontinuity is
\beq
    \centering
   \begin{tikzpicture}
   \draw (-3, 0) to (3, 0);
   \draw (0, -2) to (0, 2);
   \draw[thick, color=Blue, decorate,decoration=zigzag]  (1, 0) -- (3,0);
   \node (z) at (1,-0.5) {$1$};
   \draw[fill, color=Blue] (1, 0) circle (.5mm);
   \draw[thick] (3, .25) to (.95, .25);
   \draw[thick, Straight Barb-]  (1.9, -.25) to (3, -.25);
   \draw[thick]  (.95, -.25) to (1.95, -.25);
   \draw[thick] (.95,.25) arc[radius=7.14pt,start angle= 90,end angle=270];
   \node (z) at (2.9,1.75) {$z$};
  \draw (z.north west) -- (z.south west)  -- (z.south east);
   \end{tikzpicture}
  \nonumber
\eeq
 We can then calculate
\begin{equation}
    \text{Li}_n(z+i\epsilon)-\text{Li}_n(z-i\epsilon) = \int\limits_{1+i\epsilon}^{z+i\epsilon}\text{Li}_{n-1}(t)\,\ud\log t - \int\limits_{1-i\epsilon}^{z-i\epsilon}\text{Li}_{n-1}(t)\,\ud\log t \, ,
\end{equation}
which then implies that the discontinuity is given by
\begin{equation}
    \text{Disc}[\text{Li}_n(z)] =\int\limits_{1}^{z}  \text{Disc}[\text{Li}_{n-1}(t)]\,\ud\log t \, .
\end{equation}
This illustrates how the logarithmic singularity of the innermost integral propagates up to be the leading singularity.  We find the discontinuity
\begin{equation}
\label{eq:Lindisc}
    \text{Disc}[\text{Li}_n(z)] = \frac{2\pi i}{(n-1)!}\log^{(n-1)}(z)\, .
\end{equation}
This is consistent with the symbol derived from differentiating 
\begin{equation}
    \mathcal{S}(\text{Li}_n(z)) = -(1-z)\otimes z \dots \otimes z\, ,
\end{equation}
where the leading entry manifests the branch point at $z = 1$ and its discontinuity (\ref{eq:Lindisc}). 

\vskip4pt
We see that the symbol makes manifests two complementary pieces of information. The symbol read from left to right provides information about the branch points of a function, where the discontinuity is given by the remaining part of the symbol after deleting this entry. On the other hand, the differential of a function can be read off from the symbol starting from the right and is captured by the formula~\eqref{equ:symboldef}. In particular, the differential is given by one divided by the rightmost symbol entry times the function whose symbol is given by the remaining symbol after deleting the rightmost entry.

\subsection{Symbols from Differential Equations}
With an understanding of the properties of the symbol in general, we can now articulate how to calculate the symbol in the de Sitter limit from the finite $\varepsilon$ differential equations.  First, we need to take the appropriate $\varepsilon \to 0$ limit of the associated $A$ matrix, which we denote as $A^{(0)}$. Taking this limit simply sets certain entries in the $A$ matrix to zero.  This happens even though $\varepsilon$ is factored in front of the matrix because the functions they multiply in the differential equation begin as a Laurent expansion in $\varepsilon$ around $\varepsilon = 0$.  The functions are integrals of forms, and the power of the leading behavior in $\e$ is given by the number of twisted lines in the simplex in the form with the most twisted lines.  For example, for the function $f$ in the three-site chain, we have the integrand
\begin{equation}
	\Omega_{f} = [245]-[246]+[25\tilde 3]-[26\tilde 1]-[2\tilde 1\tilde 3]-[45\tilde 1]+[46\tilde 3]-[4\tilde 1\tilde 3]+[5\tilde 1\tilde 3]+[6\tilde 1\tilde 3] \, ,
\end{equation}
where the maximum number of twisted lines in a simplex term is two. In the $\varepsilon\to 0$ limit, we only keep entries $A_{ij}$ such that the leading pole in $\varepsilon$ of the function $f_j$ is one degree higher than that of  the function $f_i$.  In particular, all diagonal terms are set to zero.  This defines $A^{(0)}$, though there is one additional subtlety.  The functions at the bottom of our cascade of differential equations (e.g.~$g, \tilde g$, $Z$ in the three-site case) have the limit $1$ or $-1$, and the sign must be identified. This is readily done once the integrand is determined.
 
 \vskip 4pt
Given the matrix $A^{(0)}$, the symbol of the wavefunction is calculated via 
\begin{equation}
\label{eq:symbfromA}
	\mathcal{S}(\psi^{(0)}) = \sum\limits_{i_1,\dots ,i_n} A^{(0)}_{i_{n-1}i_n}\otimes A^{(0)}_{i_{n-2}i_{n-1}}\otimes \dots \otimes A^{(0)}_{1i_2} \, ,
\end{equation}
where we abuse notation by taking the entries to denote the arguments of the $\ud\log$'s in the $A$-matrix. For example, given an entry $A^{(0)}_{ij}=\ud \log (X_1+X_2)$, we would write $\cdots \otimes (X_1+X_2)\otimes \cdots$ in (\ref{eq:symbfromA}). The transcendental weight $n$ is the maximal number at which the entries are all nontrivial.  That is, the transcendental weight is of course an output of the $A$-matrix and we need not fix $n$ externally: all terms of the form in (\ref{eq:symbfromA}) will become trivial for $n$ beyond the transcendental weight of $\psi^{(0)}$.

\subsection{Symbols from Discontinuities} 
The symbol of our wavefunctions in the $\e\to0$ limit can also be constructed recursively  from knowledge of the discontinuities. The details can be found in \cite{Hillman:2019wgh}.  
In  order to derive the recursion relations, we need a few facts:
\begin{itemize}
\item 
The branch points of the wavefunction on the first sheet (the first entries in the symbol) are at the same locations as the poles of the flat-space wavefunction. 
\item The sum of all discontinuities vanishes, reflecting a residue theorem in the integrand and the vanishing behavior of the integrand at infinity.
\item At branch point, the wavefunction has the following factorization property (see Figure~\ref{fig:discrule})
\begin{equation}
	\label{eq:dsdisc}
	\text{Disc}_{E_G} \psi_{\mathcal{G}} = \sum\limits_{\{ \sigma^i_e\}} \text{Disc}_{E_{G}} \psi_{G}^{\{ +\dots +\}}\prod\limits_{g_i}\left(\prod\limits_e -\sigma^i_e\right) \psi_{g_i}^{\{ \sigma^i_e\}}\,,
\end{equation}
where the notation is in need of clarification. The full graph has been denoted by $\mathcal{G}$, while we use $G$ for the connected subgraph that the discontinuity is associated with.  The energy $E_G$ is the partial energy associated with $G$. 
 In addition to~$G$, the cutting produces various disconnected graphs $g_i$ as depicted in Figure~\ref{fig:discrule}, but with the superscripts $\{\sigma^i_e\}$ denoting the signs which with the energies of the cut edges are absorbed onto the terminating vertices.  For $G$, these signs are always positive as the energy discontinuity arises from contributions to the time integral in which $G$ is either in the past of all the $g_i$ or unordered with respect to the $g_i$.  We sum over all sign choices $\sigma_e^i$ for the graphs $g_i$. The product over $-\sigma_e^i$ ensures the correct overall sign of the contribution from the $g_i$, depending on whether the edge energy is positively or negatively absorbed (where we substitute $1$ for $\sigma_e^i$ when it is $+$ and $-1$ for $\sigma_e^i $ when it is $-$).  These signs come directly from the propagator.

\item Finally, a single vertex has discontinuity $2\pi i$.
\end{itemize}
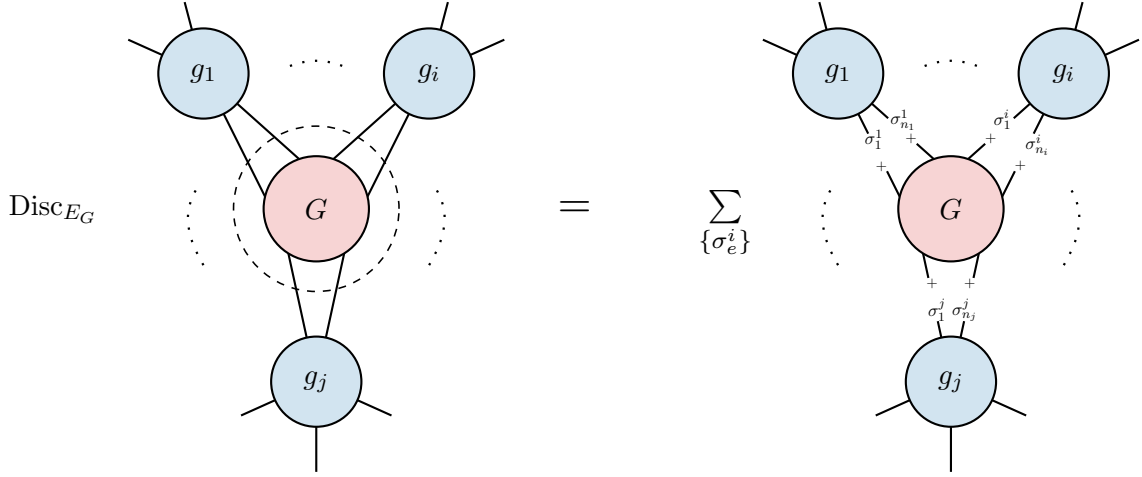
\begin{figure}
    \centering
   \begin{tikzpicture}
   \coordinate (v1) at (-1.5, 1.8);
   \coordinate (v2) at (1.5, 1.8);
   \coordinate (v3) at (0, -2.3);
   \coordinate (e11) at (-2.5, 2.25);
   \coordinate (e12) at (-1.75, 2.75);
   \coordinate (ei1) at (2.5, 2.25);
   \coordinate (ei2) at (1.75, 2.75);
   \coordinate (e31) at (-1, -2.75);
   \coordinate (e32) at (1, -2.75);
   \coordinate (e33) at (0, -3.5);
        \draw[thick] (-.4, -.4) -- (v1);
        \draw[thick] (.4, .1) -- (v1);
        \draw[thick] (.4, -.4) -- (v2);
        \draw[thick] (-.4, .1) -- (v2);
        \draw[thick] (-.5, 0) -- (v3) ;
        \draw[thick] (.5, 0) -- (v3) ;
        \draw[thick] (e11) -- (v1) -- (e12);
        \draw[thick] (ei1) -- (v2) -- (ei2);
        \draw[thick] (e31) -- (v3) -- (e32);
        \draw[thick] (e33) -- (v3);
        \draw[thick, fill=Red!20] (0, 0) circle (7mm);
        \draw[thick,fill=Blue!20] (v1) circle (6mm);
        \draw[thick,fill=Blue!20] (v2) circle (6mm);
        \draw[thick,fill=Blue!20] (v3)  circle (6mm);
        \node at (0, 0) {$G$};
        \node at (v1) {$g_1$};
        \node at (v2) {$g_i$};
        \node at (v3)  {$g_j$};
        \draw[dashed,line width=0.6pt] (0, 0) circle (11mm);
        \draw[loosely dotted,line width = .3mm] (-1.5, -.75) to [out = 120, in = -120] (-1.6, .25);
        \draw[loosely dotted,line width = .3mm] (1.5, -.75) to [out = 60, in = -60] (1.6, .25);
        \draw[loosely dotted,line width = 0.3mm] (-.35, 1.9) to [out = 20, in = 160] (.35, 1.9);
       \node at (-3.5, 0) {$\text{Disc}_{E_G}$};
\end{tikzpicture}
\hspace{0.25cm}
 \begin{tikzpicture}
   \coordinate (v1) at (-1.5, 1.8);
   \coordinate (v2) at (1.5, 1.8);
   \coordinate (v3) at (0, -2.3);
   \coordinate (v1m1) at ( $(v1) + .4*(1, -2)$) ;
   \coordinate (v1m2) at ( $(v1) + .65*(1, -2)$) ;
   \coordinate (v1n1) at ( $(v1) + .35*(1.9, -1.7)$) ;
   \coordinate (v1n2) at ( $(v1) + .55*(1.9, -1.7)$) ;
   \coordinate (v2m1) at ( $(v2) + .4*(-1, -2)$) ;
   \coordinate (v2m2) at ( $(v2) + .65*(-1, -2)$) ;
   \coordinate (v2n1) at ( $(v2) + .35*(-1.9, -1.7)$) ;
   \coordinate (v2n2) at ( $(v2) + .55*(-1.9, -1.7)$) ;
   \coordinate (v3m1) at ( $(v3) + .35*(-.5, 2.3)$) ;
   \coordinate (v3m2) at ( $(v3) + .6*(-.5, 2.3)$) ;
   \coordinate (v3n1) at ( $(v3) + .35*(.5, 2.3)$) ;
   \coordinate (v3n2) at ( $(v3) + .6*(.5, 2.3)$) ;
   \coordinate (e11) at (-2.5, 2.25);
   \coordinate (e12) at (-1.75, 2.75);
   \coordinate (ei1) at (2.5, 2.25);
   \coordinate (ei2) at (1.75, 2.75);
   \coordinate (e31) at (-1, -2.75);
   \coordinate (e32) at (1, -2.75);
        \draw[thick] (-.4, -.4) -- (v1m2);
        \draw[thick] (v1m1)--(v1);
        \draw[thick] (.4, .1) -- (v1n2);
        \draw[thick] (v1n1) -- (v1) ;
        \draw[thick] (.4, -.4) -- (v2m2);
        \draw[thick] (v2m1)--(v2);
        \draw[thick] (-.4, .1) -- (v2n2);
        \draw[thick] (v2n1) -- (v2) ;
        \draw[thick] (-.5, 0) -- (v3m2);
        \draw[thick] (v3m1) -- (v3);
        \draw[thick] (.5, 0) -- (v3n2);
        \draw[thick] (v3n1) -- (v3);
        \draw[thick] (e11) -- (v1) -- (e12);
        \draw[thick] (ei1) -- (v2) -- (ei2);
        \draw[thick] (e31) -- (v3) -- (e32);
        \draw[thick] (e33) -- (v3);
        \draw[thick,fill=Red!20] (0, 0) circle (7mm);
        \draw[thick,fill=Blue!20] (v1) circle (6mm);
        \draw[thick,fill=Blue!20] (v2) circle (6mm);
        \draw[thick,fill=Blue!20] (v3) circle (6mm);
        \node at (0, 0) {$G$};
        \node at (v1) {$g_1$};
        \node at (v2) {$g_i$};
        \node at (v3) {$g_j$};
        \draw[loosely dotted,line width = .3mm] (-1.5, -.75) to [out = 120, in = -120] (-1.6, .25);
        \draw[loosely dotted,line width = .3mm] (1.5, -.75) to [out = 60, in = -60] (1.6, .25);
        \draw[loosely dotted,line width = .3mm] (-.35, 1.9) to [out = 20, in = 160] (.35, 1.9);
        \node[scale=.5] at($(v1m2)+(-.075, .07)$) {$\small +$};
        \node[scale=.5] at($(v1n2)+(-.075, .07)$) {$\small +$};
        \node[scale=.5] at($(v2n2)+(.075, .075)$) {$\small +$};
        \node[scale=.5] at($(v2m2)+(.075, .075)$) {$\small +$};
        \node[scale=.5] at($(v3n2)+(-.05, -.075)$) {$\small +$};
        \node[scale=.5] at($(v3m2)+(.05, -.075)$) {$\small +$};
         \node[scale=.6] at($(v1m1)+(.075, -.075)$) {$\small \sigma^1_1$};
        \node[scale=.6] at($(v1n1)+(.2, -.02)$) {$\small \sigma^1_{n_1}$};
        \node[scale=.6] at($(v2n1)+(-.15, -.02)$) {$\small \sigma^i_1$};
        \node[scale=.6] at($(v2m1)+(.05, -.15)$) {$\small \sigma^i_{n_i}$};
        \node[scale=.6] at($(v3n1)+(.01, .15)$) {$ \sigma^j_{n_j}$};
        \node[scale=.6] at($(v3m1)+(-.01, .15)$) {$\sigma^j_1$};
        \node[scale=1.15] at (-3, -0.2) {$\sum\limits_{ \{\sigma_e^i \}}$};
        \node[scale=1.5] at (-5, 0) {$=$};
\end{tikzpicture}
\caption{Figure depicting the expression of the discontinuity associated with the partial energy of a connected subgraph $G$ going to zero, and a sum over products of graphs.  Every term of the sum is a product over all the resultant graphs disconnected by the dashed cut, and the \textit{discontinuity} of the resultant $G$ graph (which is $G$ with all cut edge energy absorbed positively at incident vertices), summed over ways of absorbing cut edge energy. The  graph $G$ with all incident edge energies positively absorbed is in every term of the sum, and the dashed line means the total energy of this resultant graph.  The sum is over all ways of absorbing positive or negative cut edge energy into the graphs $g_i$. These signs are denoted by the~$\sigma^i_e$'s. For this to be useful as a recursion, we note that the total energy discontinuity is minus the sum over all other discontinuities of a graph. }
\label{fig:discrule}
\end{figure}

\noindent
Equation (\ref{eq:dsdisc}) is a statement about the factorization of the wavefunction at a branch point.  Leveraging the shuffle product, this statement about functions is immediately converted to the associated statement about symbols.  Notice that this sum over signs reflects the structure of the propagators, and indeed is readily seen in the time-integral representation: contributions to the integrand with partial energy $E_G$ only come from terms in which $G$ is to the past of or unordered with respect to the $g_i$.  Below, we illustrate the application of this recursion to the two- and three-site chains.

\subsection{Examples}

In the following, we derive the symbol of the two-site chain in two different ways: 1) From the matrix $A^{(0)}$ of the differential equation, and 2) from the symbol recursion. We will show explicitly that the two results agree.

\paragraph{Two-site chain:} Taking the limit $\e \to 0$, we find the $A^{(0)}$-matrix of the two-site chain simply by deleting the diagonal entries of the $A$-matrix for the general case. This gives
\beq
A^{(0)} = \left[\begin{array}{cccc}
\phantom{\hskip 3pt} 0\phantom{\hskip 3pt} & \phantom{\hskip 3pt} \ell_3- \ell_1 \phantom{\hskip 3pt}&  \phantom{\hskip 3pt}\ell_4- \ell_2\phantom{\hskip 3pt} & 0 \\[5pt]
 0 & 0 & 0 &  \ell_5- \ell_2 \\[5pt]
 0 & 0 &  0 & \phantom{\hskip 3pt} \ell_5- \ell_1 \phantom{\hskip 3pt}\\[5pt]
 0 & 0 & 0 & 0
\end{array}\right],
\eeq
with the $\ell_i$'s defined in (\ref{eq:dlogell}). We see by application of (\ref{eq:symbfromA}) that the symbol is 
\begin{equation}
	{\cal S}_{\text{2-site, diff}} = \frac{X_1+X_2}{X_1+Y}\otimes \frac{X_2-Y}{X_2+Y}+\frac{X_1+X_2}{X_2+Y}\otimes \frac{X_1-Y}{X_1+Y}\,.
	\label{eq:s2diff}
\end{equation}
Next, we compare this calculation to the symbol derived recursively from factorization at singularities.  From the above prescription, we find the discontinuities associated with $X_1+Y$ and $X_2+Y$. For example, the discontinuity of $X_1+Y$ involves summing over two cut graphs
\be
\raisebox{2.5pt}{
 \begin{tikzpicture}[baseline=(current  bounding  box.center)]
\draw[thick] (4, 0) -- (5, 0);
\draw[color=Red, fill=Red] (4, 0) circle (.5mm);
\draw[color=Blue, fill=Blue] (5, 0) circle (.5mm);
\draw[line width=0.6pt, dashed] (4, 0) circle (2mm);
\end{tikzpicture}}
\,\, \ =\ \raisebox{2.5pt}{
 \begin{tikzpicture}[baseline=(current  bounding  box.center)]
\draw[thick] (5, 0) -- (5.5, 0);
\draw[thick] (4, 0) -- (4.5, 0);
 \draw[color=Red, fill=Red] (4, 0) circle (.5mm);
 \draw[color=Blue, fill=Blue] (5.5, 0) circle (.5mm);
\node[color=Blue] at (4.85, .0) {\tiny$-$};
\end{tikzpicture}}\hspace{1.5mm}
\,-\,\raisebox{2.5pt}{
 \begin{tikzpicture}[baseline=(current  bounding  box.center)]
 \draw[thick] (4, 0) -- (4.5, 0);
  \draw[color=Red, fill=Red] (4, 0) circle (.5mm);
\draw[thick] (5, 0) -- (5.5, 0);
\draw[color=Blue, fill=Blue] (5.5, 0) circle (.5mm);
\node[color=Blue] at (4.85,0) {\tiny$+$};
\end{tikzpicture}}
\,\,,
\ee
where the circled vertices denote the trivial discontinuity of the single vertex, which is $2\pi i$. We emphasize that at the level of functions the two terms above are each products, with the small plus and minus signs denoting the absorption of edge energy $Y$ for the rightmost graph.  From this, we can read off the symbol of the discontinuity
\be
{\cal S}(
\text{Disc}_{X_1+Y}[\psi^{(0)}])= \frac{X_2-Y}{X_2+Y}\,.
\ee
The $X_2+Y$ discontinuity is obtained by symmetry (interchanging $X_1\leftrightarrow X_2$) and the total energy discontinuity is minus the sum of the others,
\be 
 \raisebox{2pt}{\begin{tikzpicture}[baseline=(current  bounding  box.center)]
\draw[fill] (4, 0) circle (.5mm);
\draw[fill] (5, 0) circle (.5mm);
\draw[thick] (4, 0) -- (5, 0);
\node [
        draw, color=black, line width=0.6pt, dashed,
        rounded rectangle,
        minimum height = 1.1em,
        minimum width = 4.7em,
        rounded rectangle arc length = 180,
    ] at (4.5,0)
    {};
\end{tikzpicture}}
\,\, \ =\  - \, \raisebox{2pt}{\begin{tikzpicture}[baseline=(current  bounding  box.center)]
\draw[thick] (4, 0) -- (5, 0);
\draw[color=Red,fill=Red] (4, 0) circle (.5mm);
\draw[color=Blue,fill=Blue] (5, 0) circle (.5mm);
\draw[line width=0.6pt, dashed] (4, 0) circle (2mm);
\end{tikzpicture}}
\,\,- \, \raisebox{2pt}{\begin{tikzpicture}[baseline=(current  bounding  box.center)]
\draw[thick] (4, 0) -- (5, 0);
\draw[color=Red,fill=Red] (4, 0) circle (.5mm);
\draw[color=Blue,fill=Blue] (5, 0) circle (.5mm);
\draw[line width=0.6pt, dashed] (5, 0) circle (2mm);
\end{tikzpicture}}
\,\,,
\ee
which implies
\be 
{\cal S}(\text{Disc}_{X_1+X_2}[\psi^{(0)}]) = -\frac{X_2-Y}{X_2+Y}-\frac{X_1-Y}{X_1+Y}\,.
\label{eq:totalEdisc}
\ee
From this information it is straightforward to reconstruct the full symbol itself. Recall that the symbol, read from left to right, gives the sequence of discontinuities of the function. We see that there are two leading entries in the symbol ($X_1+Y$ and $X_2+Y$), which each come divided by the total energy $X_1+X_2$, so that the discontinuity of $X_1+X_2$ is minus the others as in~\eqref{eq:totalEdisc}. Putting this together, we obtain
\begin{equation}
	{\cal S}_{\text{2-site, disc}} = \frac{X_1+Y}{X_1+X_2}\otimes \frac{X_2-Y}{X_2+Y}+\frac{X_2+Y}{X_1+X_2}\otimes \frac{X_1-Y}{X_1+Y}\,,
\end{equation}
which is manifestly the same as~\eqref{eq:s2diff} obtained from the differential picture (up to an overall sign that depends on conventions). The consistency of the two methods for deriving the symbol is highly nontrivial.

\vskip 4pt
From the point of view of a second-order ordinary differential equation, it is not clear why one local picture need imply the other, but the kinematic flow makes it clear in the two-site case.  The first-order differentials are local to a vertex, and the only term remaining for the final symbol entry is itself local at the remaining vertex according to the next tier of the differential equation.  Explicitly, we can think of the tubing picture of letters in the $A$-matrix as implying the following schematic structure of the symbol 
\begin{equation}
	\label{eq:blobsym}
	\mathcal{S} = \frac{B_1(X_1)}{B_{\text{tot}}(X_1, X_2)}\otimes D_2(X_2)+\frac{B_2(X_2)}{B_{\text{tot}}(X_1, X_2)}\otimes D_1(X_1)\,,
\end{equation}
where $B_{i},D_{i}$ and $B_{\rm tot}$ can be arbitrary (rational) functions of their arguments. The crucial aspect of the tubing picture is that
 the rightmost entries depend only on a single kinematic variable $X_1$ or $X_2$, but not both.
The same structure is  implied by the symbol recursion point of view, independent of the specific functions $B_i,D_i$, because this structure is determined by the possible kinematic dependencies allowed by the underlying graph given a tubing picture.  This fact of course implies a differential equation by applying (\ref{equ:transcdef}) which takes the form 
\begin{equation}
	\label{eq:blobdiff}
	\partial_{X_1} \frac{D_1}{\partial_{X_1} D_1}\,\partial_{X_1}\psi = -\frac{\partial_{X_1} B_{\text{tot}}}{B_{\text{tot}}}\,.
\end{equation}
Hence, as the symbol, read from left to right, is the sequence of discontinuities of the wavefunction, locality of the time evolution implies locality of the singularities and vice versa.   This implication is highly nontrivial, and in general it is rather difficult to see explicitly. The two-site example is sufficiently small that two methods generate exactly the same symbol, so one can trivially check compatibility of the pictures. For more complicated examples, the presentation of the symbol from the two methods of derivation is typically very different, as we will now see at three sites.

\paragraph{Three-site chain:} We now state the result of the calculation of the three-site chain from the differential equation construction and from the discontinuity recursion.  From the recursive calculation, the symbol is naturally handed to us in the following form 
\begin{align}
	{\cal S}_{\text{3-site, disc}} &= \frac{X_1^+}{X_{123}} \otimes \bigg[\frac{X_3^+}{X_{23}^-}\otimes  \frac{X_2^{--}}{X^{-+}_2}+\frac{X_2^{-+}}{X_{23}^-}\otimes  \frac{X_3^-}{X_3^+}-\frac{X_3^+}{X_{23}^+}\otimes  \frac{X_2^{+-}}{X_2^{++}}-\frac{X_2^{++}}{X_{23}^+}\otimes  \frac{X_3^-}{X_3^+}\bigg] \nonumber\\
    &+\frac{X_3^+}{X_{123}}\otimes \bigg [ \frac{X_1^+}{X_{12}^-}\otimes  \frac{X_2^{--}}{X_2^{+-}}+\frac{X_2^{+-}}{X_{12}^-}\otimes  \frac{X_1^-}{X_1^+}-\frac{X_1^+}{X_{12}^+}\otimes  \frac{X_2^{-+}}{X_2^{++}}-\frac{X_2^{++}}{X_{12}^+}\otimes  \frac{X_1^-}{X_1^+}\bigg] \nonumber\\
	&+ \frac{X_2^{++}}{X_{123}} \otimes \left[\frac{X_1^-}{X_1^+}\otimes \frac{X_3^-}{X_3^+}+ \frac{X_3^-}{X_3^+}\otimes \frac{X_1^-}{X_1^+} \right] \\ 
	&+\frac{X_{12}^+}{X_{123}}\otimes \bigg[\frac{X_1^-}{X_1^+}\otimes \frac{X_3^-}{X_3^+}+\frac{X_3^-}{X_3^+}\otimes \frac{X_1^-}{X_1^+}+\frac{X_2^{-+}}{X_2^{++}}\otimes \frac{X_3^-}{X_3^+}+ \frac{X_3^-}{X_3^+}\otimes \frac{X_2^{-+}}{X_2^{++}} \bigg] \nonumber\\
	&+ \frac{X_{23}^+}{X_{123}}\otimes \bigg[\frac{X_3^-}{X_3^+}\otimes \frac{X_1^-}{X_1^+}+\frac{X_1^-}{X_1^+}\otimes \frac{X_3^-}{X_3^+}+ \frac{X_2^{+-}}{X_2^{++}}\otimes \frac{X_1^-}{X_1^+}+ \frac{X_1^-}{X_1^+}\otimes \frac{X_2^{+-}}{X_2^{++}} \bigg] \nonumber\, ,
\end{align}
whereas, from the differential equations, we find 
\begin{align}
	{\cal S}_{\text{3-site, diff}} &= \frac{X_{1\phantom{23}}^-}{X_{123}}\otimes \frac{X_{23\phantom{1}}^+}{X_{3\phantom{12}}^+}\otimes \frac{X_{2\phantom{13}}^{+-}}{X_{2\phantom{13}}^{++}}+\frac{X_{1\phantom{23}}^-}{X_{123}}\otimes \frac{X_{23\phantom{1}}^+}{X_{2\phantom{13}}^{++}}\otimes \frac{X_{3\phantom{12}}^-}{X_{3\phantom{12}}^+}+\frac{X_{3\phantom{12}}^-}{X_{123}}\otimes \frac{X_{12\phantom{3}}^+}{X_{1\phantom{23}}^+}\otimes \frac{X_{2\phantom{13}}^{-+}}{X_{2\phantom{13}}^{++}} \nonumber \\
	&\vspace{5mm}+\frac{X_{3\phantom{12}}^-}{X_{123}}\otimes \frac{X_{12\phantom{3}}^+}{X_{2\phantom{13}}^{++}}\otimes \frac{X_{1\phantom{23}}^-}{X_{1\phantom{23}}^+}\nonumber+\frac{X_{123}}{X_{1\phantom{23}}^+}\otimes \frac{X_{3\phantom{12}}^-}{X_{23\phantom{1}}^-}\otimes \frac{X_{2\phantom{13}}^{-+}}{X_{2\phantom{13}}^{++}}
	+\frac{X_{123}}{X_{1\phantom{23}}^+}\otimes \frac{X_{23\phantom{1}}^-}{X_{3\phantom{12}}^+}\otimes \frac{X_{2\phantom{13}}^{--}}{X_{2\phantom{13}}^{++}}\nonumber\\&+\frac{X_{123}}{X_{1\phantom{23}}^+}\otimes \frac{X_{23\phantom{1}}^-}{X_{2\phantom{13}}^{++}}\otimes \frac{X_{3\phantom{12}}^-}{X_{3\phantom{12}}^+}
	-\frac{X_{1\phantom{23}}^-}{X_{1\phantom{23}}^+}\otimes \frac{X_{23\phantom{1}}^+}{X_{3\phantom{12}}^+}\otimes \frac{X_{2\phantom{13}}^{+-}}{X_{2\phantom{13}}^{++}}-\frac{X_{1\phantom{23}}^-}{X_{1\phantom{23}}^+}\otimes \frac{X_{23\phantom{1}}^+}{X_{2\phantom{13}}^{++}}\otimes \frac{X_{3\phantom{12}}^-}{X_{3\phantom{12}}^+}\nonumber\\
	\begin{split}
	&-\frac{X_{12\phantom{3}}^+}{X_{1\phantom{23}}^+}\otimes \frac{X_{3\phantom{12}}^-}{X_{3\phantom{12}}^+}\otimes \frac{X_{2\phantom{13}}^{-+}}{X_{2\phantom{13}}^{++}}-\frac{X_{12\phantom{3}}^+}{X_{1\phantom{23}}^+}\otimes \frac{X_{2\phantom{13}}^{-+}}{X_{2\phantom{13}}^{++}}\otimes \frac{X_{3\phantom{12}}^-}{X_{3\phantom{12}}^+}+\frac{X_{123}}{X_{3\phantom{12}}^+}\otimes \frac{X_{12\phantom{3}}^-}{X_{1\phantom{23}}^+}\otimes \frac{X_{2\phantom{13}}^{--}}{X_{2\phantom{13}}^{++}}\\
	&+\frac{X_{123}}{X_{3\phantom{12}}^+}\otimes \frac{X_{1\phantom{23}}^-}{X_{12\phantom{3}}^-}\otimes \frac{X_{2\phantom{13}}^{+-}}{X_{2\phantom{13}}^{++}}+\frac{X_{123}}{X_{3\phantom{12}}^+}\otimes \frac{X_{12\phantom{3}}^-}{X_{2\phantom{13}}^{++}}\otimes \frac{X_{1\phantom{23}}^-}{X_{1\phantom{23}}^+}-\frac{X_{23\phantom{1}}^+}{X_{3\phantom{12}}^+}\otimes \frac{X_{1\phantom{23}}^-}{X_{1\phantom{23}}^+}\otimes \frac{X_{2\phantom{13}}^{+-}}{X_{2\phantom{13}}^{++}}\\
	\end{split}\\
	&-\frac{X_{23\phantom{1}}^+}{X_{3\phantom{12}}^+}\otimes \frac{X_{2\phantom{13}}^{+-}}{X_{2\phantom{13}}^{++}}\otimes \frac{X_{1\phantom{23}}^-}{X_{1\phantom{23}}^+}-\frac{X_{3\phantom{12}}^-}{X_{3\phantom{12}}^+}\otimes \frac{X_{12\phantom{3}}^+}{X_{1\phantom{23}}^+}\otimes \frac{X_{2\phantom{13}}^{-+}}{X_{2\phantom{13}}^{++}}
	-\frac{X_{3\phantom{12}}^-}{X_{3\phantom{12}}^+}\otimes \frac{X_{12\phantom{3}}^+}{X_{2\phantom{13}}^{++}}\otimes \frac{X_{1\phantom{23}}^-}{X_{1\phantom{23}}^+}\nonumber\\
	&+\frac{X_{123}}{X_{2\phantom{13}}^{++}}\otimes \frac{X_{1\phantom{23}}^-}{X_{1\phantom{23}}^+}\otimes \frac{X_{3\phantom{12}}^-}{X_{3\phantom{12}}^+}+\frac{X_{123}}{X_{2\phantom{13}}^{++}}\otimes \frac{X_{3\phantom{12}}^-}{X_{3\phantom{12}}^+}\otimes \frac{X_{1\phantom{23}}^-}{X_{1\phantom{23}}^+}-\frac{X_{23\phantom{1}}^+}{X_{2\phantom{13}}^{++}}\otimes \frac{X_{1\phantom{23}}^-}{X_{1\phantom{23}}^+}\otimes \frac{X_{3\phantom{12}}^-}{X_{3\phantom{12}}^+} \nonumber\\
	&-\frac{X_{23\phantom{1}}^+}{X_{2\phantom{13}}^{++}}\otimes \frac{X_{3\phantom{12}}^-}{X_{3\phantom{12}}^+}\otimes \frac{X_{1\phantom{23}}^-}{X_{1\phantom{23}}^+}-\frac{X_{12\phantom{3}}^+}{X_{2\phantom{13}}^{++}}\otimes \frac{X_{1\phantom{23}}^-}{X_{1\phantom{23}}^+}\otimes \frac{X_{3\phantom{12}}^-}{X_{3\phantom{12}}^+}-\frac{X_{12\phantom{3}}^+}{X_{2\phantom{13}}^{++}}\otimes \frac{X_{3\phantom{12}}^-}{X_{3\phantom{12}}^+}\otimes \frac{X_{1\phantom{23}}^-}{X_{1\phantom{23}}^+}\nonumber\, .
\end{align}
The compact notation used in these expressions is defined in \eqref{threesiteX}. One can check that ${\cal S}_{\text{3-site, disc}} = {\cal S}_{\text{3-site, diff}}$.  This of course had to be the case, but given the two distinct presentations it appears somewhat remarkable.  The most salient difference between these two equivalent representations of the symbol are the manifestation of leading singularities.  By construction, ${\cal S}_{\text{3-site, disc}} $ only has singularities associated with partial energy sums.  On the other hand, ${\cal S}_{\text{3-site, diff}}$ naively has many folded singularities on the first sheet, which of course are spurious.  This is simply the price we pay for manifesting the differential properties via the $A$-matrix in this particular basis.


\clearpage
\phantomsection
\addcontentsline{toc}{section}{References}
\bibliographystyle{utphys}
{\linespread{1.075}
\bibliography{DE-Refs}

\providecommand{\href}[2]{#2}\begingroup\raggedright\begin{thebibliography}{100}

\bibitem{Maldacena:2002vr}
J.~Maldacena, ``{Non-Gaussian Features of Primordial Fluctuations in
  Single-Field Inflationary Models},''
  \href{http://dx.doi.org/10.1088/1126-6708/2003/05/013}{{\em JHEP} {\bfseries
  05} (2003) 013}, \href{http://arxiv.org/abs/astro-ph/0210603}{{\ttfamily
  arXiv:astro-ph/0210603}}.

\bibitem{Anninos:2014lwa}
D.~Anninos, T.~Anous, D.~Z. Freedman, and G.~Konstantinidis, ``{Late-time
  Structure of the Bunch-Davies De Sitter Wavefunction},''
  \href{http://dx.doi.org/10.1088/1475-7516/2015/11/048}{{\em JCAP} {\bfseries
  11} (2015) 048}, \href{http://arxiv.org/abs/1406.5490}{{\ttfamily
  arXiv:1406.5490 [hep-th]}}.

\bibitem{Ghosh:2014kba}
A.~Ghosh, N.~Kundu, S.~Raju, and S.~Trivedi, ``{Conformal Invariance and the
  Four Point Scalar Correlator in Slow-Roll Inflation},''
  \href{http://dx.doi.org/10.1007/JHEP07(2014)011}{{\em JHEP} {\bfseries 07}
  (2014) 011}, \href{http://arxiv.org/abs/1401.1426}{{\ttfamily arXiv:1401.1426
  [hep-th]}}.

\bibitem{Arkani-Hamed:2017fdk}
N.~Arkani-Hamed, P.~Benincasa, and A.~Postnikov, ``{Cosmological Polytopes and
  the Wavefunction of the Universe},''
  \href{http://arxiv.org/abs/1709.02813}{{\ttfamily arXiv:1709.02813
  [hep-th]}}.

\bibitem{Albayrak:2018tam}
S.~Albayrak and S.~Kharel, ``{Towards the Higher-Point Holographic Momentum
  Space Amplitudes},'' \href{http://dx.doi.org/10.1007/JHEP02(2019)040}{{\em
  JHEP} {\bfseries 02} (2019) 040},
  \href{http://arxiv.org/abs/1810.12459}{{\ttfamily arXiv:1810.12459
  [hep-th]}}.

\bibitem{Albayrak:2019yve}
S.~Albayrak and S.~Kharel, ``{Towards the Higher-Point Holographic Momentum
  Space Amplitudes. Part II. Gravitons},''
  \href{http://dx.doi.org/10.1007/JHEP12(2019)135}{{\em JHEP} {\bfseries 12}
  (2019) 135}, \href{http://arxiv.org/abs/1908.01835}{{\ttfamily
  arXiv:1908.01835 [hep-th]}}.

\bibitem{Hillman:2019wgh}
A.~Hillman, ``{Symbol Recursion for the dS Wave Function},''
  \href{http://arxiv.org/abs/1912.09450}{{\ttfamily arXiv:1912.09450
  [hep-th]}}.

\bibitem{Baumann:2020dch}
D.~Baumann, C.~Duaso~Pueyo, A.~Joyce, H.~Lee, and G.~L. Pimentel, ``{The
  Cosmological Bootstrap: Spinning Correlators from Symmetries and
  Factorization},'' \href{http://dx.doi.org/10.21468/SciPostPhys.11.3.071}{{\em
  SciPost Phys.} {\bfseries 11} (2021) 071},
  \href{http://arxiv.org/abs/2005.04234}{{\ttfamily arXiv:2005.04234
  [hep-th]}}.

\bibitem{Goodhew:2020hob}
H.~Goodhew, S.~Jazayeri, and E.~Pajer, ``{The Cosmological Optical Theorem},''
  \href{http://dx.doi.org/10.1088/1475-7516/2021/04/021}{{\em JCAP} {\bfseries
  04} (2021) 021}, \href{http://arxiv.org/abs/2009.02898}{{\ttfamily
  arXiv:2009.02898 [hep-th]}}.

\bibitem{Cespedes:2020xqq}
S.~C\'espedes, A.-C. Davis, and S.~Melville, ``{On the Time Evolution of
  Cosmological Correlators},''
  \href{http://dx.doi.org/10.1007/JHEP02(2021)012}{{\em JHEP} {\bfseries 02}
  (2021) 012}, \href{http://arxiv.org/abs/2009.07874}{{\ttfamily
  arXiv:2009.07874 [hep-th]}}.

\bibitem{Baumann:2021fxj}
D.~Baumann, W.-M. Chen, C.~Duaso~Pueyo, A.~Joyce, H.~Lee, and G.~L. Pimentel,
  ``{Linking the Singularities of Cosmological Correlators},''
  \href{http://dx.doi.org/10.1007/JHEP09(2022)010}{{\em JHEP} {\bfseries 09}
  (2022) 010}, \href{http://arxiv.org/abs/2106.05294}{{\ttfamily
  arXiv:2106.05294 [hep-th]}}.

\bibitem{Goodhew:2021oqg}
H.~Goodhew, S.~Jazayeri, M.~H. Gordon~Lee, and E.~Pajer, ``{Cutting
  Cosmological Correlators},''
  \href{http://dx.doi.org/10.1088/1475-7516/2021/08/003}{{\em JCAP} {\bfseries
  08} (2021) 003}, \href{http://arxiv.org/abs/2104.06587}{{\ttfamily
  arXiv:2104.06587 [hep-th]}}.

\bibitem{Jazayeri:2021fvk}
S.~Jazayeri, E.~Pajer, and D.~Stefanyszyn, ``{From Locality and Unitarity to
  Cosmological Correlators},''
  \href{http://dx.doi.org/10.1007/JHEP10(2021)065}{{\em JHEP} {\bfseries 10}
  (2021) 065}, \href{http://arxiv.org/abs/2103.08649}{{\ttfamily
  arXiv:2103.08649 [hep-th]}}.

\bibitem{Hillman:2021bnk}
A.~Hillman and E.~Pajer, ``{A Differential Representation of Cosmological
  Wavefunctions},'' \href{http://dx.doi.org/10.1007/JHEP04(2022)012}{{\em JHEP}
  {\bfseries 04} (2022) 012}, \href{http://arxiv.org/abs/2112.01619}{{\ttfamily
  arXiv:2112.01619 [hep-th]}}.

\bibitem{Meltzer:2021zin}
D.~Meltzer, ``{The Inflationary Wavefunction from Analyticity and
  Factorization},'' \href{http://dx.doi.org/10.1088/1475-7516/2021/12/018}{{\em
  JCAP} {\bfseries 12} no.~12, (2021) 018},
  \href{http://arxiv.org/abs/2107.10266}{{\ttfamily arXiv:2107.10266
  [hep-th]}}.

\bibitem{Bittermann:2022nfh}
N.~Bittermann and A.~Joyce, ``{Soft Limits of the Wavefunction in Exceptional
  Scalar Theories},'' \href{http://dx.doi.org/10.1007/JHEP03(2023)092}{{\em
  JHEP} {\bfseries 03} (2023) 092},
  \href{http://arxiv.org/abs/2203.05576}{{\ttfamily arXiv:2203.05576
  [hep-th]}}.

\bibitem{Bonifacio:2022vwa}
J.~Bonifacio, H.~Goodhew, A.~Joyce, E.~Pajer, and D.~Stefanyszyn, ``{The
  Graviton Four-Point Function in de Sitter Space},''
  \href{http://dx.doi.org/10.1007/JHEP06(2023)212}{{\em JHEP} {\bfseries 06}
  (2023) 212}, \href{http://arxiv.org/abs/2212.07370}{{\ttfamily
  arXiv:2212.07370 [hep-th]}}.

\bibitem{Salcedo:2022aal}
S.~Salcedo, M.~H.~G. Lee, S.~Melville, and E.~Pajer, ``{The Analytic
  Wavefunction},'' \href{http://arxiv.org/abs/2212.08009}{{\ttfamily
  arXiv:2212.08009 [hep-th]}}.

\bibitem{Lee:2022fgr}
H.~Lee and X.~Wang, ``{Cosmological Double-Copy Relations},''
  \href{http://dx.doi.org/10.1103/PhysRevD.108.L061702}{{\em Phys. Rev. D}
  {\bfseries 108} no.~6, (2023) L061702},
  \href{http://arxiv.org/abs/2212.11282}{{\ttfamily arXiv:2212.11282
  [hep-th]}}.

\bibitem{Stefanyszyn:2023qov}
D.~Stefanyszyn, X.~Tong, and Y.~Zhu, ``{Cosmological Correlators Through the
  Looking Glass: Reality, Parity, and Factorisation},''
  \href{http://arxiv.org/abs/2309.07769}{{\ttfamily arXiv:2309.07769
  [hep-th]}}.

\bibitem{Albayrak:2023hie}
S.~Albayrak, P.~Benincasa, and C.~Duaso~Pueyo, ``{Perturbative Unitarity and
  the Wavefunction of the Universe},''
  \href{http://arxiv.org/abs/2305.19686}{{\ttfamily arXiv:2305.19686
  [hep-th]}}.

\bibitem{Cespedes:2023aal}
S.~C\'espedes, A.-C. Davis, and D.-G. Wang, ``{On the IR Divergences in de
  Sitter Space: Loops, Resummation and the Semi-Classical Wavefunction},''
  \href{http://arxiv.org/abs/2311.17990}{{\ttfamily arXiv:2311.17990
  [hep-th]}}.

\bibitem{Chen:2009zp}
X.~Chen and Y.~Wang, ``{Quasi-Single Field Inflation and Non-Gaussianities},''
  \href{http://dx.doi.org/10.1088/1475-7516/2010/04/027}{{\em JCAP} {\bfseries
  04} (2010) 027}, \href{http://arxiv.org/abs/0911.3380}{{\ttfamily
  arXiv:0911.3380 [hep-th]}}.

\bibitem{Maldacena:2011nz}
J.~Maldacena and G.~Pimentel, ``{On Graviton Non-Gaussianities during
  Inflation},'' \href{http://dx.doi.org/10.1007/JHEP09(2011)045}{{\em JHEP}
  {\bfseries 09} (2011) 045}, \href{http://arxiv.org/abs/1104.2846}{{\ttfamily
  arXiv:1104.2846 [hep-th]}}.

\bibitem{Mata:2012bx}
I.~Mata, S.~Raju, and S.~Trivedi, ``{CMB from CFT},''
  \href{http://dx.doi.org/10.1007/JHEP07(2013)015}{{\em JHEP} {\bfseries 07}
  (2013) 015}, \href{http://arxiv.org/abs/1211.5482}{{\ttfamily arXiv:1211.5482
  [hep-th]}}.

\bibitem{Creminelli:2012ed}
P.~Creminelli, J.~Nore\~na, and M.~Simonovi\'c, ``{Conformal Consistency
  Relations for Single-Field Inflation},''
  \href{http://dx.doi.org/10.1088/1475-7516/2012/07/052}{{\em JCAP} {\bfseries
  07} (2012) 052}, \href{http://arxiv.org/abs/1203.4595}{{\ttfamily
  arXiv:1203.4595 [hep-th]}}.

\bibitem{Assassi:2012zq}
V.~Assassi, D.~Baumann, and D.~Green, ``{On Soft Limits of Inflationary
  Correlation Functions},''
  \href{http://dx.doi.org/10.1088/1475-7516/2012/11/047}{{\em JCAP} {\bfseries
  11} (2012) 047}, \href{http://arxiv.org/abs/1204.4207}{{\ttfamily
  arXiv:1204.4207 [hep-th]}}.

\bibitem{Hinterbichler:2013dpa}
K.~Hinterbichler, L.~Hui, and J.~Khoury, ``{An Infinite Set of Ward Identities
  for Adiabatic Modes in Cosmology},''
  \href{http://dx.doi.org/10.1088/1475-7516/2014/01/039}{{\em JCAP} {\bfseries
  01} (2014) 039}, \href{http://arxiv.org/abs/1304.5527}{{\ttfamily
  arXiv:1304.5527 [hep-th]}}.

\bibitem{Kundu:2014gxa}
N.~Kundu, A.~Shukla, and S.~P. Trivedi, ``{Constraints from Conformal Symmetry
  on the Three Point Scalar Correlator in Inflation},''
  \href{http://dx.doi.org/10.1007/JHEP04(2015)061}{{\em JHEP} {\bfseries 04}
  (2015) 061}, \href{http://arxiv.org/abs/1410.2606}{{\ttfamily arXiv:1410.2606
  [hep-th]}}.

\bibitem{Arkani-Hamed:2015bza}
N.~Arkani-Hamed and J.~Maldacena, ``{Cosmological Collider Physics},''
  \href{http://arxiv.org/abs/1503.08043}{{\ttfamily arXiv:1503.08043
  [hep-th]}}.

\bibitem{Lee:2016vti}
H.~Lee, D.~Baumann, and G.~L. Pimentel, ``{Non-Gaussianity as a Particle
  Detector},'' \href{http://dx.doi.org/10.1007/JHEP12(2016)040}{{\em JHEP}
  {\bfseries 12} (2016) 040}, \href{http://arxiv.org/abs/1607.03735}{{\ttfamily
  arXiv:1607.03735 [hep-th]}}.

\bibitem{Arkani-Hamed:2018kmz}
N.~Arkani-Hamed, D.~Baumann, H.~Lee, and G.~L. Pimentel, ``{The Cosmological
  Bootstrap: Inflationary Correlators from Symmetries and Singularities},''
  \href{http://dx.doi.org/10.1007/JHEP04(2020)105}{{\em JHEP} {\bfseries 04}
  (2020) 105}, \href{http://arxiv.org/abs/1811.00024}{{\ttfamily
  arXiv:1811.00024 [hep-th]}}.

\bibitem{Baumann:2019oyu}
D.~Baumann, C.~Duaso~Pueyo, A.~Joyce, H.~Lee, and G.~L. Pimentel, ``{The
  Cosmological Bootstrap: Weight-Shifting Operators and Scalar Seeds},''
  \href{http://dx.doi.org/10.1007/JHEP12(2020)204}{{\em JHEP} {\bfseries 12}
  (2020) 204}, \href{http://arxiv.org/abs/1910.14051}{{\ttfamily
  arXiv:1910.14051 [hep-th]}}.

\bibitem{Sleight:2019mgd}
C.~Sleight, ``{A Mellin Space Approach to Cosmological Correlators},''
  \href{http://dx.doi.org/10.1007/JHEP01(2020)090}{{\em JHEP} {\bfseries 01}
  (2020) 090}, \href{http://arxiv.org/abs/1906.12302}{{\ttfamily
  arXiv:1906.12302 [hep-th]}}.

\bibitem{Sleight:2019hfp}
C.~Sleight and M.~Taronna, ``{Bootstrapping Inflationary Correlators in Mellin
  Space},'' \href{http://dx.doi.org/10.1007/JHEP02(2020)098}{{\em JHEP}
  {\bfseries 02} (2020) 098}, \href{http://arxiv.org/abs/1907.01143}{{\ttfamily
  arXiv:1907.01143 [hep-th]}}.

\bibitem{Pajer:2020wxk}
E.~Pajer, ``{Building a Boostless Bootstrap for the Bispectrum},''
  \href{http://dx.doi.org/10.1088/1475-7516/2021/01/023}{{\em JCAP} {\bfseries
  01} (2021) 023}, \href{http://arxiv.org/abs/2010.12818}{{\ttfamily
  arXiv:2010.12818 [hep-th]}}.

\bibitem{DiPietro:2021sjt}
L.~Di~Pietro, V.~Gorbenko, and S.~Komatsu, ``{Analyticity and Unitarity for
  Cosmological Correlators},''
  \href{http://dx.doi.org/10.1007/JHEP03(2022)023}{{\em JHEP} {\bfseries 03}
  (2022) 023}, \href{http://arxiv.org/abs/2108.01695}{{\ttfamily
  arXiv:2108.01695 [hep-th]}}.

\bibitem{Hogervorst:2021uvp}
M.~Hogervorst, J.~a. Penedones, and K.~S. Vaziri, ``{Towards the
  Non-Perturbative Cosmological Bootstrap},''
  \href{http://dx.doi.org/10.1007/JHEP02(2023)162}{{\em JHEP} {\bfseries 02}
  (2023) 162}, \href{http://arxiv.org/abs/2107.13871}{{\ttfamily
  arXiv:2107.13871 [hep-th]}}.

\bibitem{McFadden:2009fg}
P.~McFadden and K.~Skenderis, ``{Holography for Cosmology},''
  \href{http://dx.doi.org/10.1103/PhysRevD.81.021301}{{\em Phys. Rev. D}
  {\bfseries 81} (2010) 021301},
  \href{http://arxiv.org/abs/0907.5542}{{\ttfamily arXiv:0907.5542 [hep-th]}}.

\bibitem{Bzowski:2012ih}
A.~Bzowski, P.~McFadden, and K.~Skenderis, ``{Holography for Inflation using
  Conformal Perturbation Theory},''
  \href{http://dx.doi.org/10.1007/JHEP04(2013)047}{{\em JHEP} {\bfseries 04}
  (2013) 047}, \href{http://arxiv.org/abs/1211.4550}{{\ttfamily arXiv:1211.4550
  [hep-th]}}.

\bibitem{Sleight:2021plv}
C.~Sleight and M.~Taronna, ``{From dS to AdS and back},''
  \href{http://dx.doi.org/10.1007/JHEP12(2021)074}{{\em JHEP} {\bfseries 12}
  (2021) 074}, \href{http://arxiv.org/abs/2109.02725}{{\ttfamily
  arXiv:2109.02725 [hep-th]}}.

\bibitem{Bzowski:2022rlz}
A.~Bzowski, P.~McFadden, and K.~Skenderis, ``{A Handbook of Holographic
  Four-Point Functions},''
  \href{http://dx.doi.org/10.1007/JHEP12(2022)039}{{\em JHEP} {\bfseries 12}
  (2022) 039}, \href{http://arxiv.org/abs/2207.02872}{{\ttfamily
  arXiv:2207.02872 [hep-th]}}.

\bibitem{Baumgart:2019clc}
M.~Baumgart and R.~Sundrum, ``{De Sitter Diagrammar and the Resummation of
  Time},'' \href{http://dx.doi.org/10.1007/JHEP07(2020)119}{{\em JHEP}
  {\bfseries 07} (2020) 119}, \href{http://arxiv.org/abs/1912.09502}{{\ttfamily
  arXiv:1912.09502 [hep-th]}}.

\bibitem{Gorbenko:2019rza}
V.~Gorbenko and L.~Senatore, ``{$\lambda \phi^4$ in dS},''
  \href{http://arxiv.org/abs/1911.00022}{{\ttfamily arXiv:1911.00022
  [hep-th]}}.

\bibitem{Cohen:2020php}
T.~Cohen and D.~Green, ``{Soft de Sitter Effective Theory},''
  \href{http://dx.doi.org/10.1007/JHEP12(2020)041}{{\em JHEP} {\bfseries 12}
  (2020) 041}, \href{http://arxiv.org/abs/2007.03693}{{\ttfamily
  arXiv:2007.03693 [hep-th]}}.

\bibitem{Cohen:2021fzf}
T.~Cohen, D.~Green, A.~Premkumar, and A.~Ridgway, ``{Stochastic Inflation at
  NNLO},'' \href{http://dx.doi.org/10.1007/JHEP09(2021)159}{{\em JHEP}
  {\bfseries 09} (2021) 159}, \href{http://arxiv.org/abs/2106.09728}{{\ttfamily
  arXiv:2106.09728 [hep-th]}}.

\bibitem{Cohen:2021jbo}
T.~Cohen, D.~Green, and A.~Premkumar, ``{A Tail of Eternal Inflation},''
  \href{http://dx.doi.org/10.21468/SciPostPhys.14.5.109}{{\em SciPost Phys.}
  {\bfseries 14} no.~5, (2023) 109},
  \href{http://arxiv.org/abs/2111.09332}{{\ttfamily arXiv:2111.09332
  [hep-th]}}.

\bibitem{Pimentel:2022fsc}
G.~L. Pimentel and D.-G. Wang, ``{Boostless Cosmological Collider Bootstrap},''
  \href{http://dx.doi.org/10.1007/JHEP10(2022)177}{{\em JHEP} {\bfseries 10}
  (2022) 177}, \href{http://arxiv.org/abs/2205.00013}{{\ttfamily
  arXiv:2205.00013 [hep-th]}}.

\bibitem{Qin:2022fbv}
Z.~Qin and Z.-Z. Xianyu, ``{Helical Inflation Correlators: Partial
  Mellin-Barnes and Bootstrap Equations},''
  \href{http://dx.doi.org/10.1007/JHEP04(2023)059}{{\em JHEP} {\bfseries 04}
  (2023) 059}, \href{http://arxiv.org/abs/2208.13790}{{\ttfamily
  arXiv:2208.13790 [hep-th]}}.

\bibitem{Wang:2022eop}
D.-G. Wang, G.~L. Pimentel, and A.~Ach\'ucarro, ``{Bootstrapping Multi-Field
  Inflation: Non-Gaussianities from Light Scalars Revisited},''
  \href{http://dx.doi.org/10.1088/1475-7516/2023/05/043}{{\em JCAP} {\bfseries
  05} (2023) 043}, \href{http://arxiv.org/abs/2212.14035}{{\ttfamily
  arXiv:2212.14035 [astro-ph.CO]}}.

\bibitem{Penedones:2023uqc}
J.~Penedones, K.~Salehi~Vaziri, and Z.~Sun, ``{Hilbert Space of Quantum Field
  Theory in de Sitter Spacetime},''
  \href{http://arxiv.org/abs/2301.04146}{{\ttfamily arXiv:2301.04146
  [hep-th]}}.

\bibitem{Loparco:2023akg}
M.~Loparco, J.~Qiao, and Z.~Sun, ``{A Radial Variable for de Sitter Two-Point
  Functions},'' \href{http://arxiv.org/abs/2310.15944}{{\ttfamily
  arXiv:2310.15944 [hep-th]}}.

\bibitem{Loparco:2023rug}
M.~Loparco, J.~Penedones, K.~Salehi~Vaziri, and Z.~Sun, ``{The
  K\"all\'en-Lehmann Representation in de Sitter Spacetime},''
  \href{http://arxiv.org/abs/2306.00090}{{\ttfamily arXiv:2306.00090
  [hep-th]}}.

\bibitem{Qin:2023nhv}
Z.~Qin and Z.-Z. Xianyu, ``{Nonanalyticity and On-Shell Factorization of
  Inflation Correlators at All Loop Orders},''
  \href{http://arxiv.org/abs/2308.14802}{{\ttfamily arXiv:2308.14802
  [hep-th]}}.

\bibitem{DuasoPueyo:2023viy}
C.~Duaso~Pueyo and E.~Pajer, ``{A Cosmological Bootstrap for Resonant
  Non-Gaussianity},'' \href{http://arxiv.org/abs/2311.01395}{{\ttfamily
  arXiv:2311.01395 [hep-th]}}.

\bibitem{Xianyu:2023ytd}
Z.-Z. Xianyu and J.~Zang, ``{Inflation Correlators with Multiple Massive
  Exchanges},'' \href{http://arxiv.org/abs/2309.10849}{{\ttfamily
  arXiv:2309.10849 [hep-th]}}.

\bibitem{Bzowski:2013sza}
A.~Bzowski, P.~McFadden, and K.~Skenderis, ``{Implications of Conformal
  Invariance in Momentum Space},''
  \href{http://dx.doi.org/10.1007/JHEP03(2014)111}{{\em JHEP} {\bfseries 03}
  (2014) 111}, \href{http://arxiv.org/abs/1304.7760}{{\ttfamily arXiv:1304.7760
  [hep-th]}}.

\bibitem{Bzowski:2015pba}
A.~Bzowski, P.~McFadden, and K.~Skenderis, ``{Scalar Three-Point functions in
  CFT: Renormalisation, Beta Functions and Anomalies},''
  \href{http://dx.doi.org/10.1007/JHEP03(2016)066}{{\em JHEP} {\bfseries 03}
  (2016) 066}, \href{http://arxiv.org/abs/1510.08442}{{\ttfamily
  arXiv:1510.08442 [hep-th]}}.

\bibitem{Bzowski:2017poo}
A.~Bzowski, P.~McFadden, and K.~Skenderis, ``{Renormalised Three-Point
  Functions of Stress Tensors and Conserved Currents in CFT},''
  \href{http://dx.doi.org/10.1007/JHEP11(2018)153}{{\em JHEP} {\bfseries 11}
  (2018) 153}, \href{http://arxiv.org/abs/1711.09105}{{\ttfamily
  arXiv:1711.09105 [hep-th]}}.

\bibitem{Bzowski:2019kwd}
A.~Bzowski, P.~McFadden, and K.~Skenderis, ``{Conformal $n$-Point Functions in
  Momentum Space},''
  \href{http://dx.doi.org/10.1103/PhysRevLett.124.131602}{{\em Phys. Rev.
  Lett.} {\bfseries 124} no.~13, (2020) 131602},
  \href{http://arxiv.org/abs/1910.10162}{{\ttfamily arXiv:1910.10162
  [hep-th]}}.

\bibitem{Bzowski:2020kfw}
A.~Bzowski, P.~McFadden, and K.~Skenderis, ``{Conformal Correlators as Simplex
  Integrals in Momentum Space},''
  \href{http://dx.doi.org/10.1007/JHEP01(2021)192}{{\em JHEP} {\bfseries 01}
  (2021) 192}, \href{http://arxiv.org/abs/2008.07543}{{\ttfamily
  arXiv:2008.07543 [hep-th]}}.

\bibitem{Dymarsky:2014zja}
A.~Dymarsky, K.~Farnsworth, Z.~Komargodski, M.~Luty, and V.~Prilepina, ``{Scale
  Invariance, Conformality, and Generalized Free Fields},''
  \href{http://dx.doi.org/10.1007/JHEP02(2016)099}{{\em JHEP} {\bfseries 02}
  (2016) 099}, \href{http://arxiv.org/abs/1402.6322}{{\ttfamily arXiv:1402.6322
  [hep-th]}}.

\bibitem{Gillioz:2018mto}
M.~Gillioz, ``{Momentum-Space Conformal Blocks on the Light Cone},''
  \href{http://dx.doi.org/10.1007/JHEP10(2018)125}{{\em JHEP} {\bfseries 10}
  (2018) 125}, \href{http://arxiv.org/abs/1807.07003}{{\ttfamily
  arXiv:1807.07003 [hep-th]}}.

\bibitem{Gillioz:2019lgs}
M.~Gillioz, ``{Conformal Three-Point Functions and the Lorentzian OPE in
  Momentum Space},'' \href{http://dx.doi.org/10.1007/s00220-020-03836-8}{{\em
  Commun. Math. Phys.} {\bfseries 379} no.~1, (2020) 227--259},
  \href{http://arxiv.org/abs/1909.00878}{{\ttfamily arXiv:1909.00878
  [hep-th]}}.

\bibitem{Gillioz:2020mdd}
M.~Gillioz, M.~Meineri, and J.~Penedones, ``{A Scattering Amplitude in
  Conformal Field Theory},''
  \href{http://dx.doi.org/10.1007/JHEP11(2020)139}{{\em JHEP} {\bfseries 11}
  (2020) 139}, \href{http://arxiv.org/abs/2003.07361}{{\ttfamily
  arXiv:2003.07361 [hep-th]}}.

\bibitem{Baumann:2022jpr}
D.~Baumann, D.~Green, A.~Joyce, E.~Pajer, G.~L. Pimentel, C.~Sleight, and
  M.~Taronna, ``{Snowmass White Paper: The Cosmological Bootstrap},'' in {\em
  {2022 Snowmass Summer Study}}.
\newblock 3, 2022.
\newblock \href{http://arxiv.org/abs/2203.08121}{{\ttfamily arXiv:2203.08121
  [hep-th]}}.

\bibitem{Arkani-Hamed:2023bsv}
N.~Arkani-Hamed, D.~Baumann, A.~Hillman, A.~Joyce, H.~Lee, and G.~L. Pimentel,
  ``{Kinematic Flow and the Emergence of Time},''
  \href{http://arxiv.org/abs/2312.05300}{{\ttfamily arXiv:2312.05300
  [hep-th]}}.

\bibitem{Bern:1993kr}
Z.~Bern, L.~Dixon, and D.~Kosower, ``{Dimensionally Regulated Pentagon
  Integrals},'' \href{http://dx.doi.org/10.1016/0550-3213(94)90398-0}{{\em
  Nucl. Phys. B} {\bfseries 412} (1994) 751--816},
  \href{http://arxiv.org/abs/hep-ph/9306240}{{\ttfamily arXiv:hep-ph/9306240}}.

\bibitem{Kotikov:1990kg}
A.~Kotikov, ``{Differential Equations Method: New Technique for Massive Feynman
  Diagrams Calculation},''
  \href{http://dx.doi.org/10.1016/0370-2693(91)90413-K}{{\em Phys. Lett. B}
  {\bfseries 254} (1991) 158--164}.

\bibitem{Remiddi:1997ny}
E.~Remiddi, ``{Differential Equations for Feynman Graph Amplitudes},''
  \href{http://dx.doi.org/10.1007/BF03185566}{{\em Nuovo Cim. A} {\bfseries
  110} (1997) 1435--1452},
  \href{http://arxiv.org/abs/hep-th/9711188}{{\ttfamily arXiv:hep-th/9711188}}.

\bibitem{Gehrmann:1999as}
T.~Gehrmann and E.~Remiddi, ``{Differential Equations for Two-Loop Four-Point
  Functions},'' \href{http://dx.doi.org/10.1016/S0550-3213(00)00223-6}{{\em
  Nucl. Phys. B} {\bfseries 580} (2000) 485--518},
  \href{http://arxiv.org/abs/hep-ph/9912329}{{\ttfamily arXiv:hep-ph/9912329}}.

\bibitem{Henn:2013pwa}
J.~Henn, ``{Multiloop Integrals in Dimensional Regularization Made Simple},''
  \href{http://dx.doi.org/10.1103/PhysRevLett.110.251601}{{\em Phys. Rev.
  Lett.} {\bfseries 110} (2013) 251601},
  \href{http://arxiv.org/abs/1304.1806}{{\ttfamily arXiv:1304.1806 [hep-th]}}.

\bibitem{Henn:2014qga}
J.~Henn, ``{Lectures on Differential Equations for Feynman Integrals},''
  \href{http://dx.doi.org/10.1088/1751-8113/48/15/153001}{{\em J. Phys. A}
  {\bfseries 48} (2015) 153001},
  \href{http://arxiv.org/abs/1412.2296}{{\ttfamily arXiv:1412.2296 [hep-ph]}}.

\bibitem{Abreu:2022mfk}
S.~Abreu, R.~Britto, and C.~Duhr, ``{The SAGEX Review on Scattering Amplitudes,
  Chapter 3: Mathematical Structures in Feynman Integrals},''
  \href{http://arxiv.org/abs/2203.13014}{{\ttfamily arXiv:2203.13014
  [hep-th]}}.

\bibitem{Parra-Martinez:2020dzs}
J.~Parra-Martinez, M.~Ruf, and M.~Zeng, ``{Extremal Black Hole Scattering at
  $\mathcal{O}(G^3)$: Graviton Dominance, Eikonal Exponentiation, and
  Differential Equations},''
  \href{http://dx.doi.org/10.1007/JHEP11(2020)023}{{\em JHEP} {\bfseries 11}
  (2020) 023}, \href{http://arxiv.org/abs/2005.04236}{{\ttfamily
  arXiv:2005.04236 [hep-th]}}.

\bibitem{Kalin:2020fhe}
G.~K\"alin, Z.~Liu, and R.~Porto, ``{Conservative Dynamics of Binary Systems to
  Third Post-Minkowskian Order from the Effective Field Theory Approach},''
  \href{http://dx.doi.org/10.1103/PhysRevLett.125.261103}{{\em Phys. Rev.
  Lett.} {\bfseries 125} no.~26, (2020) 261103},
  \href{http://arxiv.org/abs/2007.04977}{{\ttfamily arXiv:2007.04977
  [hep-th]}}.

\bibitem{De:2023xue}
S.~De and A.~Pokraka, ``{Cosmology Meets Cohomology},''
  \href{http://arxiv.org/abs/2308.03753}{{\ttfamily arXiv:2308.03753
  [hep-th]}}.

\bibitem{Arkani-Hamed:2017mur}
N.~Arkani-Hamed, Y.~Bai, S.~He, and G.~Yan, ``{Scattering Forms and the
  Positive Geometry of Kinematics, Color and the Worldsheet},''
  \href{http://dx.doi.org/10.1007/JHEP05(2018)096}{{\em JHEP} {\bfseries 05}
  (2018) 096}, \href{http://arxiv.org/abs/1711.09102}{{\ttfamily
  arXiv:1711.09102 [hep-th]}}.

\bibitem{LectureNotes}
D.~Baumann and A.~Joyce, {\em {Lectures on Cosmological Correlations (to
  appear)}}.

\bibitem{Benincasa:2022gtd}
P.~Benincasa, ``{Amplitudes Meet Cosmology: A (Scalar) Primer},''
  \href{http://arxiv.org/abs/2203.15330}{{\ttfamily arXiv:2203.15330
  [hep-th]}}.

\bibitem{CARR20062155}
M.~Carr and S.~Devadoss, ``{Coxeter Complexes and Graph Associahedra},''
  \href{http://dx.doi.org/https://doi.org/10.1016/j.topol.2005.08.010}{{\em
  Topology and its Applications} {\bfseries 153} no.~12, (2006) 2155--2168}.

\bibitem{Chen:2011zf}
X.~Chen, ``{Primordial Features as Evidence for Inflation},''
  \href{http://dx.doi.org/10.1088/1475-7516/2012/01/038}{{\em JCAP} {\bfseries
  01} (2012) 038}, \href{http://arxiv.org/abs/1104.1323}{{\ttfamily
  arXiv:1104.1323 [hep-th]}}.

\bibitem{Chen:2014cwa}
X.~Chen, M.~H. Namjoo, and Y.~Wang, ``{Models of the Primordial Standard
  Clock},'' \href{http://dx.doi.org/10.1088/1475-7516/2015/02/027}{{\em JCAP}
  {\bfseries 02} (2015) 027}, \href{http://arxiv.org/abs/1411.2349}{{\ttfamily
  arXiv:1411.2349 [astro-ph.CO]}}.

\bibitem{Mizera:2019ose}
S.~Mizera, ``{Status of Intersection Theory and Feynman Integrals},''
  \href{http://dx.doi.org/10.22323/1.383.0016}{{\em PoS} {\bfseries MA2019}
  (2019) 016}, \href{http://arxiv.org/abs/2002.10476}{{\ttfamily
  arXiv:2002.10476 [hep-th]}}.

\bibitem{Brunello}
G.~Brunello and P.~Mastrolia, {\em {unpublished}}.

\bibitem{aomoto1975vanishing}
K.~Aomoto, ``{On Vanishing of Cohomology Attached to Certain Many Valued
  Meromorphic Functions},'' {\em Journal of the Mathematical Society of Japan}
  {\bfseries 27} no.~2, (1975) 248--255.

\bibitem{mastrolia2019feynman}
P.~Mastrolia and S.~Mizera, ``{Feynman Integrals and Intersection Theory},''
  {\em Journal of High Energy Physics} {\bfseries 2019} no.~2, (2019) 1--25.

\bibitem{Benincasa:2019vqr}
P.~Benincasa, ``{Cosmological Polytopes and the Wavefunction of the Universe
  for Light States},'' \href{http://arxiv.org/abs/1909.02517}{{\ttfamily
  arXiv:1909.02517 [hep-th]}}.

\bibitem{Muller-Stach:2012tgj}
S.~M\"uller-Stach, S.~Weinzierl, and R.~Zayadeh, ``{Picard-Fuchs Equations for
  Feynman Integrals},'' \href{http://dx.doi.org/10.1007/s00220-013-1838-3}{{\em
  Commun. Math. Phys.} {\bfseries 326} (2014) 237--249},
  \href{http://arxiv.org/abs/1212.4389}{{\ttfamily arXiv:1212.4389 [hep-ph]}}.

\bibitem{Agostini:2022cgv}
D.~Agostini, C.~Fevola, A.-L. Sattelberger, and S.~Telen, ``{Vector Spaces of
  Generalized Euler Integrals},''
  \href{http://arxiv.org/abs/2208.08967}{{\ttfamily arXiv:2208.08967
  [math.AG]}}.

\bibitem{Lairez:2022zkj}
P.~Lairez and P.~Vanhove, ``{Algorithms for Minimal Picard-Fuchs Operators of
  Feynman Integrals},'' \href{http://arxiv.org/abs/2209.10962}{{\ttfamily
  arXiv:2209.10962 [hep-th]}}.

\bibitem{Arkani-Hamed:2017tmz}
N.~Arkani-Hamed, Y.~Bai, and T.~Lam, ``{Positive Geometries and Canonical
  Forms},'' \href{http://dx.doi.org/10.1007/JHEP11(2017)039}{{\em JHEP}
  {\bfseries 11} (2017) 039}, \href{http://arxiv.org/abs/1703.04541}{{\ttfamily
  arXiv:1703.04541 [hep-th]}}.

\bibitem{Kuhne:2022wze}
L.~K\"uhne and L.~Monin, ``{Faces of Cosmological Polytopes},''
  \href{http://arxiv.org/abs/2209.08069}{{\ttfamily arXiv:2209.08069
  [math.CO]}}.

\bibitem{Juhnke-Kubitzke:2023nrj}
M.~Juhnke-Kubitzke, L.~Solus, and L.~Venturello, ``{Triangulations of
  Cosmological Polytopes},'' \href{http://arxiv.org/abs/2303.05876}{{\ttfamily
  arXiv:2303.05876 [math.CO]}}.

\bibitem{Arkani-Hamed:2023lbd}
N.~Arkani-Hamed, H.~Frost, G.~Salvatori, P.-G. Plamondon, and H.~Thomas, ``{All
  Loop Scattering as a Counting Problem},''
  \href{http://arxiv.org/abs/2309.15913}{{\ttfamily arXiv:2309.15913
  [hep-th]}}.

\bibitem{Arkani-Hamed:2018bjr}
N.~Arkani-Hamed and P.~Benincasa, ``{On the Emergence of Lorentz Invariance and
  Unitarity from the Scattering Facet of Cosmological Polytopes},''
  \href{http://arxiv.org/abs/1811.01125}{{\ttfamily arXiv:1811.01125
  [hep-th]}}.

\bibitem{Workshop}
{\em Cosmological Correlators}, 2020.
\newblock \url{https://indico.cern.ch/event/943614/}.
\newblock Online Workshop.

\bibitem{Simons}
{\em Amplitudes Meet Cosmology}, 2022.
\newblock
  \url{https://www.simonsfoundation.org/event/amplitudes-meet-cosmology-2022/}.

\bibitem{hilbert1952geometry}
D.~Hilbert and S.~Cohn-Vossen, {\em Geometry and the Imagination}.
\newblock Chelsea scientific books. Chelsea Publishing Company, 1952.

\bibitem{hartshorne1967foundations}
R.~Hartshorne, {\em Foundations of Projective Geometry}.
\newblock Harvard University. Lecture notes. W. A. Benjamin, 1967.

\bibitem{richter2011perspectives}
J.~Richter-Gebert, {\em Perspectives on Projective Geometry: A Guided Tour
  Through Real and Complex Geometry}.
\newblock Springer Berlin Heidelberg, 2011.

\bibitem{aomoto2011theory}
K.~Aomoto, M.~Kita, T.~Kohno, and K.~Iohara, {\em Theory of Hypergeometric
  Functions}.
\newblock Springer, 2011.

\bibitem{yoshida2013hypergeometric}
M.~Yoshida, {\em Hypergeometric Functions, My Love}.
\newblock Springer, 2013.

\bibitem{Cacciatori:2021nli}
S.~Cacciatori, M.~Conti, and S.~Trevisan, ``{Cohomology of Differential Forms
  and Feynman Diagrams},''
  \href{http://dx.doi.org/10.3390/universe7090328}{{\em Universe} {\bfseries 7}
  no.~9, (2021) 328}, \href{http://arxiv.org/abs/2107.14721}{{\ttfamily
  arXiv:2107.14721 [hep-th]}}.

\bibitem{Weinzierl:2022eaz}
S.~Weinzierl, \href{http://dx.doi.org/10.1007/978-3-030-99558-4}{{\em {Feynman
  Integrals}}}.
\newblock Springer-Verlag, Berlin, 1, 2022.
\newblock \href{http://arxiv.org/abs/2201.03593}{{\ttfamily arXiv:2201.03593
  [hep-th]}}.

\bibitem{brown2015feynman}
F.~Brown, ``{Feynman Amplitudes and Cosmic Galois Group},''
  \href{http://arxiv.org/abs/1512.06409}{{\ttfamily arXiv:1512.06409}}.

\bibitem{Mastrolia:2018uzb}
P.~Mastrolia and S.~Mizera, ``{Feynman Integrals and Intersection Theory},''
  \href{http://dx.doi.org/10.1007/JHEP02(2019)139}{{\em JHEP} {\bfseries 02}
  (2019) 139}, \href{http://arxiv.org/abs/1810.03818}{{\ttfamily
  arXiv:1810.03818 [hep-th]}}.

\bibitem{Abreu:2019wzk}
S.~Abreu, R.~Britto, C.~Duhr, E.~Gardi, and J.~Matthew, ``{From Positive
  Geometries to a Coaction on Hypergeometric Functions},''
  \href{http://dx.doi.org/10.1007/JHEP02(2020)122}{{\em JHEP} {\bfseries 02}
  (2020) 122}, \href{http://arxiv.org/abs/1910.08358}{{\ttfamily
  arXiv:1910.08358 [hep-th]}}.

\bibitem{Caron-Huot:2021xqj}
S.~Caron-Huot and A.~Pokraka, ``{Duals of Feynman Integrals. Part I.
  Differential Equations},''
  \href{http://dx.doi.org/10.1007/JHEP12(2021)045}{{\em JHEP} {\bfseries 12}
  (2021) 045}, \href{http://arxiv.org/abs/2104.06898}{{\ttfamily
  arXiv:2104.06898 [hep-th]}}.

\bibitem{zaslavsky1975facing}
T.~Zaslavsky, {\em Facing up to Arrangements: Face-Count Formulas for
  Partitions of Space by Hyperplanes: Face-count Formulas for Partitions of
  Space by Hyperplanes}.
\newblock American Mathematical Society, 1975.

\bibitem{stanley2004introduction}
R.~Stanley {\em et~al.}, ``{An Introduction to Hyperplane Arrangements},'' in
  {\em Lecture notes, IAS/Park City Mathematics Institute}, Citeseer.
\newblock 2004.

\bibitem{orlik1992arrangements}
P.~Orlik and H.~Terao, {\em {Arrangements of Hyperplanes}}.
\newblock Springer, 1992.

\bibitem{brylawski1972decomposition}
T.~Brylawski, ``{A Decomposition for Combinatorial Geometries},'' {\em
  Transactions of the American Mathematical Society} {\bfseries 171} (1972)
  235--282.

\bibitem{goncharov2013simple}
A.~Goncharov, ``{A Simple Construction of Grassmannian Polylogarithms},''
  \href{http://arxiv.org/abs/0908.2238}{{\ttfamily arXiv:0908.2238 [math.AG]}}.

\bibitem{Goncharov:2010jf}
A.~Goncharov, M.~Spradlin, C.~Vergu, and A.~Volovich, ``{Classical
  Polylogarithms for Amplitudes and Wilson Loops},''
  \href{http://dx.doi.org/10.1103/PhysRevLett.105.151605}{{\em Phys. Rev.
  Lett.} {\bfseries 105} (2010) 151605},
  \href{http://arxiv.org/abs/1006.5703}{{\ttfamily arXiv:1006.5703 [hep-th]}}.

\bibitem{Duhr:2011zq}
C.~Duhr, H.~Gangl, and J.~Rhodes, ``{From Polygons and Symbols to
  Polylogarithmic Functions},''
  \href{http://dx.doi.org/10.1007/JHEP10(2012)075}{{\em JHEP} {\bfseries 10}
  (2012) 075}, \href{http://arxiv.org/abs/1110.0458}{{\ttfamily arXiv:1110.0458
  [math-ph]}}.

\bibitem{Duhr:2019tlz}
C.~Duhr and F.~Dulat, ``{PolyLogTools \textemdash{} Polylogs for the Masses},''
  \href{http://dx.doi.org/10.1007/JHEP08(2019)135}{{\em JHEP} {\bfseries 08}
  (2019) 135}, \href{http://arxiv.org/abs/1904.07279}{{\ttfamily
  arXiv:1904.07279 [hep-th]}}.

\end{thebibliography}\endgroup
}

\end{document}